\documentclass[b5paper,twoside,openright]{book}
\usepackage[utf8x]{inputenc}
\usepackage[colorlinks=true,linkcolor=blue,citecolor=red]{hyperref}
\usepackage{amssymb,amsmath}
\usepackage{mathtools}
\usepackage{titlesec}
\usepackage{graphicx}
\usepackage{float}
\usepackage{lipsum}
\usepackage[numbers,compress]{natbib}
\usepackage{setspace}
\usepackage{fancyhdr}
\usepackage{tikz}
\usepackage{pdfpages}
\usepackage{caption}
\usepackage{xcolor}
\usepackage[title, titletoc]{appendix}
\usepackage{bigints}
\usepackage{bm}
\usepackage{subfig}
\usepackage{bookmark}

\usepackage[nottoc,notlof,notlot]{tocbibind}

\usepackage[					% page layout modifications
paper=b5paper,					% 	- use A4 paper size
includefoot,					% 	- include footer space
includemp,						% 	- include side note space
bindingoffset=0.5cm,			% 	- binding correction
top=2.5cm,						% 	- total body: top margin
left=2.4cm,						% 	- total body: left margin (odd pages)
right=0.05cm,					% 	- total body: right margin (odd pages)
bottom=1.25cm,					% 	- total body: bottom margin
footskip=2cm				% 	- footer skip size
]{geometry}

\definecolor{titlec}{rgb}{0,0.332,0.59}
\definecolor{titlepagecolor}{rgb}{0,0.332,0.59}
\definecolor{pc}{rgb}{0,0.332,0.59}

\fancypagestyle{thesisStyle}{
	\fancyhf{} % clear all header and footer fields
	\fancyhead[EL]{ \leftmark} % except the center
	\fancyhead[OR]{ \rightmark} % except the center
	\fancyfoot[OR,EL]{\thepage} % except the center
}

\fancypagestyle{pdfStyle}{
	\fancyhf{} % clear all header and footer fields
	\fancyfoot[EL]{\vspace*{-1mm}\hspace*{-1.9cm}{\large{(\thepage})}} % except the center
	\fancyfoot[OR]{\vspace*{-1mm}\hspace*{13.5cm}{\large{(\thepage})}}
}

\def\mypart#1#2{%
	\cleardoublepage % Page break
	\thispagestyle{plain} % page style
	\vspace*{8em} % Vertical shift
	\refstepcounter{part}% Next part
	\addcontentsline{toc}{part}{\thepart \;\; #1}
	{\centering\huge \textbf{ Part \thepart}\par}% 
	\vskip .05\vsize % Vertical shift
	{\centering\Huge \textbf{ #1}\par}% 
	\vskip .05\vsize % Vertical shift 
	#2
	\vfill\break\thispagestyle{empty} % Fill the end of page and page break
}

\titlespacing*{\chapter}{0pt}{0pt}{6ex}
\newcommand{\hsp}{\hspace{10pt}}
\titleformat{\chapter}[hang]{\huge\bfseries}{\thechapter\hsp\textcolor{titlec}{\Huge\textbar\huge}\hsp}{0pt}{\huge\bfseries}
\titlespacing*{\section}{0pt}{2.3ex}{1ex}
\titlespacing*{\subsection}{0pt}{1.476ex}{0.7ex}
\newcommand{\kB}{k_{\rm B}}
\newcommand{\bx}{\textbf{x}}
\newcommand{\pd}[1]{\frac{\partial}{\partial #1}}
\newcommand{\bp}{\mathbf{p}}

\begin{document}

\begin{titlepage}
\begin{tikzpicture}[overlay]
\node at (7.3,-9.1){\includegraphics[width = 4 \linewidth]{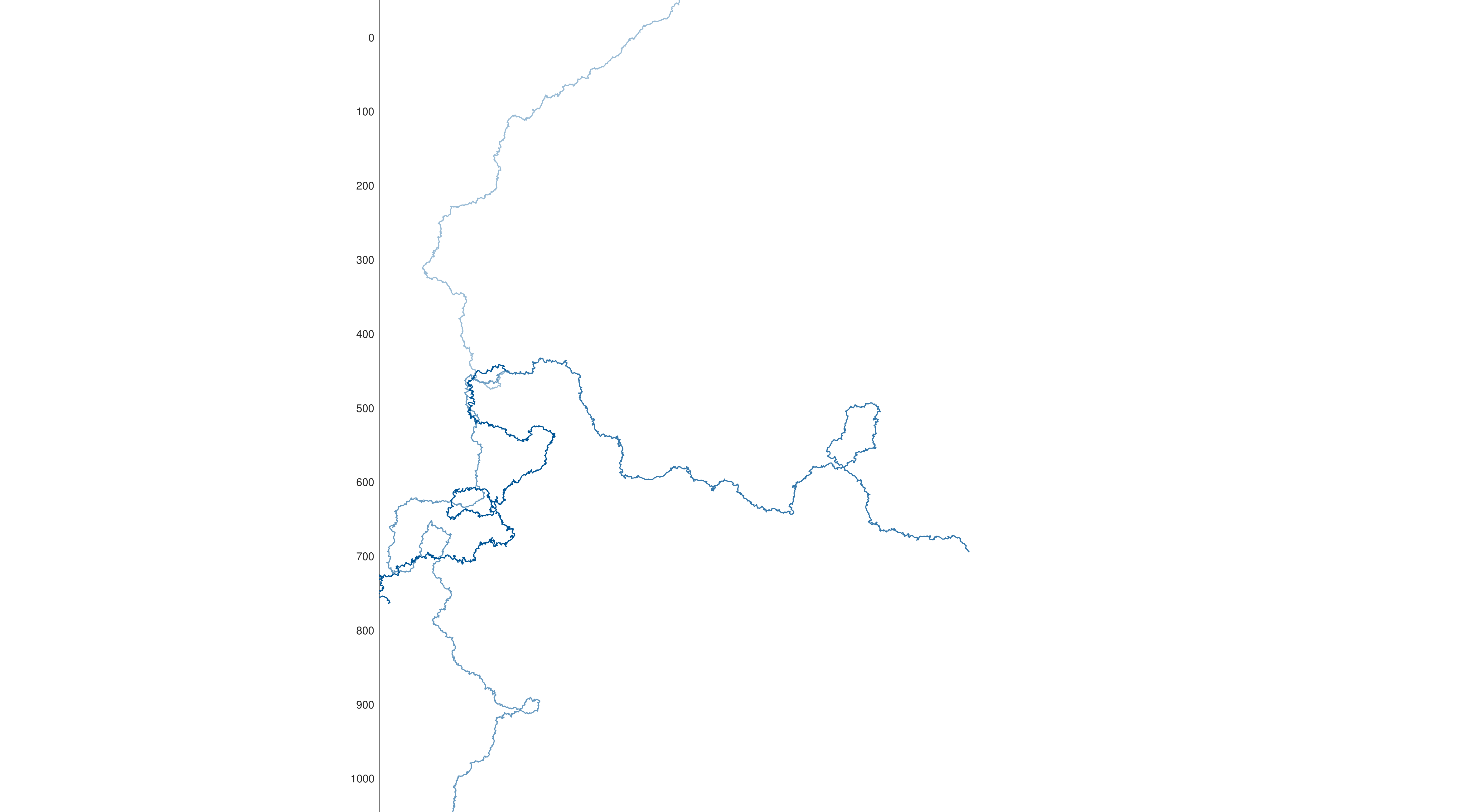}};
\end{tikzpicture}
\begin{flushright}
\includegraphics[width = 0.5 \linewidth]{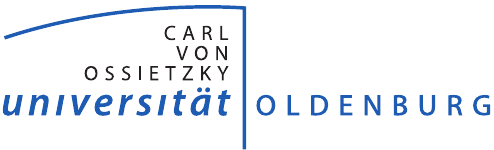}
\end{flushright}
\vspace{1.0cm}
%\noindent \Large \textcolor{titlec}{\textbf{Dissertation}}

\noindent\LARGE\textbf{Coupled and Hidden Degrees of Freedom in Stochastic Thermodynamics} 
\vfill

\noindent\normalsize Von der Fakultät für Mathematik und Naturwissenschaften der Carl von \mbox{Ossietzky} Universität
Oldenburg zur Erlangung des Grades und Titels eines \newline 

\noindent\large \textbf{Doktors der Naturwissenschaften \mbox{(Dr. rer. nat.)}} \newline

\noindent\normalsize angenommene Dissertation \newline

\vspace{-0.6cm}
von Herrn \noindent\large \textbf{Jannik Ehrich, M. Sc.} \newline

\vspace{-0.6cm}
\noindent\normalsize geboren am 17.01.1993 in Oldenburg \newline
\end{titlepage}

\newpage
\mbox{}
\vfill
\begin{minipage}[t]{3.5cm}
Gutachter:\\\\[0.4cm]
Weitere Gutachter:\\\\[0.4cm]
%Externer Gutachter:\\\\[0.4cm]
\end{minipage}
\begin{minipage}[t]{0.6\linewidth}
\textbf{Prof. Dr. Andreas Engel}\\Carl von Ossietzky Universität Oldenburg\\[0.39cm]
\textbf{Prof. Dr. Martin Holthaus}\\Carl von Ossietzky Universität Oldenburg\\[0.39cm]
\textbf{Prof. Dr. Bart Cleuren}\\Universiteit Hasselt\\[0.39cm]
\end{minipage}\\
\begin{minipage}[t]{3.5cm}
	Tag der Disputation:
\end{minipage}
\begin{minipage}[t]{0.6\linewidth}
	14. Februar 2020
\end{minipage}

\thispagestyle{empty}

\frontmatter

\section*{Danksagungen}
Die zurückliegenden drei Jahre meiner Promotion waren außergewöhnlich inspirierend, motivierend und befriedigend. Wissenschaft spielt sich allerdings nicht im stillen Kämmerlein ab. Daher sollten jene Personen und Institutionen nicht unerwähnt bleiben, die maßgeblich zu meinem Erfolg beigetragen haben.

Allen voran möchte ich meinem Doktorvater \textbf{Andreas Engel} für die sehr gute Betreuung während Bachelor- und Masterarbeit und Promotion danken. Seine Präzision und Herangehensweise an physikalische Fragen beeindruckt mich sehr und hat mich nachhaltig geprägt. Außerdem bin ich dankbar für den Freiraum den er mir gegeben hat, um eigene Ideen und Forschungsfragen zu verfolgen und seine Unterstützung abseits von fachlichen Belangen, wie die Ermöglichung vieler Tagungsbesuche, eines Forschungsaufenthaltes in Madrid und darüber hinaus für die zahlreichen Gutachten, die so vieles für mich ermöglichten.

Außerdem möchte ich \textbf{Martin Holthaus} für die Übernahme des Zweitgutachtens dieser Arbeit danken und für seinen Vorlesungskanon 2012-2014 in theoretischer Physik. Wehmütig blicke ich auf seine außerordentlich klaren Vorlesungen und die gemein-kniffligen Übungsaufgaben zurück, die mich dazu bewogen haben, aus dem Ingenieursstudium in die theoretische Physik zu wechseln.

Further, I would like to thank \textbf{Juan Parrondo} for the time he took for me during my research stay in Madrid. I am also thankful for his cheerful attitude towards research and his encouragement.

Additionally, I thank \textbf{Luis Dinis} for many interesting discussions, scientific and otherwise. ¡Muchas gracias!

I'm grateful to \textbf{Bart Cleuren} for agreeing to be an external reviewer for my thesis.

Ganz besonderer Dank gilt meinem Kollegen und Freund \textbf{Marcel Kahlen} für die unzähligen Diskussionen, die fruchtbare Kollaboration und seine Unterstützung für meine Science Slams. Außerdem bin ich dankbar für seinen mathematischen Scharfsinn und seine Geduld mit vielen unausgegorenen und naiven Ideen, ohne die so manches Resultat nie zustande gekommen wäre.

Danke auch an \textbf{Stefan Landmann} und \textbf{Marcel Kahlen} für das kritisches Lesen der Dissertation, das wesentlich zu ihrer Qualität beigetragen hat.

Für die freundliche Arbeitsatmosphäre und diverse Spiele- sowie Cocktailabende möchte ich der Arbeitsgruppe ``Statistische Physik'', also \textbf{Sebastian}, \textbf{Marcel}, \textbf{Stefan}, \textbf{Mattes} und \textbf{Axel}, danken. Vielen Dank außerdem an \textbf{Sören} und \textbf{Christoph} für die Organisation der ``Physikerfrühstücke'' und an \textbf{Christoph}, \textbf{Sören}, \textbf{Cornelia} und \textbf{Age} und die anderen Mitglieder der ``Theoretischen Laufgruppe''; Sport macht gemeinsam einfach mehr Spaß!

Außerdem hatte ich während meiner Promotionszeit die Ehre drei außerordentlich motivierte und talentierte Studierende bei ihren Bachelorarbeiten zu begleiten. Vielen Dank daher an \textbf{Axel}, \textbf{Lea-Christin} und \textbf{Juliana} für ihr Engagement und ihren Enthusiasmus.

Danken möchte ich zudem \textbf{Frauke Arens} und \textbf{Stefan Krautwald} für ihre administrative Unterstützung.

Darüber hinaus möchte ich den vielen ehrenamtlichen Organisator*innen verschiedener Science Slams dafür danken, dass sie Wissenschaft für alle verständlich auf die Bühne bringen und immer ein offenes Ohr für die Nervöseren unter den Slammer*innen haben.

Vielen Dank außerdem an die Lindau-Stiftung und die Wilhelm und Else Heraeus-Stiftung dafür, dass sie meine Teilnahme an der Lindauer Nobelpreisträgertagung ermöglicht haben, welche meine Motivation dafür, Wissenschaft zu betreiben, nachhaltig gestärkt hat.

Dank geht auch an \textbf{Benedikt}, \textbf{Daniela} und \textbf{Malte} für die schöne Zeit innerhalb und außerhalb der Universität (wenn Martins Übungsaufgaben es zuließen) und für unsere, teilweise erschreckend tiefgründigen, Diskussionen.

Auch meinen Großeltern \textbf{Helga}, \textbf{Christa} und \textbf{Robert} und meiner Großtante \textbf{Greta} möchte ich danken für ihre Ermutigung, für mehr Kekse und Kuchen als ich jemals essen kann und ihr Verständnis dafür, dass ich ihnen weniger Zeit schenken konnte, als sie verdient haben.

Insbesondere danke ich meinen Eltern \textbf{Doris} und \textbf{Michael} dafür, dass sie mich schon früh gefördert, unterstützt und meine eigenen Entscheidungen respektiert haben. Außerdem bin ich dankbar für die finanzielle Unterstützung während meines Studiums und ihr Interesse an meiner Arbeit.

Zuletzt geht mein größter Dank an meine Frau \textbf{Charlotte} für ihre selbstlose Unterstützung, ihre Abenteuerlust und ihre verständnisvolle Art; außerdem für ihre Fähigkeit, eine andere Perspektive zu vermitteln und ihre Geduld mit mir: Du hast mich zu dem Menschen gemacht, der ich heute bin!

\begin{flushright}
Jannik Ehrich\\
Oldenburg, Dezember 2019
\end{flushright}

\newpage

\section*{Abstract}
Stochastic thermodynamics extends thermodynamics to mesoscopic scales. Building on the mathematical framework of stochastic processes, one can consider the prevailing role of thermal fluctuations in small systems and assign heat and work to single stochastic trajectories. By carefully defining probabilities for the occurrence of a specific trajectory and its time-reversed version, one can present the second law as an equality, the so-called fluctuation theorem. By including an information-theoretic notion of entropy à la Shannon, one can show that information is a thermodynamic resource just like heat, work, and energy. This solves the long-standing mystery around Maxwell's demon once and for all. 

This thesis investigates an important generalization by considering the interactions of different degrees of freedom of one joint system. First, a comprehensive introduction to the subjects of stochastic processes, information theory and the theory of stochastic thermodynamics is given, thereby highlighting the key results.

In the second part, systems with interacting degrees of freedom are considered. This allows one to investigate the thermalization properties of collisional baths, i.e. particles at equilibrium interacting with a localized system via collisions. It is shown that the interactions between system and bath must be reversible to ensure thermalization of the system. Moreover, the role of information in thermodynamics is presented and interpreted in the context of interacting subsystems. Using the concept of causal conditioning, a framework is developed for finding entropy productions that capture the mutual influence of coupled subsystems. This framework is applied to diverse setups which are usually studied separately in information thermodynamics.

The third part covers the special case of important system variables being hidden from observation. The problem is motivated by presenting a model of a microswimmer and showing that its movement can be approximated by active Brownian motion. However, its energy dissipation is massively underestimated by this procedure. It is shown that this is a consequence of the fact that the swimming mechanism is a hidden variable. Subsequently, different methods of effective description are discussed and applied to a simple model system with which the impact of hidden slow degrees of freedom on fluctuation theorems is studied. Finally, a setting is investigated in which it is possible to give bounds for the hidden entropy production from only partial observation of the system dynamics by fitting an underlying hidden Markov process to the observable data.

\newpage

\section*{Zusammenfassung}
Die stochastische Thermodynamik ist eine Erweiterung der Thermodynamik hin zu mikroskopischen Skalen. Die auf diesen Skalen dominierenden thermischen Fluktuationen werden aufbauend auf dem mathematischen Grundgerüst der stoch\-astischen Prozesse mit in die Beschreibung einbezogen. Damit kann einzelnen stochastischen Trajektorien Wärme und Arbeit zugeordnet werden. Indem Wahr\-scheinlichkeiten für bestimmte Trajektorien und ihre zeitumgekehrten Versionen definiert werden, kann der zweite Hauptsatz als Gleichung formuliert werden, das sogenannte Fluktuationstheorem. Mithilfe der Informationstheorie kann Information auf gleiche Weise wie Wärme, Arbeit und Energie als thermodynamische Ressource aufgefasst werden. Damit wird das Rätsel um den Maxwellschen Dämon endgültig gelöst.

Diese Dissertation behandelt eine wichtige Verallgemeinerung, da die Wechselwirkung zwischen verschiedenen Freiheitsgraden eines größeren Verbundsystems betrachtet wird. Zunächst wird in die Gebiete der stochastischen Prozesse, der Informationstheorie und der stochastischen Thermodynamik eingeführt. Dabei werden die wichtigsten Resultate herausgestellt.

Im zweiten Teil werden Systeme mit wechselwirkenden Freiheitsgraden betrachtet. Dies erlaubt die Untersuchung der Thermalisierungseigenschaften von soge\-nannten Kollisionsbädern, bestehend aus Teilchen im Gleichgewicht, welche durch Stöße mit einem lokalisierten System interagieren. Es wird gezeigt, dass die Wechselwirkung zwischen System und Bad reversibel sein muss, um die Thermalisierung des Systems sicherzustellen. Zudem wird die thermodynamische Rolle von Information behandelt und im Kontext miteinander wechselwirkender Systeme interpretiert.  Mithilfe des Konzepts des ``causal conditioning'' wird ein System zur Definition von Entropieproduktionsmaßen entwickelt. Diese bilden den wechselseitigen Einfluss gekoppelter Subsysteme aufeinander ab. Das System wird auf verschiedenartige Anordnungen angewandt, welche bisher in der Literatur getrennt voneinander untersucht wurden.

Der dritte Teil behandelt den Spezialfall, in dem wichtige Systemvariablen verborgen sind. Das Problem wird veranschaulicht, indem ein Modell eines Mikroschwimmers vorgestellt wird, dessen Dynamik durch aktive Brownsche Bewegung abgebildet werden kann. Seine Energiedissipation wird durch dieses Vorgehen jedoch grob unterschätzt. Man kann zeigen, dass dies daran liegt, dass der Schwimmmechanismus eine verborgene Variable ist. Anschließen werden verschiedene effektive Beschreibungen des sichtbaren Teils des Systems besprochen und auf ein einfaches Modell angewandt. Dieses Modell erlaubt die Untersuchung des Einflusses verborgener langsamer Freiheitsgrade auf Fluktuationstheoreme. Schließlich wird eine Situation untersucht, in der es möglich ist, durch nur teilweise Beobachtung Schranken für die vollständige, jedoch verborgene, Entropieproduktion zu bestimmen, indem ein zugrundeliegender Markov-Prozess an die beobachtbaren Daten angepasst wird.

\begin{spacing}{0.81}
\tableofcontents
\end{spacing}

\mainmatter

\chapter{Introduction}
\emph{Thermodynamics} is the theory of the exchange of heat, work, and matter between systems and environments. Its great power lies in its generality: The second law, for instance, states that in any process a quantity called entropy must be growing, which leads, e.g., to fundamental limits on the efficiency of heat engines.

\emph{Statistical physics} provides the rigorous derivation of the laws of thermodynamics from an underlying microscopic reality. The microscopic world is composed of countless individual particles, extremely tiny and way too numerous to ever be perceived without sophisticated instruments. Importantly, it is exactly this key fact about the microscopic world that is exploited by statistical mechanics: It is possible to very accurately describe the \emph{average behavior} of many particles without having to know the exact state of any one of them.

The last two decades have seen tremendous experimental progress in the manipulation of small-scale systems; mainly in the study of biological machinery like molecular motors. In essence, many of these systems are \emph{microscopic engines}. It is thus natural to ask whether thermodynamics can be applied to these objects.

Crucially, the main tenets of statistical mechanics do not apply to these systems: Firstly, they are not composed of near infinitely many constituent parts. Secondly, the magnitude of energy exchanges becomes comparable to thermal fluctuations which makes the naive application of thermodynamics in the field of biophysics impossible, e.g., when considering efficiencies of molecular motors or the energetics of the folding and unfolding of biopolymers such as DNA and RNA.

Instead, thermal fluctuations must be explicitly built into the description of these systems, e.g., by utilizing \emph{stochastic processes} to capture their dynamics. Fueled by experimental advances over the last 20 years, this approach has led to a significant honing of theoretical tools applicable to the thermodynamics of small-scale systems. Thus has emerged the theory of \emph{stochastic thermodynamics} which allows the study of small machines in much the same terms as thermodynamics enabled the analysis of their large counterparts.

Although rigorously verified by experiments, two consequences of the theory stick out and seem strange to the novice in this field. These are: (1) The fact that the second law does not hold always for microscopic processes. Instead, it holds only when averaging over many realizations of the process. This, at first sight astonishing, fact is perfectly captured by the so-called \emph{fluctuation theorems}. (2) The fact that \emph{information} enters as a \emph{thermodynamic resource} like heat and work, which was already anticipated by Maxwell, Szil\'ard, Bennett, and others and led to the apparent paradox of \emph{Maxwell's demon}. Importantly, within the framework of stochastic thermodynamics this difficult point can finally be discussed in the necessary detail.

This thesis addresses, among others, both of the mentioned aspects. It extends our understanding of small-scale (information) thermodynamics towards systems in which multiple constituent parts interact with each other. Moreover it addresses situations in which some of the degrees of freedom making up a system might be hidden from observation, as it is the case in many experimental situations.

\section{Outline of the thesis}
This thesis is based on the four included peer-reviewed publications (Refs.~\cite{Ehrich2017, Kahlen2018, Ehrich2019, Ehrich2019a}). It is written in such a way as to provide a cohesive flow of the presented material. Therefore, the publications are included into the appropriate chapters of the thesis irrespective of their chronological order. Some results which are not yet published are included as well.

The thesis is grouped into three parts. \textbf{Part I} deals with the fundamental mathematical and physical concepts pertaining to stochastic processes (\textbf{chaper~\ref{chap_basics_stochProc}}), information theory (\textbf{chapter~\ref{chap_basics_infoTheory}}), and the theory of stochastic thermodynamics (\textbf{chapter~\ref{chap_stochasticThermodynamics}}). While the second and third chapters contain material that can readily be found in textbooks on the respective subjects, the fourth chapter on stochastic thermodynamics represents a selection of key results that are mostly re-derived with regard to the stochastic methods presented before. This chapter is aimed at keeping the thesis as self-contained as possible.

\textbf{Part II} presents original research on the role of interacting degrees of freedom in stochastic thermodynamics. Particularly, it discusses a viewpoint which treats the thermal reservoir as one subsystem of a larger joint system. Special emphasis is put on collisional baths (\textbf{chapter~\ref{chap_int_reservoirSubs}}). Further, it contains a detailed discussion of the role of information in thermodynamics (\textbf{chapter~\ref{chap_informationThermodynamics}}) and shows how it results from the interplay of several interacting subsystems. It presents new results on how this interaction can be disentangled in such a way as to preserve their causal influence on each other. 

\textbf{Part III} covers important results on the role of unobserved slow degrees of freedom in stochastic thermodynamics. It discusses the energy requirements of microswimmers (\textbf{chapter~\ref{chap_hid_microswimmers}}) and shows that they serve as a testbed to study how the presence of hidden degrees of freedom leads to an underestimation of energy dissipation. Moreover, it discusses different effective descriptions for the observed dynamics and analyzes the impact of hidden slow degrees of freedom on fluctuation theorems (\textbf{chapter~\ref{sec_hid_hiddenDegreesInSTD}}). Finally, it presents a method with which one can construct tight bounds for the real hidden entropy production from partial information about the system dynamics (\textbf{chapter~\ref{chap_hid_inferringDiss}}).

\section{Notation}
The notation in this thesis follows the usual standards established in the field of stochastic thermodynamics. However, some remarks on the peculiarities are in order:
\begin{itemize}
	\item We usually do not differentiate between \emph{probabilities} and \emph{probability densities} where it is clear from context. When the identification is ambiguous, we use the Greek letter $\rho$ for probability densities.
	\item Moreover, we do not differentiate between a random variable $X$ and the value $x$ that it takes, i.e., we write $p(x)$ for the probability of outcome $x$ where other authors might use the notation $p(X=x)$.
	\item Similarly, whenever it is clear from context, we denote by $p(x)$ and $p(y)$ \emph{different functions}, distinguished by their argument. Other authors might use $p_x(x)$ and $p_y(y)$.
	\item We denote averages using angled brackets, which translate as follows:
	\begin{align*}
	\left\langle f(x)\right\rangle = \int dx\, p(x)\, f(x).
	\end{align*}
	Whenever it is not clear from context which variables are included in the average, a subscript is added, i.e.,
	\begin{align*}
	\left\langle g(x,y) \right\rangle_{p(x|y)} = \int dx\,p(x|y) \,g(x,y).
	\end{align*}
\end{itemize}

\mypart{The basics}{
	The topics in this part establish the mathematical and physical basis for the research presented in later chapters. It is assumed that the reader is familiar with the basics of probability theory and statistics, as it is presented, e.g., in the book by Feller~\cite{Feller1968}. In particular, this part aims at
	\begin{itemize}
		\item Presenting the topic of stochastic processes (especially Markov processes) in a pedagogical way.
		\item Establishing the basics of information theory.
		\item Reviewing the key progress in the field of stochastic thermodynamics (excluding information thermodynamics).
\end{itemize}}

\chapter{Stochastic processes}\label{chap_basics_stochProc}
This first chapter is a collection of fundamental concepts relevant to the topic of the thesis. It is based on the books by Risken~\cite{Risken1996}, Gardiner~\cite{Gardiner2004}, and van~Kampen~\cite{vanKampen2007}, which have proven very useful while conducting research.

In this thesis we will consider systems which evolve probabilistically in time, meaning they are described by some time-dependent random variable $X(t)$. This is called a \emph{stochastic process}. It is clear that any stochastic process can only be defined with regard to its statistic properties, e.g., the probability of a certain outcome $X(t)$. Assume that we measure the states $x_1, x_2,...,x_N$ of the process at different times $t_1 < t_2<...<t_N$. Then, a characterization of the statistics is the joint probability
\begin{align}\label{eqn_basics_JointProb}
p(x_1,t_1;\,x_2,t_2;\,...;\,x_N,t_N)
\end{align}
of the different measurement outcomes. This joint probability describes the probability of measuring $x_1$ at time $t_1$, $x_2$ at time $t_2$ and so on.

In the simplest case, e.g. a repeated coin toss, all measurements are independent, thus implying
\begin{align}
p(x_1,t_1;\,x_2,t_2;...;\,x_N,t_N) = \prod\limits_{i=1}^N p(x_i,t_i).
\end{align}
This means that no predictions of future values based on past or current values of the process are possible. 

In a common setting, the immediate future of the process is dependent on the past, while it is independent of the future, as demanded by causality:
\begin{align}\label{eqn_basics_CondCausal}
p(x_i,t_i|x_1,t_1;...;x_{i-1},t_{i-1};\,x_{i+1},t_{i+1};...;x_{N},t_N) = p(x_i,t_i|x_1,t_1;...;x_{i-1},t_{i-1}).
\end{align}

\section{Markov processes}
An important special case are \emph{Markov processes} for which the future only depends on the most recent value of the process, i.e., the conditional probability in Eq.~\eqref{eqn_basics_CondCausal} reads:
\begin{align}\label{eqn_basics_transitionProbability}
p(x_i,t_i|x_1,t_1;...;x_{i-1},t_{i-1}) = p(x_i,t_i|x_{i-1},t_{i-1}).
\end{align}
It is often called \emph{transition probability} since it describes a transition from state $x_i$ at time $t_i$ given the process was in state $x_{i-1}$ at time $t_{i-1}$.

The joint probability in Eq.~\eqref{eqn_basics_JointProb} can then be expanded into
\begin{align}
p(x_1,t_1;\,x_2,t_2;\,...;\,x_N,t_N) = p(x_1,t_1)\, \prod\limits_{i=2}^N p(x_i,t_i|x_{i-1},t_{i-1}).
\end{align}

Two comments on different aspects of \emph{continuity} in stochastic processes are in order. The first concerns the question about the \emph{state space} of the stochastic process: We can deal with continuous state spaces, e.g. the price of some share on the stock market, and discrete state spaces, e.g. the number of people in an elevator at a given time. Most calculations are similar in both versions: Often, one only has to switch integrals to sums or vice versa. Additionally, whenever it is clear from context, we use the word \emph{probability} to mean either \emph{probability density} or probability in the proper meaning of the word.

The second aspect concerns how the state of a process can change in time: continuously or at discrete times. It would, e.g., be economical not to include the time evolution between floors of a stochastic process describing the occupancy of an elevator. A discrete-time Markov process is usually called a \emph{Markov chain}. Both kinds of time evolution are used in this thesis. As a general rule throughout, a discrete-time process is marked by a lower-index as in $x_t$, while a continuous-time process is represented by a bracketed notation as in $x(t)$. However, sometimes a continuous trajectory can be \emph{discretized}, thereby switching from one notation to the other.

\section{Master equation}\label{sec_basics_masterEq}
The transition probabilities $p(x_i;t_i|x_{j},t_{j})$ [Eq.~\eqref{eqn_basics_transitionProbability}] of a Markov process fulfill the \emph{Chapman-Kolmogorov equation}: For $t_3>t_2>t_1$
\begin{align}
p(x_3,t_3|x_1,t_1) &= \int dx_2\, p(x_2,t_2;x_3,t_3|x_1,t_1) \\
&= \int dx_2\, p(x_3,t_3|x_1,t_1;x_2,t_2)\,p(x_2,t_2|x_1,t_1)\\
&= \int dx_2\, p(x_3,t_3|x_2,t_2)\,p(x_2,t_2|x_1,t_1), \label{eqn_basics_CKE}
\end{align}
where we used the marginalization property in the first line and the Markov property [Eq.~\eqref{eqn_basics_transitionProbability}] in the third line.

The interpretation is straightforward: Since in a Markov process the future only depends on the current state, the transition probability from one state $x_1$ at time $t_1$ to another state $x_3$ at time $t_3$ can be unraveled via the sum of all probabilities to go to \emph{all possible} intermediate states $x_2$ at time $t_2$ multiplied by the probability to go from there to the final state.

Because both sides of the equation can be calculated from sampled transition probabilities, the Chapman-Kolmogorov equation can be used to check for the Markov property of a given stochastic process. However, we have to point out a common problem: The Markov property in Eq.~\eqref{eqn_basics_transitionProbability} has to hold true for \emph{all} conditional probabilities. Therefore, one cannot really infer that a process is Markovian from a few (possibly sampled) conditional probabilities. Nevertheless, a violation of the Chapman-Kolmogorov equation is enough to rule out the Markovian nature of a certain process. Similarly, knowing that a process is Markovian, one can characterize the entire process from the transition probabilities. See Sec.~IV.1~of~Ref.~\cite{vanKampen2007} for more details.

The Chapman-Kolmogorov equation has a more intuitive version which concerns \emph{changes in probabilities}:
\begin{align}
p(x,t_3|x'',t_1) - p(x,t_2|x'',t_1) &= \int dx'\Big[ p(x,t_3|x',t_2)\, p(x',t_2|x'',t_1)\nonumber\\
&\qquad\qquad\quad- p(x',t_3|x,t_2)\,p(x,t_2|x'',t_1)  \Big], \label{eqn_basics_masteEQ1}
\end{align}
where we used the fact that the transition probability is normalized:
\begin{align}
\int dx'\, p(x',t_3|x,t_2) = 1.
\end{align}
Equation~\eqref{eqn_basics_masteEQ1} has a straightforward interpretation in terms of \emph{conservation of probability}: The change in probability at $x$ from one time to another is given by the \emph{influx} of probability from all other states $x'$ minus the \emph{outflux} out of state $x$ into all other states $x'$.

The Chapman-Kolmogorov equation concerns finite time intervals $t_3-t_2$. Next, let us instead derive a differential equation. To this end, we expand the transition probability for small time differences $dt$:
\begin{align}\label{eqn_basics_expandTransitionProb}
p(x',t+dt|x,t) = \delta(x-x') + \tilde W(x'|x;t)\, dt + \mathcal{O}(dt^2).
\end{align}
Due to normalization, one finds
\begin{align}\label{eqn_basics_integralWtilde}
\int dx'\, \tilde W(x'|x;t) = 0.
\end{align} 
It is thus convenient to define:
\begin{align}\label{eqn_basics_WtildeDef}
\tilde{W}(x'|x;t) := W(x'|x;t) - r(x,t)\, \delta(x-x').
\end{align}
Here, $W(x'|x;t)$ denotes the \emph{transition rate}, i.e., the transition probability per unit time from state $x$ to $x'$, and
\begin{align}\label{eqn_basics_exitRate}
r(x,t) := \int dx'\, W(x'|x;t)
\end{align}
is the total \emph{exit rate} out of state $x$.

Inserting this expression into Eq.~\eqref{eqn_basics_masteEQ1} for $t_1=t''$, $t_3-t_2 = dt$ and $t_2 = t$ yields:
\begin{align}
p(x,&t+dt|x'',t'') - p(x,t|x'',t'')\nonumber\\ 
&= \int dx'\,\Big[\big( (1-dt\,r(x'))\,\delta(x-x') + dt\, W(x|x';t)\big)\,p(x',t|x'',t'') \nonumber\\
&\qquad\qquad\qquad\quad -\big( (1-dt\,r(x))\,\delta(x-x')) + dt\, W(x'|x;t)\big)\,p(x,t|x'',t'')\Big]\nonumber\\
&= dt\int dx'\,\Big[ W(x|x';t)\,p(x',t|x'',t'') -  W(x'|x;t)\,p(x,t|x'',t'')\Big].
\end{align}
In the limit $dt\rightarrow 0$ we thus find the (differential) \emph{master equation}
\begin{align}\label{eqn_basics_differentialMasterEquation}
\frac{\partial}{\partial t} p(x,t|x'',t'') = \int dx' \,\Big[W(x|x';t)\,p(x',t|x'',t'') - W(x'|x;t)\,p(x,t|x'',t'')\Big].
\end{align}

Multiplying Eq.~\eqref{eqn_basics_differentialMasterEquation} by $p(x'',t'')$ and integrating over $x''$, one gets
\begin{align}\label{eqn_basics_simplifiedMasterEquation}
\frac{\partial}{\partial t} p(x,t) = \int dx' \,\Big[W(x|x';t)\,p(x',t) - W(x'|x;t)\,p(x,t)\Big],
\end{align}
which is the most commonly used version of the master equation. It describes how the probability function $p(x,t)$ changes with time. It hides, however, important information about the underlying process: It is possible to find a master equation of the form of Eq.~\eqref{eqn_basics_simplifiedMasterEquation} for non-Markovian processes, as we did, e.g., in Ref.~\cite{Kahlen2018}, which might seem confusing at first sight.

If one is willing to use $\tilde{W}(x|x';t)$ instead of $W(x|x';t)$, the master equation can even be written in the following form:
\begin{align}\label{eqn_basics_verySimpleME}
\frac{\partial}{\partial t} p(x,t) = \int dx' \,\tilde W(x|x';t)\,p(x',t),
\end{align}
which is especially appealing for discrete state spaces, for which one has a \emph{transition rate matrix} $\tilde{\mathcal{W}}(t)$ and a vector $\mathbf{p}(t)$ of probabilities: 
\begin{align}\label{eqn_basics_MEMatrixNotation}
\dot{\mathbf{p}}(t) = \tilde{\mathcal{W}}(t)\,\mathbf{p}(t).
\end{align}
The diagonal elements $\tilde W_{ii}$ of the transition matrix must then be chosen negative such that the column sums vanish: $\sum_i \tilde W_{ij} = 0$. In this case they are the negative of the total exit rate from state $i$.

\subsection{Example Markov jump process}\label{sec_basics_exampleJumpProcess}
\begin{figure}[ht]
	\centering
	\raisebox{-0.5\height}{\includegraphics[height=2.5cm]{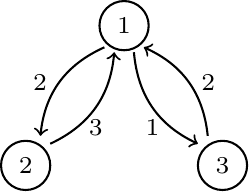}}
	\hspace*{5mm}
	\raisebox{-0.5\height}{\includegraphics[height=5cm]{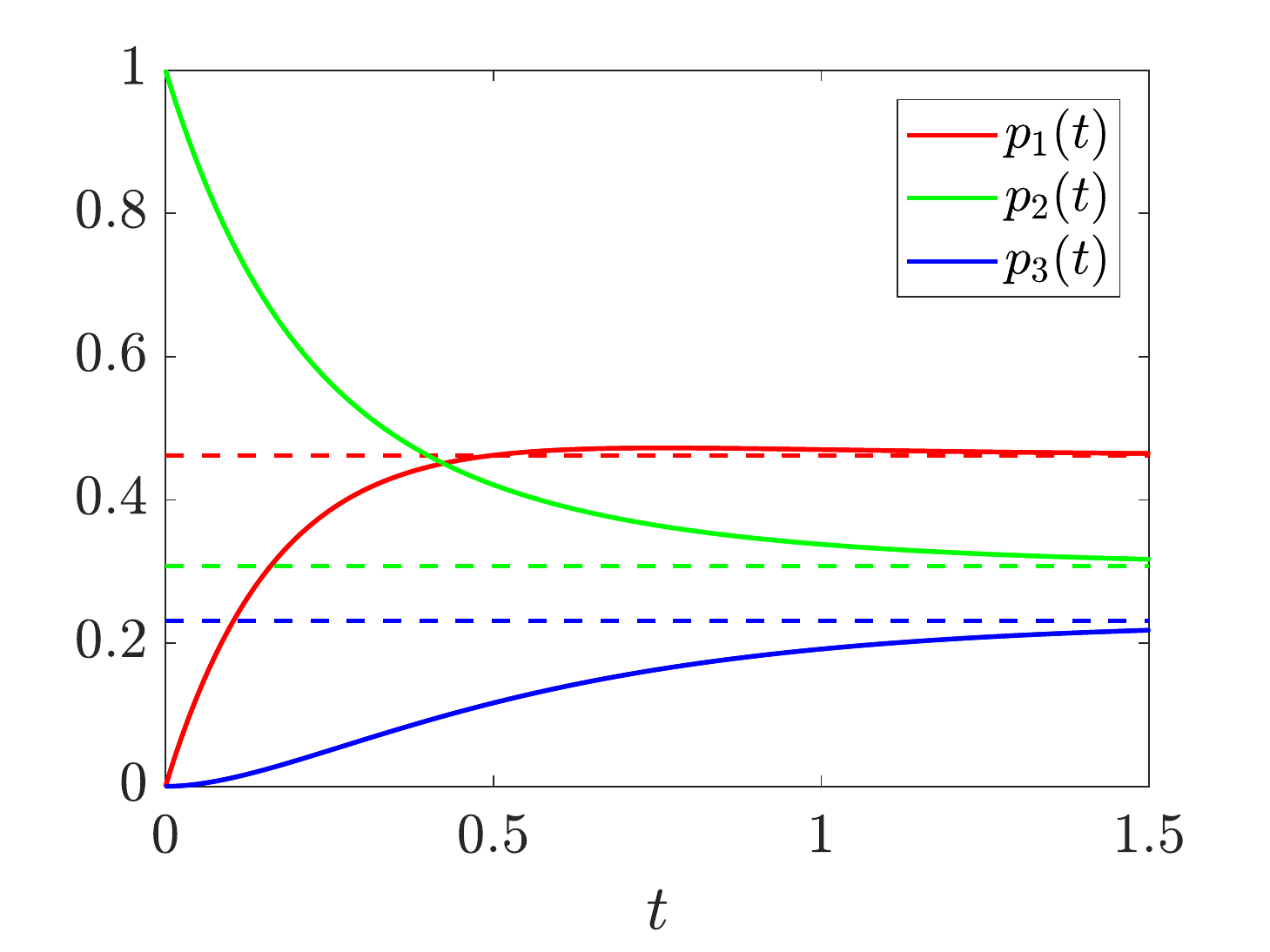}}
	\caption{Example Markov jump process. Left: Transition rates between the three states. Right: Solution of the Master equation for the initial condition $p_i(0) = \delta_{i,2}$. The dashed lines indicate the stationary solution.}
	\label{fig_basics_exampleJumpProcess}
\end{figure}

Let us consider a Markovian jump process with three states $1$, $2$, and $3$ and the following transition rate matrix depicted in the left half of Fig.~\ref{fig_basics_exampleJumpProcess}:
\begin{align}\label{eqn_basics_exampleTransitionRates}
\tilde{\mathcal{W}} = \begin{pmatrix}
-3 & 3 & 2\\
2 & -3 & 0\\
1 & 0 & -2
\end{pmatrix}
\end{align}

Given an initial probability density $\mathbf{p}_{0}$, the solution of the Master equation is given by:
\begin{align}
\mathbf{p}(t) = \exp{\left(\tilde{\mathcal{W}}\,t\right)}\,\mathbf{p}_0.
\end{align}
The solution for $\mathbf{p}_0 = (0,1,0)^T$ is displayed in the right half of Fig.~\ref{fig_basics_exampleJumpProcess}. We see that starting from state $2$, on average, the time evolution first populates state $1$ and then state $3$. The probability distribution eventually relaxes towards the \emph{stationary state} 
\begin{align}\label{eqn_basics_exampleSteadyStateProbs}
\mathbf{p}_{\rm st} = \left(\frac{6}{13},\frac{4}{13},\frac{3}{13}\right)^T
\end{align}
set by
\begin{align}
0 = \tilde{\mathcal{W}}\,\mathbf{p}_{\rm st}. 
\end{align}

\section{Kramers-Moyal expansion}
Consider a continuous-time stochastic process with a continuous state space. Assuming that its trajectory $x(t)$ is a continuous function of $t$, we expect that its transition rate $W(x|x';t)$ does not allow jumps, i.e., in a certain sense it has a sharp peak around $x=x'$ with a rapid decay. Then, a Taylor expansion around small jump lengths $\delta := x-x'$ is useful. We first rewrite the transition rate: $W(x|x';t) =: W(x',\delta;t)$. Inserting in the master equation~\eqref{eqn_basics_differentialMasterEquation} yields:
\begin{align}
\frac{\partial}{\partial t} p(x,t|x'',t'') &= \int d\delta \,\Big[W(x-\delta,\delta;t)\,p(x-\delta,t|x'',t'') \nonumber\\
&\qquad\qquad\qquad\qquad\qquad- W(x,-\delta ;t)\,p(x,t|x'',t'')\Big].
\end{align}

Next, we expand the first line for small jumps,
\begin{align}
&W(x-\delta,\delta;t)\,p(x-\delta,t|x'',t'') = W(x,\delta;t)\,p(x,t|x'',t'')\nonumber\\
&\qquad\quad- \frac{\partial}{\partial x} W(x,\delta;t)\,p(x,t|x'')\, \delta +\frac{1}{2} \frac{\partial^2}{\partial x^2} W(x,\delta;t)\,p(x,t|x'')\, \delta^2 -...,
\end{align}
thus obtaining
\begin{align}
\frac{\partial}{\partial t} p(x,t|x'',t'') &= \int d\delta \Bigg[W(x,\delta;t) \sum\limits_{n=1}^\infty \frac{\delta^n}{n!}\left(-\frac{\partial}{\partial x}\right)^n W(x,\delta;t)\Bigg]\, p(x,t|x'',t'') + \nonumber\\
&\qquad\qquad\qquad\qquad\qquad\qquad\qquad-\int d\delta \,W(x,\delta;t)\,p(x,t|x'',t'')\nonumber\\
&=\sum\limits_{n=1}^\infty \left(-\frac{\partial}{\partial x}\right)^n D^{(n)}(x,t)\, p(x,t|x'',t''), \label{eqn_basics_KramersMoyalExpansion}
\end{align}
where we substituted $-\delta$ in the second line, changed the integration limits, and introduced the coefficient functions
\begin{align}
D^{(n)}(x,t) := \int d\delta\, \frac{\delta^n}{n!}\,W(x,\delta;t).
\end{align}
The above procedure is known as \emph{Kramers-Moyal expansion}~\cite{Kramers1940,Moyal1949}. 

Equation~\eqref{eqn_basics_KramersMoyalExpansion} is useful because the expansion can be truncated to obtain estimates for the evolution equation. Furthermore, a theorem by Pawula~\cite{Pawula1967} states that the expansion either stops after the first or second term, or otherwise, never. 

\subsection{Fokker-Planck equation}
The expansion with two terms yields the \emph{Fokker-Planck equation}:
\begin{align}\label{eqn_basics_FPE}
\frac{\partial}{\partial t}\, p(x,t|x'',t'') &= \left[-\frac{\partial}{\partial x} D^{(1)}(x,t) + \frac{\partial^2}{\partial x^2} D^{(2)}(x,t) \right]\,p(x,t|x'',t''),
\end{align}
which is an extremely useful version of the master equation in the context of diffusive processes as we will see in the following section.

Equation~\eqref{eqn_basics_FPE} can also be written as a continuity equation with the probability current
\begin{align}\label{eqn_basics_defProbCurrentFPE}
j(x,t|x'',t'')  = \left[D^{(1)}(x,t) - \frac{\partial}{\partial x} D^{(2)}(x,t) \right]\,p(x,t|x'',t''),
\end{align}
yielding
\begin{align}\label{eqn_basics_continuityEq}
\frac{\partial}{\partial t}\, p(x,t|x'',t'') = -\frac{\partial}{\partial x} j(x,t|x'',t'').
\end{align}

\section{Brownian motion}
In this subsection we shall briefly depart from our mathematical treatment of stochastic processes and turn to their origin. In fact the whole subject of stochastic thermodynamics, and with it the contents of this thesis, can be regarded as a continuation (albeit with slightly different goals) of the analysis of Brownian motion spearheaded by Einstein, Smoluchowski, and Langevin. For this reason we will give a somewhat historical treatment of \emph{Brownian motion}.

\begin{figure}[ht]
	\centering
	\includegraphics[width = 0.6 \linewidth]{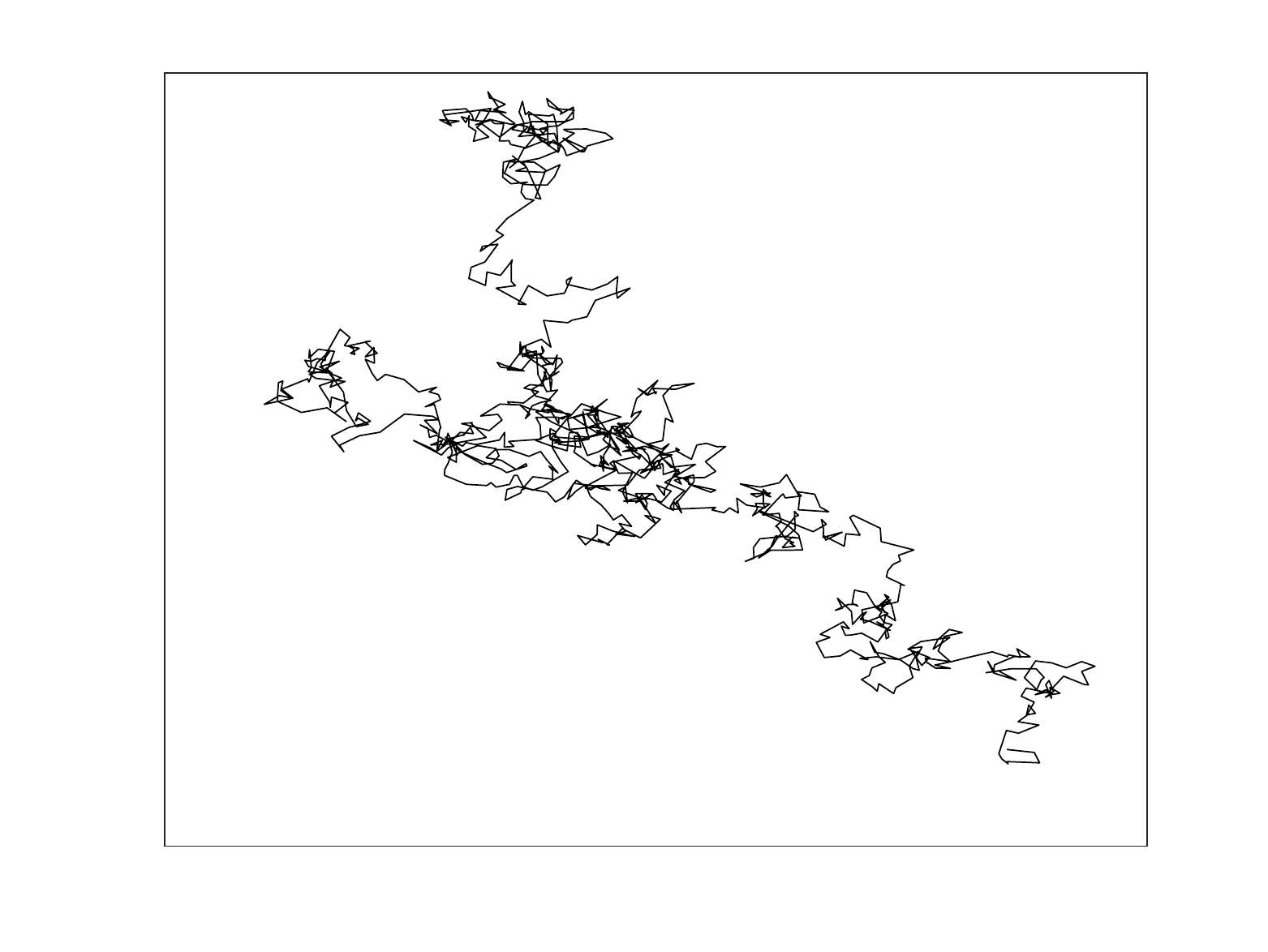}
	\caption{Trajectory of the simulation of a point-particle undergoing Brownian motion.}
	\label{fig_basics_brownianMotion}
\end{figure}

In 1827 the botanist Robert Brown observed under his microscope that small grains of pollen suspended in water perform an irregular, jittery motion, which is illustrated in Fig.~\ref{fig_basics_brownianMotion}. He ruled out that this movement was life-related but was unable to explain its origin. An explanation arrived much later, in 1905, with the theoretical treatments by Einstein~\cite{Einstein1905} and Smoluchowski~\cite{Smoluchowski1906}. Following Einstein's derivation, one encounters many of the concepts introduced previously.

The observed motion is due to very frequent collisions between the molecules of the suspension and the pollen grain. Because one cannot accurately describe the motion of this many molecules, the collisions can only be treated \emph{statistically}. Let us restrict the discussion to a one-dimensional setup. The density of particles per unit volume is given by $n(x,t)$. In some time interval $dt$, sufficiently small with respect to the observation times, each particle will experience a shift $\Delta$ due to the collisions. The probability $p(\Delta)$ of a certain shift shall be independent for every particle, independent from the past, symmetric, $p(-\Delta)=p(\Delta)$, and have a sharp peak around $\Delta =0$. The density for the time $t+dt$ then reads:
\begin{align} \label{eqn_basics_EinsteinsCKE}
n(x,t+ dt) = \int d\Delta \, n(x-\Delta,t)\,p(\Delta).
\end{align}
This is similar to the \emph{Chapman-Kolmogorov equation}~\eqref{eqn_basics_CKE} derived from assuming that collisions have no memory.

Next, one expands the LHS of Eq.~\eqref{eqn_basics_EinsteinsCKE} for small $dt$:
\begin{align}
n(x,t+dt) = n(x,t) + dt\,\frac{\partial n}{\partial t},
\end{align}
which is a crucial step in deriving the master equation. From assuming a sharp peak in $p(\Delta)$, Einstein continues with his version of the \emph{Kramers-Moyal-expansion}:
\begin{align}
n(x-\Delta,t) = n(x,t) - \Delta\,\frac{\partial n}{\partial x} + \frac{\Delta^2}{2}\frac{\partial^2 n}{\partial x^2}-...
\end{align}

Inserting into Eq.~\eqref{eqn_basics_EinsteinsCKE}, using the symmetry and the normalization of $p(\Delta)$, he obtains the equivalent of the \emph{Fokker-Planck equation}~\eqref{eqn_basics_FPE}:
\begin{align}\label{eqn_basics_diffusionEq}
\frac{\partial n}{\partial t} = D\, \frac{\partial^2 n}{\partial x^2},
\end{align}
thereby recovering Fick's law of diffusion~\cite{Fick1855} with the \emph{diffusion coefficient}
\begin{align}
D := \frac{1}{dt}\int d\Delta\,\frac{\Delta^2}{2}p(\Delta).
\end{align}

In essence, Einstein envisions a \emph{random walk} of the pollen grain with infinitely small steps and shows how this relates to diffusion (see, e.g., Sec. 3.8.2 of Ref.~\cite{Gardiner2004}). The most important contribution, however, is Einstein's and Smoluchowski's linking of the diffusion coefficient $D$ with the temperature $T$ and the coefficient of Stokes's friction $\gamma$ that a spherical particle in solution experiences:
\begin{align}\label{eqn_basics_EinsteinSmoluchowskiRelation}
D = \frac{\kB T}{\gamma},
\end{align}
where $k_B$ is Boltzmann's constant. We will prove this relation at a later point using a different method. This identification allows estimates of microscopic information like Boltzmann's constant or, equivalently, the Avogadro number $N_{\rm A} = R/\kB$, (where $R$ is the gas constant) from the experimentally accessible mean squared displacement of small particles in solution. 

\subsection{Langevin equation}\label{sec_basics_LengevinEquation}
The approach to explaining Brownian motion that is most used today is due to Langevin~\cite{Langevin1908}. He set up an equation of motion for the position $x$ of a \emph{colloidal particle}, i.e., a particle suspended in a medium and subject to collisions from its surroundings:
\begin{align}\label{eqn_basics_originalLangevinEq}
m\, \ddot{x} = -\gamma \,\dot{x} + \sqrt{2D}\,\gamma\, \xi(t).
\end{align}
The first term on the RHS describes the effect of Stokes's friction on the particle, while the second term represents some other \emph{fluctuating force} that is due to the incessant collisions with the molecules of the suspension. The proportionality factor $\sqrt{2D}\, \gamma$ is included for convenience, in fact, $D$ will later turn out to be the diffusion constant.

Langevin was able to recover Einstein's result with very little assumptions about the fluctuating force $\xi(t)$: It needs to be zero on average, 
\begin{align}\label{eqn_basics_meanNoise}
\langle \xi(t) \rangle=0
\end{align}
and must be completely uncorrelated with itself: 
\begin{align}\label{eqn_basics_noiseCorrFunction}
\left\langle \xi(t) \xi(t') \right\rangle = \delta(t-t')
\end{align}
(although the second assumption is only implicit in his original work). In the following, we use the approach by Uhlenbeck and Ornstein~\cite{Uhlenbeck1930} to derive the Einstein-Smoluchowsky relation [Eq.~\eqref{eqn_basics_EinsteinSmoluchowskiRelation}].

With the initial condition $\dot{x}(t=0)=v_0$, we can solve Eq.~\eqref{eqn_basics_originalLangevinEq} for the velocity:
\begin{align}\label{eqn_basics_UhlenbeckSolution}
v(t) = v_0\,e^{-\gamma t/m} + \frac{\sqrt{2D}\,\gamma}{m} e^{-\gamma t/m} \int\limits_0^t dt'\, e^{\gamma t'/m} \xi(t').
\end{align}
With this, we calculate the second moment
\begin{align}
\left\langle v^2(t)\right\rangle &= v_0^2 \,e^{-2\gamma t/m} + \frac{2D\,\gamma^2}{m^2} \int\limits_0^t dt'  \int\limits_0^t dt''\, e^{\gamma (t'+t'')/m}\left\langle \xi(t') \xi(t'')\right\rangle\\
&= v_0^2\, e^{-2\gamma t/m} + \frac{D\,\gamma}{m} \left( 1-e^{-2 \gamma t/m} \right),
\end{align}
where we used the fact that $\xi(t)$ is delta-correlated. This means that for small particles the mean-squared velocity is quickly decaying towards $\left\langle v_\infty^2\right\rangle = D\gamma/m$. Now, $D$ can be identified from the fact that the mean kinetic energy must obey the equipartition theorem:
\begin{align}
\frac{1}{2}\kB T \overset{!}{=} \frac{1}{2} m \langle v_\infty^2\rangle = \frac{1}{2}D\gamma  \Rightarrow D = \frac{\kB T}{\gamma},
\end{align}
which recovers Eq.~\eqref{eqn_basics_EinsteinSmoluchowskiRelation}. This, however, does not yet prove the equivalence of Einstein's and Langevin's approaches, which can be achieved by integrating Eq.~\eqref{eqn_basics_UhlenbeckSolution} and computing the mean squared displacement $\left\langle(x(t)-x_0)^2\right\rangle$. 

The \emph{Einstein-Smoluchowski} relation has important conceptual implications: It states that the energy dissipation due to the friction and the energy intake due to the fluctuations are related. This is because they both result from collisions with the surrounding molecules. It is a manifestation of the far more versatile \emph{fluctuation-dissipation theorem} which yields similar relations in other settings, e.g., current fluctuations in resistors and thermal radiation. For a review see Ref.~\cite{Marconi2008}.

Einstein's treatment of Brownian motion does not refer to velocities at all. Instead, only shifts in position are considered. This works because of the \emph{overdamped} limit in which Brownian motion usually takes place. As can be seen from Eq.~\eqref{eqn_basics_UhlenbeckSolution}, velocity correlations decay extremely quickly for small particles: The characteristic timescale for a spherical colloid is $m/\gamma$ = $m/(6\pi \mu R)$, where $m$ is the particle's mass, $\mu$ the dynamic viscosity of the medium, and $R$ the radius. For small pollen, Langevin estimated a correlation time of about $10^{-8}\,\text{s}$. This means that one will almost never see the particle's velocity fluctuations. Instead, we see the integrated effect of small collisions as a shift in the particle's position. 

With this in mind, one can simplify Langevin's approach by neglecting the inertia term $m \ddot{x}$ in Eq.~\eqref{eqn_basics_originalLangevinEq}, thus obtaining the \emph{overdamped Langevin equation}: 
\begin{align}\label{eqn_basics_overdampedLENoDrift}
\dot{x} = \sqrt{2D}\,\xi(t).
\end{align}
This equation describes a Markov process whereas the \emph{underdamped} equation~\eqref{eqn_basics_originalLangevinEq} does not. Instead, the underdamped diffusion is a \emph{bi-variate} Markov process, namely in position $x$ \emph{and} velocity $v = \dot{x}$.

In many interesting scenarios there is also a deterministic force $f(x,t)$ acting on the particle which, when included in the equation of motion, leads to the overdamped Langevin equation, which we will use throughout this thesis:
\begin{align}\label{eqn_basics_overdampedLangevinEquation}
\dot{x} = \nu f(x,t) + \sqrt{2D}\,\xi(t),
\end{align}
where $\nu := \gamma^{-1}$ is the \emph{mobility} of the particle.

Finally, we need to fully specify the stochastic force $\xi(t)$ by supplementing one final condition: $\xi(t)$ shall be \emph{Gaussian}, i.e., all cumulants higher than two vanish. Then, it is completely defined by Eqs.~\eqref{eqn_basics_meanNoise}~and~\eqref{eqn_basics_noiseCorrFunction}. However, the force $\xi(t)$ does not exist as a function, i.e., we could not plot one specific realization of it. Rather, it needs to be understood as the limiting case of a stochastic force with a correlation time tending towards zero. In any case this is more realistic from a physics standpoint since collisions between particle and medium are not completely uncorrelated, but, due to the vast number of degrees of freedom, their correlation time is very short.

The peculiar properties of $\xi(t)$ cause an issue when \emph{stochastic integrals}, i.e., integrals of the form $\int dt\, g(x)\xi(t)$, arise. This happens when one studies trajectories generated by Eq.~\eqref{eqn_basics_overdampedLangevinEquation} in which $D$ is not constant. Then, the equation needs to be supplemented with a \emph{discretization rule}, i.e., a prescription of how to interpret the RHS. We will always use the \emph{Stratonovich} interpretation, which means that a Langevin equation of the form
\begin{align}\label{eqn_basics_generalOverdampedLangevin}
\dot{x} = \nu f(x,t) + g(x)\,\xi(t)
\end{align}
needs to be discretized into:
\begin{align}\label{eqn_basics_StratonivichDiscretization}
x(t+dt)-x(t) &= dt\cdot \nu  f\left(\frac{x(t)+x(t+dt)}{2},t+\frac{dt}{2}\right) \nonumber\\
&\qquad\qquad\qquad+ g\left(\frac{x(t)+x(t+dt)}{2}\right)\int\limits_{t}^{t+dt} dt' \xi(t').
\end{align}

We choose this interpretation because, as alluded to above, in physical settings there exists a non-vanishing correlation time of the stochastic noise, making it \emph{colored} rather than \emph{white}. It was shown by Wong and Zakai~\cite{Wong1965} that in the white-noise limit, solutions of the Langevin equation with colored noise obey the Stratonovich form. Section~4.3 of Ref.~\cite{Gardiner2004} gives a good overview on the different interpretations and their consequences.

Next, we show that there is a relation between the Langevin equation for \emph{individual} trajectories $x(t)$ and the diffusion (or Fokker-Planck) equation for the evolution of an \emph{ensemble} of systems. This is to be expected as both the Langevin and Einstein treatments concern the same physical process.

Starting from the discretized Langevin equation~\eqref{eqn_basics_StratonivichDiscretization}, we obtain the shift $\Delta := x(t+dt)-x(t)$:
\begin{align}\label{eqn_basics_deltaLangevin}
\Delta = dt\cdot \nu\, f\left(x+\frac{\Delta}{2},t+\frac{dt}{2}\right) + g\left(x+\frac{\Delta}{2}\right)\int\limits_{t}^{t+dt} dt' \xi(t').
\end{align}
This is not a closed-form equation for $\Delta$. Therefore, we only keep terms up to first order in $dt$. The integral gives us the \emph{Wiener process}, which is the random quantity
\begin{align}
dW := \int\limits_{t}^{t+dt} dt' \xi(t').
\end{align}
Due to the properties of $\xi(t)$ [Eqs.~\eqref{eqn_basics_meanNoise}~and~\eqref{eqn_basics_noiseCorrFunction}], it is a Gaussian with mean and variance given by
\begin{align}
\left\langle dW \right\rangle = 0 \qquad \text{and}\qquad \left\langle dW^2\right\rangle = dt.
\end{align}
Therefore, the $dW$ is $ \mathcal{O}(\sqrt{dt})$. We thus need to take $f$ to zeroth order and expand $g$ to order of $\sqrt{dt}$:
\begin{align}
g\left(x+\frac{\Delta}{2}\right) &= g(x) + g'(x) \frac{\Delta}{2} + \mathcal{O}(\Delta^2)\\
&= g(x) + \frac{1}{2} g'(x)g(x)\, dW + \mathcal{O}(dt),
\end{align}
where $g'(x)$ denotes the derivative of $g(x)$ and we inserted Eq.~\eqref{eqn_basics_deltaLangevin} recursively. We thus obtain
\begin{align}
\Delta = dt\, \nu\, f(x,t) + g(x)\, dW + \frac{1}{2}g'(x) g(x)\, dW^2 + \mathcal{O}\left(dt^\frac{3}{2}\right)
\end{align}
which defines the probability distribution $p(\Delta|x,t)$ of the shift with the moments
\begin{align}
\langle\Delta\rangle &= dt\,\nu\,f(x,t) + \frac{1}{2} g'(x) g(x)\, dt \qquad \text{and} \label{eqn_basics_Stratonovich_shift_1}\\
\langle\Delta^2\rangle &= g^2(x)\, dt, \label{eqn_basics_Stratonovich_shift_2}
\end{align}
which, since the process described by the overdamped Langevin equation is Markovian, can be used in the Chapman-Kolmogorov equation~\eqref{eqn_basics_CKE} [cf.~also~eq.~\eqref{eqn_basics_EinsteinsCKE}]:
\begin{align}
p(x,t+dt) &= \int d\Delta\,p(x-\Delta,t)\,p(\Delta|x-\Delta;t)\\
&= p(x,t) - \frac{\partial}{\partial x}\, p(x,t)\left(\int d\Delta\,\Delta\,p(\Delta|x;t)\right) \nonumber\\
&\qquad\qquad\qquad\qquad + \frac{\partial^2}{\partial x^2}  p(x,t)\left(\int d\Delta\,\frac{\Delta^2}{2}\,p(\Delta|x;t)\right)-...\\
&= p(x,t) - dt\,\frac{\partial}{\partial x}\, p(x,t)\left(\nu f(x,t) + \frac{1}{2}g'(x)g(x) \right)\nonumber\\
&\qquad\qquad\qquad\qquad+ dt\,\frac{\partial^2}{\partial x^2} p(x,t)\,\frac{g^2(x)}{2}-...\,.
\end{align}

Therefore, the overdamped Langevin equation~\eqref{eqn_basics_generalOverdampedLangevin} corresponds to the Fokker-Planck equation
\begin{align}\label{eqn_basics_FokkerPlanckCorrespondingLangevin}
\frac{\partial}{\partial t}\, p(x,t) &= -\frac{\partial}{\partial x}\left(\nu f(x,t) + \frac{1}{2} g'(x)g(x)\right)\,p(x,t) +\frac{1}{2} \frac{\partial^2}{\partial x^2}\,g^2(x)\,p(x,t).
\end{align}
For $f\equiv 0$ and $g(x) \equiv \sqrt{2D}$ this proves the equivalence of the overdamped Langevin equation~\eqref{eqn_basics_overdampedLENoDrift} and the diffusion equation~\eqref{eqn_basics_diffusionEq}. Thus, $D$ is indeed the diffusion constant. Therefore, Einstein's and Langevin's treatments of Brownian motion are indeed equivalent.

\section{Path probabilities}\label{sec_basics_pathProbabilities}
In stochastic thermodynamics the concept of \emph{path probabilities} plays an important role. Simply put, it is the probability of a certain \emph{trajectory} of the stochastic process. We already encountered a (coarse-grained) path probability in Eq.~\eqref{eqn_basics_JointProb}: The joint probability
\begin{align}
p(x_1,t_1;\,x_2,t_2;\,...;\,x_N,t_N)
\end{align}
gives the probability of an entire sequence of measurements, which is a path through the state space of the process. 

The concept is fairly clear for discrete-time processes. For continuous-time processes the interpretation as a joint probability distribution becomes less intuitive, since it necessitates a infinitely fine-grained sampling of the process. Nonetheless, this limit exists in a rigorous mathematical sense. Dispensing with the measure-theoretic details, one gets a probability density in function space. Thus, the probability density of a certain trajectory $x(\cdot)$ is a functional $p[x(\cdot)]$ with an appropriate normalization given by:
\begin{align}\label{eqn_basics_pathIntegralNoBoundaryTerms}
\int \mathcal{D}x(\cdot)\,p[x(\cdot)] = 1.
\end{align}
The notation $x(\cdot)$ indicates that we mean the \emph{entire} function instead of a distinct value $x(t)$. The integral $\int \mathcal{D}x(\cdot)$ is a \emph{path integral} indicating integration over the whole function space.

We proceed to derive the path probability for a stochastic trajectory generated by the overdamped Langevin equation~\eqref{eqn_basics_overdampedLangevinEquation}. As before, we need to agree on a discretization rule, in this case both for the stochastic differential equation and for the interpretation of the resulting path integral representation. We opt for the mid-point dicretization, i.e., for a constant diffusion coefficient we get:
\begin{align}\label{eqn_basics_pathProbLangevin}
x_{i} - x_{i-1} = \nu dt\, f\left(\frac{x_{i}+x_{i-1}}{2},t_i+\frac{dt}{2}\right) + \sqrt{2D}\,dW.
\end{align}
This generates a discretized version $\{x_0,x_1,...,x_N\}$ of the trajectory $x(\cdot)$ of length $T$ starting at $x(0) = x_0$ and ending at $x(T)=x_T$. Thus, $T = N\cdot dt$ and $x_N= x_T$. 

Since the diffusion process described by the Langevin equation is Markovian, we find the joint probability for the discretized trajectory in terms of the transition probability $p(x_{i+1}|x_{i})$, which is most easily calculated from the \emph{Wiener process} by a transformation of variables:
\begin{align}
p_x(x_{i}|x_{i-1}) &= p_W\left(dW(x_{i})\right)\, \left|\frac{d dW}{d x_i} \right|\\
&= \frac{1}{\sqrt{2\pi dt}}\exp{\left(-\frac{dW^2(x_i)}{2dt}\right)} \frac{1}{\sqrt{2 D}} \left(1-\frac{1}{2} \nu dt f'_i + \mathcal{O}\left(dt^2\right)  \right)\\
&= \frac{1}{\sqrt{4\pi D dt}}\exp{\left[- \frac{dt}{4D}\left(\frac{x_i-x_{i-1}}{dt} - \nu f_i\right)^2 -\frac{1}{2} \nu dt f'_i \right]} + \mathcal{O}\left( dt^{3/2}\right) ,\label{eqn_basics_transitionProbSmallTimes}
\end{align}
where $f^{(n)}_i := \frac{\partial^n}{\partial x_i^n} f\left(\frac{x_i + x_{i-1}}{2},t+\frac{dt}{2}\right)$.

Stringing these probabilities together, we get the joint trajectory probability:
\begin{align}
p(x_1,...,x_N|x_0) dx_1...dx_N &= \prod\limits_{i=1}^N p(x_i|x_{i-1}) dx_i\\
&\approx \frac{dx_1...dx_N}{(4 \pi D dt)^{N/2}} \exp{\left[- \frac{dt}{4D} \sum\limits_{i=1}^N \left(\frac{x_i-x_{i-1}}{dt} - \nu f_i\right)^2 \right]} \nonumber\\
&\qquad\qquad\qquad\qquad\qquad \times \exp{\left(-\frac{dt}{2}\sum\limits_{i=1}^N \nu f'_i\right)}.
\end{align}

In the limit $dt\rightarrow 0$ and $N \rightarrow \infty$, we finally obtain the path probability
\begin{align}
p(x_1,...,x_N|x_0) dx_1...dx_N \longrightarrow p[x(\cdot)|x_0] \mathcal{D}x(\cdot).
\end{align}
Here, $\frac{dx_1...dx_N}{(4 \pi D dt)^{N/2}} \rightarrow \mathcal{D}x(\cdot)$ is the differential trajectory element. The trajectory probability reads
\begin{align}\label{eqn_basics_trajectoryProbLangevin}
p[x(\cdot)|x_0] = \exp{\left\{ - S[x(\cdot)]\right\}},
\end{align}
and is defined in terms of the \emph{action}
\begin{align}\label{eqn_basics_actionLangevin}
S[x(\cdot)] := \int\limits_0^T dt \left(\frac{1}{4D}\left[\dot{x}(t)-\nu f(x,t)\right]^2 + \frac{\nu}{2} f'(x,t) \right)
\end{align}
associated to the trajectory.

Normalization follows from the \emph{path integral representation} of the transition probability
\begin{align}
p(x_T,T|x_0,0) = \int\limits_{(x_0,0)}^{(x_T,T)} \mathcal{D}x(\cdot)\, p[x(\cdot)|x_0],
\end{align}
which gives rise to an intuitive interpretation: To get the probability of a transition from $x_0$ at time $0$ to $x_T$ at time $T$, one has to sum over the probabilities of \emph{all} possible paths connecting the two states. More information on path integrals can be found in the book by Chaichian and Demichev~\cite{Chaichian2001}.

To keep the notation light, we will in the following denote the probability of a complete path including the starting point simply by $p[x(\cdot)]$:
\begin{align}\label{eqn_basics_notationCompletePathProb}
p[x(\cdot)] := p[x(\cdot)|x_0]\, p(x_0,0).
\end{align}
The normalization given in Eq.~\eqref{eqn_basics_pathIntegralNoBoundaryTerms} then needs to be resolved as
\begin{align}
1 = \int \mathcal{D}x(\cdot)\,p[x(\cdot)] = \iint dx_T dx_0 \int\limits_{(x_0,0)}^{(x_T,T)} \mathcal{D}x(\cdot)\, p[x(\cdot)|x_0] \, p(x_0,0).
\end{align}

\section{Stationarity}
A \emph{stationary stochastic process} has a joint probability distribution that is invariant under arbitrary time-shifts $\Delta t$:
\begin{align}
p(x_1,t_1+\Delta t;\, x_2,t_2+\Delta t;...; x_N,t_N+\Delta t) = p(x_1,t_1;\, x_2,t_2;...;\,x_N,t_N).
\end{align}
One-point measures like the simple probability $p(x_i)$ then become independent of time and multi-point statistics only depend on time differences. Stationary Markov processes are generated by a time-independent transition probability (rate): $W(x|x')$. The stationary solution $p_{\rm st}(x)$ must then obey the \emph{stationary master equation}:
\begin{align}\label{eqn_basics_stationaryME}
0 = \int dx' \,\Big[W(x|x')\,p_{\rm st}(x') - W(x'|x)\,p_{\rm st}(x)\Big].
\end{align}

Stationary diffusion-type processes described by a Langevin equation~\eqref{eqn_basics_overdampedLangevinEquation} or Fokker-Planck equation~\eqref{eqn_basics_FokkerPlanckCorrespondingLangevin} have time-independent drifts $f(x)$ and diffusion coefficients $D$ and their stationary probability current obeys:
\begin{align}
0 = \frac{\partial}{\partial x} j_{\rm st}(x),
\end{align}
where $j_{\rm st}(x) = \left[-\nu f(x) + D \frac{\partial}{\partial x}\right] p_{\rm st}(x)$. Without periodic boundary conditions, a stationary solution $p_{\rm st}(x)$ can only be achieved when the probability current vanishes, thus:
\begin{align}
D \frac{\partial}{\partial x} \ln p_{\rm st}(x) = \nu f(x).
\end{align}
Therefore, the stationary solution reads:
\begin{align}
p_{\rm st}(x) = \frac{1}{Z}\,\exp{\left[-\varphi(x)\right]},
\end{align}
where
\begin{align}
\varphi(x) := -\frac{\nu}{D} \int dx\,f(x)\quad \text{and}\quad Z := \int dx\, \exp{\left[-\varphi(x)\right]}.
\end{align}

For an overdamped Langevin equation of the form of Eq.~\eqref{eqn_basics_overdampedLangevinEquation} where the force results from a potential $f(x) = - \frac{\partial}{\partial x} V(x)$, we consequently find:
\begin{align}
p_{\rm st}(x) = \exp{\left(-\frac{V(x)-F}{\kB T }\right)},
\end{align}
where $F := -\kB T\ln\left[\int dx\,\exp{\left(-V(x)/\kB T\right)}\right]$ is the free energy, thus recovering the Boltzmann distribution indicating that the stationary distribution is the equilibrium distribution, as it should be.

We recover the equilibrium distribution because the above diffusion process obeys \emph{detailed balance}, i.e., the frequency of transitions $x' \rightarrow x$ is statistically balanced by the frequency of the opposite transitions $x \rightarrow x'$. For a general Markov process detailed balance necessitates the integrand in Eq.~\eqref{eqn_basics_stationaryME} to vanish:
\begin{align}
W(x|x')\,p_{\rm st}(x') = W(x'|x)\,p_{\rm st}(x).
\end{align}
Refer to Secs.~\ref{sec_basics_detailedBalance}~and~\ref{sec_basics_brokenDetailedBalance} for more details. In anticipation of the thermodynamic interpretation in chapter~\ref{chap_stochasticThermodynamics}, we refer to stationary processes with detailed balance as \emph{equilibrium processes} and to those without as \emph{nonequilibrium stationary (or steady) states}. 

\section{Higher dimensions}

Most concepts from the above sections can straightforwardly be generalized to higher dimensions. The overdamped Langevin equation~\eqref{eqn_basics_overdampedLangevinEquation}, for example, becomes a vector equation. Usually, the noise terms $\{\xi_i(t)\}
$ affecting individual degrees of freedom are independent, so that we obtain
\begin{align}
\dot{\bx} = \mathbf{F}(\bx,t) + \mathbf{G}(\bx,t)\, \bm{\xi}(t),
\end{align}
where $\mathbf{G}(\bx,t)$ is the \emph{diffusion matrix} and $\bm{\xi}(t)$ is the vector collecting the individual noise terms obeying $\left\langle \xi_i(t) \xi_j(t')\right\rangle =\delta_{i j}\, \delta(t-t')$. For this process the equivalent Fokker-Planck equation in Stratonovich interpretation  reads (cf.~Sec.~4.3.6 of~Ref.~\cite{Gardiner2004}):
\begin{align}\label{eqn_basics_multiDimensionalFokkerPlanck}
\frac{\partial}{\partial t}p(\bx,t) = \left[- \sum\limits_i \frac{\partial}{\partial x_i} F_i(\bx,t) + \frac{1}{2} \sum_{i j k} \frac{\partial}{\partial x_i} G_{ik}(\bx,t) \frac{\partial}{\partial x_j} G_{jk}(\bx,t) \right] p(\bx,t).
\end{align}

Since every force $f(x,t)$ depending on one spatial coordinate can be integrated to a potential, nonequilibrium steady states in diffusion-type processes necessitate higher dimensions. They are then characterized by constant cyclical (divergence-free) probability currents through the state space:
\begin{align}
0 = \pd{t} p(\bx,t) =  -\nabla \cdot \mathbf{j}_{st}(\bx).
\end{align}

\section{Numerics: generating stochastic trajectories}\label{sec_basics_numerics}
We close this chapter by presenting some algorithms which allow numerical simulation of a Markov process. First, for discrete-time processes, the transition probability $p_i(x|x')$ already prescribes an algorithm:
\begin{enumerate}
	\item Draw a random initial state from the initial probability distribution $p_1(x_1)$.
	\item Repeatedly draw a next state from the transition probability $p_i(x_{i+1}|x_i)$.
\end{enumerate}

Continous-time jump processes can be simulated using the \emph{Gillespie algorithm}~\cite{Gillespie1977}: Once the process reached a state $i$ at time $t$ it will stay in this state for a random \emph{waiting time} $\tau$. This waiting time needs to be drawn from a probability distribution $p_i(\tau,t)$ which can be inferred from the matrix-form master equation~\eqref{eqn_basics_MEMatrixNotation}. Setting $p_j(0)=\delta_{i j}$, we obtain:
\begin{align}
\frac{d}{d\tau}p_i(\tau,t) = \tilde W_{i i}(t+\tau)\, p_i(\tau,t) \Longrightarrow p_i(\tau,t) \propto \exp{\left[\int\limits_0^\tau d\tau' \,\tilde W_{i i}(t+\tau')\right]}.
\end{align}
Notice that $\tilde W_{ii}(t) = -\sum\limits_{j\neq i} \tilde W_{ji}(t) < 0$.

After drawing a random waiting time, adding it to the simulation time and checking whether it exceeds the allotted total simulation time, the next state is chosen by drawing a random number from the normalized transition rate vector excluding the current state:
\begin{align}
p\big(x(t+\tau) = j\big|x(t)=i\big) = \frac{W_{ji}(t+\tau)}{\sum\limits_{k\neq i} W_{ki}(t+\tau)}.
\end{align}
The algorithm then proceeds with drawing the next waiting time.

Finally, diffusion-type processes governed by a Langevin equation~\eqref{eqn_basics_generalOverdampedLangevin} can be straightforwardly simulated using the Euler integration scheme with time step $dt$. Care needs to be taken when the diffusion coefficient $g(x)$ is not constant, since then the choice of discretization becomes important. We use Eqs.~\eqref{eqn_basics_Stratonovich_shift_1}~and~\eqref{eqn_basics_Stratonovich_shift_2} derived previously to obtain:
\begin{align}
x(t+dt) = x(t) + dt\,\nu f\big(x(t),t\big) + \frac{dt}{2} g'\big(x(t)\big) g\big(x(t)\big) + \sqrt{dt}\, g\big(x(t)\big)\, \hat{\xi},
\end{align}
where $\hat{\xi}$ is a zero-mean Gaussian random number with unit variance:
\begin{align}
\hat{\xi} \sim \mathcal{N}(0,1).
\end{align}

This concludes the introductory chapter on stochastic processes. We will now expand from this basis towards key results of information theory and eventually towards physics by presenting the basics of stochastic thermodynamics.

\chapter{Information theory}\label{chap_basics_infoTheory}
In 1948 Claude Shannon published a groundbreaking paper: \emph{A Mathematical Theory of Communication}~\cite{Shannon1948}. Having worked on cryptography during World War II, he considered the question of how best to encode a message transmitted through a noisy communication channel and thereby prone to signal corruption. He found that the ultimate data compression of any message, and therefore its information content, is given by its \emph{entropy}.

The field he established was later called \emph{information theory} and plays an important role in many areas of science. We will focus on its relevance in thermodynamics. There, the links are deeper than just the similarity in names: In stochastic thermodynamics, information enters as a proper thermodynamic resource like heat and work, as we will see in Chap.~\ref{chap_informationThermodynamics}. A good introduction to the field of information theory is given in the book by Cover and Thomas~\cite{Cover2006} on which the first four sections of this chapter are based.

\section{Entropy}\label{sec_basics_entropy}
Consider the outcome $x$ of a random event with a discrete probability distribution $p(x)$. We seek to quantify the \emph{information content} or \emph{surprisal} (or \emph{news-worthiness}) of this outcome in such a way that an expected result has a low surprisal value while an unexpected result has a high surprisal. A handy measure is the \emph{entropy}
\begin{align}\label{eqn_basics_defEntropy}
s(x) := -\ln{p(x)},
\end{align}
since it is a monotonically decreasing positive function of the probability of the event. A sure event occurs with probability one and thus has zero entropy. Additionally, for two unrelated events $x_1$ and $x_2$ the entropy is additive, since
\begin{align}
s(x_1,x_2)  = - \ln{[p(x_1)\,p(x_2)]} =  -\ln{p(x_1)} -\ln{p(x_2)} = s(x_1) + s(x_2).
\end{align}

Beyond the information content of a single random event, one often wants to quantify the \emph{average information content} or \emph{average entropy}:
\begin{align}\label{eqn_basics_defAverageEntropy}
S[x] := -\sum\limits_x p(x) \ln{p(x)}.
\end{align}
Notice the slight inconsistency in notation as the average entropy is actually a \emph{functional} of the entire probability distribution $p(\cdot)$. However, it is useful to think of it as the \emph{average entropy associated to the variable $x$}.

Consider, e.g., a lottery with $n$ tickets. Learning that a given lottery ticket is not the winning one is unsurprising (if there are many tickets) and consequently it has a very low entropy: $s = \ln{\left[n/(n-1)\right]}\approx 0$. In contrast, learning that a given ticket is the winning one is a high-entropy event: $s~=~\ln{n}\gg~0$. The average entropy of the lottery ticket outcome is:
\begin{align}
S = \frac{n-1}{n}\ln\frac{n}{n-1}+ \frac{1}{n}\ln{n},
\end{align}
which vanishes for $n=1$ (Where is the surprise then?). It has its maximum at $n=2$ and then decays towards $\ln{(n)}/n$: The vast probability of losing simply outweighs the contribution to the average entropy from the winning ticket. Therefore, winning the lottery is surprising but playing it must be expected to be unsurprising.

In the literature one can often find different logarithm bases for the entropy definition. We choose the natural logarithm due to its relevance in physics.

\section{Kullback-Leibler distance}
The \emph{Kullback-Leibler distance} or \emph{relative entropy} is a distance measure in the space of probability distributions. It is defined for two probability distributions $p(x)$ and $q(x)$ with the same support:
\begin{align}\label{eqn_basics_defKLDistance}
D_{\rm KL}[p||q] := \sum\limits_x p(x) \ln\frac{p(x)}{q(x)}.
\end{align}
From its definition it is evident that the Kullback-Leibler distance is not a true distance since it is not symmetric and does not fulfill the triangle inequality. However, $D_{\rm KL}$ is nonnegative as can be proven by applying Jensen's inequality (cf. Sec.~2.6 of Ref.~\cite{Cover2006}) or one of its derivatives, the log sum inequality: For nonnegative numbers $\{a_i\}$ and $\{b_i\}$ with $a = \sum_i a_i$ and $b = \sum_i b_i$ one has (cf. Sec.~2.7 of Ref.~\cite{Cover2006}):
\begin{align}
\sum\limits_i a_i \ln\frac{a_i}{b_i} \geq a\,\ln\frac{a}{b},
\end{align}
with equality holding for $a_i/b_i = \rm const.$

Since both $p(x)$ and $q(x)$ are normalized probability distributions, this proves $D_{\rm KL}[p||q] \geq 0$ with equality holding when $p(x)\equiv q(x)$.

\section{Joint and conditional entropy and mutual information}\label{sec_basics_jointCondMutualEntropy}
The situation becomes slightly more involved when there are several correlated random variables. For two random variables $x$ and $y$ we define the \emph{joint entropy} as the entropy of the combined outcome
\begin{align}
s(x,y) := -\ln{p(x,y)}
\end{align}
and thus
\begin{align}
S[x,y] := -\sum\limits_{x,y} p(x,y) \ln{p(x,y)}.
\end{align}

Additionally, we define the \emph{conditional entropy} as the entropy of one variable (say, $x$) that is left upon learning the value of the other (say, $y$):
\begin{align}
s(x|y) := - \ln{p(x|y)}.
\end{align}
For the \emph{average conditional entropy} we have to take the average with respect to the joint probability distribution:
\begin{align}\label{eqn_basics_defAverageConditionalEntropy}
S[x|y] := - \sum\limits_{x,y} p(x,y)\ln{p(x|y)}.
\end{align}

This gives rise to the \emph{chain rule of entropy}:
\begin{align}
s(x,y) = s(x|y) + s(y) \quad \text{and} \quad S[x,y] = S[x|y] + S[y].
\end{align}

Finally, we require a measure of the information that is \emph{shared} between two random variables, i.e., their correlation. This is given by the \emph{mutual information}:
\begin{align}\label{eqn_basics_defMutualInformation}
i(x,y) &:= -\ln{p(x)}-\ln{p(y)} + \ln{p(x,y)}.
\end{align}
This definition is symmetric and leads to the following relations between it and the joint and conditional entropies:
\begin{subequations}\label{eqn_basics_relationsMutualInformation}
	\begin{align}
	i(x,y) &= s(x) + s(y) - s(x,y)\\
	&= s(x) - s(x|y)\\
	&= s(y) - s(y|x).
	\end{align}
\end{subequations}
Obviously, it is zero for uncorrelated random variables. The \emph{average mutual information} is consequently given by:
\begin{align}\label{eqn_basics_defAverageMutualInformation}
I[x,y] := -\sum\limits_{x,y} p(x,y)\ln\frac{p(x)\, p(y)}{p(x,y)}
\end{align}
with similar relations to the average entropies as in Eqs.~\eqref{eqn_basics_relationsAverageMutualInformation_a}~to~\eqref{eqn_basics_relationsAverageMutualInformation_c}:
\begin{subequations}
	\begin{align}
	I(x,y) &= S[x] + S[y] - S[x,y]\label{eqn_basics_relationsAverageMutualInformation_a}\\
	&= S[x] - S[x|y]\label{eqn_basics_relationsAverageMutualInformation_b}\\
	&= S[y] - S[y|x]\label{eqn_basics_relationsAverageMutualInformation_c}.
	\end{align}
\end{subequations}

Comparing Eqs.~\eqref{eqn_basics_defKLDistance} and \eqref{eqn_basics_defAverageMutualInformation}, we see that the average mutual information is the Kullback-Leibler distance between the joint distribution $p(x,y)$ and the product of the marginal distributions $p(x)\,p(y)$, which makes its interpretation as a correlation measure apparent.

\section{Differential entropy}
The concept of entropy can be extended to continuous random variables with a probability density $\rho(x)$\footnote{Here, we use the greek letter $\rho$ to avoid confusion.}:
\begin{align}\label{eqn_basics_defDifferentialEntropy}
s(x) := -\ln{\rho(x)}.
\end{align} 
It is then called \emph{differential entropy}. Correspondingly, the average differential entropy is defined as:
\begin{align}\label{eqn_basics_defAvgDifferentialEntropy}
S[x] := -\int dx\, \rho(x)\ln{\rho(x)}.
\end{align}
Importantly, the differential entropy can become negative, since the probability density $\rho(x)$ is not bounded by one. Nonetheless, all the previously discussed quantities straightforwardly generalize to continuous variables by changing sums to integrals.

Some comments on the more nasty peculiarities of the definition in Eq.~\eqref{eqn_basics_defDifferentialEntropy} are in order. Firstly, the differential entropy is not the result of an infinitely fine-grained discrete entropy. Let's say we sample the continuous variable $x$ in discrete bins $x_i$ of width $\Delta$. Then, $\rho(x_i)\,\Delta = p(x_i)$, where $\rho(x_i)$ on the LHS is a probability density and $p(x_i)$ on the RHS is a probability. The discrete average entropy $S_\Delta$ is then given by:
\begin{align}
S_\Delta[x_i] &= -\sum\limits_i p(x_i)\ln{p(x_i)} = -\sum\limits_i \rho(x_i) \Delta \ln{\rho(x_i)} - \sum\limits_i \rho(x_i)\Delta\ln\Delta\\
&\rightarrow -\int dx\,\rho(x)\ln{\rho(x)} - \ln{\Delta},
\end{align}
as $\Delta \rightarrow 0$. It thus differs from the differential entropy by an infinite offset.

Secondly, a related issue concerns a change of variables, in the easiest case this is a simple scaling: $y := a x$. One would expect the entropy not to change, however, since $\rho_y(y) = \rho_x(x/a)/|a|$,
\begin{align}
S[y] &= -\int dy\, \rho_y(y)\ln{\rho_y(y)} = -\int dx\, \rho_x(x)\ln{\left(\rho_x(x)/|a|\right)}\\
&= S[x] + \ln|a|.
\end{align}
For a physicist, these problems are immediately apparent from asking what units enter the logarithm. Since the probability density $\rho(x)$ has the inverse units of $x$, it is clear that differential entropy needs to be defined relative to a \emph{unit volume}. Then, in Shannon's words, ``the scale of measurements sets an arbitrary zero corresponding to a uniform distribution over [this] unit volume"~\cite{Shannon1948}.

In contrast, the Kullback-Leibler distance defined analogous to Eq.~\eqref{eqn_basics_defKLDistance} does not suffer from the same shortcomings. Consequently, the mutual information is also invariant under coordinate transformations. Luckily, we will not be concerned by the aforementioned problems even when considering differential entropies proper, as we will only study entropy \emph{differences} in which the potential offsets cancel.

\section{Stochastic entropy production}\label{sec_basics_stochasticEP}
In this section we will combine the content of this and the previous chapter and assign entropies to stochastic processes. The purpose is to show the general new quantities and relations resulting from such a procedure. We will mostly follow the strategy employed by Seifert in Ref.~\cite{Seifert2012}. In the next chapter we will give the thermodynamic interpretation in the context in which most of these relations have originally been discovered.

In the following we require a \emph{dual} or \emph{conjugated} process for every stochastic process. We will denote the conjugated process with an overbar: $\bar{x}$. The conjugation shall fulfill $\bar{\bar{x}} = x$. The most prominent choice is time reversal, i.e., the time is mirrored in all distribution functions: $\bar{t} := T-t$, where $T$ is the final time.

Consider a given trajectory $x(\cdot)$ of a stochastic process. The following derivations hold mostly unmodified for discrete and continuous-time processes. While the original process results in a certain path probability $p[x(\cdot)]$, the conjugated process has a different path probability $\bar{p}[x(\cdot)]$.

In the spirit of the definition of entropy in Eq.~\eqref{eqn_basics_defEntropy}, we define the \emph{trajectory entropy production}:
\begin{align}\label{eqn_basics_defTrajectoryEP}
\sigma[x(\cdot)] := \ln\frac{p[x(\cdot)]}{\bar{p}[\bar{x}(\cdot)]},
\end{align}
where $\bar{x}(\cdot)$ denotes the conjugated trajectory. It is required that if a given trajectory occurs in the original process, it must also occur with non-zero probability in the conjugated process, which somewhat restricts the choice of dual dynamics. It can pose problems even for time-reversed dynamics, e.g, for reset-processes~\cite{Evans2011} in which a stochastic process is randomly reset, the reverse of which never occurs~\cite{Murashita2014,Pal2017}.

Equation~\eqref{eqn_basics_defTrajectoryEP} reveals what \emph{entropy production} means in this context: It is a measure of how typical a given trajectory $x(\cdot)$ is for the original process compared to the conjugated one. The bigger the entropy production, the more certain one can be that a trajectory was generated by the original process. A negative entropy production indicates that the given trajectory is more likely to have been generated by the conjugated process, which is also reflected in the fact that entropy production has odd parity under conjugation:
\begin{align}
\bar{\sigma}[\bar{x}(\cdot)] = \ln\frac{\bar{p}[\bar{x}(\cdot)]}{p[x(\cdot)]} = - \sigma[x(\cdot)].
\end{align}

\subsection{Fluctuation theorems}\label{sec_basics_fluctuationTheorems}
The entropy production fulfills some interesting symmetry relations which are called \emph{fluctuation theorems}. The first fluctuation theorem was discovered by Evans \emph{et al.}~\cite{Evans1993,Evans1994} for shear-driven fluids and later proven by Gallavotti and Cohen~\cite{Gallavotti1995, Gallavotti1995a}, Kurchan~\cite{Kurchan1998} and Lebowitz and Spohn~\cite{Lebowitz1999} for different kinds of driven dynamics. Related theorems, called \emph{nonequilibrium} work relations, were found by Jarzynski~\cite{Jarzynski1997,Jarzynski1997a} and Crooks~\cite{Crooks1999,Crooks2000}. We will now derive prototype fluctuation theorems from our general definition of entropy production in Eq.~\eqref{eqn_basics_defTrajectoryEP}.

With $x(\cdot)$ also $\sigma[x(\cdot)]$ is a random variable. Consider therefore the probability $p(\sigma)$ of observing a given entropy production $\sigma$ which follows from a transformation of probabilities:
\begin{align}
p(\sigma) = \int \mathcal{D}x(\cdot)\, p\left[ x(\cdot) \right]\,\delta\left(\sigma - \ln\frac{p[x(\cdot)]}{\bar{p}[\bar{x}(\cdot)]} \right).
\end{align}

Let us denote by $\bar{p}(\sigma)$ the probability to find an entropy production $\sigma$ in the conjugated process. Then, we obtain
\begin{align}
\bar{p}(-\sigma) &= \int \mathcal{D}\bar{x}(\cdot)\, \bar{p}[\bar{x}(\cdot)]\,\delta\left(\sigma + \ln\frac{\bar{p}[\bar{x}(\cdot)]}{p[x(\cdot)]}\right)\\
&= \int \mathcal{D}\bar{x}(\cdot)\, \exp{\left\{-\ln{\frac{p[x(\cdot)]}{\bar{p}[\bar{x}(\cdot)]}}\right\}} p[x(\cdot)]\,\delta\left(\sigma - \ln\frac{p[x(\cdot)]}{\bar{p}[\bar{x}(\cdot)]}\right)\label{eqn_basics_derivationCrooks1}\\
&= e^{-\sigma}\,\int \mathcal{D}x(\cdot)\, p\left[ x(\cdot) \right]\,\delta\left(\sigma - \ln\frac{p[x(\cdot)]}{\bar{p}[\bar{x}(\cdot)]} \right)\label{eqn_basics_derivationCrooks2}\\
&= e^{-\sigma}\,p(\sigma), \label{eqn_basics_derivationCrooks3}
\end{align}
where we used the properties of the delta function and changed the integration variable from $\bar{x}(\cdot)$ to $x(\cdot)$ in the third line. Equation~\eqref{eqn_basics_derivationCrooks3} is the \emph{Crooks-type fluctuation theorem}\footnote{Some authors use the term \emph{detailed fluctuation theorem} which we reserve for a special variant of the \emph{Crooks-type fluctuation theorem} (see Sec.~\ref{sec_basics_detailedFT}).}:
\begin{align}\label{eqn_basics_CrooksTypeFT}
\ln\frac{p(\sigma)}{\bar{p}(-\sigma)} = \sigma,
\end{align}
so named because of its similarity to the theorem found by Crooks~\cite{Crooks1999}.

It immediately implies the weaker \emph{integral fluctuation theorem}:
\begin{align}\label{eqn_basics_IFT}
\left\langle e^{-\sigma} \right\rangle = 1,
\end{align}
since
\begin{align}
\int d\sigma \, p(\sigma)\, e^{-\sigma} = \int d\sigma \, \bar{p}(-\sigma) = 1.
\end{align}
The integral fluctuation theorem has the property that it can be evaluated without having to know the statistics of the conjugated process. This makes it a useful tool for situations in which one does not have access to the conjugated dynamics.

As mentioned previously, the fluctuation theorems are interesting symmetry relations constraining the statistics of the entropy production $\sigma$. One example is a lower bound on the frequency of negative-entropy-production trajectories~\cite{Jarzynski2008}:
\begin{align}
p(\sigma \leq -\alpha) \leq e^{-\alpha},
\end{align}
where $\alpha > 0$. It can be proven as follows:
\begin{align}
p(\sigma \leq -\alpha) = \int\limits_{-\infty}^{-\alpha} d\sigma\, p(\sigma) \leq \int\limits_{-\infty}^{-\alpha} d\sigma\, p(\sigma)\,e^{-\alpha -\sigma} \leq e^{-\alpha} \int\limits_{-\infty}^{\infty} d\sigma\, p(\sigma)\,e^{-\sigma} = e^{-\alpha}.
\end{align}
This means that negative-entropy trajectories are exponentially unlikely. This effectively prohibits large violations of the second law.

Even though fluctuation theorems originate from the statistical physics of nonequilibrium processes for small-scale systems, from this very general derivation it is clear that the concept of entropy production and fluctuation theorems is applicable beyond small-scale thermodynamics. Examples are found in Bayesian statistics~\cite{Ahlers2008,Favaro2015}, in gambling~\cite{Hirono2015,Vinkler2016}, and in the Markov analysis of turbulent flows~\cite{Nickelsen2013,Reinke2016} and rogue ocean waves~\cite{Hadjihosseini2014}, where the identification of extreme rogue waves with negative entropies promises a new way to estimate the frequency of their occurrence~\cite{Hadjihoseini2018}.

\chapter{Stochastic thermodynamics}\label{chap_stochasticThermodynamics}
We will deviate slightly from the presentation style of the previous sections and use the following section to first show how small-scale thermodynamics differs from its classical, macroscopic counterpart. In the following, we will use the insight into stochastic processes and information theory from the previous chapters to derive the key results of stochastic thermodynamics. Most of the material presented can be found in the pedagocial reviews by Jarzynski~\cite{Jarzynski2011} and Van den Broeck~\cite{VandenBroeck2013} and the comprehensive one by Seifert~\cite{Seifert2012}. Our presentation's emphasis on stochastic dynamics is most in line with the latter one. More recent reviews include Refs.~\cite{Seifert2018,Seifert2019} by the same author.

One very appealing aspect of the theory is the area of \emph{information thermodynamics}. It has been postponed until Chap.~\ref{chap_informationThermodynamics}, since it fits nicely within the framework of interacting subsystems which simplifies the discussion.

\section{Macroscopic and microscopic}\label{sec_basics_microscopicMacroscopic}
\begin{figure}[ht]
	\centering
	\begin{minipage}{0.5\linewidth}
		\centering
		\includegraphics[width = 0.8 \linewidth]{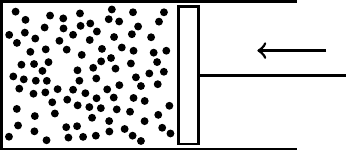}\\
		\vspace*{8pt}
		\includegraphics[width = 1 \linewidth]{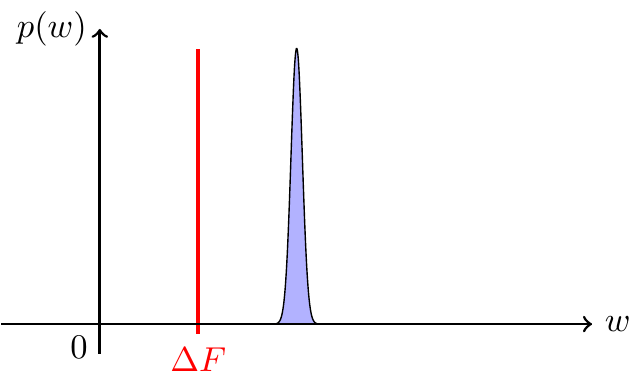}
	\end{minipage}%
	\begin{minipage}{0.5\linewidth}
		\centering
		\includegraphics[width = 0.8 \linewidth]{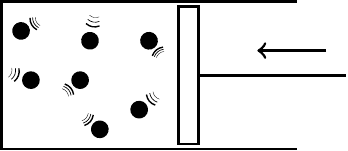}\\
		\vspace*{8pt}
		\includegraphics[width = 1 \linewidth]{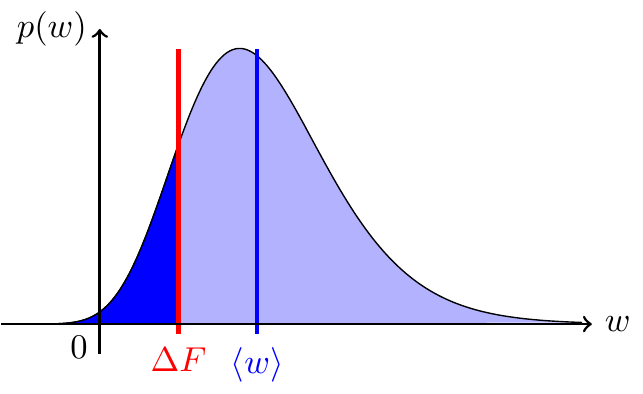}
	\end{minipage}%
	\caption{Top: Irreversible isothermal compression of an ideal gas of a large number (say $10^{23}$) of particles (left) and a small number (7) of particles (right): Bottom: Exemplary distribution of work values $w$ compared with the free energy difference $\Delta F$. The dark blue region indicates seeming violations of the second law.}
	\label{fig_basics_compressionExample}
\end{figure}

Classical thermodynamics deals with exchanges of energy and matter between macroscopic systems. Due to the large system size, fluctuations of interesting quantities around their averages are comparatively small and can therefore safely be neglected. A classic thought experiment is the compression of a macroscopic amount of ideal gas (about $10^{23}$ particles) in a cylinder with adiabatic walls as depicted in Fig.~\ref{fig_basics_compressionExample}. During the compression we need to perform work on the gas molecules. The second law tells us that this work is at least as big as the free energy difference between the uncompressed and compressed state, and performing the process so slow that the gas inside the cylinder remains in equilibrium throughout will saturate the bound.

Repeating the experiment a few times, we would find that we need the same amount of work (at least within the tolerance of our measurement) in every repetition. If we, however, go to very small system sizes, say seven particles, the situation changes dramatically. The work needed to compress the gas will be of the order of a few $k_{\rm B} T$ and (ignoring the difficulty to measure such small energies) we would see significant fluctuations from one realization to the next. This is because, for such few particles, their individual velocities become important: Sometimes we get lucky and a large proportion of the molecules will move away from the piston, which means almost no work is needed, and at other times most of the particles will move towards the piston, which necessitates a large amount of work. Once in a while we might even measure a work value that is less than the free energy difference, thereby seemingly violating the second law of thermodynamics!

The fact that it is in principle possible to violate the second law is not terribly surprising, since there exist sound microscopic trajectories of the particles that lead to such small work values. However, for large system sizes they are so rare that we can never observe them while they are more frequent for small system sizes. We should thus reformulate the second law from a statement holding for every trajectory to one that addresses \emph{averages}: We cannot beat the second law \emph{on average}. For our isothermal compression example it would therefore read:
\begin{align}
\left\langle w \right\rangle \geq \Delta F,
\end{align}
where $w$ is the work done on the system, the angled brackets indicate an average over many realizations, and $\Delta F$ is the free energy difference.

Stochastic thermodynamics is the extension of thermodynamics to small scales. It has become necessary due to advances in experimental equipment for measuring small amounts of energy and manipulating small-scale objects, which allows the study of microscopic (e.g., biological) machinery. It therefore became possible to study microscopic machines like molecular motors in the same way as macroscopic machines were studied in the nineteenth century.

The systems considered in the framework of stochastic thermodynamics are colloids, molecular motors, and biopolymers which are heavily influenced by thermal fluctuations. Stochastic processes thus provide a good model for their dynamics.  The  embedding aqueous solution in biological systems provides a thermal reservoir with a well defined temperature. The distinction between the environment degrees of freedom and the system degrees of freedom is achieved by a time-scale separation between the two: The environment degrees of freedom evolve on much faster time-scales than the system degrees of freedom. This provides Markovian dynamics for the system as it ensures that the environment is in a conditional stationary state with respect to the system state.

All biological machinery must be in a \emph{nonequilibrium} state. For small systems this fact becomes very obvious. The \emph{energy currency} of biological systems is the chemical compound ATP (\emph{Adenosine Triposphate}). Upon hydrolysis at body temperature to ADP it very quickly releases energy of the order of 10 $k_{\rm B}T$. If thermal fluctuations are visible for a given system, an energy input this large over a short time span will surely drive it out of equilibrium.

Three distinct nonequilibrium situations can be differentiated. Firstly, the dynamics can be \emph{driven} by an explicit dependence on time like in the example of a compression of gas above. Secondly, an ensemble of systems can be prepared in an initial nonequilibrium state which results in a subsequent \emph{relaxation} towards equilibrium. Thirdly, a system may be in a \emph{nonequilibrium steady state} where the dynamics are time-independent but there exist constant cyclical probability currents in the state space. 

Apart from external time-dependent control parameters or external flows, nonequilibrium can also be reached by coupling to several thermal and/or chemical reservoirs. In those cases care has to be taken to correctly identify the energy flows and attribute them to the correct reservoir. We will mostly restrict ourselves to \emph{isothermal} situations such that the thermodynamic potential appearing in the relations we derive is the (Helmholtz) free energy $F$.

In the following we will explore the nonequilibrium thermodynamics of these stochastic systems by using the frameworks of stochastic processes and information theory laid out in the first chapters. To make the discussion simpler, let us assign a \emph{potential energy} $V(x)$ to each state. That way we may keep track of changes in internal energy as the system progresses through its state space. Secondly, instead of an explicit time dependence, we use a \emph{work parameter} $\lambda$ (think of the position of the piston in the example above) which is time-dependent. In so doing we can identify a potential function $V(x,\lambda)$.

\section{Stochastic energetics for Langevin equation}\label{sec_basics_stochasticEnergetics}
The key to a thermodynamic interpretation of stochastic processes is to identify the usual thermodynamic quantities like heat and work. The whole field of stochastic thermodynamics would be very different \emph{if we were able to measure tiny amounts of energy directly} (of the order of a few $k_{\rm B} T$). Since we cannot, we are forced to deduce energy exchanges from the \emph{stochastic dynamics} of the systems under study. This enables one, e.g., to use one small system as a calorimeter for another. In short: we need a thermodynamic interpretation of the stochastic processes that govern small-scale systems.

Before considering a general Markov process, we will first turn to the simpler thermodynamic interpretation for diffusive dynamics described by an overdamped Langevin equation. This was accomplished by Sekimoto~\cite{Sekimoto1998} (see also the book by the same author~\cite{Sekimoto2010} with many thermodynamic applications of the Langevin equation). 

Consider the position $x$ of a colloidal particle diffusing in a potential $V(x,\lambda)$. Its dynamics shall be described by an overdamped Langevin equation of the form in Eq.~\eqref{eqn_basics_overdampedLangevinEquation}:
\begin{align}\label{eqn_basics_overdampedLangevinEquatuationPotential}
\dot{x} = -\nu \pd{x}V(x,\lambda(t)) + \sqrt{2D}\,\xi(t).
\end{align}

The principal insight is that the internal energy of a diffusing colloid is given by its position in the potential landscape. Changes in internal energy can have two causes: Firstly, the position of the particle can change by an amount $dx$ through thermal fluctuations. Since this displacement is induced by the thermal reservoir, these changes are identified as \emph{heat}:
\begin{align}
dq := \pd{x}V(x,\lambda)\, dx.
\end{align}

Secondly, the internal energy can change by modifying the potential landscape through the explicit dependence on $\lambda$ in $V(x,\lambda)$. This must be accomplished from the \emph{outside} (e.g., by an experimenter) and the energy must be externally supplied or extracted. Therefore, these changes represent \emph{work done on the system}:
\begin{align}\label{eqn_basics_workLangevinBasic}
dw := \pd{\lambda}V(x,\lambda)\, d\lambda .
\end{align}

\begin{figure}[ht]
	\centering
	\includegraphics[width =0.33 \linewidth]{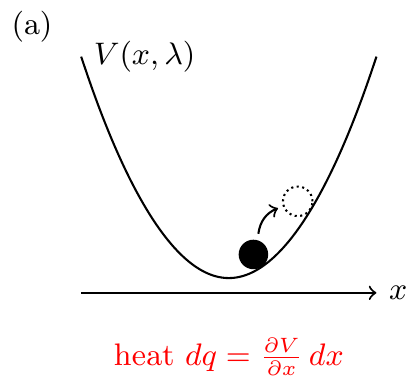}\includegraphics[width =0.33 \linewidth]{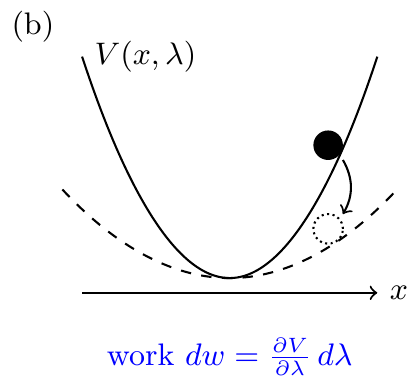}\includegraphics[width =0.33 \linewidth]{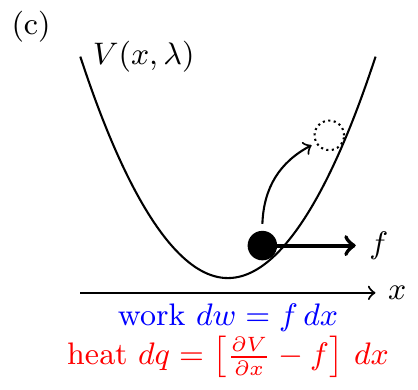}
	\caption{Illustration of the three contributions to the energy balance of a colloidal particle: (a) Fluctuations in position change the potential energy. They are driven by the thermal environment and thus contribute to the \emph{heat}. (b) Changing the potential energy function directly requires one to do \emph{work} on the system. (c) A non-conservative force also performs \emph{work} on the system but if the injected energy surpasses the changes in potential energy, the excess energy must be dissipated as \emph{heat}.}
	\label{fig_basics_Langevin_stochasticEnergetics}
\end{figure}

In some situations there is also an \emph{external `non-conservative'} force $f(x,t)$ acting on the system. Usually it is clear from the context that this force does not result from a potential or the thermal environment. An example is an \emph{active colloid} (see Sec.~\ref{sec_hid_activeBrownianMotion}) that has a constant self-propulsion force pushing it along. In this case the work done by this force must be added to the total work:
\begin{align}
dw = \pd{\lambda}V(x,\lambda)\, d\lambda + f(x,t) \, dx.
\end{align}
However, if the work done by this force exceeds the corresponding change in potential energy, the excess energy must be dissipated as heat. From the point of view of the system this is an energy loss to the environment. Thus, the total dissipated heat then reads:
\begin{align}
dq &= \left[ \pd{x}V(x,\lambda(t)) - f(x,t)\right] \, dx\label{eqn_basics_heatLangevinForce}\\
&= -F(x,t) \, dx,
\end{align}
where we identified the \emph{total force} $F(x,t) := - \pd{x}V(x,\lambda(t)) +f(x,t)$ acting on the system. Figure~\ref{fig_basics_Langevin_stochasticEnergetics} illustrates the three contributions to the energy balance.

Obviously, these relations can also be integrated to obtain the heat $q[x(\cdot)]$ exchanged and work $w[x(\cdot)]$ done for \emph{individual trajectories} of length $T$:
\begin{subequations}
	\begin{align}
	q[x(\cdot)] &= \int\limits_0^T dt\, \dot{x} \circ \left[\pd{x}V(x,\lambda(t))-f(x,t)\right]\label{eqn_basics_SekimotoTrajectoryHeat_a} \\
	&= -\int\limits_0^T dt\,\dot{x} \circ F(x,t)  \label{eqn_basics_SekimotoTrajectoryHeat_b}\\
	w[x(\cdot)] &= \int\limits_0^T dt\, \left[\,\dot{\lambda}\pd{\lambda}V(x,\lambda(t))+ \dot{x} \circ f(x,t) \right],\label{eqn_basics_SekimotoTrajectoryWork}
	\end{align}
\end{subequations}
where the stochastic integral is to be interpreted in the Stratonovich sense, according to our convention, and $\dot{\lambda}$ is the time-derivative of the work parameter.

We thus get the \emph{first law} for individual trajectories which expresses the \emph{energy balance} for the stochastic system:
\begin{align}
q[x(\cdot)] + w[x(\cdot)] &= \int\limits_0^T dt\, \left[\dot{x} \circ \pd{x}V(x,\lambda(t))+\dot{\lambda}\,\pd{\lambda}V(x,\lambda(t)) \right]\\
&= V(x(T),\lambda(T)) - V(x(0),\lambda(0)).
\end{align}

Turning to systems coupled to multiple reservoirs, the identification of heat and work is more intricate. One difficulty lies in the fact that state transitions $x~\rightarrow~x'$ can be caused by \emph{different mechanisms} pertaining to the different reservoirs. One particularly hopeless situation is given by a diffusive process described by a Langevin equation with several noise sources, e.g., a simultaneous coupling to two heat baths. Based on the model we can make predictions about the average heat that is exchanged between the two reservoirs and the system. However, it is impossible to assign heat flows to an individual trajectory. To which bath should we attribute a change in internal energy?

\section{Thermodynamics of master equation systems}
We discuss the setting in the context of a generic master equation encountered before in Sec.~\ref{sec_basics_masterEq}. We assume that we can assign a reservoir to each individual transition. Note that the following formalism applies to all Markov processes, not only jump processes, although it is mostly used in that context.

\subsection{Detailed balance}\label{sec_basics_detailedBalance}
\emph{Equilibrium and nonequilibrium} are defined by means of the \emph{probability fluxes}:
\begin{align}
j(x \to x';t) := W(x'|x;\lambda(t))\,p(x,t) - W(x|x';\lambda(t))\,p(x',t).
\end{align}
In the same way as macroscopic nonequilibrium is characterized by an energy or a matter flux, we use a nonvanishing probability flux as a definition of microscopic nonequilibrium. This is also used in experiments to infer whether some process is an equilibrium process or not (see, e.g., Ref.~\cite{Battle2016}). Given a set of transition rates $W(x'|x;\lambda)$, equilibrium is thus characterized by \emph{detailed balance}:
\begin{align}\label{eqn_basics_detailedBalance}
W(x'|x;\lambda)\,p_{\rm eq}(x;\lambda) &= W(x|x';\lambda)\,p_{\rm eq}(x';\lambda).
\end{align}
For systems coupled to a heat bath at temperature $T$, the equilibrium distribution is the Boltzmann distribution,
\begin{align}\label{eqn_basics_BoltzmannDist}
p_{\rm eq}(x;\lambda) = \exp{\left(-\frac{V(x;\lambda)-F(\lambda)}{\kB T }\right)},
\end{align}
where $F(\lambda)$ is the free energy. Detailed balance implies the following relation for the transition rates:
\begin{align}\label{eqn_basics_relationTransitionRates}
\ln\frac{W(x|x';\lambda)}{W(x'|x;\lambda)} &= \ln\frac{p_{\rm eq}(x;\lambda)}{p_{\rm eq}(x';\lambda)} = \frac{V(x',\lambda)-V(x,\lambda)}{k_{\rm B} T}.
\end{align}

Thus, just like in the previous section, if a transition $x \rightarrow x'$ changes the internal energy and is mediated by the heat bath, it should be identified as heat. If detailed balance holds, this change is given by
\begin{align}\label{eqn_basics_identificationHeatMasterEquationPreliminary}
\Delta q(x\to x';\lambda) := k_{\rm B} T\, \ln\frac{W(x|x';\lambda)}{W(x'|x;\lambda)}.
\end{align}

The identification of work is simpler. By changing the work parameter $\lambda$, one changes the potential energy and thus does work on the system:
\begin{align}\label{eqn_basics_identificationWorkMasterEquation}
\Delta w(\lambda \to \lambda';x) := V(x,\lambda') - V(x,\lambda).
\end{align}

\subsection{Broken detailed balance and multiple reservoirs}\label{sec_basics_brokenDetailedBalance}
We will now turn to situations in which detailed balance does not hold, i.e., there are probability currents in the system, even in the stationary state:
\begin{align}
W(x'|x;\lambda)\,p_{\rm st}(x;\lambda) &\neq W(x|x';\lambda)\,p_{\rm st}(x';\lambda).
\end{align}

The key insight is that this is only possible if the system is coupled to several reservoirs that are not in equilibrium with each other. Otherwise, if it were only interacting with one reservoir, the system would reach equilibrium. Note that the term \emph{reservoir} also includes work reservoirs and chemical reservoirs, etc. We already encountered one example: the non-conservative force $f(x,t)$ in Sec.~\eqref{sec_basics_stochasticEnergetics}. It represents a work reservoir that is breaking detailed balance and constantly delivering energy to the system, which it must dissipate to the heat bath.

Let us assume that transitions between states $x'$ and $x$ can have several mechanisms indexed by $\nu$ related to the different reservoirs. The transition rate $W(x|x';\lambda)$ is therefore the sum of all individual transition rates $W^{(\nu)}(x|x';\lambda)$:
\begin{align}
W(x|x';\lambda) = \sum\limits_\nu W^{(\nu)}(x|x';\lambda).
\end{align}

The different reservoirs try to impose different equilibria $p_{\rm eq}^{(\nu)}(x;\lambda)$ on the system which is encoded in the \emph{local detailed balance} relation:
\begin{align}\label{eqn_basics_localDetailedBalance}
W^{(\nu)}(x'|x;\lambda)\,p_{\rm eq}^{(\nu)}(x;\lambda) &= W^{(\nu)}(x|x';\lambda)\,p^{(\nu)}_{\rm eq}(x';\lambda).
\end{align}
This means that if the system were only coupled to one of the reservoirs, it would relax towards equilibrium with that reservoir. Since it is not, it can only relax to a nonequilibrium stationary state. 

The local detailed balance relation can also be justified by assuming that the system is repeatedly interacting with a memoryless reservoir and that this interaction is micro-reversible (see Sec.~\ref{sec_int_detailedBalanceFromRepeatedInt}).

Using Eq.~\eqref{eqn_basics_localDetailedBalance}, we can identify the heat that is exchanged between the system and the reservoirs. Let us assume that there are \emph{heat and particle} reservoirs; we thus also allow particle transport. The equilibrium distribution is then dictated by the grand canonical ensemble:
\begin{align}\label{eqn_basics_eqGrandCanonical}
p_{\rm eq}(x;\lambda) = \exp{\left(-\frac{V(x,\lambda) - \mu\,n(x)-G(\lambda,\mu)}{\kB T }\right)},
\end{align}
where
\begin{align}
G(\lambda,\mu) := -\kB T\ln\left[\int dx\,\exp{\left(-\frac{V(x,\lambda) - \mu\,n(x)}{\kB T}\right)}\right]
\end{align}
is the associated thermodynamic  potential, $\mu$ is the chemical potential, and $n(x)$ is the particle number associated to system state $x$.

Each transition $x \rightarrow x'$ via the pathway $\nu$ is now accompanied by an exchange of \emph{chemical work} done on the system by the particle reservoir,
\begin{align}\label{eqn_basics_chemicalWork}
\Delta w^{(\nu)}(x\rightarrow x') := \mu^{(\nu)} \left[ n(x') - n(x) \right],
\end{align}
and by an exchanged \emph{heat} flowing from the heat reservoir into the system,
\begin{align}\label{eqn_basics_heatWithChemicalWork}
\Delta q^{(\nu)}(x\to x') := V(x',\lambda) - V(x,\lambda) - \Delta w^{(\nu)}(x\rightarrow x').
\end{align}
Notice the similarity to Eq.~\eqref{eqn_basics_heatLangevinForce}: The heat entering the system is given by the change in potential energy minus the chemical work. The energy done by the `non-conservative' chemical work in excess of the potential energy difference is dissipated as heat.

With Eq.~\eqref{eqn_basics_localDetailedBalance} we thus find:
\begin{align}
\Delta q^{(\nu)}(x\to x') &= V(x',\lambda) - V(x,\lambda) - \mu^{(\nu)} \left[ n(x') - n(x) \right]\\
&= \ln\frac{p_{\rm eq}^{(\nu)}(x;\lambda)}{p_{\rm eq}^{(\nu)}(x';\lambda)}\\
&=  k_{\rm B} T\, \ln\frac{W^{(\nu)}(x|x';\lambda)}{W^{(\nu)}(x'|x;\lambda)} \label{eqn_basics_heatMasterEquationFinal},
\end{align}
which is the same as Eq.~\eqref{eqn_basics_identificationHeatMasterEquationPreliminary} for only one reservoir.

\begin{figure}[ht]
	\centering
	\begin{tabular}{c|c|c}
		\includegraphics[width = 0.31\linewidth]{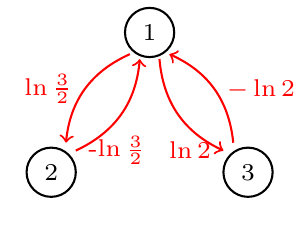} &
		\includegraphics[width = 0.31\linewidth]{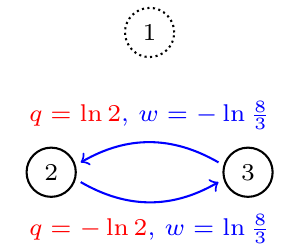} &
		\includegraphics[width = 0.31\linewidth]{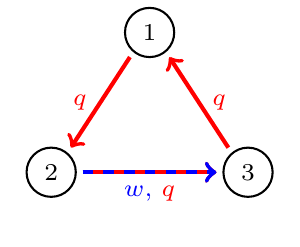} \\
		\includegraphics[width = .31 \linewidth]{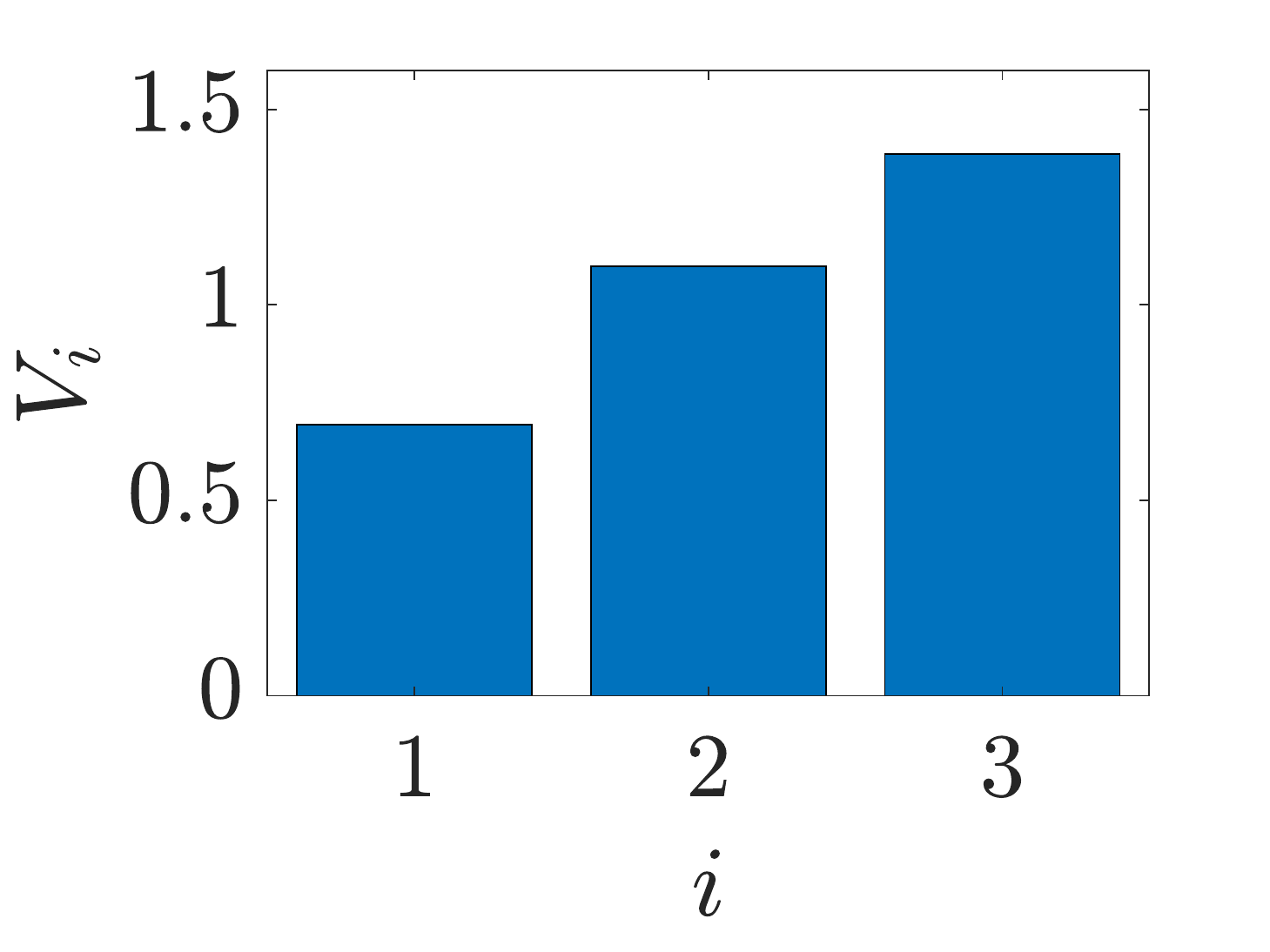} &
		\includegraphics[width = .31\linewidth]{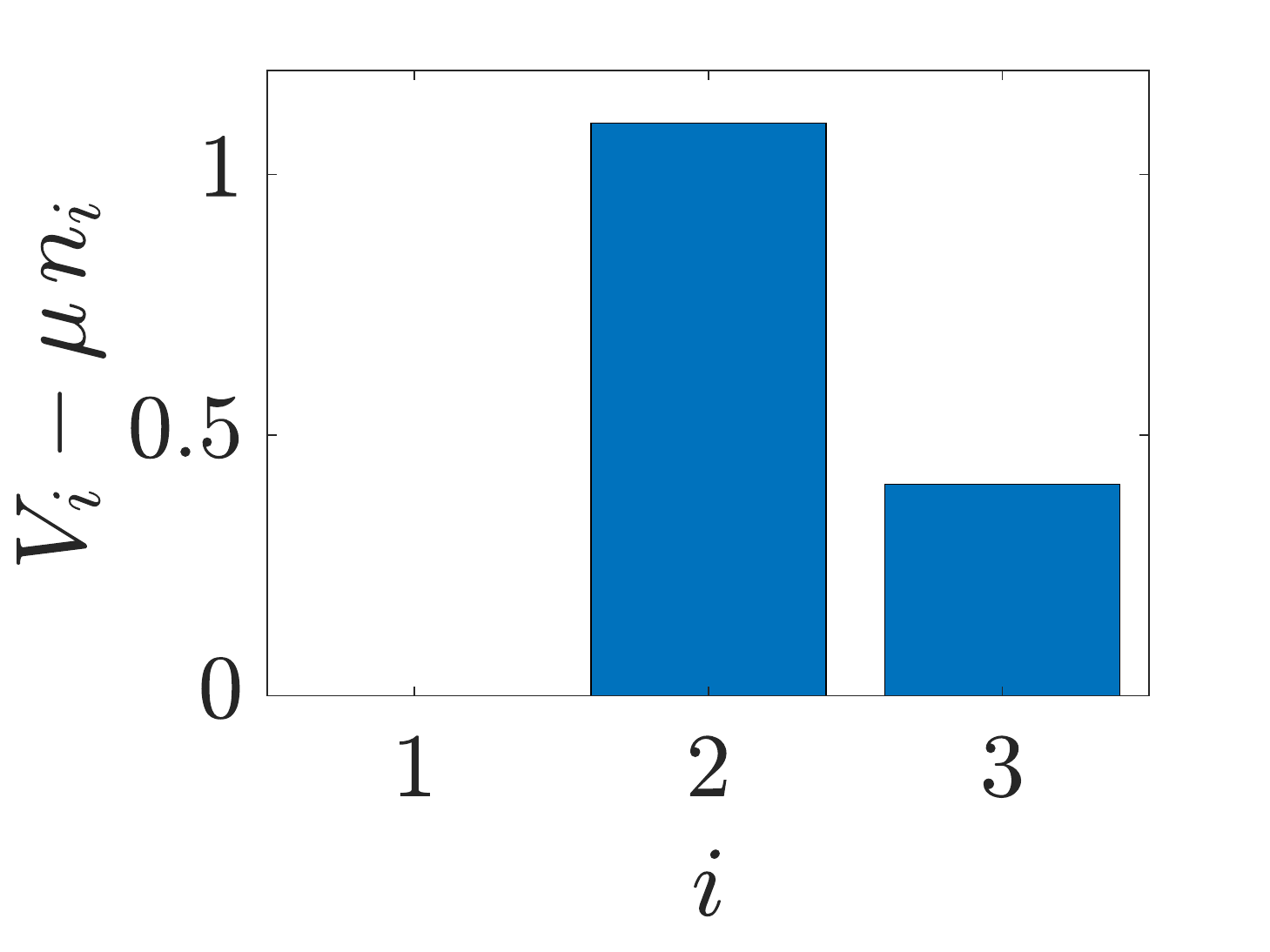} & \\
		\includegraphics[width =.31 \linewidth]{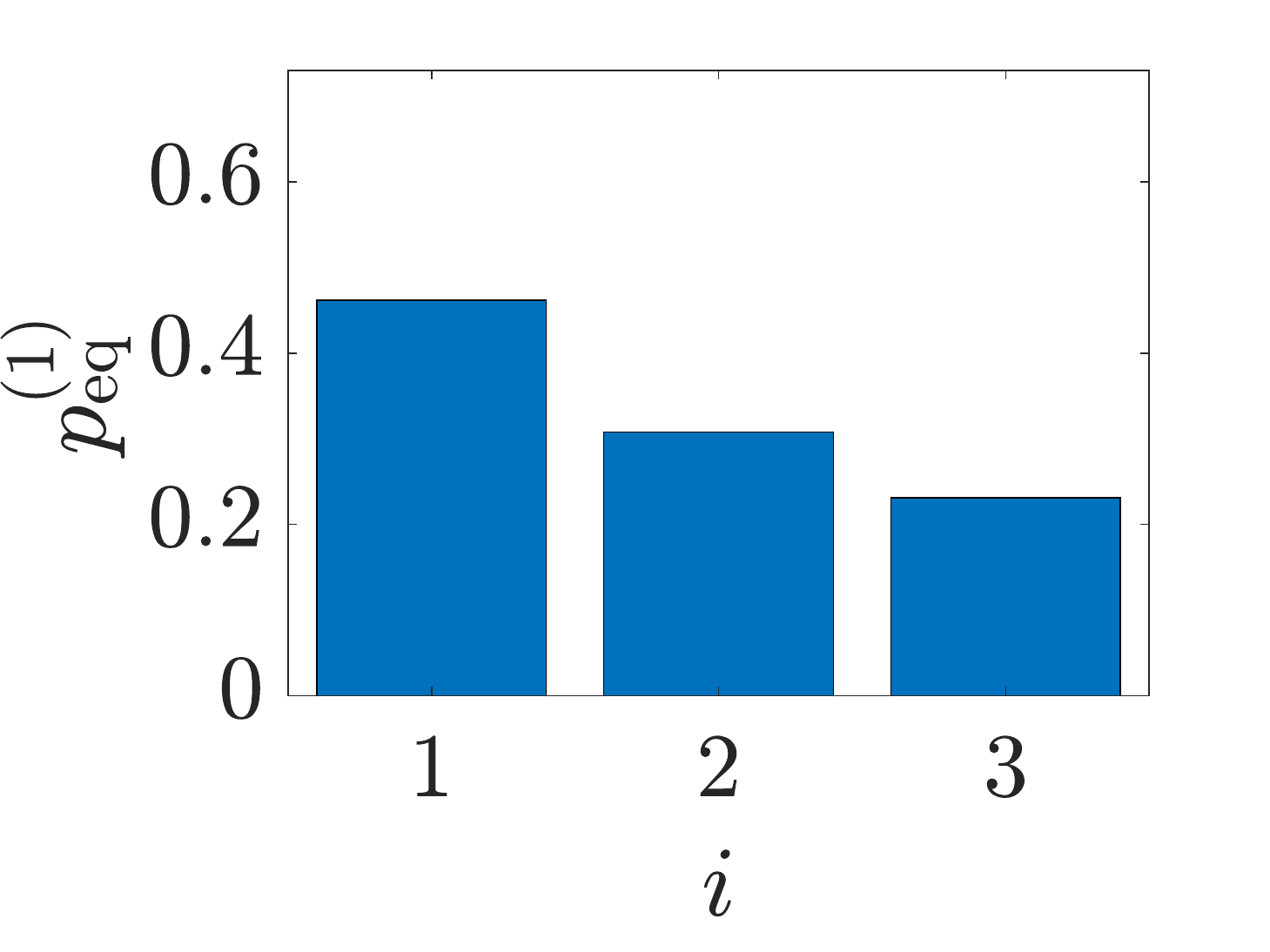} &
		\includegraphics[width =.31 \linewidth]{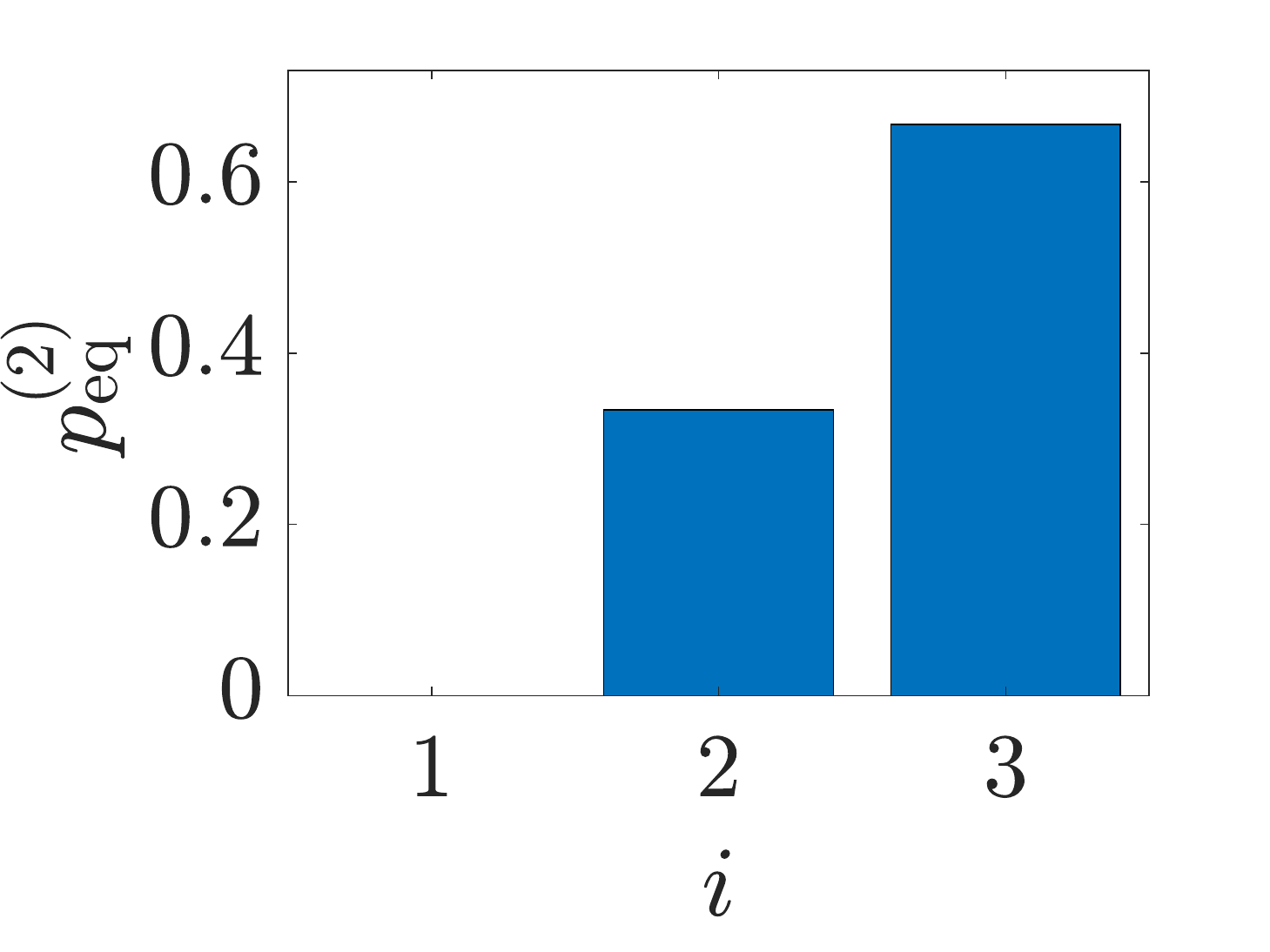} &
		\includegraphics[width =.31 \linewidth]{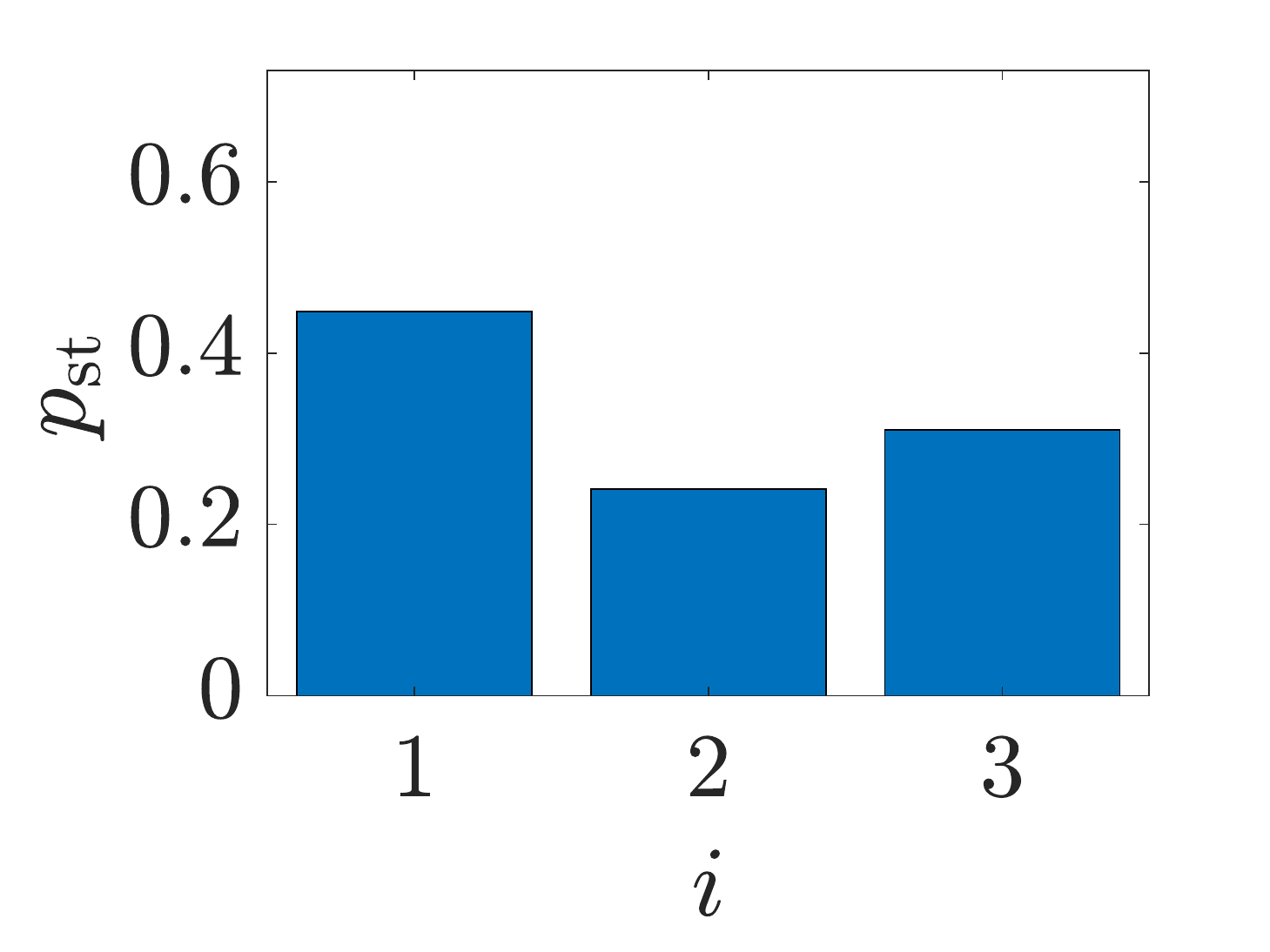}
	\end{tabular}
	\caption{Left column: Example jump process from Sec.~\ref{sec_basics_exampleJumpProcess} now interpreted in the thermodynamic sense with $\kB T = 1$. It is augmented with the heat exchanges for all transitions (top), a possible potential landscape (middle), and the corresponding equilibrium distribution (bottom). Center column: Secondary process mediated by coupling to a particle reservoir resulting in a grand canonical potential landscape (middle) that favors state $3$ over state $2$ (bottom) against the potential gradient. Right column: Resulting combined process that constantly takes up chemical work and dissipates heat (top) in a nonequilibrium steady state (bottom).}
	\label{fig_basics_heatExampleJumpProcess}
\end{figure}

Let us consider a simple example of a system displaying broken detailed balance. Take the jump process from Sec.~\ref{sec_basics_exampleJumpProcess}. It is easy to check that the transition rate matrix
\begin{align}
\tilde{W}^{(1)}_{ij} = \begin{pmatrix}
-3 & 3 & 2\\
2 & -3 & 0\\
1 & 0 & -2
\end{pmatrix},
\end{align}
already given in Eq.~\eqref{eqn_basics_exampleTransitionRates} together with the steady-state probabilities in Eq.~\eqref{eqn_basics_exampleSteadyStateProbs}, fulfills detailed balance. Imagine the transitions result from coupling to a heat bath at temperature $\kB T := 1$. One can then find a potential $V_i$ which generates these dynamics. One possibility is given by $V_1 = \ln{2}$, $V_2 = \ln{3}$, and $V_3 = \ln{4}$. Equation~\eqref{eqn_basics_BoltzmannDist} then fixes the equilibrium probabilities $\mathbf{p}^{(1)}_{\rm{eq}}$.  Using Eqs.~\eqref{eqn_basics_relationTransitionRates}~and~\eqref{eqn_basics_identificationHeatMasterEquationPreliminary} we can calculate the exchanged heat for every transition $i\to j$ and check consistency with the transition rates $\tilde{w}^{(1)}_{ij}$. The result is displayed in the left column of Fig.~\ref{fig_basics_heatExampleJumpProcess}. 

Now imagine that the system is also coupled to a particle reservoir such that it exchanges a particle during a transition between states $2$ and $3$. The particle number is such that $n_2 = 0$ and $n_3 = 1$ (see the center column of Fig.~\ref{fig_basics_heatExampleJumpProcess}). Setting $\mu := \ln{\left(8/3\right)}$, we obtain: $V_3 - \mu = \ln{(3/2)}$ from which we find the equilibrium distribution $\mathbf{p}_{\rm{ eq}}^{(2)}$ with Eq.~\eqref{eqn_basics_eqGrandCanonical}. We may calculate the transition rate matrix $\tilde{\mathcal{W}}^{(2)}$ using Eq.~\eqref{eqn_basics_localDetailedBalance}:
\begin{align}
\tilde{\mathcal{W}}^{(2)} = r \begin{pmatrix}
0 & 0 & 0\\
0 & -2 & 1\\
0 & 2 & -1
\end{pmatrix},
\end{align}
where $r$ is a free constant that is determined by the timescale of the process. We conveniently set it to $r := 1$. We can calculate the chemical work during the transitions with Eq.~\eqref{eqn_basics_chemicalWork} and the heat with Eq.~\eqref{eqn_basics_heatWithChemicalWork} and check consistency with Eq.~\eqref{eqn_basics_heatMasterEquationFinal}. See the center column of Fig.~\ref{fig_basics_heatExampleJumpProcess} for the results.

Finally, we can consider the total system with all transition rates given by:
\begin{align}
\tilde{\mathcal{W}} := \tilde{\mathcal{W}}^{(1)} + \tilde{\mathcal{W}}^{(2)} = \begin{pmatrix}
-3 & 3 & 2\\
2 & -5 & 1\\
1 & 2 & -3
\end{pmatrix}.
\end{align}
We can solve for the stationary state $\mathbf{p}_{\rm{st}}$,
\begin{align}
0 = \tilde{\mathcal{W}}\, \mathbf{p}_{\rm{st}},
\end{align}
yielding $\mathbf{p}_{\rm{st}} = (13/29, 7/29, 9/29)^T$ (see the right column of Fig.~\ref{fig_basics_heatExampleJumpProcess}), which is not an equilibrium state as it does not fulfill global detailed balance [Eq.~\eqref{eqn_basics_detailedBalance}]. The secondary process breaks detailed balance as it \emph{pumps} the system from state $2$ to $3$ against the potential gradient. Consequently, there is a constant cyclical probability current
\begin{align}
j_{\rm st} &= \tilde{W}_{21}\, p_{\rm{st},1} -\tilde{W}_{12}\, p_{\rm{st},2} = \tilde{W}_{32}\, p_{\rm{st},2} -\tilde{W}_{23}\, p_{\rm{st},3} = \tilde{W}_{13}\, p_{\rm{st},3} -\tilde{W}_{31}\, p_{\rm{st},1}\\
&= \frac{5}{29}.
\end{align}

\section{Entropy production}\label{sec_basics_entropyProductionSTD}
We proceed to apply the definition of stochastic entropy production from Sec.~\ref{sec_basics_stochasticEP} to stochastic thermodynamics. Recall that different choices for the conjugated dynamics are possible. For now we stick to a simple time-reversal, i.e., in the conjugated process all driving is reversed and in the conjugated trajectory odd variables under time-reversal (e.g., velocity) are negated. Like in the previous section, we start with overdamped Langevin dynamics, i.e., the system trajectory is the position $x(\cdot)$ which is even under time reversal.

\subsection{Entropy production for Langevin equation}\label{eqn_basics_EPLangevin}

Consider a single trajectory $x(\cdot)$ starting at $t=0$ and ending at $t=T$ of a colloidal system described by the overdamped Langevin equation
\begin{align}\label{eqn_basics_LangevinWithTotalForce}
\dot{x} = \nu F(x,t) + \sqrt{2D}\,\xi(t).
\end{align}
Inserting Eq.~\eqref{eqn_basics_notationCompletePathProb} into the definition~\eqref{eqn_basics_defTrajectoryEP} of trajectory entropy production, we obtain:
\begin{align}\label{eqn_basics_stochasticEntropyLanghevin1}
\sigma[x(\cdot)] = \ln\frac{p(x_0,0)}{p(x_T,T)} + \ln\frac{p[x(\cdot)|x_0]}{\bar{p}[\bar{x}(\cdot)|x_T]}.
\end{align}
The first term is immediately identified as the \emph{system entropy change} since, with Eq.~\eqref{eqn_basics_defEntropy},
\begin{align}\label{eqn_basics_defSystemEntropyChange}
\Delta s := s[x_T] - s[x_0] = \ln\frac{p(x_0,0)}{p(x_T,T)}.
\end{align}
We will address this definition later.

For the second term, we use Eqs.~\eqref{eqn_basics_trajectoryProbLangevin}~and~\eqref{eqn_basics_actionLangevin} for the two trajectory probabilities
\begin{subequations}
	\begin{align}
	p[x(\cdot)|x_0] &= \exp{\Bigg[-\int\limits_0^T dt \left(\frac{1}{4D}\left[\dot{x}(t)-\nu F(x,t)\right]^2 + \frac{\nu}{2} F'(x,t) \right)\Bigg]} \;\;\text{and}\\
	\bar p[\bar x(\cdot)|x_T] &= \exp \Bigg[-\int\limits_0^T dt \Big(\frac{1}{4D}\left[-\dot{x}(T-t)-\nu F(x(T-t),T-t)\right]^2\nonumber\\
	&\qquad\qquad\qquad\qquad\qquad\qquad\qquad+ \frac{\nu}{2} F'(x(T-t),T-t) \Big)\Bigg].\label{eqn_basics_reverseLangevinTrajectoryProb}
	\end{align}
\end{subequations}
Notice that there is a minus sign in front of the velocity $\dot{x}$ in Eq.~\eqref{eqn_basics_reverseLangevinTrajectoryProb} due to time reversal. 

Inserting these into the second term of Eq.~\eqref{eqn_basics_stochasticEntropyLanghevin1}, we obtain
\begin{align}
\ln\frac{p[x(\cdot)|x_0]}{\bar{p}[\bar{x}(\cdot)|x_T]} = -\int\limits_0^T dt \left(\frac{1}{4D}\left[-4 \nu\,\dot{x}(t)\circ F(x,t)\right]\right),
\end{align}
and with Eq.~\eqref{eqn_basics_SekimotoTrajectoryHeat_b} and the Einstein-Smoluchowski relation in Eq.~\eqref{eqn_basics_EinsteinSmoluchowskiRelation} we see that this term is the \emph{heat exchanged} with the reservoir up to a prefactor:
\begin{align}
q[x(\cdot)] = -\int dt\,\dot{x} \circ F(x,t) = -k_{\rm B} T\,\ln\frac{p[x(\cdot)|x_0]}{\bar{p}[\bar{x}(\cdot)|x_T]}.
\end{align}

We thus get the following decomposition of the trajectory entropy production:
\begin{align}\label{eqn_basics_decompositionEntropyProduction}
\sigma[x(\cdot)] = \Delta s - q[x(\cdot)]/k_{\rm B}T,
\end{align}
which already hints at the second law of thermodynamics, as $\sigma[x(\cdot)]$ looks like the difference between both sides of the Clausius inequality.

Even though identifying the heat is formally more involved, the arguably bigger leap is defining the system entropy as $s(t) := -\ln{p(x(t),t)}$. Firstly, we can observe that its \emph{ensemble average} recovers the usual Gibbs entropy known from macroscopic thermodynamics up to rescaling with $k_{\rm B}$:
\begin{align}
S(t) := \left\langle s(t) \right\rangle = - \sum\limits_x p(x,t)\ln p(x,t),
\end{align}
which we simply postulate to be a valid entropy measure even in nonequilibrium situations and refer the skeptics to the multitude of successful applications of this definition.

The novel aspect is assigning a state-dependent entropy that changes along the system trajectory\footnote{Note that this quantity still contains information about the whole ensemble through $p(x,t)$, which follows from solving the Fokker-Planck equation corresponding to the Langevin equation.}. The differentiated version of Eq.~\eqref{eqn_basics_decompositionEntropyProduction} is thus the evolution equation of the system entropy~\cite{Seifert2005}:
\begin{align}
\dot{s} = \dot{q}/k_{\rm B}T + \dot{\sigma}.
\end{align}

\subsection{Entropy production for jump processes}

We proceed by briefly laying out the corresponding identification for Markovian jump processes (such as the example discussed in Secs.~\ref{sec_basics_exampleJumpProcess}~and~\ref{sec_basics_brokenDetailedBalance}). Let us assume for simplicity that the states $x$ of the process are even under time reversal. The trajectories are not continuous but a series of jumps among a discrete set of states with intermediate holding times inside of the current state. Such a process is often employed to model enzyme dynamics like the progress of molecular motors (see, e.g., Sec.~9.4~of~Ref.~\cite{Seifert2012}).

\begin{figure}[ht]
	\centering
	\includegraphics[width =0.55 \linewidth]{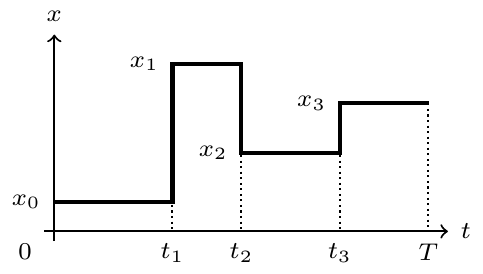}
	\caption{Example trajectory of a Markovian jump process containing three jumps.}
	\label{fig_basics_jumpTrajectory}
\end{figure}

A trajectory of length $T$ with $N$ jumps is thus specified by the sequence of states $x_i$ visited and the corresponding \emph{jump times} $t_i$, i.e., the time at which the process jumped from state $x_{i-1}$ to the next state $x_{i}$:
\begin{align}
x(\cdot) = \left\{ x_0 \rightarrow (x_1,t_1) \rightarrow (x_2,t_2) \rightarrow ... \rightarrow (x_N,t_N)\right\}.
\end{align}
Figure~\ref{fig_basics_jumpTrajectory} shows an example trajectory.

From the time-dependent transition rates $W(x|x';t)$ appearing in the master equation~\eqref{eqn_basics_differentialMasterEquation}, we find the path probability as follows: The probability of the initial state follows from the initial condition $p(x_0,0) = p_0(x_0)$. Next, we require the probability that a jump occurs at time $t_1$, which we already calculated when deriving the \emph{Gillespie algorithm} in Sec.~\ref{sec_basics_numerics}:
\begin{align}
p(t_1|x_0) = \tilde{\mathcal{N}} \exp{\left(- \int\limits_0^{t_1} dt\, r(x_0,t)\right)},
\end{align}
where $\tilde{\mathcal{N}}$ is a normalization constant and $r(x_0,t)$ is the \emph{exit rate} out of state $x_0$ defined in Eq.~\eqref{eqn_basics_exitRate}. The probability of the subsequent jump to state $x_1$ is then proportional to the rate $W(x_1|x_0;t_1)$. Continuing this procedure, one arrives at:
\begin{align}
p[x(\cdot)] = \mathcal{N}\, p(x_0,0)\,\left[\prod\limits_{i=0}^{N-1} \exp{\left(- \int\limits_{t_{i}}^{t_{i+1}} dt\, r(x_i,t)\right)} \, W(x_{i+1}|x_{i};t_{i+1})\right]\nonumber\\
\times\exp{\left(- \int\limits_{t_{N}}^{T} dt\, r(x_N,t)\right)},
\end{align}
where we set $t_0 := 0$ and $\mathcal{N}$ follows from normalization.

Next, we consider the probability of the time-reversed trajectory $\bar{x}(\cdot)$ in the time-reversed process. Its probability is given by:
\begin{align}
\bar p[\bar x(\cdot)] = \mathcal{N}\, p(x_N,T)\,&\left[\prod\limits_{i=0}^{N-1} \exp{\left(- \int\limits_{T-t_{i+1}}^{T-t_{i}} dt\, r(x_i,T-t)\right)} \, W(x_{i}|x_{i+1};t_{i+1})\right]\nonumber\\
&\qquad\qquad\qquad\qquad\times\exp{\left( -\int\limits_{T-t_1}^{T} dt\, r(x_0,T-t)\right)},
\end{align}
where $p(x_N,T)$ follows from solving the master equation and inserting the final state $x_N$.

Inserting both path probabilities into Eq.~\eqref{eqn_basics_defTrajectoryEP}, the contributions from the holding times cancel after substitution, and we obtain:
\begin{align}\label{eqn_basics_trajectoryEntropyMasterEquation}
\sigma[x(\cdot)] = \ln\frac{p(x_0,0)}{p(x_N,T)} + \sum_{i=0}^{N-1}\ln\frac{W(x_{i+1}|x_i;t_{i+1})}{W(x_{i}|x_{i+1};t_{i+1})},
\end{align}
which, with Eq.~\eqref{eqn_basics_defSystemEntropyChange}, allows us to again identify the system entropy change $\Delta s$ and with Eq.~\eqref{eqn_basics_heatMasterEquationFinal} we again find the heat absorbed by the system as the sum of the individual heat contributions from all jumps:
\begin{align}
\sigma[x(\cdot)] = \Delta s - q[x(\cdot)]/k_{\rm B}T,
\end{align}
which is the same as Eq.~\eqref{eqn_basics_decompositionEntropyProduction}. 

\subsection{Adiabatic and nonadiabatic entropy production}\label{sec_basics_adiabaticNonadabaticEP}
One very useful dissection of the entropy production splits it into two contributions each responsible for different nonequilibrium aspects of the dynamics. It was put forward by Esposito and Van den Broeck in a series of papers~\cite{Esposito2010,Esposito2010a,VandenBroeck2010} after having been proposed in Ref.~\cite{Esposito2007}.

From Eq.~\eqref{eqn_basics_trajectoryEntropyMasterEquation} we see that the entropy production can be split in the following way:
\begin{align}
\sigma[x(\cdot)] = \sigma_{\rm na}[x(\cdot)] + \sigma_{\rm a}[x(\cdot)],
\end{align}
where
\begin{align}
\sigma_{\rm na}[x(\cdot)] := \ln\frac{p(x_0,0)}{p(x_N,T)} + \sum_{i=0}^{N-1}\ln\frac{p_{\rm st}(x_{i+1};\lambda(t_{i+1}))}{p_{\rm st}(x_{i};\lambda(t_{i+1}))}\label{eqn_basics_nonadiabaticEP}
\end{align}
is the so-called \emph{nonadiabatic entropy production} and
\begin{align}
\sigma_{\rm a}[x(\cdot)] :=  \sum_{i=0}^{N-1}\ln\frac{W(x_{i+1}|x_i;t_{i+1})\,p_{\rm st}(x_{i};\lambda(t_{i+1}))}{W(x_{i}|x_{i+1};t_{i+1})\,p_{\rm st}(x_{i+1};\lambda(t_{i+1}))}
\end{align}
is called \emph{adiabatic entropy production}. Here, $p_{\rm st}(x_{i};\lambda(t_{i}))$ denotes the instantaneous stationary probability distribution that would be reached by letting the system relax at the current values of the transition rates $W(x|x';t_{i+1})$.

The naming of the two contributions can be justified as follows: Assume that the system obeys detailed balance at all times. Then, $p_{\rm st}(x_{i};\lambda(t_{i})) = p_{\rm eq}(x_{i};\lambda(t_{i}))$ and with Eq.~\eqref{eqn_basics_detailedBalance} we find: $\sigma_{\rm a}[x(\cdot)] \equiv 0$. Contributions to the entropy production can thus only come for the driving through $\lambda(t)$: The only way to obtain nonequilibrium is by explicitly driving the system out of equilibrium, i.e., non-adiabatically.

Conversely, assume that the driving happens quasi-statically (infinitely slow, or adiabatically). Then, the system can always stay in equilibrium:
\begin{align}\label{eqn_basics_quasistaticProbs}
p(x,t)= p_{\rm st}(x;\lambda(t)).
\end{align}
Let $t_0:= 0$, then, we can rewrite the boundary term in Eq.~\eqref{eqn_basics_nonadiabaticEP} as follows:
\begin{align}
\ln\frac{p(x_0,0)}{p(x_N,T)} = \sum_{i=0}^{N-1}\ln\frac{p(x_{i},t_{i})}{p(x_{i+1},t_{i+1})}.
\end{align}
We thus have
\begin{align}
\sigma_{\rm na}[x(\cdot)] &:= \sum_{i=0}^{N-1}\ln\frac{p(x_i,t_{i})\,p_{\rm st}(x_{i+1};\lambda(t_{i+1}))}{p(x_{i+1},t_{i+1})\,p_{\rm st}(x_{i};\lambda(t_{i+1}))}\\
&= 0,
\end{align}
which follows from Eq.~\eqref{eqn_basics_quasistaticProbs} and the assumption that the protocol changes very slowly. The nonadiabatic contribution thus vanishes leaving only an adiabatic contribution.

This resonates nicely with the three kinds of nonequilibrium mentioned at the beginning of this chapter: Driving and relaxation produce nonadiabtic entropy. Nonequilibrium steady states only produce adiabatic entropy.

\section{Fluctuation theorems and the second law}
Having established the energetics of Markov processes, we will turn to the fluctuation theorems. There is an impressive collection of different fluctuation theorems with different areas of applicability. Most of them have been derived on a case-by-case basis, often with specific assumptions about the underlying dynamics. We will derive some of those from our prototype fluctuation theorems presented in Sec~\ref{sec_basics_fluctuationTheorems} and the thermodynamic entropy production established in the last two sections.

\subsection{The second law as time-series irreversibility}\label{sec_basics_secondLawDKL}
Often, fluctuation theorems are presented as \emph{refinements of the second law}. In the introductory example in Sec.~\ref{sec_basics_microscopicMacroscopic} we have already seen that the second law can be expected to only hold on average. Conveniently, the integral fluctuation theorem in Eq.~\eqref{eqn_basics_IFT},
\begin{align}\label{eqn_basics_IFT2}
\left\langle e^{-\sigma} \right\rangle = 1,
\end{align}
implies the \emph{non-negativity of the average trajectory entropy production}: Since $\exp{(x)} \geq 1 + x$, we have $1 = \left\langle \exp{(-\sigma)}\right\rangle \geq \left\langle 1-\sigma\right\rangle$ and thus:
\begin{align}\label{eqn_basics_basicSecondLaw}
\left\langle\sigma\right\rangle \geq 0 .
\end{align}
This is the most general statement of the second law inequalities we are about to derive.

In stochastic thermodynamics, we find the integral fluctuation theorem [with Eq.~\eqref{eqn_basics_decompositionEntropyProduction}],
\begin{align}
\left\langle e^{- \Delta s + q[x(\cdot)]/k_{\rm B}T} \right\rangle = 1,
\end{align}
and thus:
\begin{align}\label{eqn_basics_ClausiusInequality}
\left\langle \Delta s \right\rangle \geq \frac{\left\langle q \right\rangle}{k_{\rm B} T},
\end{align}
which is the \emph{Clausius inequality} for the average system entropy change and the average heat exchanged with the reservoir.

Let us define
\begin{align}
\Sigma := \langle \sigma \rangle = \left\langle \Delta s \right\rangle - \frac{\left\langle q \right\rangle}{k_{\rm B} T},
\end{align}
the \emph{average entropy production}. Recall that $\Delta S = \left\langle \Delta s \right\rangle$ is the average change in system entropy and $-Q := -\left\langle q \right\rangle$ is the average heat that flows out of the system into the heat bath. The heat bath is always in equilibrium such that $Q$ is the only entropy change in it. Then,
\begin{align}
\Sigma = \Delta S - \frac{Q}{\kB T}
\end{align}
is the \emph{`entropy change of the universe'} during the process we consider. However, the definition of $\sigma$ in Eq.~\eqref{eqn_basics_defTrajectoryEP} implies:
\begin{align}
\Sigma &= \langle \sigma \rangle = \int \mathcal{D}x(\cdot)\, p[x(\cdot)] \ln\frac{p[x(\cdot)]}{\bar{p}[\bar{x}(\cdot)]}\\
&= D_{\rm KL}\Big[ p[x(\cdot)]\,||\,\bar{p}[\bar{x}(\cdot)] \Big], \label{eqn_basics_DissipationKullbackLeibler}
\end{align}
where we used the definition of the Kullback-Leibler distance [Eq.~\eqref{eqn_basics_defKLDistance}] in the second line.

This is a remarkable relation, as it states that the entropy produced by a nonequilibrium process is exactly given by the difficulty to differentiate whether the process runs forwards or backwards in time! It resonates with the identification of an arrow of time dictated by the second law but also allows a quantitative statement about its `length'. 

The relation was pointed out by Kawai \emph{et al.}~\cite{Kawai2007,Parrondo2009}. Even though we derived it from stochastic dynamics in which dissipation was built in from the start, it also holds for Hamiltonian dynamics, as we will show in Sec.~\ref{sec_basics_HamiltonianDynamics}. It allows numerous applications in estimating the dissipation of small-scale systems, particularly when only limited data are available~\cite{Gomez-Marin2008,Gomez-Marin2008a,Roldan2010,Roldan2012,Martinez2019}. We will exploit this relation in Chap.~\ref{chap_hid_inferringDiss}.

\subsection{Nonequilibrium work relations}\label{eqn_basics_noneqWorkRels}
A first special case of the above equations are the so-called \emph{nonequilibrium work relations}. These apply to driven processes obeying detailed balance that start and end in equilibrium. Consider a stochastic system whose potential landscape is influenced by a \emph{work parameter}, or protocol, $\lambda$: $V(x;\lambda)$. The system is initially in equilibrium with a protocol value $\lambda_0$. We drive it for a time $T$ out of equilibrium by changing the protocol from $\lambda_0$ to $\lambda_T$, and finally let it relax to state $x_\infty$ by keeping the protocol constant at the final $\lambda_T$. 

Using the first law, $\Delta V = q + w$, we can rewrite the entropy production:
\begin{align}
\sigma[x(\cdot)] &= \Delta s - \frac{1}{k_{\rm B} T} \Big( V(x_\infty;\lambda_T)- V(x(0),\lambda_0) - w[x(\cdot)]\Big)\\
&= \frac{w[x(\cdot)]}{k_{\rm B} T} - \frac{V(x_\infty;\lambda_T)- V(x(0),\lambda_0) - k_{\rm B}T \Delta s}{k_{\rm B} T}.
\end{align}
Even though the second term looks like a fluctuating quantity, it is in fact the \emph{equilibrium free energy difference} which does not depend on individual realizations. This is because the system entropy change is evaluated for equilibrium distributions,
\begin{align}
p_{\rm eq}(x;\lambda) = \exp{\left(-\frac{V(x;\lambda)-F(\lambda)}{\kB T }\right)},
\end{align}
such that
\begin{align}
\Delta s &= \ln\frac{p_{\rm eq}(x(0);\lambda_0)}{p_{\rm eq}(x_\infty;\lambda_T)}\\
&= \frac{1}{k_{\rm B} T}\left[ -V(x(0);\lambda_0) + V(x_\infty;\lambda_T) + F(\lambda_0) - F(\lambda_T) \right].
\end{align}
We finally obtain
\begin{align}\label{eqn_basics_entropyFreeEnergyDifference}
\sigma[x(\cdot)] = \frac{w[x(\cdot)]- \Delta F}{\kB T},
\end{align}
where $\Delta F := F(\lambda_T) - F(\lambda_0)$. In the final equilibration step the protocol value stays constant, therefore no work is done on the system. This means that we do not need to trace the system during that time.

Inserting Eq.~\eqref{eqn_basics_entropyFreeEnergyDifference} into Eq.~\eqref{eqn_basics_IFT2}, we obtain the celebrated \emph{Jarzynski equality}~\cite{Jarzynski1997}:
\begin{align}\label{eqn_basics_JarzynskiEquality}
\left\langle e^{-w/\kB T} \right\rangle = e^{-\Delta F/\kB T}.
\end{align}

Inserting Eq.~\eqref{eqn_basics_entropyFreeEnergyDifference} instead in Eq.~\eqref{eqn_basics_CrooksTypeFT}, we get:
\begin{align}
\ln\frac{p_\sigma\left(\frac{w-\Delta F}{\kB T}\right)}{\bar p_\sigma\left(\frac{-w+\Delta F}{\kB T}\right)} = \frac{w-\Delta F}{\kB T},
\end{align}
and a transformation of probabilities to $p_w(w) = p_\sigma\left(\frac{w-\Delta F}{\kB T}\right)/\kB T$ yields the \emph{Crooks fluctuation theorem}~\cite{Crooks1999}:
\begin{align}\label{eqn_basics_CrooksFT}
\ln\frac{p(w)}{\bar{p}(-w)} = \frac{w-\Delta F}{\kB T}.
\end{align}
Here, $p(w)$ denotes the probability to measure a work $w$ and $\bar{p}(-w)$ denotes the probability to measure the negative of that work in the time-reversed process.

\begin{figure}[ht]
	\centering
	\includegraphics[width =0.95 \linewidth]{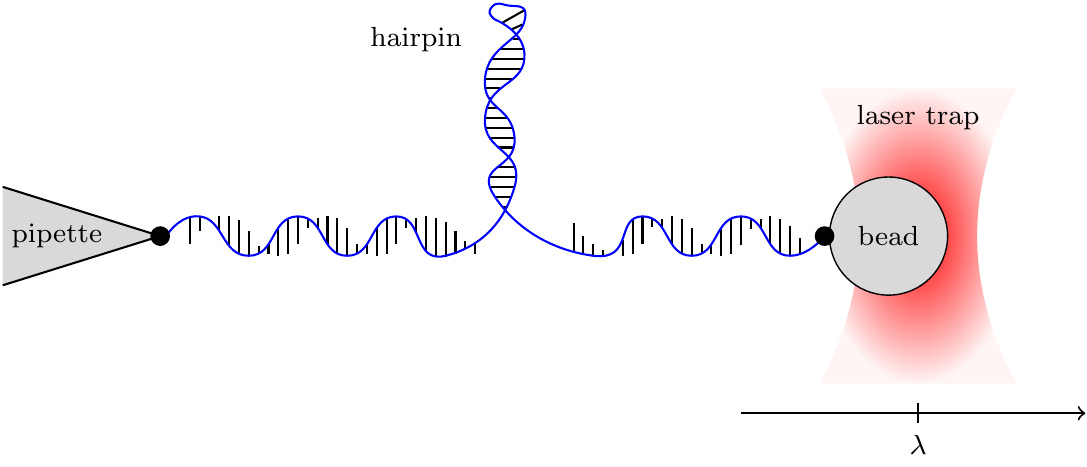}
	\caption{Optical force spectroscopy experiment: A biomolecule is held between a micropipette and a micron-sized bead in an optical trap. By moving the trap (changing $\lambda$), a force can be exerted on the molecule and work can be measured from the position of the bead inside the trap. }
	\label{fig_basics_foldingExperiment}
\end{figure}

Both the Jarzynski equality and the Crooks fluctuation theorem are very useful experimentally: They allow the measurement of the free energy, an equilibrium property, from many realizations of a nonequilibrium process. A prominent example are so-called \emph{force spectroscopy experiments}. In these experiments the system under study is subjected to a mechanical force (often a stretching force). Exerting a force on the system drives it out of equilibrium and thus makes measuring equilibrium properties difficult. Figure~\ref{fig_basics_foldingExperiment} schematically shows the setup of one example. A biomolecule (DNA or RNA) hairpin is held between a a fixed micropipette and a micron-sized bead that is in an optical trap. By moving the trap, force can be exerted on the molecule and induces an unfolding of the molecule. 

The trap position acts as a \emph{protocol} $\lambda(t)$ influencing the potential energy of the whole system. The motion of the bead can be described by an overdamped Langevin equation~\eqref{eqn_basics_overdampedLangevinEquatuationPotential}. Work can be measured by the position of the bead inside the trapping potential using Eq.~\eqref{eqn_basics_workLangevinBasic}, which effectively renders the setup of trap plus bead a microscopic calorimeter for the folding energies of the molecule. See Ref.~\cite{Alemany2011} for a review.

To get the free energy difference between the folded and unfolded state, one can perform many repetitions of the stretching experiments, record the work and evaluate Eq.~\eqref{eqn_basics_JarzynskiEquality} as done, e.g., in Ref.~\cite{Liphardt2002}. Alternatively, one can also perform the reverse experiment and record the work statistics $p(w)$ of the forward and $\bar{p}(w)$ of the reverse experiment. Equation~\eqref{eqn_basics_CrooksFT} then tells us that the free energy difference is given by the intersection point of the two histograms $p(w)$ and $\bar{p}(-w)$. This procedure has been used, e.g., in Ref.~\cite{Collin2005}.

Usually, the second option is more accurate, as the exponential average in the Jarzynski equality is dominated by the rare very negative work values. Hummer and Szabo~\cite{Hummer2001} derived a relation that allows the reconstruction of the \emph{free energy landscape} $F(\lambda)$, which has been successfully used in Refs.~\cite{Harris2007, Gupta2011}.

Let us point out some subtleties of the force spectroscopy setup that touch on the main topics of this thesis. Because the trap is moving, from the co-moving reference frame, the bead inside encounters a constant flow of the surrounding medium. This ensures the correct measurement of work values~\cite{Speck2008} but poses a problem when using two traps (one moving and one stationary) and measuring work in the stationary one~\cite{Ribezzi-Crivellari2014}. The setup thus has \emph{hidden degrees of freedom} (cf. Chap.~\ref{sec_hid_hiddenDegreesInSTD}) as one only observes one variable of the joint system of two beads and one biomolecule~\cite{Alemany2015}. One is only interested in the free energy profile of the hairpin. However, there are also fluctuations due to the bead and the rest of the DNA or RNA strain which then blur the resulting profile~\cite{Jarzynski2011a} making a deconvolution necessary~\cite{Hummer2010}.

\subsection{Detailed fluctuation theorem}\label{sec_basics_detailedFT}
In some rare cases of driven dynamics, e.g., the model system considered in Sec.~\ref{sec_hid_FTs}, the probability distribution $\bar{p}(\sigma)$ in the reverse process is the same as in the forward process. Equation~\eqref{eqn_basics_CrooksTypeFT} then becomes
\begin{align}\label{eqn_basics_detailedFT}
\ln\frac{p(\sigma)}{p(-\sigma)} = \sigma,
\end{align}
which we shall call \emph{detailed fluctuation theorem}\footnote{In the literature one can also find the name \emph{detailed fluctuation theorem} for what we called the \emph{Crooks-type fluctuation theorem}.}. Remarkably, to evaluate this equation, one only needs to sample the forward process.

A detailed fluctuation theorem can also arise in nonequilibrium steady states. Since there is no control protocol changing in time, the time-reversal only applies to the trajectory, not to the trajectory probability function. Retracing the steps that led to Eq.~\eqref{eqn_basics_derivationCrooks3}, one immediately finds Eq.~\eqref{eqn_basics_detailedFT}. In that context the detailed fluctuation theorem is often called \emph{steady-state fluctuation theorem}.

\subsection[Housekeeping heat and Hatano-Sasa relation]{Fluctuation theorem for the housekeeping heat and Hatano-Sasa relation}\label{sec_basics_housekeepingHeatFT}
Let us briefly return to the splitting of the entropy production into adiabatic and nonadiabatic contributions (cf. Sec.~\ref{sec_basics_adiabaticNonadabaticEP}). One can show~\cite{Esposito2010} that both of them fulfill Crooks-type fluctuation theorems [Eq.~\eqref{eqn_basics_CrooksTypeFT}]:
\begin{subequations}
	\begin{align}
	\ln\frac{p(\sigma_{\rm na})}{p^\dagger(-\sigma_{\rm na})} &= \sigma_{\rm na} \qquad \text{and}\label{eqn_basics_CFTNa}\\ 
	\ln\frac{p(\sigma_{\rm a})}{p^\dagger(-\sigma_{\rm a})} &= \sigma_{\rm a},\label{eqn_basics_CFTa}
	\end{align}
\end{subequations}
where the $\dagger$ indicates a special kind of conjugated process that is obtained by `\emph{twisting}' the transition rates of the underlying Markov process:
\begin{align}
W^\dagger(x|x';t) := W(x'|x;t) \frac{p_{\rm st}(x;t)}{p_{\rm st}(x';t)}.
\end{align}
Here, $p_{\rm st}(x;t)$ is the instantaneous stationary distribution. Both entropy productions can now be written as log-ratios of different path probabilities which automatically leads to Eqs.~~\eqref{eqn_basics_CFTNa}~and~\eqref{eqn_basics_CFTa} (see Ref.~\cite{Esposito2010} for details).

Equations~\eqref{eqn_basics_CFTNa}~and~\eqref{eqn_basics_CFTa} immediately yield integral fluctuation theorems:
\begin{subequations}
	\begin{align}
	\left\langle e^{-\sigma_{\rm na}}\right\rangle &= 1 \qquad \text{and}\label{eqn_basics_IFTNa}\\ 
	\left\langle e^{-\sigma_{\rm a}}\right\rangle &= 1.\label{eqn_basics_IFTa}
	\end{align}
\end{subequations}

As a special case, consider a system that is in contact with only one heat reservoir. In that setting, different names for the contributions are customary: The nonadiabatic entropy production is called \emph{excess heat} $q_{\rm ex}$ (plus system entropy change):
\begin{align}
\sigma_{\rm na}[x(\cdot)] := \Delta s -\frac{q_{\rm ex}[x(\cdot)]}{\kB T},
\end{align}
and the adiabatic entropy production is called \emph{housekeeping heat} $q_{\rm hk}$ because of its interpretation as the heat that has to be dissipated to maintain a nonequilibrium steady-state:
\begin{align}
\sigma_{\rm a}[x(\cdot)] := -\frac{q_{\rm hk}[x(\cdot)]}{\kB T}.
\end{align}

Thus, we find the \emph{fluctuation theorem for the housekeeping heat} as a special case of Eq.~\eqref{eqn_basics_IFTa}:
\begin{align}\label{eqn_basics_FTHouseKeepingHeat}
\left\langle \exp{\left(\frac{q_{\rm hk}}{\kB T}\right)} \right\rangle = 1,
\end{align}
which was first proven by Speck and Seifert~\cite{Speck2005} for systems described by a Langevin equation. The relation between the adiabatic entropy production and the housekeeping heat can be established by comparing Speck and Seifert's definition with the one by Van den Broeck and Esposito in Ref.~\cite{VandenBroeck2010} and matching with the corresponding master equation formalism in Refs.~\cite{Esposito2010,Esposito2010a}. Notice also the different sign conventions for the heat. 

With the second-law inequality [Eq.~\eqref{eqn_basics_basicSecondLaw}], Eq.~\eqref{eqn_basics_FTHouseKeepingHeat} immediately implies:
\begin{align}
\left\langle q_{\rm hk} \right\rangle \leq 0,
\end{align}
indicating that, on average, the system has to dissipate heat to maintain the nonequilibrium steady-state.

A second special case concerns transitions between \emph{different nonequilibrium steady-states}. Assume that the system starts and ends in a nonequilibrium steady-state. Then, the change in system entropy [Eq.~\eqref{eqn_basics_defSystemEntropyChange}] can be rewritten as
\begin{align}
\Delta s = \Delta \phi,
\end{align}
where $\phi(x,\lambda)$ is understood as a \emph{nonequilibrium potential} such that:
\begin{align}
p_{\rm st}(x;\lambda) = \exp{\left[-\phi(x,\lambda)\right]}.
\end{align}
Equation~\eqref{eqn_basics_IFTNa} then reads:
\begin{align}
\left\langle \exp{\left(-\Delta \phi + \frac{q_{\rm ex}}{\kB T}\right)} \right\rangle = 1,
\end{align}
which is known as the \emph{Hatano-Sasa relation}~\cite{Hatano2000}. It implies
\begin{align}
\left\langle \Delta \phi \right\rangle \geq  \left\langle q_{\rm ex}\right\rangle/\kB T,
\end{align}
which is a kind of Clausius inequality [Eq.~\eqref{eqn_basics_ClausiusInequality}] for transitions between nonequilibrium steady states.

\section{Average entropy production rate}
In some situations one is interested in the \emph{average rate} of heat, work, and entropy production. These quantities are readily available starting from a Fokker-Planck equation (see, e.g., Ref.~\cite{VandenBroeck2010}) or a master equation (see, e.g., Ref.~\cite{Esposito2010a}) upon identification of the different contributions to the system entropy change. In deriving the corresponding expressions, this strategy is less cumbersome than the one we will employ. However, since we already have the corresponding trajectory quantities, we might as well start from these. Again, we first present the continuous state-space-treatment and then the formalism for jump processes described by a Master equation.

\subsection[Fokker-Planck dynamics]{Average entropy production rate for Fokker-Planck dynamics}
Recall the expression derived in Sec.~\ref{sec_basics_stochasticEnergetics} for the trajectory-dependent heat in Eq.~\eqref{eqn_basics_SekimotoTrajectoryHeat_b}. It implies that:
\begin{align}
\dot{q} = - \dot{x} \circ F(x,t),
\end{align}
where $\circ$ indicates Stratonovich discretization and $F(x,t)$ is the total force acting on the Brownian particle.

Let us denote the average rate of heat by
\begin{align}
\dot{Q} := \left\langle \dot{q} \right\rangle.
\end{align}
As pointed out by Seifert~\cite{Seifert2005,Seifert2012}, it can be resolved as:
\begin{align}
\dot{Q} = - \left\langle \left\langle \dot{x}\right\rangle_{p(\dot{x}|x,t)}\circ F(x,t)\right\rangle_{p(x,t)},
\end{align}
where $\left\langle \dot{x}\right\rangle_{p(\dot{x}|x,t)}$ is the \emph{conditional average} of $\dot{x}(t)$ given $x(t)$. As we show in Appendix~\ref{app_conditionalVelocityAverages}, this procedure yields:
\begin{align}
\dot{Q} &= - \left\langle \frac{j(x,t)}{p(x,t)}\, F(x,t) \right\rangle_{p(x,t)}\\
&= -\int dx \, j(x,t)\,F(x,t)\\
&= - \left\langle F(x,t) \right\rangle_{j(x,t)},
\end{align}
where 
\begin{align}\label{eqn_basics_probCurrentFPELangevin}
j(x,t)= \left[\nu F(x,t)-D\pd{x}\right]p(x,t)
\end{align}
is the probability current defined in Eq.~\eqref{eqn_basics_defProbCurrentFPE} and appearing in the Fokker-Planck equation~\eqref{eqn_basics_FokkerPlanckCorrespondingLangevin},
\begin{align}
\pd{t} p(x,t) = -\pd{x} \nu F(x,t)\,p(x,t) + D \frac{\partial^2}{\partial x^2} p(x,t),
\end{align}
which corresponds to the overdamped Langevin equation~\eqref{eqn_basics_LangevinWithTotalForce}.

Similarly, with Eq.~\eqref{eqn_basics_SekimotoTrajectoryWork} the \emph{average rate of work} reads:
\begin{align}
\dot{W} &:= \left\langle \dot{w} \right\rangle\\
&= \left\langle \dot{\lambda}\pd{\lambda}V(x,\lambda(t))+ \dot{x} \circ f(x,t) \right\rangle\\
&= \dot{\lambda}\pd{\lambda} \Big\langle V(x,\lambda(t))\Big\rangle_{p(x,t)} +  \Big\langle f(x,t) \Big\rangle_{j(x,t)},
\end{align}
where $f(x,t)$ is the non-conservative part of the total force.

Finally, we can consider the \emph{average entropy production rate} $\dot{\Sigma}$, for which we find from Eq.~\eqref{eqn_basics_decompositionEntropyProduction}
\begin{align}
\dot{\Sigma} := \left\langle \dot{\sigma} \right\rangle = \dot{S} - \dot{Q}/\kB T,
\end{align}
where $S$ is the differential Shannon (or Gibbs) entropy given in Eq.~\eqref{eqn_basics_defDifferentialEntropy}.

We therefore find
\begin{align}
\dot{\Sigma} &= \int dx\,\left[- \left( \pd{t} p(x,t)\right) \ln{p(x,t)} + \frac{F(x,t) j(x,t)}{\kB T} \right]\\
&= \int dx\,\left[ \left( \pd{x} j(x,t)\right) \ln{p(x,t)} + \frac{F(x,t) j(x,t)}{\kB T} \right] \label{eqn_basics_FPEEP_2}\\
&= \int dx\,\left[-\frac{j(x,t)}{p(x,t)}\pd{x} p(x,t) + \frac{F(x,t) j(x,t)}{\kB T} \right] \label{eqn_basics_FPEEP_3}\\
&= \int dx\,\left[-\frac{\nu j(x,t) F(x,t)}{D} + \frac{j^2(x,t)}{D\,p(x,t)} + \frac{F(x,t) j(x,t)}{\kB T} \right] \label{eqn_basics_FPEEP_4}\\
&= \int dx \frac{j^2(x,t)}{D \, p(x,t)} \geq 0, \label{eqn_basics_FPEEP_5}
\end{align}
where we inserted the Fokker-Planck equation to get to Eq.~\eqref{eqn_basics_FPEEP_2}, used partial integration to obtain Eq.~\eqref{eqn_basics_FPEEP_3} and noted that the boundary terms vanish, and inserted Eq.~\eqref{eqn_basics_probCurrentFPELangevin} to get Eq.~\eqref{eqn_basics_FPEEP_4}. Finally, we used the Einstein relation $D~=~\nu \kB T$ [Eq.\eqref{eqn_basics_EinsteinSmoluchowskiRelation}] to obtain Eq.~\eqref{eqn_basics_FPEEP_5}.

Equation~\eqref{eqn_basics_FPEEP_5} lends itself to an intuitive interpretation: The rate of entropy production is nonnegative and directly linked to the existence of probability flows in the system, neatly highlighting the special role of equilibrium as a state without average probability flows.

\subsection[Master equation dynamics]{Average entropy production rate for master equation dynamics}\label{sec_basics_avgEPMasterEq}
In master equation systems, the \emph{average heat dissipation rate} $\dot{Q}$ is given by:
\begin{align}
\dot{Q} := \lim\limits_{dt \to 0} \frac{\left\langle \Delta q\right\rangle}{dt},
\end{align}
where $\Delta q$ denotes the heat flowing into the system in a time interval $dt$ so small that there is at most one state transition $x$ to $x'$ occurring. The heat exchange with the reservoir is then given by Eq.~\eqref{eqn_basics_identificationHeatMasterEquationPreliminary}, so that we may write:
\begin{align}
\left\langle \Delta q \right\rangle &=\Big\langle \Delta q(x\to x';\lambda(t))\Big\rangle\\
&= k_{\rm B} T\iint dx dx'\, p(x',t+dt|x,t)\,p(x,t) \ln\frac{W(x|x';\lambda(t))}{W(x'|x;\lambda(t))}.
\end{align}
Then, using the expansion of the transition probability [Eq.~\eqref{eqn_basics_expandTransitionProb}] and Eq.~\eqref{eqn_basics_WtildeDef}, we obtain:
\begin{align}
\left\langle \Delta q \right\rangle = k_{\rm B} T\iint dx dx'\, \Big[\delta(x-x')\,\big(1-r(x;t)\,dt\big) + W(x'|x;\lambda(t))\,dt + \mathcal{O}(dt^2)\Big] \nonumber\\
\times\, p(x,t) \ln\frac{W(x|x';\lambda(t))}{W(x'|x;\lambda(t))}.
\end{align}
Thus, we find:
\begin{align}\label{eqn_basics_averageHeatRateMasterEquation}
\dot{Q} = k_{\rm B} T\iint dx dx'\,W(x'|x;\lambda(t))\,p(x,t) \ln\frac{W(x|x';\lambda(t))}{W(x'|x;\lambda(t))}.
\end{align}

Similarly, the \emph{average rate of work} for a system coupled to one thermal reservoir is given by Eq.~\eqref{eqn_basics_identificationWorkMasterEquation}:
\begin{align}
\dot{W} = \dot{\lambda} \int dx\, p(x,t) V(x,\lambda(t)).
\end{align}

To calculate the \emph{average rate of entropy production} $\dot{\Sigma}$, note first that we can write Eq.~\eqref{eqn_basics_averageHeatRateMasterEquation} as follows:
\begin{align}
\dot{Q} = k_{\rm B} T\iint dx dx'\,\tilde W(x|x';\lambda(t))\,p(x',t) \ln\frac{\tilde W(x'|x;\lambda(t))}{\tilde W(x|x';\lambda(t))},
\end{align}
where we exchanged the integration variables and used Eq.~\eqref{eqn_basics_WtildeDef}, noting that the `diagonal' terms cancel.

Now, $\dot{\Sigma}$ follows from Eq.~\eqref{eqn_basics_decompositionEntropyProduction}:
\begin{align}
\dot{\Sigma} &:= \dot{S} - \dot{Q}/\kB T\\
&= \int dx\left[-\left( \pd{t} p(x,t)\right) \ln{p(x,t)} - \int dx'\,\tilde W(x|x';\lambda(t))\,p(x',t) \ln\frac{\tilde W(x'|x;\lambda(t))}{\tilde W(x|x';\lambda(t))}\right]\\
&= \iint dx dx'\,\Bigg[- \tilde W(x|x';\lambda(t))\,p(x',t)\ln{p(x,t)} \nonumber\\
&\qquad\qquad\qquad\qquad +\tilde W(x|x';\lambda(t))\,p(x',t) \ln\frac{\tilde W(x|x';\lambda(t))}{\tilde W(x'|x;\lambda(t))} \Bigg]\label{eqn_basics_EPRateMaster_1}\\
&= \iint dx dx'\,\Bigg[\tilde W(x|x';\lambda(t))\,p(x',t)\ln\frac{p(x',t)}{p(x,t)} \nonumber\\
&\qquad\qquad\qquad\qquad +\tilde W(x|x';\lambda(t))\,p(x',t) \ln\frac{\tilde W(x|x';\lambda(t))}{\tilde W(x'|x;\lambda(t))} \Bigg]\label{eqn_basics_EPRateMaster_2}\\
&= \iint dx dx'\,\tilde W(x|x';\lambda(t))\,p(x',t)\ln\frac{\tilde W(x|x';\lambda(t))p(x',t)}{\tilde W(x'|x;\lambda(t))p(x,t)},
\end{align}
where we used Eq.~\eqref{eqn_basics_verySimpleME} to get to Eq.~\eqref{eqn_basics_EPRateMaster_1} and Eq.~\eqref{eqn_basics_integralWtilde} to include the term $\ln p(x',t)$ in Eq.~\eqref{eqn_basics_EPRateMaster_2}.

Again, `diagonal terms' cancel, permitting us to write:
\begin{align}
\dot{\Sigma} &= \iint dx dx'\, W(x|x';\lambda(t))\,p(x',t)\ln\frac{ W(x|x';\lambda(t))p(x',t)}{ W(x'|x;\lambda(t))p(x,t)}\label{eqn_basics_averageEPMarkov}\\
&=  \frac{1}{2}\iint dx dx'\,\Big[ W(x|x';\lambda(t))\,p(x',t)- W(x'|x;\lambda(t))\,p(x,t)\Big]\nonumber\\
&\qquad\qquad\qquad\qquad\qquad\qquad\qquad\qquad\qquad \times\ln\frac{ W(x|x';\lambda(t))p(x',t)}{ W(x'|x;\lambda(t))p(x,t)}\\
&\geq 0.
\end{align}

Once again, this expression allows a transparent interpretation in terms of nonvanishing net probability fluxes $W(x|x';\lambda(t))\,p(x',t)- W(x'|x;\lambda(t))\,p(x,t)$ flowing in the system indicating that it is out of equilibrium.

\section{Hamiltonian and quantum dynamics}\label{sec_basics_HamiltonianDynamics}
So far we have presented the material using the formalism of stochastic dynamics. However, most of the central results can also be obtained by assuming Hamiltonian dynamics for the \emph{system and bath}. That is in fact how Jarzynski originally derived the work relation~\cite{Jarzynski1997} and Kawai \emph{et al.} found the identification of entropy production as time-series irreversibility~\cite{Kawai2007}.

One strong point of this approach is that it readily produces, e.g., the Jarzynski equality and thus a consistent thermodynamic interpretation for a system that is strongly coupled to a heat bath~\cite{Jarzynski2004}; an accomplishment that the stochastic approach seems not to be able to reliably reproduce, especially when it comes to the interpretation of heat (see, e.g., Ref.~\cite{Seifert2016,Talkner2016,Strasberg2017,Miller2017}). 

We will present the basic gist of the derivation for a system that is in weak contact with a reservoir. The Hamiltonian of the joint system shall read:
\begin{align}\label{eqn_basics_jointHamiltonian}
H(\Gamma,\lambda) = H_x(x,\lambda) + H_y(y) + H_i(x,y),
\end{align}
where $\Gamma = \{x,y\}$ is a point in the phase space of the joint system and $x$ and $y$ are the phase-space coordinates of the system and the heat bath, respectively. The work parameter $\lambda$ is assumed to only affect the system part of the Hamiltonian, $H_x$, which we identify as the relevant internal energy of the system\footnote{This is in fact the crucial difference between strongly and weakly coupled systems. For a strongly coupled system, one should instead use the \emph{Hamiltonian of mean force} (see, e.g, Refs.~\cite{Kirkwood1935,Gelin2009,Jarzynski2017}).}.

In an isothermal process of length $T$ in which the protocol $\lambda(t)$ runs from $\lambda_0$ to $\lambda_T$ and the system state changes from $\Gamma_0$ to $\Gamma_T$, the difference in internal energy of the system proper can then be written as~\cite{Jarzynski2004}:
\begin{align}
\Delta H_x &= H_x(x_T,\lambda_T) - H_x(x_0,\lambda_0)\\
&= \int\limits_0^T dt\, \dot{\lambda} \pd{\lambda} H_x(x(t),\lambda) + \int\limits_0^T dt\, \dot{x} \pd{x} H_x(x,\lambda(t))\\
&= \qquad\qquad w\qquad\qquad\; + \qquad q,
\end{align}
where the work $w$ results from the contribution to the energy change that comes from modifying the protocol value $\lambda$. Note that this definition agrees with the one found by Sekimoto for Langevin dynamics [Eq.~\eqref{eqn_basics_workLangevinBasic}].

Interestingly, the work is also the difference in the complete Hamiltonian:
\begin{align}
w &= \int\limits_0^T dt\, \dot{\lambda} \pd{\lambda} H_x(x(t),\lambda(t))\\
&= \int\limits_0^T dt\, \dot{\lambda} \pd{\lambda} H(\Gamma(t),\lambda(t))\\
&= \int\limits_0^T dt\, \frac{d}{dt} H(\Gamma(t),\lambda(t))\\
&= H(\Gamma_T,\lambda_T)-H(\Gamma_0,\lambda_0),
\end{align}
where we used the fact that the complete Hamiltonian only depends on $\lambda$ through $H_x$ and the fact that the total derivative of the Hamilton function along the solution of Hamilton's equations is given by its partial derivative.

The free energy difference between initial and final state is given by
\begin{align}
\Delta F &:= F(\lambda_T) - F(\lambda_0)\\
&=  \kB T \ln\frac{Z(\lambda_0)}{Z(\lambda_T)},
\end{align}
where $Z(\lambda) := \int d\Gamma \exp{\left[-{H(\Gamma ,\lambda)}/{\kB T}\right]}$ is the canonical partition function. We thus obtain the dissipation:
\begin{align}
\sigma &:= \frac{w - \Delta F}{\kB T}\\
&= \ln{\left[ \frac{1}{Z(\lambda_0)} \exp{\left(-\frac{H(\Gamma_0,\lambda_0)}{\kB T}\right)}\right]} -  \ln{\left[ \frac{1}{Z(\lambda_T)} \exp{\left(-\frac{H(\Gamma_T,\lambda_T)}{\kB T}\right)}\right]}\\
&= \ln\frac{\rho(\Gamma_0)}{\rho(\Gamma_T)},\label{eqn_basics_HamiltonianDissipation}
\end{align}
where $\rho(\Gamma_0)$ and $\rho(\Gamma_T)$ denote the initial and final equilibrium densities in phase-space, respectively\footnote{Notice the similarity to our previous definition of entropy production in Eq.~\eqref{eqn_basics_defTrajectoryEP}. Instead of weighing the probability that a specific trajectory occurs in the forward process against that of the reverse process, it measures the change in phase-space density between initial and final system state.}. Because $\Gamma_T$ follows from $\Gamma_0$ via integration of the equations of motion, irreversibility is solely encoded in the \emph{initial condition} of the process. It is important to stress that $\rho(\Gamma_T)$ is the \emph{equilibrium} phase-space density evaluated at the final (nonequilibrium) coordinates $\Gamma_T$ directly after the driving ceases and thus \emph{before} the system can relax to equilibrium.

We can now derive the Jarzynski equality~\eqref{eqn_basics_JarzynskiEquality} by noting
\begin{align}
\left\langle \exp{\left(-\frac{w-\Delta F}{\kB T}\right)} \right\rangle &= \iint d\Gamma_0 d\Gamma_T\, \rho(\Gamma_T|\Gamma_0)\,\rho(\Gamma_0)\, \exp{\left(-\sigma\right)}\\
&= \iint d\Gamma_0 d\Gamma_T\, \rho(\Gamma_T|\Gamma_0)\,\rho(\Gamma_T),
\end{align}
where $\rho(\Gamma_T|\Gamma_0) = \delta\big( \Gamma_T - \mathcal{U}(\Gamma_0)\big)$ is the transition probability, with $U(\Gamma)$ denoting the time evolution operator in the interval $[0,T]$. We thus get the Jarzynski equality:
\begin{align}
\left\langle \exp{\left(-\frac{w-\Delta F}{\kB T}\right)} \right\rangle &= \iint d\Gamma_0 d\Gamma_T\, \delta\big( \Gamma_T - \mathcal{U}(\Gamma_0)\big)\,\rho(\Gamma_T)\\
&= \iint d\Gamma_0 d\Gamma_T\, \delta\big( \Gamma_0 - \mathcal{U}^{-1}(\Gamma_T)\big)\,\left|\frac{d \Gamma_0}{d \Gamma_T} \right|\,\rho(\Gamma_T)\\
&= 1,
\end{align}
where the Jacobian is unity, $\left|\frac{d \Gamma_0}{d \Gamma_T} \right|=1$, as a consequence of Liouville's theorem.

Along similar lines we can take Eq.~\eqref{eqn_basics_HamiltonianDissipation} and note that another consequence of Liouville's theorem is that the probability $\rho(\Gamma_0)$ is the same as the probability $p[\Gamma(\cdot)]$ of the entire phase-space trajectory $\Gamma(\cdot)$ from $0$ to $T$. Similarly, since the Hamiltonian is even in the momenta, the second phase-space density can be rewritten as:
\begin{align}
\rho(\Gamma_T) = \rho(\bar\Gamma_T),
\end{align}
where $\bar{\Gamma}_T$ is the time-reversal state, i.e., all momenta are flipped. Consequently, this probability is the same as the probability $\bar{\rho}(\bar{\Gamma}(\cdot))$ of  an entire trajectory originating in $\bar{\Gamma}_T$ and terminating in $\bar\Gamma_0$ in which the protocol $\lambda(t)$ is time-reversed. We thus obtain
\begin{align}
\sigma = \ln\frac{\rho[\Gamma(\cdot)]}{\bar\rho[\bar\Gamma(\cdot)]},
\end{align}
which again leads to Eq.~\eqref{eqn_basics_DissipationKullbackLeibler} for the average dissipation~\cite{Kawai2007}:
\begin{align}\label{eqn_basics_DissipationKullbackLeiblerHamiltonian}
\Sigma = D_{\rm KL}\Big[ p[\Gamma(\cdot)]\,||\,\bar{p}[\bar{\Gamma}(\cdot)] \Big].
\end{align}
Thus, amazingly, this relation also holds for Hamiltonian dynamics when the start and endpoint of the process are in equilibrium.

Combining Eq.~\eqref{eqn_basics_DissipationKullbackLeibler}, or Eq.~\eqref{eqn_basics_DissipationKullbackLeiblerHamiltonian}, with the average of the Crooks fluctuation theorem [Eq.~\eqref{eqn_basics_CrooksFT}],
\begin{align}
\frac{\langle w \rangle-\Delta F}{\kB T} = D_{\rm KL}\Big[ p(w)\,||\,\bar{p}(-w) \Big],
\end{align}
we obtain:
\begin{align}
D_{\rm KL}\Big[ p[\Gamma(\cdot)]\,||\,\bar{p}[\bar{\Gamma}(\cdot)] \Big] = D_{\rm KL}\Big[ p(w)\,||\,\bar{p}(-w) \Big].
\end{align}
This relation shows that ``the exact dissipation is revealed by any set of variables that contains the statistical information about the work''~\cite{Gomez-Marin2008a}, which gives an insight into why even in the Hamiltonian formalism, which captures all the information about the system and the environment, we only need to measure the system variables to calculate the dissipation and evaluate fluctuation theorems. 

Let us finally mention some aspects of the ongoing effort to extend the results to the quantum regime. One conceptually challenging issue is the fact that there is no equivalent of a classical trajectory in quantum mechanics, which invalidates most of the approaches we mentioned before. This can be partly remedied by making measurements at the beginning and end of a unitary quantum evolution, or, alternatively, the dynamics of an open quantum system in which coherences are neglected. This enables, among others, the derivation of fluctuation theorems (see, e.g., Refs.~\cite{Talkner2007,Engel2007,Jarzynski2015,Funo2018}, the reviews in Refs.~\cite{Esposito2009,Campisi2011}, and a recent book~\cite{Binder2018}).

\mypart{Interacting degrees of freedom}{
	In this second part of the thesis we consider the interaction of multiple subsystems. We first treat the heat bath as one subsystem interacting with the system proper. Next, we consider stochastic systems and disentangle their mutual influences on each other. This perspective is shown to clarify the role of information in thermodynamics. The aims of this part are to:
	\begin{itemize}
		\item Show how a \emph{microscopically reversible} interaction between system and reservoir leads to proper thermalization of the system.
		\item Study the special setup of a \emph{collisional bath} from this perspective.
		\item Introduce the main features of \emph{information thermodynamics}, especially the thought experiment of \emph{Maxwell's demon}.
		\item Clarify how information enters the theory of thermodynamics.
		\item Show how one can disentangle the interaction of different subsystems using information thermodynamics.
		\item Present some of the setups illustrating the thermodynamics of information from one global perspective.
	\end{itemize}
}

\chapter{The reservoir as a subsystem}\label{chap_int_reservoirSubs}
This chapter presents a perspective already encountered in Sec.~\ref{sec_basics_HamiltonianDynamics} in the context of Hamiltonian dynamics for stochastic thermodynamics: The system under study and the thermal reservoir jointly form \emph{one big system}. Interestingly, this viewpoint provides a neat explanation of the local detailed balance relation, which is the key ingredient in the thermodynamic interpretation of stochastic processes (c.f. Secs.~\ref{sec_basics_detailedBalance}~and~\ref{sec_basics_brokenDetailedBalance}).

Any thermodynamically sound description of a physical system must include the possibility of thermalization, i.e., the system should take on the equilibrium distribution set by the temperature of its thermal environment. We already demanded thermalization when deriving the local detailed balance relations in Sec.~\ref{sec_basics_brokenDetailedBalance}.

An important component ensuring thermalization is \emph{microscopic reversibility}, i.e., the evolution equations of the \emph{system and environment} are symmetric with respect to time-reversal. Since both Hamiltonian and unitary quantum dynamics are time-reversal invariant, this warrants thermalization in these scenarios. As we will see, the situation becomes a bit obscure when considering a localized system interacting with its environment via collisions, so-called \emph{collisional baths}. These are especially relevant in semi-classical scattering setups which makes it worthwhile to study collisional baths in detail.

\section{Detailed balance in equilibrium ensembles}
A derivation of detailed balance from Hamiltonian dynamics is found in Sec.~V.6 of van Kampen's book~\cite{vanKampen2007}. We take a different approach: Assume that detailed balance holds and evaluate its consequences. In the following, we will make no assumptions about the underlying dynamics (Hamiltonian, Markovian, or quantum) and simply evaluate the consequences of the interplay between detailed balance and micro-reversibility.

Let $\Gamma$ and $\Gamma'$ be two phase-space coordinates (or normalized quantum states). Microscopic reversibility then implies:
\begin{align}\label{eqn_int_microReversibility}
\rho(\Gamma'|\Gamma;\Delta t) = \rho(\bar\Gamma|\bar\Gamma';\Delta t),
\end{align}
where $\rho(\Gamma'|\Gamma;\Delta t)$ is the conditional probability of finding the system in $\Gamma'$ when it started in $\Gamma$ and evolved for a time $\Delta t$. The bar indicates the time-reversed state. Note that in the case of Hamiltonian dynamics these probabilities are delta functions. 

The formal proof that classical Hamiltonian dynamics and unitary quantum dynamics are micro-reversible can be found in Sec.~3 of our article in Ref.~\cite{Ehrich2019a} reprinted below.

Macroscopic phenomena appear irreversible while the laws governing the microscopic dynamics are time-reversible. This fact has puzzled statistical physicists for a long time, even though Boltzmann and Maxwell understood the effect quite well. Albeit a fascinating topic by itself, we will refrain from discussing this issue and instead refer the interested reader to the insightful articles by Lebowitz~\cite{Lebowitz1993,Lebowitz1993a} and the references therein.

In our article (Ref.~\cite{Ehrich2019a}) we show that the detailed balance condition supplemented with micro-reversibility ensures that the stationary distribution $\rho(\Gamma)$ is the uniform distribution of the microcanonical ensemble.

We will in the following show how the local detailed balance relation in Eq.~\eqref{sec_basics_brokenDetailedBalance} results from micro-reversibility.

\subsection{Repeated interactions}\label{sec_int_detailedBalanceFromRepeatedInt}
The equilibrium distribution of a system in weak contact with a large reservoir is the Boltzmann distribution. The derivation can be found in standard textbooks, e.g., Chap.~16 of Ref.~\cite{Callen1985}. In large reservoirs that are separable into noninteracting parts (e.g., individual gas molecules, electromagnetic or vibrational modes, etc.), this distribution also holds for the individual parts. 

\begin{figure}[ht]
	\centering
	\includegraphics[width =0.33 \linewidth]{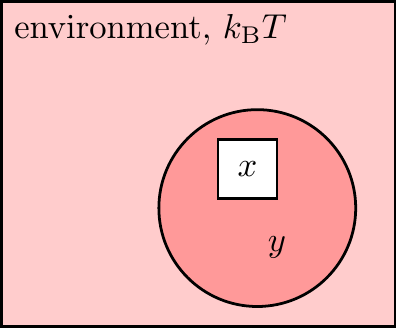}\hspace{1cm}\includegraphics[width =0.5 \linewidth]{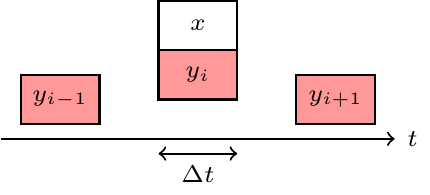}
	\caption{Left: Schematic setup of a system $x$ interacting with a part of an environment $y$. Right: The system is repeatedly interacting with fresh copies of the environment for a time $\Delta t$.}
	\label{fig_int_setupSystemEnvironment}
\end{figure}

Let us now assume that the system's degrees of freedom $x$ are only interacting with one part $y$ of the environment. This interaction shall happen repeatedly in such a way that, in every time interval $\Delta t$, the system is coupled to a fresh equilibrium copy of the environment that is uncorrelated with the system state. The schematics of the setup are depicted in Fig.~\ref{fig_int_setupSystemEnvironment}.

This setup yields Markovian dynamics for the system state as the environment is \emph{memoryless}: each fresh copy of the environment is uncorrelated with the system state. This assumption is also the underpinning of Boltzmann's \emph{Stoßzahlansatz}~\cite {Boltzmann1872} for modeling gas molecule collisions as completely uncorrelated, independent events. The \emph{repeated-interaction framework} is very versatile and also allows the inclusion of work and information reservoirs~\cite{Strasberg2017a} and helps in analyzing quantum-thermodynamic setups (see, e.g., Refs.~\cite{Scarani2002,Karevski2009,DeChiara2018}).

In Ref.~\cite{Ehrich2019a} we show how local detailed balance results from such a repeated interaction framework when the evolution conserves the total energy and the interactions are completely reversible:
\begin{align}\label{eqn_int_microReversibilityXY}
\rho(x',y'|x,y;\Delta t) = \rho(\bar x,\bar{y}|\bar{x}',\bar{y}';\Delta t).
\end{align}
We assume that the Hamiltonian is even in the momenta and reproduce the proof in the following.

The probability for a transition $x \to x'$ can be calculated from the transition probability for the combined system by marginalization:
\begin{align}
\rho(x'| x;\Delta t) = \iint dy dy'\, \rho(x',y'|x,y;\Delta t)\, \rho(y|x;t),
\end{align}
where $\rho(y|x;t)$ is the conditional probability of the environment $y$ given the system state $x$ at time $t$. However, since we assumed that the environment is initially uncorrelated with the system and in equilibrium, we can write
\begin{align}\label{eqn_int_defTransitionProbX}
\rho(x'|x;\Delta t) = \iint dy dy'\, \rho(x',y'|x,y;\Delta t)\, \rho_{\rm eq}(y).
\end{align}

Similarly, the probability for the reverse transition reads:
\begin{align}
\rho(\bar x|\bar x';\Delta t) = \iint d\bar y' d \bar y\, \rho(\bar x,\bar y|\bar x',\bar y';\Delta t)\, \rho_{\rm eq}(\bar y').
\end{align}
Using Eq.~\eqref{eqn_int_microReversibilityXY}, we can rewrite this probability as follows:
\begin{align}
\rho(\bar x|\bar x';\Delta t) = \iint d\bar y' d \bar y\, \rho(x',y'|x,y;\Delta t)\, \rho_{\rm eq}( y) \, \frac{ \rho_{\rm eq}(\bar y')}{\rho_{\rm eq}( y)}.
\end{align}

Since we know the equilibrium distribution of the environment, we find
\begin{align}
\frac{ \rho_{\rm eq}(\bar y')}{\rho_{\rm eq}( y)} = \exp{\left(-\frac{H_y(y')-H_y(y)}{\kB T}\right)},
\end{align}
because $H_y(\bar{y}) = H_y(y)$, where $H_y(y)$ is the part of the Hamiltonian belonging to the $y$-subsystem alone. Since the evolution conserves the total energy, we have:
\begin{align}
H_y(y') - H_y(y) = H_x(x) - H_x(x'),
\end{align}
where $H_x(x)$ is the Hamiltonian of the $x$-dynamics.

Compare this to the Hamiltonian used in Eq.~\eqref{eqn_basics_jointHamiltonian} which also contains an interaction term $H_i(x,y)$. Between interaction intervals this part vanishes since system and environment are decoupled.

We therefore obtain:
\begin{align}
\rho(\bar x|\bar x';\Delta t) = \exp{\left(-\frac{H_x(x)-H_x(x')}{\kB T}\right)} \, \iint d\bar y' d \bar y\, \rho(x',y'|x,y;\Delta t)\, \rho_{\rm eq}( y),
\end{align}
which, with Eq.~\eqref{eqn_int_defTransitionProbX}, leads to the detailed balance condition
\begin{align}
\frac{\rho(x'|x;\Delta t)}{\rho(\bar x|\bar x';\Delta t)} = \exp{\left(-\frac{H_x(x')-H_x(x)}{\kB T}\right)}
\end{align}
found, e.g., in Eq.~\eqref{eqn_basics_relationTransitionRates} (Note that previously we considered transition rates, did not deal with time-reversal states, and set $H(x) =: V(x;\lambda)$).

This is an alternative derivation of the detailed balance condition in which the fact that the system should relax to the thermal state does not enter \emph{a priori}, but results from a micro-reversible interaction between system and environment. However, this comes at the cost of assuming repeated interactions of the system and memoryless copies of the environment. Importantly, it shows that conservation of energy alone is not sufficient to ensure thermalization. It must be supplemented with micro-reversibility.

\section{Collisional baths}
One setup that perfectly fits the repeated-interaction framework is that of a \emph{collisional bath}: The environment is represented by a large container filled with noninteracting ideal gas molecules of mass $m$ with velocities $v$ distributed according to the Maxwell-Boltzmann distribution, i.e.,
\begin{align}
\rho_{\rm eq}(v) = \sqrt{\frac{m}{2\pi \kB T}}\, \exp{\left(-\frac{m v^2}{2 \kB T}\right)},
\end{align}
in one dimension.

\begin{figure}[ht]
	\centering
	\includegraphics[width =0.7 \linewidth]{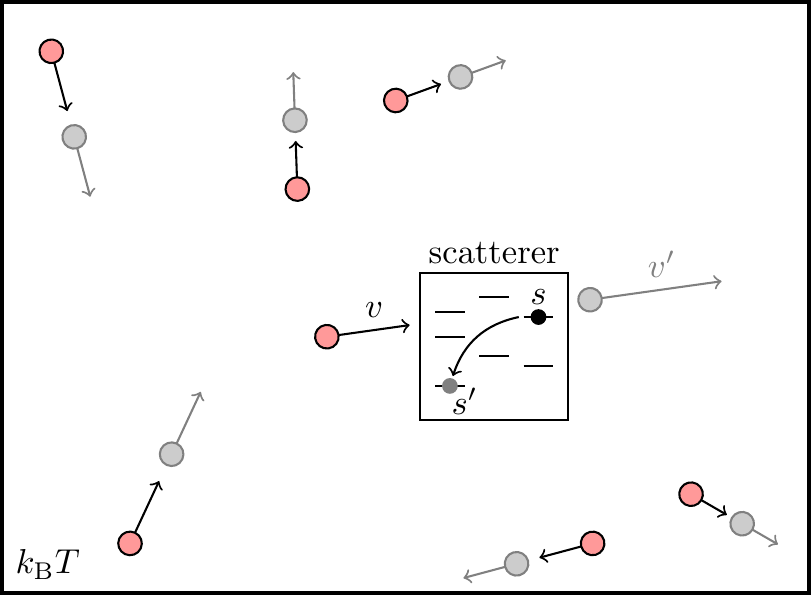}
	\caption{Setup of a system (or scatterer) in a collisional bath consisting of ideal gas particles. A snapshot of the setup after a short time interval is shown in gray.}
	\label{fig_int_setupCollisionalBath}
\end{figure}

The system, or \emph{scatterer}, is fixed in space and possesses internal degrees of freedom $s$ which can evolve upon interaction with the environment particles. By neglecting the possibility of re-collisions, one ensures that the system is only interacting with fresh, equilibrium copies of the environment. Figure~\ref{fig_int_setupCollisionalBath} shows the setup schematically.

Collisional baths have been a testbed of thermalization of classical systems, e.g., modifications of the Lorentz gas with freely rotating disks, which can exchange energy with the gas through inelastic collisions~\cite{Rateitschak2000,Mejia-Monesterio2001,Collet2009,Collet2009a,DeBievre2016}. Interestingly, some of the proposed models (such as the one in Ref.~\cite{Collet2009a}) do not thermalize in that setting. Collisional baths can also be analyzed semi-classically using Wigner distributions or wave packets (see our derivation in the appendix of the article in Ref.~\cite{Ehrich2019a} and Refs.~\cite{Littlejohn1986,Zachos2005}).

As we show in Sec.~4 of our article, a naive treatment based only on the velocities of the reservoir particles automatically leads to a violation of the micro-reversibility condition. Confusingly, the prefactor caused by this apparent violation is exactly compensated by the effusion prefactor, which is a result of the velocity distribution of the particles that are interacting with the system not being the Maxwellian distribution but the effusion distribution
\begin{align}
\rho_{\rm eff}(v) \propto |v|\,\rho_{\rm eq}(v).
\end{align}

Correctly accounting for the effusion prefactor becomes important, e.g, when one wants to simulate such a process and needs the particles to \emph{thermalize} at the wall, i.e., draw a new random velocity from that distribution which ensures that the particle velocities follow Boltzmann statistics (see Refs.~\cite{Tehver1998,Hoppenau2013} for examples).

If one accounts for the effusion prefactor, all calculations are correct; but for the wrong reasons, which is innocent enough in simple scenarios. However, it can lead to wrong conclusions in more complicated setups as we show in our article.

Ultimately, a complete analysis shows that all canonical coordinates, i.e., momenta \emph{and position} of the particles need to be considered. This restores micro-reversibility and thus ensures thermalization of the system. Applying this analysis to the model system~\cite{Collet2009a} that does not thermalize reveals that its apparently sensible collision rules violate micro-reversibility. Lastly, we show how this lacking thermalization can be exploited to build a heat pump able to transport heat against a temperature gradient, thereby illustrating that the model is unphysical.

\section{Article: [\emph{Physica A}, 122108 (2019)]}
The following article is reprinted from J. Ehrich, M. Esposito, F. Barra, and J. M. R. Parrondo, arXiv1904.07931. Journal reference [J. Ehrich, M. Esposito, F. Barra, and J. M. R. Parrondo, ``Micro-reversibility and thermalization with collisional baths,'' \emph{Physica A} \textbf{552}, 122108 (2020)] (Ref.~\cite{Ehrich2019a}).

The bracketed numbers provide a continuous pagination.
\cleardoublepage
\pagestyle{pdfStyle}
\includepdf[pages=-,pagecommand={},scale=0.97]{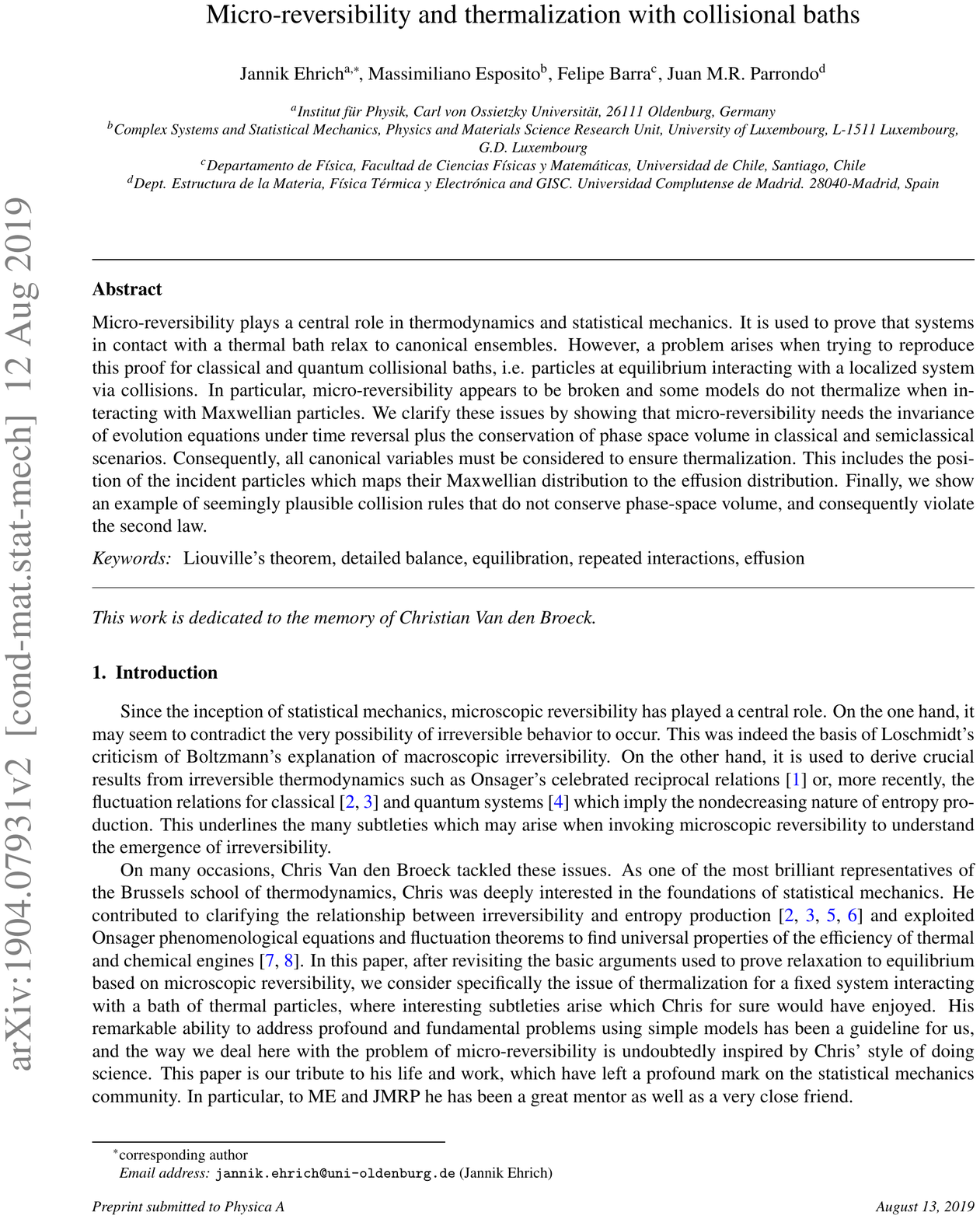}
\pagestyle{thesisStyle}

\chapter{Thermodynamics of information}\label{chap_informationThermodynamics}
We will now turn to one of the most appealing areas of stochastic thermodynamics: \emph{information thermodynamics}. It will become clear how information can be thought of as a thermodynamic resource, just like heat and work. After presenting the key results in this area, we will elucidate how information thermodynamics can be seen as stemming from the interplay of different interacting subsystems. This more global view then allows to study the multiple facets of the role information plays in thermodynamics from one unifying framework.

Much of the following material is based on the review by Maruyama \emph{et al.}~\cite{Maruyama2009} and the recent one by Parrondo \emph{et al.}~\cite{Parrondo2015}.

\section{Maxwell's demon}
The question about the role of information in statistical physics is quite old, dating back to Maxwell himself. In the last chapter of his book \emph{Theory of Heat}~\cite{Maxwell1872}, he points out a seeming problem with the second law. He envisions a container (a \emph{vessel}) filled with an ideal gas and goes on to say: ``Now let us suppose that such a vessel is divided into two portions, A and B, by a division in which there is a small hole, and that a being, who can see the individual molecules, opens and closes this hole, so as to allow only the swifter molecules to pass from $A$ to $B$, and only the slower ones to pass from $B$ to $A$. He will thus, without expenditure of work, raise the temperature of $B$ and lower that of $A$, in contradiction to the second law of thermodynamics''~\cite{Maxwell1872}.

\begin{figure}[ht]
	\centering
	\includegraphics[width =0.58 \linewidth]{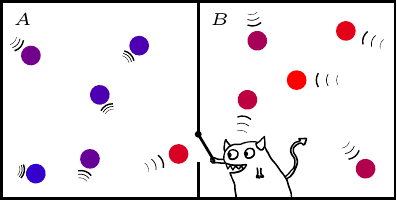}
	\caption{Maxwell's demon is a hypothetical intelligent being sitting at the boundary between two volumes $A$ and $B$ of gas. It operates a small door between them and is thus able to sort molecules according to their speeds, thereby raising the temperature in volume $B$ and lowering it in volume $A$ without expenditure of work.}
	\label{fig_int_maxwellsDemon}
\end{figure}

The \emph{being} was named \emph{intelligent demon} by William Thomson~\cite{Thomson1874} (later Lord Kelvin). Fig~\ref{fig_int_maxwellsDemon} depicts the setup as Maxwell envisioned it. The problem might at that time have seemed of purely academic interest; far away from any experimental reality due to the requirement of having to observe individual molecules. Nonetheless, this riddle has kept generations of theoreticians busy as can be gauged, e.g., from the collection of historical papers in Ref.~\cite{Leff1990} and the recent historical review by Rex~\cite{Rex2017}.

It could be argued that Maxwell overshot with his thought experiment, as a simple spring-loaded trap door would have done the trick. This was pointed out by Smoluchowski~\cite{Smoluchowski1912}. The trapdoor would act as a \emph{one-way} valve letting particles only pass in one direction and thereby raising the pressure in one compartment while lowering it in the other, apparently without expenditure of work.

Dispensing with the intelligent being smartly operating the trap door leads us to the first key towards resolving the paradox: The \emph{demon} itself must be incorporated into the energy and entropy balance of the setup. Smoluchowski observes that after a while the door would thermalize (see the previous chapter on collisional thermalization) and therefore, due to thermal fluctuations, open spontaneously, thereby letting molecules pass into the other direction, eventually equalizing the pressure. This has even been verified using computer simulations~\cite{Skordos1992}, showing that the demon could only work when heat is removed from the door, which makes the device operate akin to a conventional heat engine.

Smoluchowski also envisions a \emph{ratchet} device consisting of an axle with vanes on one side and a gearwheel on the other with a pawl that only allows it to turn one way. Gas molecules hitting the vane can then slowly wind an attached weight upwards. This device was analyzed by Feynman in his \emph{Lectures on Physics}~\cite{Feynman1963} and although his analysis is flawed~\cite{Parrondo1996}, he shows that this device, too, can only work as a heat engine, i.e., the vanes have to be kept at a higher temperature than the ratchet. Nonetheless, a slight modification of such a setup, e.g., by periodically switching on and off an asymmetric potential, allows such devices to achieve directed motion, albeit, again, without breaking the second law~\cite{Reimann2002}.

\subsection{The Szil\'ard engine}
\begin{figure}[ht]
	\centering
	\includegraphics[width =1 \linewidth]{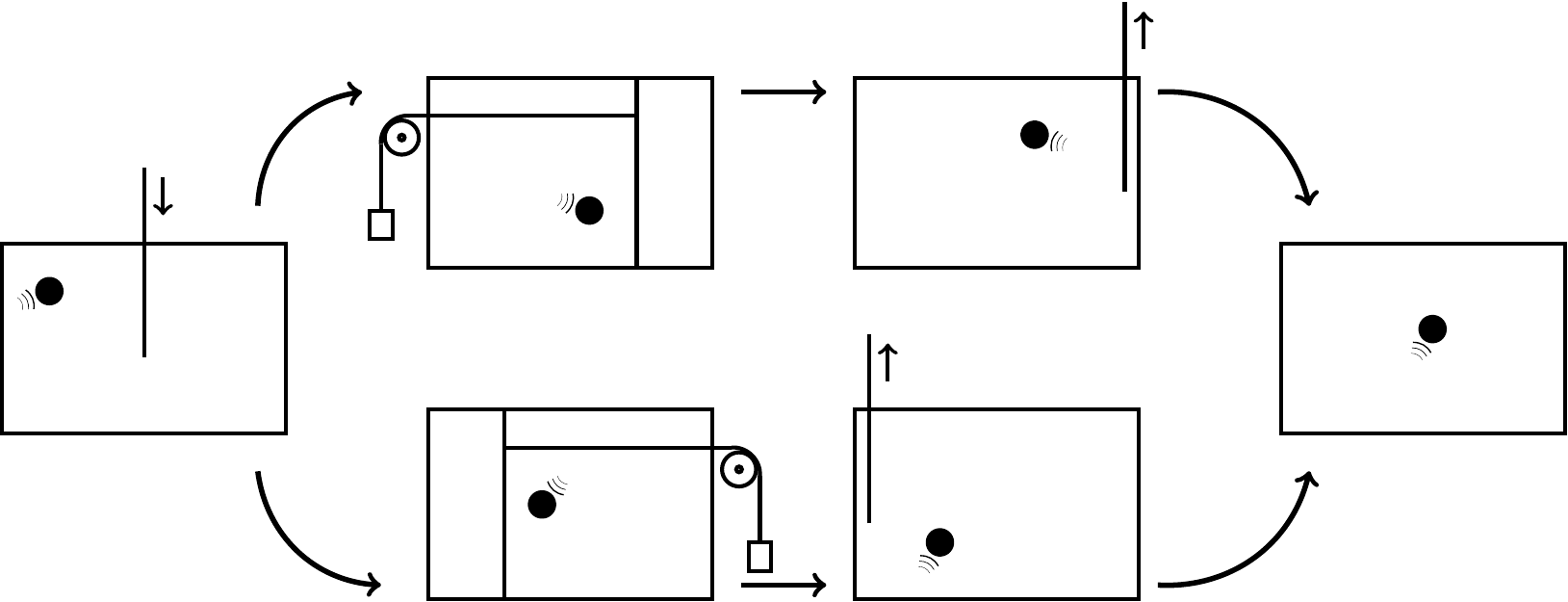}
	\caption{Schematic setup of the Szil\'ard engine. The working principle is explained in the main text.}
	\label{fig_int_szilardEngine}
\end{figure}

A simplified demon that allows to asses what kind of \emph{intelligence} is needed in order for Maxwell's original demon to work goes back to L\'eo Szil\'ard~\cite{Szilard1929}. The setup is depicted in Fig.~\ref{fig_int_szilardEngine}. Initially, a box contains one ideal gas molecule in equilibrium\footnote{\emph{Equilibrium} means that the particle's velocity follows Boltzmann statistics. One may think of it thermalizing at the walls of the box.} at some fixed temperature. Next, a wall is inserted into the container, forcing the molecule to stay on one side of the container. Then the one-molecule gas, which exerts a certain pressure on the wall via collisions, is left to isothermally and reversibly expand, thereby doing work on a weight attached to the wall via a frictionless pulley. When the expansion is finished, the slightly lifted weight is detached and the wall removed.

During the process, the engine delivers a work $W =- \kB T  \ln{2}$ because the volume of the gas doubles during expansion. Since the process is isothermal, on average, the same amount of heat will be flowing into the gas: $Q = \kB T \ln{2}$. This means that the engine perfectly converts heat into work, in violation of the second law.

Although a crude theoretical model, there is nothing wrong with the analysis \emph{in principle}: The wall can \emph{in principle} be inserted and removed without friction. Moving the weight from left to right and vice versa between different cycles does not \emph{in principle} require spending energy. Thermalizing the particle at the walls is possible \emph{in principle}. It would be bad style for a theoretician to dismiss the model on practical grounds. So what is wrong here?

The key point that is missing in the analysis is that a \emph{measurement} has to be performed after the wall has been inserted to learn on which side of the container the particle is located and to attach the weight \emph{at the correct side}. This measurement generates \emph{information} which accumulates over many cycles of engine operation. However, engines should perform \emph{cyclically}, i.e. this information has to be \emph{erased} at some point. This restores the second law, as we shall see shortly. 

Lately, the physics of Maxwell's demon and Szil\'ard's engine have come into the realm of experimental reality. Some laboratory realizations are found in Refs.~\cite{Toyabe2010,Koski2014,Koski2014a,Pekola2016}. 

\subsection{Nature of information}
In the abstract sense, information is a correlation between the states of different systems. When statistics are involved, it can be quantified in the information-theoretic sense, i.e., using mutual information (see. Sec.~\ref{sec_basics_jointCondMutualEntropy}).

From a physical perspective, information needs to be stored somewhere. That means that some physical system must be in a distinct (and stable) state associated with the state of the external world. For example: In the case of the Szil\'ard engine, a computer memory of one bit could store the measurement outcome: $0$ for left, $1$ for right.

A complete analysis of the process builds on the works of Brillouin~\cite{Brillouin1951}, Landauer~\cite{Landauer1961,Landauer1991}, and Bennett~\cite{Bennett1981}. It goes as follows: The engine must run cyclically, so the \emph{statistical state} of the engine and the memory must be the same at the beginning and end of the process. We thus begin with the memory being in one standard state, say $0$. Note that all of the following steps happen quasi\-statically, so as to ensure that a treatment using equilibrium thermodynamics is correct.

The wall gets inserted and the measurement is performed. Interestingly, this measurement can be performed reversibly, even if one has to use such strange contraptions as the one imagined by Bennett~\cite{Bennett1987}.

The measurement perfectly correlates the memory with the position of the particle, thereby doubling the phase-space volume accessible to the memory, which translates to an average entropy increase of $\ln{2}$ in the memory (cf. Sec.~\ref{sec_basics_entropy}). At the same time, the entropy in the engine is halved because, as soon as we learn the particle's position, its available phase-space volume is cut in half\footnote{This is a difficult point and perhaps a matter of philosophy rather than physics: Of course nothing is changed about the system at this point. It is only our knowledge of the \emph{statistical} state of the system that is affected. This effectively replaces the \emph{marginal entropy} [Eq.~\eqref{eqn_basics_defAverageEntropy}] of the system  with the \emph{conditional} entropy [Eq.~\eqref{eqn_basics_defAverageConditionalEntropy}] of the system given the measurement outcome. In the next section we will formalize this argument. A nice illustration of this effect can be found in Ref.~\cite{Granger2016}. There, the position $x$ of a colloidal particle trapped in a harmonic potential is measured ($y$) with Gaussian measurement errors. Using Bayes's theorem, the \emph{statistical state} $p(x|y)$ is determined and the trapping potential adjusted, such that this distribution becomes the equilibrium distribution. Here too, on average, no heat is dissipated in the process.}. The net effect thus far is zero global entropy change.

Now, work is done reversibly, the weight is lifted, heat of the amount $\kB T \ln{2}$ flows into the gas while its entropy increases by $\ln{2}$. Again, there is no net entropy change. Removing the wall has no effect on the entropy and energy balance. So far it looks as if we have converted heat into work without entropy increase.

However, now the memory must be reset to its original state $0$. This operation is \emph{logically} irreversible, i.e., no matter what the current state of the memory, the final state is $0$ and afterwards there is no way of recovering the original memory contents. Landauer argued~\cite{Landauer1961} that this process requires heat dissipation, at least as much as $\kB T$ times the entropy reduction achieved by the reset.

\begin{figure}[ht]
	\centering
	\includegraphics[width =0.85 \linewidth]{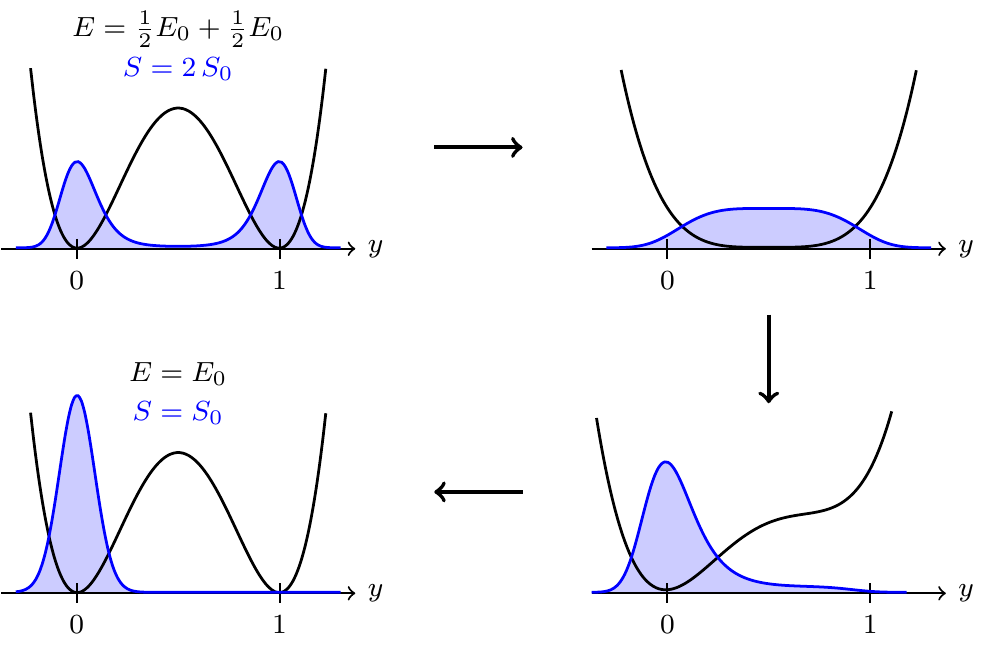}
	\caption{Example reset process. The memory is a Brownian particle in a bistable potential (black lines) with well defined metastable states. The equilibrium distribution is indicated by the blue curves. The memory is reset to the default state ($0$) by quasistatically modifying the potential landscape as indicated, thereby doing work on the particle. This results in the particle being trapped in the left well at the end of the process. The average internal energy $E$ does not change during reset, but the entropy $S$ is halved. This reset process has been studied experimentally in Refs.~\cite{Dillenschneider2009,Berut2012,Jun2014}.}
	\label{fig_int_resetprocess}
\end{figure}

We can easily see why: The phase space of the memory is halved again, decreasing the entropy by $\ln{2}$. This must be accompanied by a heat flow of at least $\kB T \ln{2}$ out of the memory as per the second law. Where should the required energy come from? The available phase-space of systems does not spontaneously halve! It can only be supplied externally as work done on the memory. Therefore, deleting a two-bit memory requires at least the conversion of $\kB T \ln{2}$ of work into heat. Fig.~\ref{fig_int_resetprocess} shows one example setup of memory erasure.

Consider now the total energy balance: We have converted at best $\kB T \ln{2}$ of heat into work and again \emph{at least} the same amount of work into heat. At best, i.e., when operating reversibly, the demon does \emph{nothing}.

The key to resolving the paradox is the so-called \emph{Landauer principle}, i.e., deleting information $I$ from a memory requires at least $\kB T\,I$ of work to be dissipated as heat. Although strange at first sight, this principle has by now been experimentally tested using memories consisting of colloidal particles~\cite{Berut2012,Jun2014,Gavrilov2016,Gavrilov2017}, classical magnetic bits~\cite{Hong2016,Martini2016}, as well as quantum spins~\cite{Peterson2016}.

\section{General picture: Measurement and feedback}\label{sec_int_generalPicture}
Let us back away from the specific realization of the Szil\'ard engine and take a global view of the process: The setup consists of the \emph{system} $x$ that is interacting with the \emph{memory} $y$. Both are coupled to a heat bath at temperature $T$. Initially system and memory are uncorrelated, so that their average \emph{joint entropy} [Eq.~\eqref{eqn_basics_defAverageEntropy}] is given by:
\begin{align}
S_0[x,y] = S_0[x] + S_0[y].
\end{align}
The operation consists of three steps:
\begin{enumerate}
	\item \emph{Measurement}: The memory becomes correlated with the system, creating the information $I_1[x,y]$ [Eq.~\eqref{eqn_basics_defAverageMutualInformation}]. Importantly, this includes the possibility of measurement errors, which have a negative effect on $I_1[x,y]$. The measurement leaves the marginal entropy $S_0[x]$ of the system unchanged. However, writing the joint entropy in terms of the mutual information [cf. Eqs.~\eqref{eqn_basics_relationsAverageMutualInformation_b}], we obtain
	\begin{align}
	S_1[x,y] &= S_0[x] + S_1[y] - I_1[x,y]\\
	&= S_1[x|y] + S_1[y].
	\end{align}
	Thus, the measurement has two effects: Firstly, it effectively updates the entropy of the system from $S_0[x]$ to $S_1[x|y]$. Secondly, it increases the entropy in the memory from $S_0[y]$ to $S_1[y]$. The second law [$\kB T \Delta S \geq \Delta Q]$ implies that a heat 
	\begin{align}
	\Delta Q_{0\to 1} &\leq \kB T \big( S_1[x,y] - S_0[x,y] \big)\\
	&= \kB T \big(S_1[y] - S_0[y] - I_1[x,y] \big)
	\end{align}
	flows into the system.
	
	It is instructive to apply this to the Szil\'ard engine: The measurement perfectly correlates the engine and the memory and therefore creates an information of $I_1[x,y]= \ln{2}$ while increasing the entropy of the memory from $S_0[y]$ to $S_1[y] = S_0[y] + \ln{2}$, at no cost of heat dissipation: $\Delta Q_{0 \to 1} = \kB T (\ln 2 - \ln 2) = 0$.
	\item \emph{Feedback}: The information in the memory can be used to influence the system. During this step, the entropy in the memory is not affected, thus, afterwards, the joint entropy reads:
	\begin{align}
	S_2[x,y] &= S_2[x] + S_1[y] - I_2[x,y].
	\end{align}
	This step does not have to be perfect and can leave correlations with the system encoded in $I_2[x,y]$. The second law requires that a heat 
	\begin{align}
	\Delta Q_{1\to 2} &\leq \kB T \big( S_2[x,y] - S_1[x,y] \big)\\
	&= \kB T \big(\Delta S_{1\to 2}[x] - \Delta I_{1\to 2}[x,y] \big)
	\end{align}
	flows into the system. Applying this analysis to the Szil\'ard engine, we see that the expansion results in $S_2[x] = S_1[x|y] + \ln{2} = S_1[x] = S_0[x]$, and that $I_2[x,y] = 0$ because in the end the memory is uncorrelated with the final position of the particle. The heat reads $\Delta Q_{1\to 2} = \kB T \ln{2}$.
	\item \emph{Reset}: Finally, everything is reset to the original state, thereby requiring a heat flow of
	\begin{align}
	\Delta Q_{2\to 0} \leq \kB T \Big( S_0[x] - S_2[x] + S_0[y] - S_2[y] + I_2[x,y] \Big).
	\end{align}
	In the case of the Szil\'ard engine this heat is bounded by the memory entropy change: $\Delta Q_{2\to 0} \leq \kB T (S_0[y] - S_2[y]) = - \kB T \ln{2}$.
\end{enumerate}
We see that the second law applies to all steps of the process. A seeming violation of the second law can thus only occur either by not accounting for the memory and only focusing on the system, or by neglecting some steps of the process (e.g., the reset).

One way of making the complete system work as an engine is by performing the erasure at some other temperature $T'< T$ which means that less work needs to be converted into heat for reset than heat is converted into work during feedback. 

Assuming that we have a (near) infinitely big memory, we can also treat the memory itself as an \emph{information reservoir}~\cite{Deffner2013,Barato2014}. An ordered, i.e., low entropy, information reservoir can than be used as \emph{fuel} to convert heat from a heat bath into work at the cost of continuously \emph{writing} information to the information reservoir, i.e., increasing its entropy. These reservoirs often come in form of a \emph{tape}~\cite{Mandal2012,Mandal2013} of binary units that are \emph{athermal}\footnote{A notable exception is presented in Ref.~\cite{Hoppenau2014}.}, only containing entropy and no energy.

\subsection{Information fluctuation relations}\label{sec_int_informationFTs}
Turning to stochastic thermodynamics leads us to consider how the methods described in the first part of the thesis can be applied to information thermodynamics. The central quantity will again be the \emph{trajectory entropy production} in Eq.~\eqref{eqn_basics_defTrajectoryEP}. However, it must be slightly modified as to include the measurement outcome.

\begin{figure}[ht]
	\centering
	\includegraphics[width =0.65 \linewidth]{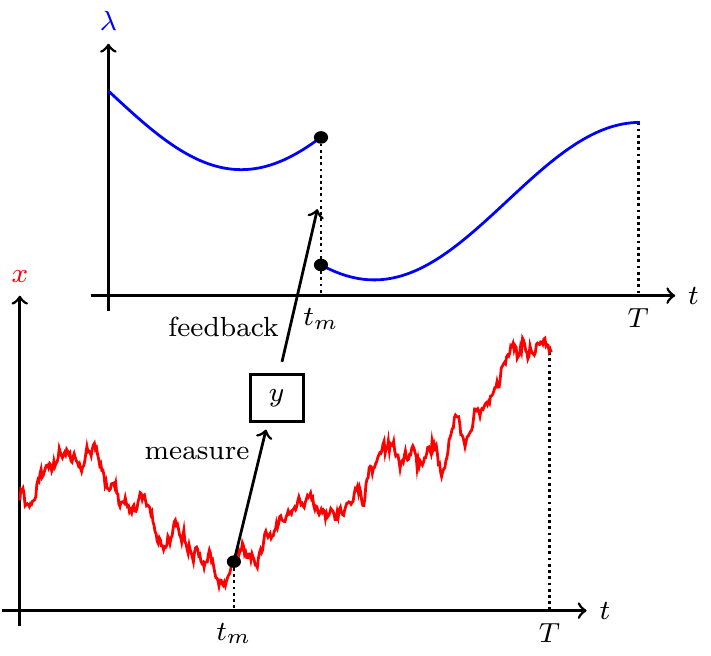}
	\caption{Example measurement-feedback process. The system state is measured at time $t_m$. The measurement is stored in the variable $y$ and used to perform feedback, i.e, change the subsequent driving protocol.}
	\label{fig_int_feedbackProcess}
\end{figure}

Consider the following setting: The stochastic evolution between time $0$ and $T$ of the \emph{system} $x$ shall be influenced by a \emph{protocol} trajectory $\lambda(\cdot)$. At time $t_m$ a measurement of the current system state $x(t_m)$ is performed and the result stored in the variable $y$. The measurement outcome is governed by the distribution $p_{m}(y|x)$. Immediately after the measurement \emph{feedback} is performed on the system, i.e., the protocol trajectory $\lambda_{t_m:T}(\cdot)$ between times $t_m$ and $T$ is chosen based on the measurement outcome (e.g., to extract work from the system). The schematics of this process are depicted in Fig.~\ref{fig_int_feedbackProcess}.

The joint probability of the whole trajectory $x(\cdot)$ and the measurement outcome $y$ is given by:
\begin{align}\label{eqn_int_forwardFeedbackTrajProb}
p\big[x(\cdot),y\big] = p_m\big(y|x(t_m)\big)\; p\big[x(\cdot);\lambda(\cdot)\big],
\end{align}
where $p[x(\cdot);\lambda(\cdot)]$ denotes the path probability of the system trajectory given the protocol. Note that this probability still depends on $y$ through the part $\lambda_{t_f:T}(\cdot)$ of the protocol.

In the reverse process we choose a special structure: \emph{No feedback} is performed. Instead, a measurement outcome is drawn from the marginal distribution
\begin{align}
p_m(y) = \int dx(t_m)\, p_m(y|x(t_m)) p(x(t_m))
\end{align}
of measurement outcomes. Then, the protocol corresponding to this measurement outcome is chosen and executed in reverse. The initial system state is drawn from the time-reverse of the final distribution. We thus obtain for the reverse process:
\begin{align}
\bar{p}\big[\bar{x}(\cdot),y\big] = p_m(y)\; \bar{p}\big[\bar x(\cdot);\bar \lambda(\cdot)\big],
\end{align}
where the overbar indicates time-reversal as in Sec.~\ref{sec_basics_entropyProductionSTD}. Notice that this prescription for the time-reversed process is in principle experimentally feasible. It simply requires that the forward experiment is performed first, so that the statistics of measurement outcomes are known before executing the reverse experiment.

Upon taking the log-ratio of forward and reverse trajectory probabilities, we obtain:
\begin{align}\label{eqn_int_probRatioInfo}
\ln\frac{p\big[x(\cdot),y\big]}{\bar{p}\big[\bar{x}(\cdot),y\big]} = \sigma[x(\cdot)] + i\big(x(t_m),y\big),
\end{align}
where the first term is the logratio of trajectory probabilities and therefore identified with the entropy production for all the reasons laid out in Sec.~\ref{sec_basics_entropyProductionSTD} and we used the definition of mutual information in Eq.~\eqref{eqn_basics_defMutualInformation},
\begin{align}
i(x,y) := \ln\frac{p(x,y)}{p(x)\,p(y)} = \ln\frac{p(y|x)}{p(y)}.
\end{align}

Since $\sigma[x(\cdot)] = \Delta s(x) - \frac{q[x(\cdot)]}{\kB T}$, Eq.~\eqref{eqn_int_probRatioInfo} implies the following integral fluctuation theorem first found by Sagawa and Ueda~\cite{Sagawa2010}:
\begin{align}\label{eqn_int_generalizedJarzynski}
\left\langle \exp{\left[-\Delta s + \frac{q}{\kB T} - i\big(x(t_m),y\big)\right]} \right\rangle = 1,
\end{align}
where the average has to be taken with respect to the system trajectories and the measurement outcomes. For $\kB T\Delta s - q = w- \Delta F$, which holds whenever the process starts and ends in equilibrium (see Sec.~\ref{eqn_basics_noneqWorkRels}), this is known as the \emph{generalized Jarzynski equality}.

Equation~\eqref{eqn_int_generalizedJarzynski} implies the following second law:
\begin{align}\label{eqn_int_generalizedSecondLaw}
\left\langle \Delta s(x) \right\rangle - \frac{\left\langle q[x(\cdot)] \right\rangle}{\kB T} \geq -I\big[x(t_m),y\big],
\end{align}
where $I[x,y] = \langle i(x,y) \rangle$ is the average mutual information [Eq.~\eqref{eqn_basics_defAverageMutualInformation}]. This neatly generalizes the second law to situations with feedback control. Since $I[x,y]\geq 0$, the RHS can become negative, which explains why a Maxwell demon can achieve the apparent violation of the second law.

From this basis the analysis of feedback-controlled systems can be extended to situations with multiple feedback loops~\cite{Cao2009,Horowitz2010,Sagawa2012} including the possibility of feedback control in both directions~\cite{Ponmurugan2010}. Importantly, the informational term $I$ gets modified in such cases to represent \emph{transfer entropy}~\cite{Schreiber2000}.  However, one still recovers the Sagawa-Ueda result in the limiting case of one measurement and feedback loop. 

A notable recent development concerns the fact that to evaluate the above theorems, one needs access to the system state \emph{and} the measurements. Of course, from an experimental standpoint it is desirable to only use the measurements. This is indeed possible as shown in Ref.~\cite{Potts2018}.

\section{Disentangling interacting systems}
We have seen that information thermodynamics is relevant in setups of two subsystems which are interacting with each other. Consider the three steps of measurement, feedback and reset described in Sec.~\ref{sec_int_generalPicture}: To see what is actually going on in the system, we had to disentangle the energy and entropy balance of the individual subsystems, $x$ and $y$, and their interaction.

Arguably, the most illuminating perspective on information thermodynamics comes from considering subsystems interacting with each other; thereby not only producing an energy flow but also an \emph{information flow}.

\subsection{Information flow}
For simplicity, we restrict the discussion to systems with two interacting subsystems, i.e., the state $z := \{x,y\}$ of the joint system is composed of the individual states $x$ and $y$. Further, we require a \emph{bipartite structure}, i.e., a change of state can only affect \emph{one subsystem at a time}. This is a sensible assumption: Ref.~\cite{Horowitz2014} lists four examples relevant to all kinds of systems studied in stochastic thermodynamics. Moreover, it is a \emph{necessary} assumption: Changes in state are the origin of entropy production and heat flow. We need to be able to assign state changes to individual subsystems if we want to disentangle their dynamics.

As an example, let us consider the average rate of entropy production for a bipartite Markovian jump process [Eq.~\eqref{eqn_basics_averageEPMarkov}] as done, e.g., in Refs.~\cite{Horowitz2014,Barato2013,Barato2013a,Diana2014}:
\begin{align}
\dot{\Sigma}_{xy} = \iint dx dx' dy dy'\, W(x,y|x',y') p(x',y';t) \ln\frac{W(x,y|x',y')\, p(x',y';t)}{W(x',y'|x,y)\, p(x,y;t)},
\end{align}
where we suppressed the time-dependence of the transition rates $W(x,y|x',y';t)$ for notational brevity. The bipartite assumption allows us to write the transition rate in the following way:
\begin{align}\label{eqn_int_bipartiteTransitionRates}
W(x,y|x',y') = \begin{cases}
W_x(x,y|x',y) & x\neq x', y = y' \\
W_y(x,y|x,y') & y\neq y', x = x'\\
0 & \rm{otherwise}.
\end{cases}
\end{align}

Because of this splitting, the probability flow $ W(x,y|x',y') p(x',y';t)$ consists of two terms, one due to the $x$-transitions, the other due to the $y$-transitions, which motivates the following splitting of the average rate of entropy production:
\begin{align}
\dot{\Sigma}_{xy} = \dot{\Sigma}_x +\dot{\Sigma}_y,
\end{align}
with 
\begin{subequations}
	\begin{align}\label{eqn_int_partialEPRates_a}
	\dot{\Sigma}_x &:= \iint dx dx' dy \, W_x(x,y|x',y) p(x',y;t) \ln\frac{W_x(x,y|x',y)\, p(x',y;t)}{W_x(x',y|x,y)\, p(x,y;t)} \geq 0\\
	\dot{\Sigma}_y &:= \iint dx dy dy' \, W_y(x,y|x,y') p(x,y';t) \ln\frac{W_y(x,y|x,y')\, p(x,y';t)}{W_y(x,y'|x,y)\, p(x,y;t)} \geq 0, \label{eqn_int_partialEPRates_b}
	\end{align}
\end{subequations}
the \emph{partial} entropy production rates.

To enable a thermodynamic interpretation, we assume that \emph{local detailed balance relations} hold for the individual transition rates [Eq.~\eqref{eqn_basics_localDetailedBalance}]:
\begin{subequations}
	\begin{align}
	W_x(x,y|x',y)\,p_{\rm eq}(x';y) &= W_x(x',y|x,y)\,p_{\rm eq}(x;y)\\
	W_y(x,y|x,y')\,p_{\rm eq}(y';x) &= W_y(x,y'|x,y)\,p_{\rm eq}(y;x),
	\end{align}
\end{subequations}
where $p_{\rm eq}(x|y)$ denotes the equilibrium distribution of $x$ when $y$ is kept fixed and vice versa for $p_{\rm eq}(y;x)$. This implies that we can identify heat flows, because:
\begin{subequations}
	\begin{align}
	\ln\frac{W_x(x,y|x',y)}{W_x(x',y|x,y)} &= -\frac{H_x(x;y)-H_x(x';y)}{\kB T_x}\\
	\ln\frac{W_y(x,y|x,y')}{W_y(x,y'|x,y)} &= -\frac{H_y(y;x)-H_y(y';x)}{\kB T_y},
	\end{align}
\end{subequations}
where $H_x(x;y)$ is the Hamiltonian (or potential) governing the $x$-dynamics and $H_y(y;x)$ the one governing the $y$-dynamics. Note that, depending on the physical setup, these Hamiltonians do not necessarily have to be the same. Similarly, we can allow for different temperatures (or chemical potentials, etc.) pertaining to the different transitions.

We may thus rewrite the partial entropy production rates in Eqs.~\eqref{eqn_int_partialEPRates_a} and~\eqref{eqn_int_partialEPRates_b} in the following way:
\begin{subequations}
	\begin{align}
	\dot{\Sigma}_x = \frac{d}{dt} S[x] - \frac{\dot{Q}_x}{\kB T_x} - \dot{I}_x \geq 0 \label{eqn_basics_AvergaeEPxWithInfoFlow}\\
	\dot{\Sigma}_y = \frac{d}{dt} S[y] - \frac{\dot{Q}_y}{\kB T_y} - \dot{I}_y \geq 0 ,
	\end{align}
\end{subequations}
where
\begin{align}
\frac{d}{dt} S[x] &:= -\frac{d}{dt}\int dx \, p(x,t)\ln p(x,t) \\
&= \iint dx dx' dy \,W_x(x,y|x',y) p(x',y;t) \ln\frac{p(x',t)}{p(x,t)}
\end{align}
is the rate of entropy change of $x$ (and similarly for $\frac{d}{dt} S[y]$),
\begin{align}
\dot{Q}_x := k_{\rm B} T_x\iint dx dx' dy\,W_x(x,y|x',y)\,p(x',y;t) \ln\frac{W_x(x',y|x,y)}{W_x(x,y|x',y)}
\end{align}
is the \emph{heat flow} into system $x$ (and similarly for $\dot{Q}_y$), and
\begin{subequations}
	\begin{align}
	\dot{I}_x := \iint dx dx' dy\, W_x(x,y|x',y)\,p(x',y;t) \ln\frac{p(y|x;t)}{p(y|x';t)}\\
	\dot{I}_y := \iint dx dy dy'\, W_y(x,y|x,y')\,p(x,y';t) \ln\frac{p(x|y;t)}{p(x|y';t)}
	\end{align}
\end{subequations}
are the \emph{information flows} that ``[...] quantify how information sloshes between the two subsystems: When [$\dot{I}_x > 0$], an $x$ jump on average increases the information $I$. In this way, $x$ is learning about or measuring $y$; vice versa, [$\dot{I}_x < 0$] signifies that $x$ is decreasing correlations [...]"~\cite{Horowitz2014}.

Using the bipartite assumption in Eq.~\eqref{eqn_int_bipartiteTransitionRates}, we find that the sum of the information flows gives the rate change in average mutual information~[Eq.~\eqref{eqn_basics_defAverageMutualInformation}]:
\begin{align}
\dot{I}_x + \dot{I}_y &= \iint dx dx' dy dy'\, \delta(y-y')\,W(x,y|x',y')\,p(x',y';t) \ln\frac{p(y|x;t)}{p(y|x';t)}\nonumber\\
&\qquad+ \iint dx dx' dy dy'\,\delta(x-x')\, W(x,y|x',y')\,p(x',y';t) \ln\frac{p(x|y;t)}{p(x|y';t)}\\
&= \iint dx dx' dy dy'\, \delta(y-y')\,W(x,y|x',y')\,p(x',y';t) \ln\frac{p(y'|x;t)}{p(y'|x';t)}\nonumber\\
&\qquad+ \iint dx dx' dy dy'\,\delta(x-x')\, W(x,y|x',y')\,p(x',y';t) \ln\frac{p(x|y;t)}{p(x|y';t)}\\
&= \iint dx dx' dy dy'\,W(x,y|x',y')\,p(x',y';t)\ln\frac{p(x,y;t)\,p(x',t)p(y',t)}{p(x,t)p(y,t)\,p(x',y';t)}\\
&= \frac{d}{dt} I[x,y](t).
\end{align}

Eq.~\eqref{eqn_basics_AvergaeEPxWithInfoFlow} can be thought of as a dynamic version of the generalized second law in Eq.~\eqref{eqn_int_generalizedSecondLaw}:
\begin{align}
\frac{d}{dt} S[x] - \frac{\dot{Q}_x}{\kB T_x} &= \iint dx dx' dy \,W_x(x,y|x',y) p(x',y;t) \ln\frac{W_x(x',y|x,y)\, p(x',t)}{W_x(x,y|x',y)\, p(x,t)} \nonumber\\
&\geq \dot{I}_x. \label{eqn_int_dynamicGeneralizedSecondLaw}
\end{align}
This allows an interpretation in terms of Maxwell's demon: When $\dot{I}_x$ is negative, it can appear as if the second law were violated. However, what is actually happening is that the subsystem $x$ is \emph{using up} correlations which either have been established previously, or which the subsystem $y$ is continuously building up. This is particularly clear when the system is in a nonequilibrium steady state, such that $S[x] = \text{const.}$ and $\frac{d}{dt} I[x,y](t) = 0 \Leftrightarrow \dot{I}_x = -\dot{I}_y$: The information that $x$ continuously consumes to break the second law (positive $\dot{Q}_x$) must steadily be supplied by $y$. More examples can be found in Ref.~\cite{Horowitz2014}.

\subsection{Thermodynamics of sensing}
What we called information flow in line with Refs.~\cite{Parrondo2015,Horowitz2014,Allahverdyan2009,Horowitz2015,Rosinberg2016} is called \emph{nostalgia} (although only for discrete-time processes) in Refs.~\cite{Still2012,Quenneville2018}, and \emph{learning rate} in Refs.~\cite{Hartich2014,Barato2014a,Bo2015,Hartich2016,Brittain2017,Chetrite2019}. Especially the latter two terms have been used in the context of the \emph{thermodynamics of sensing}~\cite{Mehta2012}.

In this setup a system $x$ (the \emph{sensor}) is influenced by an external \emph{stochastic signal} $y$, e.g., a chemical concentration. In this situation Eq.~\eqref{eqn_int_dynamicGeneralizedSecondLaw},
\begin{align}\label{eqn_int_dynamicGeneralizedSecondLaw_2}
\frac{d}{dt} S[x] - \frac{\dot{Q}_x}{\kB T_x} \geq \dot{I}_x,
\end{align}
needs to be interpreted differently: $\dot{I}_x$ is positive, since the sensor should build up correlations with the external signal (learn the signal), while the signal will change over time thereby decorrelating with the sensor ($\dot{I}_y < 0$). The generalized second law therefore gives a lower bound on how much dissipation must occur. The more correlations the system builds up, the more it must dissipate to do so.

Again, numerous generalizations are possible~\cite{Hartich2017}: One can study a \emph{sensor efficiency}~\cite{Barato2014a} by taking the ratio of the RHS and LHS of Eq.~\eqref{eqn_int_dynamicGeneralizedSecondLaw_2}. Additionally, one can interpret the RHS as quantifying how much of the correlation between system and signal (the memory) is predictive of the future of the external signal~\cite{Still2012,Ehrich2016}. Moreover, instead of only taking the instantaneous information flow, similar relations involving the \emph{rate of transfer entropy}~\cite{Schreiber2000} can be derived~\cite{Hartich2014,Bo2015,Ito2015}. Then, not only the efficiency plays a role but also a quantity termed \emph{sensory capacity}~\cite{Hartich2016} measuring how much information the current state of the sensor has compared to the entire trajectory of the sensor. An overview over the different inequalities involving various informational terms is given in Ref.~\cite{Horowitz2014a}.

One related result was obtained by Sartori \emph{et al.}~\cite{Sartori2014}. Their methods are more in line with what we presented in Sec.~\ref{sec_int_generalPicture}. They apply Eq.~\eqref{eqn_int_generalizedSecondLaw} to a process with a sudden external signal variation and find
\begin{align}\label{eqn_int_costAdaptation}
\Delta S[x] - \frac{{Q}_x}{\kB T_x} \geq \Delta I,
\end{align}
which they call the \emph{cost of sensory adaptation}: To build up a correlation of $\Delta I$, the sensor must at least dissipate $\kB T_x \Delta I$.

\section{Detached path probabilities}\label{sec_int_detachedPathProbs}
Having presented the splitting of the second law using information flows, it is natural to ask whether there are also corresponding fluctuation relations involving information flows. In Ref.~\cite{Ehrich2017} (reprinted below) we studied the above example setups (measurement-feedback schemes and sensors) from one general perspective.

Our contribution is twofold: Firstly, it is instructive to see that the different setups can be distinguished by their \emph{causal structure}, i.e., the question of whether there is feedback in both directions between the subsystems. Secondly, we want to show that many (variants) of the above results can be readily obtained from such a global formalism.

Of course this effort requires some degree of generalization: The main ingredient comes from information theory and goes under the name of \emph{causal conditioning}.

\subsection{Causal conditioning}
Let us return to Sec.~\ref{sec_int_informationFTs}, specifically to Eq.~\eqref{eqn_int_forwardFeedbackTrajProb}. Recall that the \emph{joint probability} of the trajectory $x(\cdot)$ and the measurement outcome $y$ in a measurement-feedback process is given by:
\begin{align}
p\big[x(\cdot),y\big] = p_m\big(y|x(t_m)\big)\; p\big[x(\cdot);\lambda(\cdot)\big].
\end{align}
We have not pointed it out at the time, but this splitting is non-trivial. The way it is written suggests that $p\big[x(\cdot);\lambda(\cdot)\big]$ is the marginal probability of the trajectory $x(\cdot)$. However, this is not true because the second part (after the measurement time $t_m$) of the protocol $\lambda(\cdot)$ depends on the measurement outcome $y$ (cf. Fig.~\ref{fig_int_feedbackProcess}).

Neither is it a conditional probability since $x(\cdot)$ is still needed (namely, at $x(t_m)$) to determine the statistics of $y$. Rewriting the above equation (thereby dropping the dependency on the protocol $\lambda(\cdot)$) clarifies the causal structure:
\begin{align}
p\big[x(\cdot),y\big] = q_x\big[x(\cdot);y\big]\; q_y\big(y;x(t_m)\big),
\end{align}
where
\begin{align}
q_x\big[x(\cdot);y\big] := p(x_0)\; p_x\big[x_{0:t_m}(\cdot)|x_0\big] \; p_x\big[x_{t_m:T}(\cdot)|x(t_m);y\big]
\end{align}
and
\begin{align}
q_y\big[y;x(\cdot)\big] := p_m\big(y|x(t_m)\big)
\end{align}
are the \emph{causally-conditioned probabilities} and $p_x\big[x_{0:t_m}(\cdot)\big]$ denotes the path probability of the first part (without feedback) while $p_x\big[x_{t_m:T}(\cdot);y\big]$ denotes the path probability of the second part, for which the protocol depends on the measurement outcome.

Causal conditioning allows one to assess \emph{causation} and not only \emph{correlation}~\cite{Marko1973,Massey1990,Jiao2013}. It is used implicitly by Ito and Sagawa~\cite{Ito2013} and is the center piece of Crook's and Still's dissection of entropy productions~\cite{Crooks2016,Crooks2019}. Because of the effect it has on the causal structure, effectively splitting (or \emph{detaching}) the individual sub-processes from each other (see Fig. 2 of Ref.~\cite{Ehrich2017}), we named the resulting probabilities \emph{detached path probabilities}. Ref.~\cite{Ehrich2017} contains a detailed elaboration on the differences between conditional and detached probabilities.

We apply the formalism to a bipartite Markov chain and recover, among other results, the Sagawa-Ueada fluctuation relation [Eq.~\eqref{eqn_int_generalizedJarzynski}] for measurement-feedback systems (albeit with \emph{information flow} due to the repeated feedback) and the cost of sensory adaptation [Eq.~\eqref{eqn_int_costAdaptation}]. Additionally, we apply it to the causal structure of a hidden Markov model.

\subsection{Detached entropy production}
We define \emph{detached entropy productions} using log-ratios of forward and time-reversed detached path probabilities. These can be thought of as the trajectory equivalents of the \emph{partial entropy production rates} in Eqs.~\eqref{eqn_int_partialEPRates_a}~and~\eqref{eqn_int_partialEPRates_b}. 

They have a special property that was pointed out by Ito and Sagawa \cite{Ito2013} and Crooks and Still~\cite{Crooks2019} but is perhaps not sufficiently appreciated in the literature: When a subsystem is in contact with a thermal reservoir, its detached entropy production equals the usual thermodynamic entropy production of system entropy change minus heat flow. The reason for this is that we can apply the local detailed balance relation to the detached probabilities but not to other, e.g., coarse-grained (see Sec.~\ref{sec_hid_coarsegraining}) probabilities. 

Thus: If one wants to define heat flow for a system that is strongly interacting with other (sub-) systems, the detached entropy production delivers the correct result. This could be a guiding principle in answering the question of how heat should be defined for systems strongly coupled to their environment~{\cite{Seifert2016,Talkner2016,Strasberg2017,Miller2017}}. The problem is, however, that the detached entropy productions are not coarse-grained entropies because they still depend on the trajectory traversed by the other process. That means that for a system strongly interacting with its environment one has to monitor the environment degrees of freedom to calculate heat flow, which is not feasible. It might therefore be advantageous to resort to coarse-graining instead, as done in Ref.~\cite{Strasberg2017}.

\subsection{Article: [\emph{Phys. Rev. E} \textbf{96}, 042129 (2017)]}
The following article is reprinted from J. Ehrich and A. Engel, arXiv:1707.07434. Journal reference: [J. Ehrich and A. Engel, ``Stochastic thermodynamics of interacting degrees of freedom: Fluctuation theorems for detached path probabilities,'' \emph{Phys. Rev. E} \textbf{96}, 042129 (2017)] (Ref.~\cite{Ehrich2017}).

The bracketed numbers provide a continuous pagination.
\cleardoublepage
\pagestyle{pdfStyle}
\includepdf[pages=-,pagecommand={},scale=1]{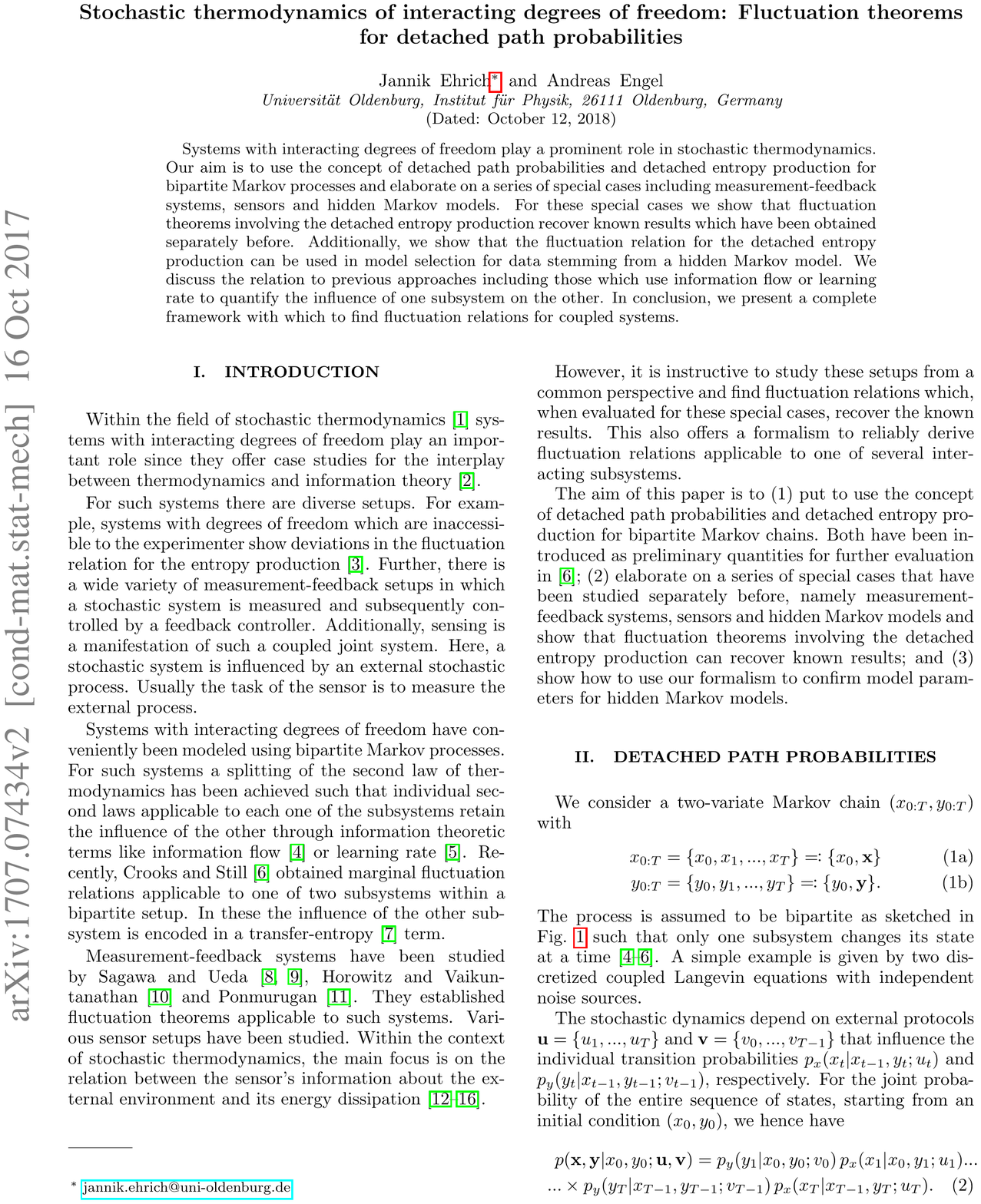}
\pagestyle{thesisStyle}

\mypart{Hidden degrees of freedom}{
	The last part of the thesis is dedicated to the study of interacting subsystems of which one is hidden from observation. In particular the aims are to:
	\begin{itemize}
		\item Show by the example of a microswimmer model how neglecting relevant degrees of freedom in the thermodynamic description of a system results in underestimating its entropy production.
		\item Formalize and organize the different effective descriptions of the visible parts of systems with hidden slow degrees of freedom.
		\item Show how an effective Markovian description of the visible parts results in violations of the standard fluctuation relations.
		\item Present an approach with which one can find bounds for the hidden, complete entropy production for masked Markovian jump networks.
	\end{itemize}
}

\chapter{Dissipation of Microswimmers}\label{chap_hid_microswimmers}
We will start with an example illustrating how neglecting some degrees of freedom in the stochastic description of the dynamics of a certain system leads one to underestimate entropy production and, therefore, energy dissipation. Nonetheless, apart from showing the mentioned effect, the example system we propose possesses some interesting properties by itself. It represents a very simple model of a \emph{microswimmer}.

\section{Microswimmers}
Locomotion is a central accomplishment in biological evolution. Motility is useful for microorganisms in their search for food, for avoiding poison, and reacting to light, etc. Biological and artificial small-scale objects with a self-propulsion mechanism are collectively called \emph{microswimmers}, and studying their motion and collective behavior is an active area of physics (for a recent review, see Ref.~\cite{Elgeti2015}).

Due to their size, they swim at very low Reynolds numbers, much like a person swimming in thick syrup. An important consequence pointed out by Purcell~\cite{Purcell1977} is that a reciprocal swimming motion, i.e., one that consists of a stroke that is subsequently retraced, does not achieve any net displacement.

Taking the limit of low Reynolds numbers in the Navier-Stokes equation leads to the Stokes equation (see, e.g. Chap.~II,~§20 of Ref.~\cite{Landau1987}):
\begin{align}
\mu \,\nabla^2 \bf{v} - \nabla\, p = 0, \qquad \nabla \cdot \mathbf{v} = 0,
\end{align}
where $\mu$ is the fluid viscosity and $\bf{v}$ the fluid velocity. The Stokes equation possesses no time-dependence and is linear in $\bf v$ and thus invariant under time-reversal\footnote{This is the basis for many jaw-dropping experiments, e.g., the reversible mixing and unmixing of tracer particles in a Taylor-Couette flow~\cite{Heller1960,movieLaminarFlow,movieLaminarFlowSmarterEveryDay}.}. Moreover, if we imagine a swimming strategy that consists of periodically modifying the shape of the swimmer (which is essentially what swimming means), even the speed of the shape modification is irrelevant. This is the underpinning of Purcell's \emph{Scallop theorem}, which he sums up as follows:

``So, if [an] animal tries to swim by a reciprocal motion, it \emph{can't go anywhere}. Fast or slow, it exactly retraces its trajectory and it's back where it started. A good example of that is a scallop. You know, a scallop opens its shell slowly and closes its shell fast, squirting out water. The moral of this is that the scallop at low Reynolds number is no good. It can't swim because it only has one hinge, and if you have only one degree of freedom in configuration space, you are bound to make a reciprocal motion"~\cite{Purcell1977}.

\begin{figure}[ht]
	\centering
	\includegraphics[width =0.60 \linewidth]{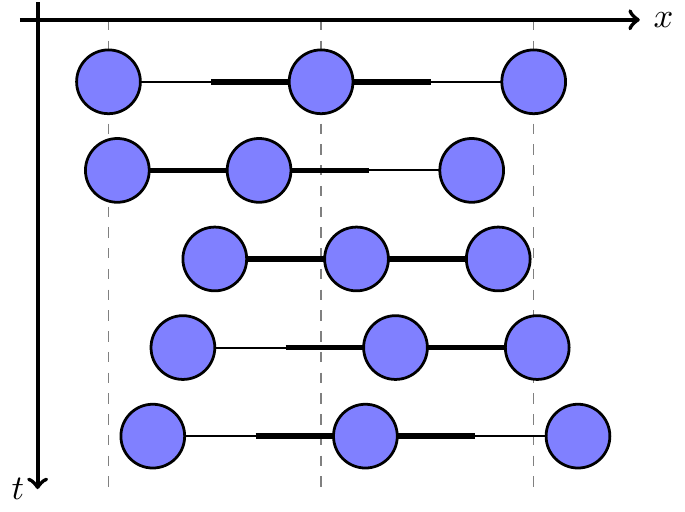}
	\caption{Three sphere swimmer by Najafi and Golestanian. Modified after Ref.~\cite{Najafi2004}. The three connected spheres move forward as a consequence of the time-asymmetric change of the length of the rod connecting them.}
	\label{fig_hid_threeSphereSwimmer}
\end{figure}

The kind of motion needed to break time-reversal symmetry is exemplified by the three-sphere swimmer envisioned by Najafi and Golestanian~\cite{Najafi2004} and depicted in Fig~\ref{fig_hid_threeSphereSwimmer}. It swims in one dimension by modifying the length of two rigid rods between three spheres in a non-reciprocal manner. This model has also been realized experimentally~\cite{Leoni2009}.

\section{Active Brownian motion}\label{sec_hid_activeBrownianMotion}
Often, the propulsion mechanism of the swimmer is hidden from observation. This is the case, e.g., when considering the trajectories traced out by microswimmers. A simple model describing their dynamics is that of \emph{active Brownian motion}, i.e., Brownian motion in two or three dimensions with a constant propulsion force that undergoes rotational diffusion. In two dimensions this process is described by the following set of overdamped Langevin equations (cf. Sec.~\ref{sec_basics_LengevinEquation}):
\begin{subequations}
	\begin{align}
	\dot{x} &= \nu f \, \cos \phi + \sqrt{2 \nu T}\, \xi_x(t)\label{eqn_hid_ABM_x}\\
	\dot{y} &= \nu f \, \sin \phi + \sqrt{2 \nu T}\, \xi_y(t)\label{eqn_hid_ABM_y}\\
	\dot{\phi} &= \sqrt{2 D_\phi}\, \xi_\phi(t),\label{eqn_hid_ABM_phi}
	\end{align}
\end{subequations}
where $x$ and $y$ are the position coordinates of the swimmer, $\phi$ is the angle of the constant force $f$, and $\nu$ is the mobility. The swimmer is affected by thermal fluctuations due to it being submersed in an aqueous solution at temperature $T$. This also makes it undergo rotational diffusion with some strength $D_\phi$, which depends on the geometry of the swimmer.

One hallmark of active Brownian motion is its enhanced diffusion. Transfoming the above system of Langevin equations to the corresponding Fokker-Planck equation (cf. Eq.~\eqref{eqn_basics_multiDimensionalFokkerPlanck}), and solving it with initial conditions $x(0) = y(0) = 0$ and a uniformly distributed initial angle $\phi(0)$, yields the following expression for the mean-squared displacement~\cite{Bechinger2016}:
\begin{align}
\mathrm{MSD}(t) &:= \big\langle x^2(t)+y^2(t)\big\rangle = \left( 4 \nu T + \frac{2 \nu^2f^2}{D_\phi} \right)t - \frac{2\nu^2f^2}{D_\phi^2}\big[ 1-\exp{\left( -D_\phi t\right) } \big], 
\end{align}
such that asymptotically $\mathrm{MSD}(t)\approx \left( 4 \nu T + \frac{2 \nu^2f^2}{D_\phi} \right)t$. We see that the diffusion is enhanced by a term proportional to the square of the propulsion force and inversely proportional to the rotational diffusivity. Some sample trajectories are shown on the cover of this very thesis with  the propulsion force increasing from darker colors to lighter colors.

The description of active Brownian motion is usually used for \emph{active particles}, e.g., Janus colloids, i.e., colloidal particles with different coatings on either hemisphere able to generate propulsion through chemical reactions or by producing thermal gradients by localized absorption of light~\cite{Bechinger2016}. However, the term \emph{microswimmer} is often applied to mean \emph{active colloid}. We will in the following only use the term microswimmer for a system that is utilizing some swimming motion for propulsion.

\subsection{Entropy production for active matter}
Recently, there have been efforts to apply the formalism of stochastic thermodynamics to active matter systems like microswimmers and active colloids~\cite{Speck2016}. The general consensus is that active matter is out of equilibrium because locomotion is produced by the conversion of stored (in the case of microswimmers) or externally supplied (in the case of active colloids) fuel. The question remaining is: How far from equilibrium is active matter~\cite{Fodor2016}?

While a comparison of trajectory weights as done in Sec.~\ref{eqn_basics_EPLangevin} yields an entropy production that can be used as a measure of dissipation for individual trajectories~\cite{Ganguly2013, Chaudhuri2014, Puglisi2017, Mandal2017, Caprini2018, Mandal2018, Dabelow2019, Szamel2019,Caprini2019}, it underestimates the true dissipation of active particles because the position data is only a coarse-grained description of the real process~\cite{Pietzonka2018,Seifert2019}: From the trajectory alone we cannot distinguish whether a change in position was due to a thermal fluctuation or the active process. This is a general problem of coarse-grained descriptions (see Sec.~\ref{sec_hid_2effDescriptions}).

We want to demonstrate this effect and show the underlying problem in assessing the dissipation: The propulsion mechanism that generates the active force itself is coarse-grained away in the active Brownian motion description.

\section{Microswimmer model}
We take one step towards including the propulsion mechanism: We model the system as consisting of two colloidal particles at positions $(x_1,y_1)$ and $(x_2,y_2)$ with tunable mobilities $\nu_1(t)$ and $\nu_2(t)$. The particles are interacting via a harmonic potential which defines an equilibrium distance $l$. We then apply a periodic protocol changing the equilibrium distance and the mobilities in a non-reciprocal manner (cf. Fig. 1 of our article in Ref.~\cite{Ehrich2019} reprinted below). The model is completely analytically solvable by using a Gaussian ansatz for the Fokker-Planck equation resulting from the coupled overdamped Langevin equations (see Appendix~\ref{app_linearLangevin} for details).

The model is inspired by the three-sphere swimmer~\cite{Najafi2004} presented above and close to the model termed \emph{Pushmepullyou} by Avron \emph{et al.}~\cite{Avron2005}. For simplicity, we do not model the system's hydrodynamics. Instead, we use a stochastic description using overdamped Langevin equations, which makes the system comparable to a model introduced by Amb\'ia and H\'ijar~\cite{Ambia2016,Ambia2017}, although their system resembles a feedback ratchet. Notice that we, too, apply a protocol that breaks time-reversal symmetry as required by the linear Stokes friction term in the overdamped Langevin equation.

We show that the center of mass movement of the swimmer is exactly described by active Brownian motion given in Eqs.~\eqref{eqn_hid_ABM_x} to~\eqref{eqn_hid_ABM_phi} when the driving protocol becomes very fast. Using Sekimoto's definition of heat and work for Langevin systems (cf. Sec.~\ref{sec_basics_stochasticEnergetics}), we can calculate the average dissipation for the real swimmer and for the approximation using active Brownian motion.

Comparing both dissipation measures, one sees that the dissipation assigned to active Brownian motion can massively underestimate the real dissipation. The reason for this is that, to calculate the full dissipation, one needs to resolve the relative distance $r~:=~\sqrt{(x_2-x_1)^2 + (y_2-y_1)^2}$ between the particles which is a \emph{hidden variable} when only considering the center of mass trajectories.

A discussion of the definition of microswimmer efficiency which we used in our article is found in Appendix~\ref{app_efficiency}.

\subsection{Article: [\emph{Phys. Rev. E} \textbf{99}, 012118 (2019)]}
The following article is reprinted from J. Ehrich and M. Kahlen, arXiv:1809.07235. Journal reference: [J. Ehrich and M. Kahlen, ``Approximating microswimmer dynamics by active Brownian motion: Energetics and efficiency,'' \emph{Phys. Rev. E} \textbf{99}, 012118 (2019)] (Ref.~\cite{Ehrich2019}).

The bracketed numbers provide a continuous pagination.
\cleardoublepage
\pagestyle{pdfStyle}
\includepdf[pages=-,pagecommand={},scale=1]{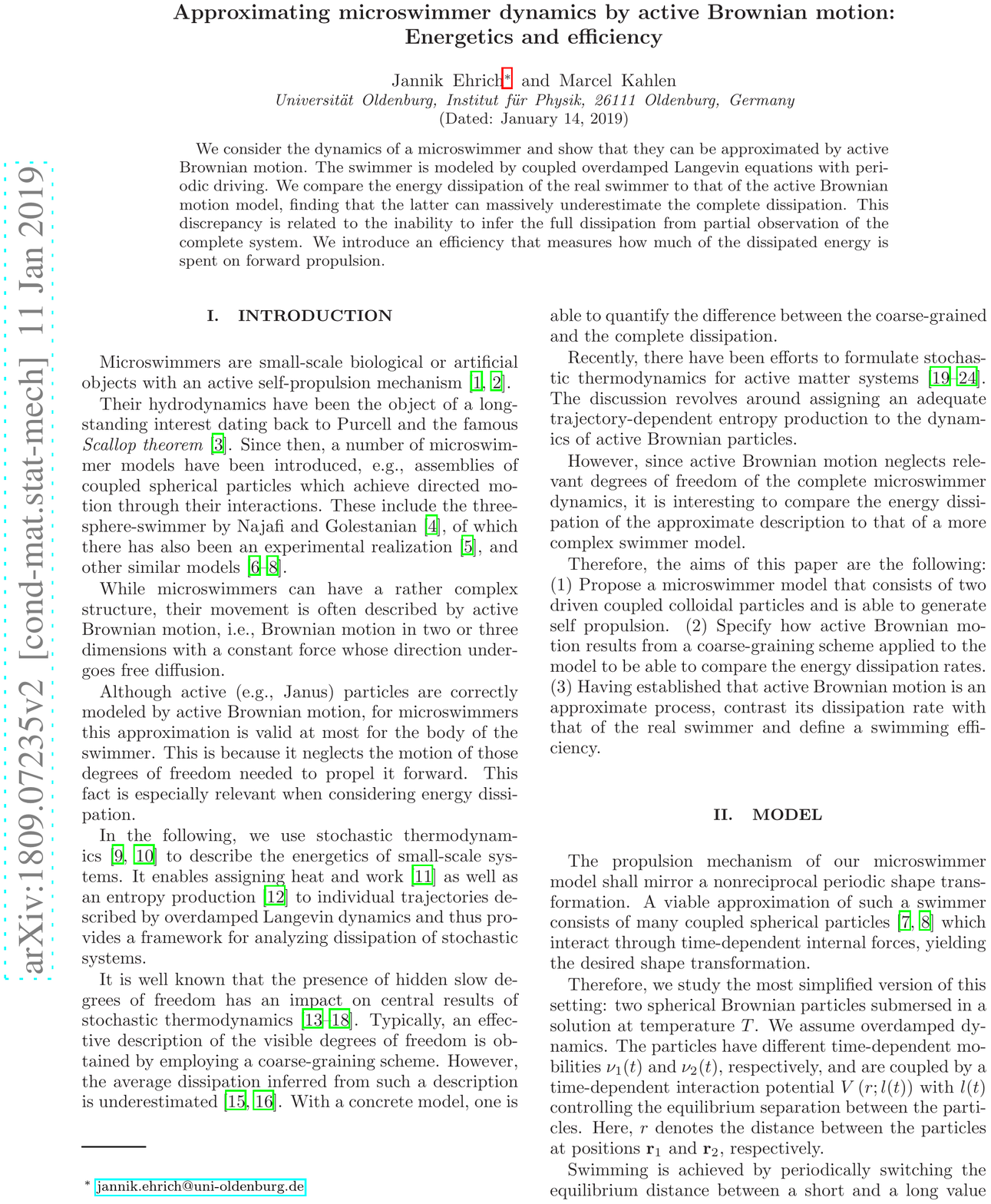}
\pagestyle{thesisStyle}

\chapter{Hidden degrees of freedom in stochastic thermodynamics}\label{sec_hid_hiddenDegreesInSTD}
In the previous chapter we have seen by example that ignoring relevant degrees of freedom in the thermodynamic description results in an underestimate of the entropy production. In this chapter we want to formalize this notion and present different effective descriptions for systems with hidden degrees of freedom. Additionally, we show the impact of hidden degrees of freedom on fluctuation theorems.

\begin{figure}[ht]
	\centering
	\includegraphics[width =0.85 \linewidth]{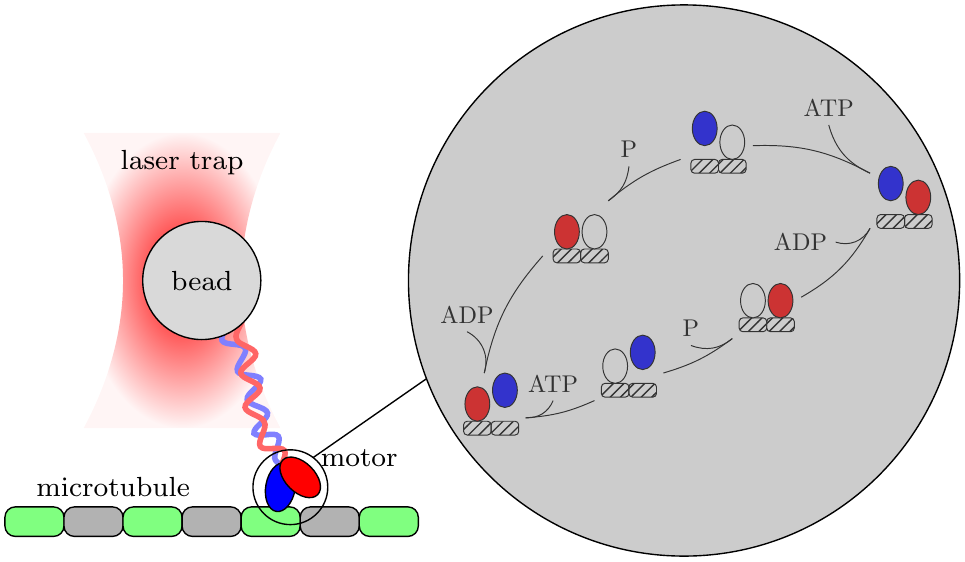}
	\caption{Schematics of a molecular motor experiment. One observes the motor dynamics via an attached bead that is in a laser trap so as to exert forces on the motor. The motor position, the linking of motor and bead and the chemical reactions inside the motor are all hidden variables in such a setup.}
	\label{fig_hid_motorChemNet}
\end{figure}

When introducing stochastic thermodynamics in Sec.~\ref{sec_basics_microscopicMacroscopic}, we made one crucial assumption that seemed innocent enough: We assumed that we measure \emph{all relevant degrees of freedom} of the system and that there is a time-scale separation between the degrees of freedom of the system and those of the environment. Actually, this is a rare situation in experiments. Consider, e.g., molecular motor experiments~\cite{Kolomeisky2013} like the one depicted in Fig~\ref{fig_hid_motorChemNet}. Often the motor dynamics are resolved via a colloidal bead that is attached to the motor. Motor and bead then form one joint stochastic system of which many important parts are not resolved, e.g., the \emph{actual} motor position, the fluctuations of the linking between motor and bead, and the chemical reactions occurring inside the motor.

Thus arises the question of what can still be learned about the entropy production in such settings and whether there is an effective thermodynamic description for the visible degrees of freedom.

\section{Setup}\label{sec_hid_setupHiddenProcess}
To formalize the setting, we assume that there exists an underlying (possibly multi-variate) \emph{complete dynamics} $z(t)$ and that we observe a \emph{visible process} $x(t)$. The visible process could consist of a subset of the variables that make up the complete process, such that one can split the variables $z=\{x,y\}$ into a visible set $x$ and a hidden set $y$. This is the case for the microswimmer considered before: Of the complete set $\{X,r\}$ of variables, we can only observe the center of mass position $X$. 

\begin{figure}[ht]
	\centering
	\includegraphics[width =0.6 \linewidth]{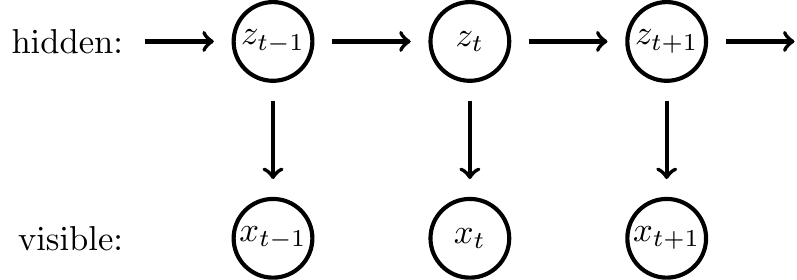}
	\caption{Causal diagram of a hidden Markov model. A Markov process $z(t)$ is observed via the visible process $x(t)$.}
	\label{fig_hid_HMMCausalDiagram}
\end{figure}

However, such a setting is too restrictive. Consider, for example, the masked jump processes discussed in the next chapter. There, we cannot make such a distinction between visible and hidden variables. The most complete setup is therefore that of a \emph{hidden Markov model} (see, e.g., Chap.~13.2 of Ref.~\cite{Bishop2006}), which we have already encountered in Sec.~\ref{sec_int_detachedPathProbs} and in the article in Ref.~\cite{Ehrich2017}. The causal structure is depicted in Fig.~\ref{fig_hid_HMMCausalDiagram}. It consists of a hidden Markov chain and observations that form the visible process.

This structure is even more general than is needed: The so-called \emph{emission probabilites} governing how $x$ results from $z$ represent a simple \emph{many-to-one} mapping using delta functions, as the hidden state completely defines the visible state. Importantly, one could generalize the setup by loosening this restriction thus including measurement errors.

Crucially, we already see the main feature of the visible dynamics: They are \emph{not Markovian}, i.e., there are (perhaps complex) memory effects that cannot be unraveled via two-point statistics.

\section{Two effective descriptions}\label{sec_hid_2effDescriptions}
We start by presenting two different strategies of effective description that have been proposed in the literature. For simplicity, it is assumed that all states $z$ and $x$ are even under time-reversal.

\subsection{Coarse-graining and apparent entropy production}\label{sec_hid_coarsegraining}
The first method goes, somewhat ambiguously, under the name of \emph{coarse-graining}. It is effectively a \emph{mapping to a Markov model}. Importantly, this would be the description used if one were unaware of the presence of hidden slow degrees of freedom, since then, naturally, one would assume Markovian dynamics for the system.

Of course we expect some level of consistency of the effective description. It should reproduce the correct \emph{marginal distribution} $p(x,t)$ of the visible process for all times $t$. Furthermore, we require that the transition rates $\tilde W(x|x';t)$ of the visible process coincide with the observed transition rates.

As pointed out by Esposito~\cite{Esposito2012}, this is accomplished by integrating out the hidden variables in the master equation~\eqref{eqn_basics_verySimpleME} governing the complete dynamics:
\begin{align}\label{eqn_hid_completeME}
\frac{\partial}{\partial t} p(z,t) = \int dz' \,W(z|z';t)\,p(z',t),
\end{align}
where $W(z|z';t)$ is the transition rate for the underlying dynamics (notice that we dropped the tilde, which we reserve for the coarse-grained transition rates).

The hidden probability can now be decomposed in the following way:
\begin{align}\label{eqn_hid_decompositionHiddenState}
p(z,t) = p_c(z|x;t)\,p(x,t),
\end{align}
where $p_c(z|x;t)$ denotes the conditional probability of the complete hidden state $z$ given the visible state $x$. Inserting this decomposition into Eq.~\eqref{eqn_hid_completeME} yields
\begin{align}
\left(\frac{\partial}{\partial t} p_c(z|x;t)\right)&p(x,t) + p_c(z|x;t) \pd{t}p(x,t)\nonumber\\
&= \int dx' \left(\int d(z'|x') \,W(z|z';t)\,p_c(z'|x';t) \right) p(x,t),
\end{align}
where $\int d(z'|x')$ denotes integration over all hidden states $z'$ belonging to the visible state $x'$. Integrating over all hidden states belonging to $x$ yields:
\begin{align}\label{eqn_hid_effectiveMasterEQ}
\pd{t}p(x,t) = \tilde{W}(x|x';t)\,p(x,t),
\end{align}
with the \emph{effective} (or \emph{coarse-grained}) rates
\begin{align}\label{eqn_hid_defEffectiveRates}
\tilde{W}(x|x';t) := \iint d(z|x) d(z'|x') \,W(z|z';t)\,p_c(z'|x';t).
\end{align}

This coarse-graining scheme has been used, e.g., in Refs.~\cite{Strasberg2017,Esposito2012,Bo2014} for Markovian jump processes and in Ref.~\cite{Herpich2019} for Fokker-Planck dynamics. One example is given by a Brownian particle in a two-dimensional potential $V(x,y;t)$. In our article in Ref.~\cite{Kahlen2018} (reprinted below) we show that coarse-graining away the $y$-coordinate results in an \emph{effective potential} $\tilde V(x;t)$ for the variable $x$.

Importantly, it is necessary to first solve the complete dynamics in order to calculate $p_c(z'|x';t)$. Moreover, the effective description is not unique since it depends on the choice of initial distribution $p_c(z|x;0)$ of hidden variables. However, in realistic settings one would either choose the equilibrium distribution or the stationary distribution of a nonequilibrium steady-state, depending on the physical context. We also need to mention that the effective transition rates $\tilde{W}(x|x';t)$ would be a measured quantity in an experiment, computed by \emph{counting} the number of transitions per time interval.

Also note that Eq.~\eqref{eqn_hid_effectiveMasterEQ} does not imply that the $x$-dynamics are Markovian. We have explicitly used the one-time distribution $p(x,t)$ and not a generic transition probability. It is generally impossible to find a corresponding Master equation for $\pd{t} p(x,t|x'';t'')$ for all $x$ and $x''$ and all $t$ and $t''$ since that would imply the Markov property of the visible process (see Sec.~IV.1~of~Ref.~\cite{vanKampen2007}).

Applying the formalism laid out in Secs.~\ref{sec_basics_stochasticEP}~and~\ref{sec_basics_entropyProductionSTD} leads to the following definition of an entropy production for the coarse-grained process:
\begin{align}\label{eqn_hid_defApparentEP}
\tilde\sigma[x(\cdot)] := \ln\frac{\tilde{p}[x(\cdot)]}{\bar{\tilde{p}}[\bar x(\cdot)]},
\end{align}
where $\tilde{p}[x(\cdot)]$ denotes the path probability of the visible trajectory $x(\cdot)$ if it were generated using the effective Master equation with the rates $\tilde{W}(x|x';t)$ (or Langevin equation with the corresponding force and diffusion) and similarly for the reverse probability $\bar{\tilde{p}}[\bar x(\cdot)]$ of the time-reversed trajectory.

We call the quantity $\tilde\sigma[x(\cdot)]$ defined in Eq.~\eqref{eqn_hid_defApparentEP} \emph{apparent entropy production}  in accordance with Refs.~\cite{Kahlen2018,Mehl2012,Uhl2018}. As shown in Sec.~\ref{sec_basics_avgEPMasterEq}, the corresponding \emph{average entropy production rate} [Eq.~\ref{eqn_basics_averageEPMarkov}] $\dot{\tilde{\Sigma}} $ reads
\begin{align}
\dot{\tilde{\Sigma}} &= \iint dx dx'\, \tilde{W}(x|x';t)\,p(x',t)\ln\frac{ \tilde W(x|x';t)\,p(x',t)}{\tilde  W(x'|x;t)\,p(x,t)}.
\end{align}

It is natural to ask how this average entropy production rate relates to the real average entropy production rate 
\begin{align}
\dot{\Sigma} = \iint dz dz'\, W(z|z';t)\,p(z',t)\ln\frac{ W(z|z';t)\,p(z',t)}{W(z'|z;t)\,p(z,t)}.
\end{align}
As shown by Esposito~\cite{Esposito2012}, the average apparent entropy production never overestimates the real entropy production rate. To show why this is the case, let us rewrite the entropy production rates. First the apparent entropy production rate
\begin{align}
\dot{\tilde{\Sigma}} &= \iint dx dx'\left(\iint d(z|x) d(z'|x') \,W(z|z')\,p_c(z'|x')\right)p(x')\ln\frac{ \tilde W(x|x')p(x')}{\tilde  W(x'|x)\,p(x)},
\end{align}
where we used Eq.~\eqref{eqn_hid_defEffectiveRates} and dropped the time-dependency. The total average entropy production rate can also be rewritten:
\begin{align}
\dot{{\Sigma}} = \iint dx dx' \iint d(z|x) d(z'|x') \,W(z|z')\,p_c(z'|x')p(x')\ln\frac{ W(z|z')p_c(z'|x')p(x')}{W(z'|z)\,p_c(z|x)p(x)},
\end{align}
where we used the decomposition in Eq.~\eqref{eqn_hid_decompositionHiddenState}.

We then find:
\begin{align}
\dot{{\Sigma}} - \dot{\tilde{\Sigma}} &=  \iint dx dx' \tilde{W}(x|x')p(x')\iint d(z|x) d(z'|x') \frac{W(z|z')p_c(z'|x')}{\tilde{W}(x|x')}\nonumber\\
&\qquad\qquad\qquad\qquad\qquad\times\ln\left[\frac{W(z|z')p_c(z'|x')}{\tilde{W}(x|x')}\,\frac{\tilde{W}(x'|x)}{W(z'|z)p_c(z|x)} \right]\\
&= \iint dx dx' \tilde{W}(x|x')p(x')\, D_{\rm KL}\big[q(z'\to z| x'\to x)\big|\big|q(z\to z'| x\to x')\big]\label{eqn_hid_diffEPKullbackLeibler}\\
&\geq 0,
\end{align}
where we defined the conditional hidden transition rate
\begin{align}\label{eqn_hid_hiddenConditionalTransitionRate}
q(z'\to z| x'\to x) := \frac{W(z|z')p_c(z'|x')}{\tilde{W}(x|x')}
\end{align}
normalized according to Eq.~\eqref{eqn_hid_defEffectiveRates} such that:
\begin{align}
\iint d(z|x) d(z'|x')\,q(z'\to z| x'\to x) = 1.
\end{align}
Additionally, we identified the Kullback-Leibler distance defined in Eq.~\eqref{eqn_basics_defKLDistance} which is nonnegative and thus proves the inequality.

Both entropy productions therefore become equal when there are no hidden degrees of freedom, i.e., when $p_c(z|x) = \delta(z-x)$ and $W(z|z') = \tilde{W}(x|x')$. However, 
Eq.~\eqref{eqn_hid_diffEPKullbackLeibler} also reveals another possibility: They become equal when $q(z'\to z| x'\to x) = q(z\to z'| x\to x')$, implying that the hidden transitions $z' \to z$ are \emph{reversible} given the visible transitions.

Apart from the average entropy production rate, one can also study the fluctuations of the apparent entropy production which we will discuss in Sec.~\ref{sec_hid_FTs}.

Turning back to the microswimmer model introduced in the previous chapter, we now understand from a formal perspective why dissipation was underestimated by the coarse-grained description: The center of mass dynamics lacked crucial information about the underlying time-reversal behavior.

\subsection{Marginal process and marginal entropy production}\label{sec_hid_marginalProcess}
The second effective description utilizes the full marginal path probability of the visible trajectories. Then, the \emph{marginal entropy production} reads:
\begin{align}\label{eqn_hid_pathMarginalization}
\sigma_x[x(\cdot)] := \ln\frac{p[x(\cdot)]}{\bar{p}[\bar{x}(\cdot)]},
\end{align}
where the probabilities result from marginalizing the complete trajectory probabilities:
\begin{align}\label{eqn_hid_pathMarginalizationReverse}
p[x(\cdot)] = \int \mathcal{D}z(\cdot)\, p\big[x(\cdot) |z(\cdot)\big]\, p[z(\cdot)]
\end{align}
and
\begin{align}
\bar p[\bar x(\cdot)] = \int \mathcal{D}\bar z(\cdot)\, p\big[\bar x(\cdot) |\bar z(\cdot)\big]\, \bar p[\bar z(\cdot)],
\end{align}
where $p[x(\cdot) |z(\cdot)]$ encodes the mapping of $z(\cdot)$ to $x(\cdot)$ which is time-reversal-symmetric: $p[x(\cdot) |z(\cdot)]=p[\bar x(\cdot) |\bar z(\cdot)]$.

By definition (see Sec.~\ref{sec_basics_stochasticEP}), this marginal entropy production fulfills an integral and a Crooks-type fluctuation theorem. We can also study the \emph{average marginal entropy production} $\Sigma_x$ [using Eq.~\eqref{eqn_basics_DissipationKullbackLeibler}]:
\begin{align}
\Sigma_x[x(\cdot)] &= \langle \sigma_x[x(\cdot)] \rangle = \int \mathcal{D}x(\cdot)\, p[x(\cdot)] \ln\frac{p[x(\cdot)]}{\bar{p}[\bar{x}(\cdot)]}= D_{\rm KL}\Big[ p[x(\cdot)]\,||\,\bar{p}[\bar{x}(\cdot)] \Big],
\end{align}
which is the irreversibility of the visible process. We can prove that this entropy production, too, gives a lower bound for the real entropy production. We first show that the complete average entropy production of the hidden, complete process is the same as the \emph{joint entropy production} of the visible and hidden process together:
\begin{align}
\Sigma[z(\cdot)] &= \int \mathcal{D}z(\cdot)\, p[x(\cdot)] \ln\frac{p[z(\cdot)]}{\bar{p}[\bar{z}(\cdot)]}\\
&=\iint \mathcal{D}z(\cdot)\mathcal{D}x(\cdot)\, p[x(\cdot),z(\cdot)] \ln\frac{p[z(\cdot)]\,p\big[x(\cdot) |z(\cdot)\big]}{\bar{p}[\bar{z}(\cdot)]\,p\big[\bar x(\cdot) |\bar z(\cdot)\big]}\\
&= \iint \mathcal{D}z(\cdot)\mathcal{D}x(\cdot)\, p[x(\cdot),z(\cdot)] \ln\frac{p[x(\cdot),z(\cdot)]}{\bar{p}[\bar{x}(\cdot),\bar{z}(\cdot)]}\\
&=: \Sigma[x(\cdot),z(\cdot)],
\end{align}
where we used the fact that the mapping probabilities $p[x(\cdot)|z(\cdot)]$ are symmetric under time-reversal in the second line.

However, when we compare this to the marginal entropy production, we find:
\begin{align}
\Sigma[x(\cdot),z(\cdot)] &= \int \mathcal{D}x(\cdot)\, p[x(\cdot)] \ln\frac{p[x(\cdot)]}{\bar{p}[\bar{x}(\cdot)]} \left(1 + \int \mathcal{D}z(\cdot) p\big[ z(\cdot)|x(\cdot)\big] \ln \frac{p\big[ z(\cdot)|x(\cdot)\big]}{\bar{p}\big[ \bar z(\cdot)|\bar x(\cdot)\big]} \right)\\
&= \int \mathcal{D}x(\cdot)\, p[x(\cdot)] \ln\frac{p[x(\cdot)]}{\bar{p}[\bar{x}(\cdot)]} \Bigg(1 + D_{\rm KL} \Big[p\big[ z(\cdot)|x(\cdot)\big]\Big|\Big|\bar{p}\big[ \bar z(\cdot)|\bar x(\cdot)\big]\Big] \Bigg)\\
&\geq \Sigma_x[x(\cdot)].\label{eqn_hid_marginalEPUnderestimates}
\end{align}
Thus $\Sigma_x[x(\cdot)] \leq \Sigma[x(\cdot),z(\cdot)] = \Sigma[z(\cdot)]$. This makes sense since the visible process alone contains less information about time-irreversibility than the hidden, complete process. Once the hidden trajectory is known, the observed one is redundant information. This reasoning was first employed by Gomez-Marin \emph{et al.}~\cite{Gomez-Marin2008a} to argue how incomplete information about a system can yield lower bounds on its average dissipation.

One big caveat of this description is that to compute the marginal entropy production, one has to know the \emph{trajectory statistics} $p[x(\cdot)]$ which is difficult in experiments, to say the least. It seems impossible to sample sufficiently good statistics of entire continuous trajectories in diffusive processes. Nonetheless, for discrete-time jump processes, the multi-point statistics for short trajectories can be obtained and supplemented with some clever extrapolation techniques as shown by Rold\'an and Parrondo~\cite{Roldan2010,Roldan2012,Roldan2013}. Recently, Mart\'inez \emph{et al.}~\cite{Martinez2019} have shown how to compute the marginal entropy production for continuous-time jump processes from the non-Poissonian distribution of dwell times inside the individual states. We will further investigate this approach in Sec.~\ref{sec_hid_estTimeIrreversibility}.

One big advantage also pointed out by Mart\'inez \emph{et al.}~\cite{Martinez2019} is that the marginal entropy production makes it able to infer time-irreversibility, even when there are no probability flows. The reason for this is that, by definition, it is sensitive to correlations beyond the two-point statistics captured by fitting a Markov process to the available data. No other proposed lower bound for the entropy production can achieve the same.

However, the question remains under what circumstances one can recover the complete entropy production from only the data of the visible process. There exist some results about the missing part of the entropy production, e.g., that it fulfills a fluctuation theorem~\cite{Kawaguchi2013}. Yet, this seems to be of dubious utility for the problem at hand. In Chap.~\ref{chap_hid_inferringDiss} we show one possible scenario in which one can constrain the hidden entropy production.

\section{Separation of time-scales}
An important and illuminating limiting case is obtained when there is a \emph{time-scale separation} between the visible and the hidden degrees of freedom such that the unobserved variables evolve on much faster time scales then the observed ones. Then, as shown in Refs.~\cite{Esposito2012,Bo2014}, the fast degrees of freedom can be adiabatically eliminated yielding Markovian dynamics for the visible process: Since the hidden degrees of freedom evolve so quickly, they relax (infinitely) quickly towards a stationary state $p_c(z|x;t) = p_{\rm st}(z|x;t)$, where the instantaneous stationary state is reached by keeping $x$ and $t$ fixed. This means that the hidden transition probability can be approximated by:
\begin{align}
W(z|z';t) \approx \tilde{W}(x|x';t)\, p_{\rm st}(z|x;t),
\end{align}
which depends on $z'$ only through the visible state $x'$ belonging to it.

Given that there exists detailed balance among the hidden transitions at fixed visible transitions, such that no nonequilibrium steady state can emerge, we can replace the conditional \emph{stationary} distribution by a conditional \emph{equilibrium} distribution: $p_{\rm eq}(z|x;t)$. Then the conditional hidden transition rate defined in Eq.~\eqref{eqn_hid_hiddenConditionalTransitionRate} reads:
\begin{align}
q(z'\to z| x'\to x) &= \frac{\tilde{W}(x|x';t)\, p_{\rm eq}(z|x;t)\,p_{\rm eq}(z'|x';t)}{\tilde{W}(x|x';t)}\\
&= p_{\rm eq}(z|x;t)\,p_{\rm eq}(z'|x';t)\\
&= q(z\to z'| x\to x'),
\end{align}
which implies that the difference between the real average entropy production rate $\dot{\Sigma}$ and the apparent entropy production rate $\dot{\tilde{\Sigma}}$ vanishes\footnote{The fact that there is detailed balance in the hidden part of the system is crucial. Otherwise we would miss a contribution to the entropy production~\cite{Bo2014}.}:
\begin{align}
\dot{\tilde{\Sigma}} = \dot{{\Sigma}}.
\end{align}
Thus, within time-scale separation, coarse-graining delivers a consistent thermodynamic interpretation~\cite{Esposito2012}, which is what we expected since this forms the basis of our thermodynamic interpretation.

Since the Markovian description becomes exact in the limit of time-scale separation, the path probabilities from the effective master equation coincide with the ones from the marginal process in that limit:
\begin{align}
\tilde{p}[x(\cdot)] = p[x(\cdot)]
\end{align}
and similarly for the reversed probabilities. 
This means that for the trajectory entropy productions, we find:
\begin{align}
\tilde{\sigma}[x(\cdot)] = \sigma_x[x(\cdot)].
\end{align}
Moreover, the hidden path probabilities and the marginal path probabilities differ only by conditional equilibrium fluctuations which are time-reversible:
\begin{align}
p[z(\cdot)] = p_{\rm eq}[z(\cdot)|x(\cdot)]\,p[x(\cdot)],
\end{align}
and equally for the reversed probabilities. Therefore, we find
\begin{align}
\sigma[z(\cdot)] &= \ln\frac{p_{\rm eq}[z(\cdot)|x(\cdot)]}{\bar p_{\rm eq}[\bar z(\cdot)|\bar x(\cdot)]} + \sigma_x[x(\cdot)] \\
&=\sigma_x[x(\cdot)].
\end{align}

Thus, in the limit of time-scale separation and when the detailed balance holds for those hidden transitions that do not change the observed state, all entropy productions agree:
\begin{align}
\sigma[z(\cdot)] = \sigma_x[x(\cdot)] = \tilde\sigma[x(\cdot)]
\end{align}
and thus
\begin{align}
\dot{\Sigma} = \dot{\Sigma}_x = \dot{\tilde{\Sigma}}.
\end{align}

\section{Hidden slow degrees of freedom in fluctuation theorems}\label{sec_hid_FTs}
We already established that the marginal entropy production $\sigma_x[x(\cdot)]$ fulfills the standard fluctuation theorems. Considering, however, the apparent entropy production, predicting how the fluctuation theorems must be modified is not trivial. For instance, take the integral fluctuation theorem:
\begin{align}
\left\langle e^{-\tilde{\sigma}} \right\rangle &= \int \mathcal{D}x(\cdot)\,p[x(\cdot)] \frac{\bar{\tilde{p}}[\bar x(\cdot)]}{\tilde{p}[x(\cdot)]}\neq 1,
\end{align}
where we used Eq.~\eqref{eqn_hid_defApparentEP}. The standard integral fluctuation theorem is violated because the integrand is not normalized. Without any input about the underlying physical process no further statements seem to be possible (e.g., under which conditions the exponential average is less or more than one). 

Rahav and Jarzynski~\cite{Rahav2007} addressed this question in an early study by performing perturbation theory on a coarse-grained system about the limiting case of time-scale separation, finding that the fluctuation relations hold approximately for small entropy productions.

Mehl \emph{et al.}~\cite{Mehl2012} have studied the problem experimentally using two magnetically coupled Brownian particles driven into a nonequilibrium steady state. Using the coarse-graining scheme outlined above and inferring entropy production from the trajectory of only one of the particles, they found that the detailed fluctuation theorem (see Sec.~\ref{sec_basics_detailedFT}) for the apparent entropy production becomes modified such that the slope is less than one:
\begin{align}\label{eqn_hid_modifiedDetailedFT}
\ln\frac{p(\sigma)}{p(-\sigma)} = \alpha\,\sigma,\quad \text{with } \alpha < 1.
\end{align}

A similar experiment on two coupled driven RC-circuits was carried out by Chiang \emph{et al.}~\cite{Chiang2016}. Their apparent detailed fluctuation theorem also shows a modified slope. An attempt at explaining these result was made in Ref.~\cite{Uhl2018} using a large-deviation analysis of a similar system. The authors find a distinctly nonlinear fluctuation theorem and argue that the experimentally observed linear behavior is the result of a lack of good statistics for large entropy productions: By definition the detailed fluctuation theorem is an antisymmetric function and every smooth antisymmetric function is linear around the origin.

\subsection{Model system}
Our contribution to the analysis of the effect of hidden variables on fluctuation relations is twofold: Firstly, we present a model system that is simple enough to be analytically tractable\footnote{Like in our analysis of the microswimmer model, we exploit the fact that linearly coupled Langevin equations transform to a Fokker-Planck equation that is solved by a Gaussian as detailed in Appendix~\ref{app_linearLangevin}.}. This enables us to exactly calculate the distribution of entropy productions. It comes at the cost of only showing a linear apparent fluctuation theorem, as the distribution of entropy productions is Gaussian. Secondly, using this model, we are able to contrast the coarse-grained with the marginal description. To do this, we exactly calculate the marginal entropy production for this model which is possible because the marginalizations in Eqs.~\eqref{eqn_hid_pathMarginalization} and~\eqref{eqn_hid_pathMarginalizationReverse} each involve a Gaussian path integral. As expected, this marginal entropy production fulfills the fluctuation theorems.

\begin{figure}[ht]
	\centering
	\raisebox{0\height}{\includegraphics[width =0.4 \linewidth]{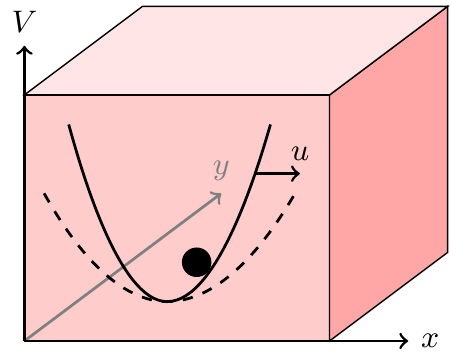}}%
	\hspace{0.5cm}
	\raisebox{-0.1\height}{\includegraphics[width =0.5 \linewidth]{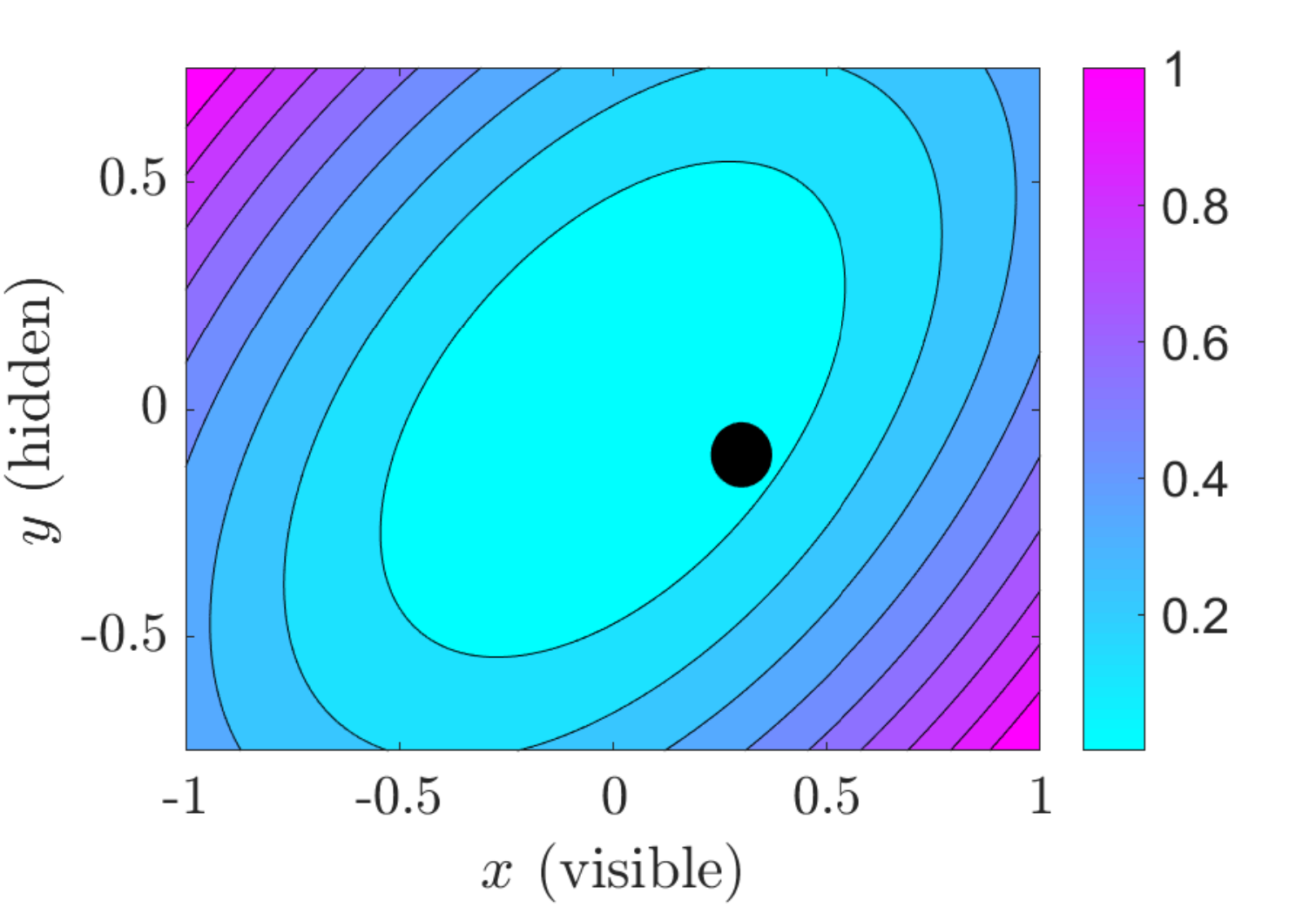}}
	\caption{Left: Setup of the model system with which we study the impact of hidden degrees of freedom on fluctuation theorems. A Brownian particle is trapped in a harmonic potential which is pulled at constant velocity $u$ in the positive $x$-direction. Coarse-graining away the $y$-degree of freedom creates an effective trapping potential in the $x$-direction (dashed curve). Right: Contourplot of the trapping potential illustrating how the potential creates correlations between the visible and the hidden coordinates.}
	\label{fig_hid_setupModelSystem}
\end{figure}

Our model consists of a Brownian particle in a two-dimensional harmonic trapping potential which is moved at a constant velocity $u$ in the positive $x$-direction. The trapping potential contains a linear coupling term between the visible ($x$) and the hidden ($y$) degree of freedom. The setup is depicted in Fig.~\ref{fig_hid_setupModelSystem}.

We show that the coarse-graining procedure with $y$ being the hidden variable results in an effective trapping potential $\tilde{V}(x;t)$ for the visible coordinate. Evaluating the work done on the particle using the effective potential and the visible trajectories results in an entropy production that breaks the fluctuation theorem. The reason is the non-Markovianity of the visible process which is not reflected in the apparent entropy production $\tilde{\sigma}[x(\cdot)]$. We show that, when the hidden degree of freedom evolves on much faster timescales than the visible one, the usual fluctuation relations are restored, as expected.

In appendix~\ref{app_CompareDiss} we additionally present the average entropy production rates of the different effective descriptions for a model system with an additional velocity component $v$ in the $y$-direction.

\subsection{Article: [\emph{J. Stat. Mech.}, 063204 (2018)]}\label{sec_hid_articleHiddenDegrees}
The following article is reprinted from M. Kahlen and J. Ehrich, arXiv:1803.04740. Journal reference: [M. Kahlen and J. Ehrich, ``Hidden slow degrees of freedom and fluctuation theorems: an analytically solvable model,'' \emph{J. Stat. Mech.}, 063204 (2018)] (Ref.~\cite{Kahlen2018}).

The bracketed numbers provide a continuous pagination.
\cleardoublepage
\pagestyle{pdfStyle}
\includepdf[pages=-,pagecommand={},scale=1]{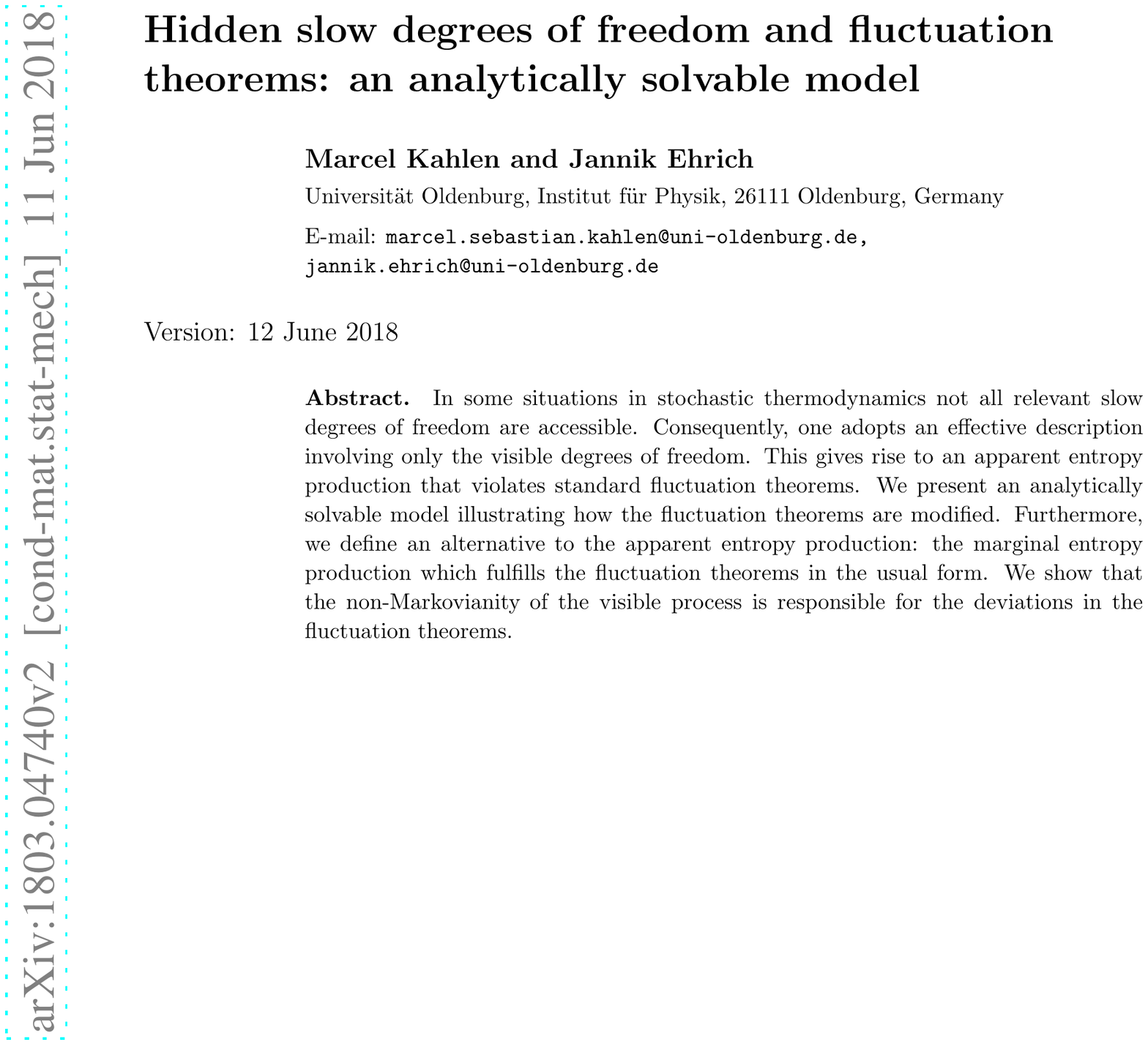}
\pagestyle{thesisStyle}

\chapter{Inferring dissipation with hidden Markov models}\label{chap_hid_inferringDiss}
In this final chapter of the thesis we turn to the question whether, in a setting with hidden slow degrees of freedom, one can reconstruct the complete hidden entropy production from the visible data. In the last chapter we have seen that using the full non-Markovian statistics of the visible trajectories allows the construction of an entropy measure that fulfills standard fluctuation theorems. This motivates the question of how else one can use the full statistics: We will now show that they can be used to bound the dissipation occurring in the complete system. 

It should be clear that this procedure can only work in a limited setting. For diffusive processes with a continuous state space, we would have to sample the statistics of entire trajectories which is not feasible in practice. However, for the setting of a \emph{masked Markov network} the necessary statistics are in the realm of experimental possibility as we will see.

We will in the following consider \emph{discrete-time} Markov chains. Instead of transition rates, we will be dealing with \emph{transition probabilities}. Similarly, entropy production \emph{rates} will thus become entropy productions \emph{per time-step}.

\section{Masked Markov networks}
We consider a stochastic system modeled by a discrete-time Markov chain $z_t$ on a finite set of states labeled $1$ through $K$. The probability of the system being in each state at time $t$ is given by the vector $\bp_t$. Its evolution is described by the discrete-time master equation (in contrast to the continous-time Master equation given in Eq.~\eqref{eqn_basics_MEMatrixNotation}):
\begin{align}
\bp_{t+1} = \mathcal{A}\,\bp_{t},
\end{align}
where $\mathcal{A}$ is the $K\times K$ \emph{transition probability matrix}, which we assume to be time-independent. Its entry $a_{ij}$ in row $i$ and column $j$ gives the probability to go from state $j$ to state $i$ in one time step:
\begin{align}
a_{ij} = p(z_{t+1}=i|z_t=j).
\end{align}
Due to probability conservation, the columns of $\mathcal{A}$ sum to $1$.

We assume that the system is in the \emph{steady-state} described by the steady-state probabilities $\bm{\pi}$ which satisfy
\begin{align}
\bm{\pi} = \mathcal{A}\, \bm{\pi}.
\end{align}
Additionally, we demand that every transition is not \emph{absolutely irreversible}, such that $a_{ij} \neq 0 \Leftrightarrow a_{ji} \neq 0$.

\begin{figure}[ht]
	\centering
	\includegraphics[width = 0.6\linewidth]{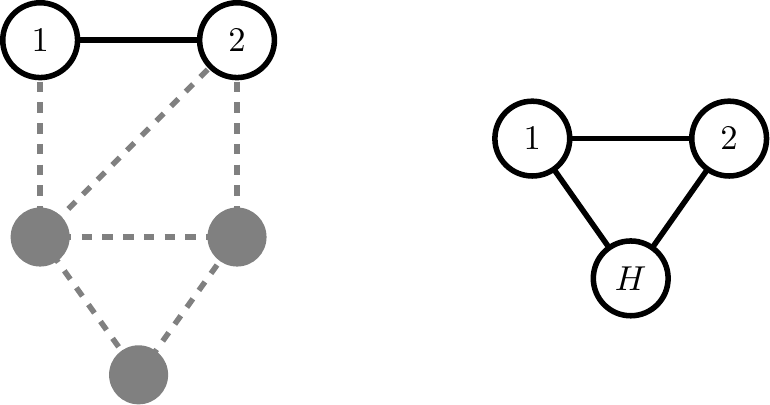}
	\caption{Left: Network of states of the complete process. The labeled states 1 and 2 are observed. All other states (grey) are not observed. Right: Observed network of states. All unobserved states are lumped together into one \emph{hidden} state $H$.}
	\label{fig_hid_setupMaskedNetwork}
\end{figure}

When some states are \emph{masked}, i.e. not observable, an observer will conclude the process to be in a generic hidden state $H$ whenever they do not see it. Therefore, from the point of view of an observer, all unobserved states are lumped together. Fig.~\ref{fig_hid_setupMaskedNetwork} shows a schematic setup. In the following the network of states is shown as a graph whose vertices are the states and whose edges indicate transitions (in both direction) between these states.

To keep the discussion simple, we will focus on cases where only two states (labeled 1 and 2) are visible. An observer thus sees a three-state network (1, 2, and $H$) and monitors three possible transitions. It will become clear that a generalization is straightforward.

\section{Entropy productions}
Let us consider some entropy productions for this setting.

\subsection{Complete hidden entropy production}
The complete \emph{average entropy production per time step} $\Delta\Sigma$ is constant in this setting. Generalizing from Eq.~\eqref{eqn_basics_averageEPMarkov} to the discrete-time setting, it is given by:
\begin{align}
\Delta \Sigma = \sum\limits_{ij} a_{ij} \pi_j \ln\frac{a_{ij}\pi_j}{a_{ji}\pi_i} \geq 0,
\end{align}
where $\pi_i = [\bm{\pi}]_i$ is the stationary probability of state $i$.

From our discussion in Sec.~\ref{sec_basics_adiabaticNonadabaticEP} it is clear that the entropy production corresponds to the \emph{adiabatic entropy production}, which means that it is proportional to the energy spent as \emph{house-keeping heat} in order to maintain non-zero probability currents (see also Sec.~\ref{sec_basics_housekeepingHeatFT}) in the steady state. This implies that we can equivalently write:
\begin{align}\label{eqn_hid_EPProdPerTimeStep}
\Delta \Sigma = -\frac{\Delta Q}{\kB T} = \sum\limits_{ij} a_{ij} \pi_j \ln\frac{a_{ij}}{a_{ji}} \geq 0.
\end{align}

\subsection{Partial entropy productions}
Two similar strategies have been proposed to infer dissipation in masked Markov networks. These are the \emph{partial entropy productions} of Shiraishi and Sagawa~\cite{Shiraishi2015} and Polettini and Esposito~\cite{Polettini2017}, who proposed to only use the net current flowing on the edge $1-2$ to estimate entropy production. 

A detailed discussion of these measures and their relation can be found in Ref.~\cite{Bisker2017}. The authors conclude that the so-called \emph{informed} partial entropy production of Polettini and Esposito is closer to the complete hidden entropy production than the partial entropy production by Shiraishi and Sagawa. We will not discuss this issue further and instead focus on the \emph{apparent entropy} production obtained from mapping a Markov process to the observed data. This method is similar in spirit to the partial entropy productions.

\subsection{Apparent entropy production}
This strategy consists of counting transitions between the states $1$, $2$, and $H$ and thus mapping the dynamics to a Markov model with three states. In accordance with our discussion in the previous chapter we call this procedure \emph{coarse-graining}. It produces a lower-bound entropy production estimate $\Delta \tilde\Sigma$, the \emph{apparent entropy production}:
\begin{align}\label{eqn_hid_defAppEP}
\Delta\tilde\Sigma = \sum\limits_{ij} \tilde a_{ij} p_j \ln\frac{\tilde a_{ij}}{\tilde a_{ji}},
\end{align}
where $i$ and $j$ run over all visible states \emph{including} the hidden state $H$, and the \emph{coarse-grained} transition probabilities can be calculated from counting the number $n[j~\rightarrow~i]$ of transitions from visible state $j$ to state $i$ in a long trajectory:
\begin{align}
\tilde{a}_{ij} = \frac{n[j\rightarrow i]}{\sum\limits_i n[j \rightarrow i]}.
\end{align}

Crucially, the trajectory-dependent counterparts of the partial entropy productions both fulfill fluctuation theorems~\cite{Shiraishi2015,Bisker2017} while the apparent entropy production does not, as we have seen already in Sec.~\ref{sec_hid_FTs}.

\subsection{Marginal entropy production}
Next, we can consider the \emph{marginal entropy production} which amounts to estimating the time-irreversibility of the observed trajectories (see Sec.~\ref{sec_hid_marginalProcess}). This is because the average complete entropy production is given by the Kullback-Leibler divergence between the full trajectories and their time-reverse (see Sec.~\ref{sec_basics_secondLawDKL}).

The \emph{average marginal entropy production per time step} is defined as
\begin{align}\label{eqn_hid_EPDKL}
\Delta \Sigma_x = \lim\limits_{N \rightarrow \infty} \frac{1}{N} \sum\limits_{x_{1:N}} p(x_{1:N}) \ln\frac{p(x_{1:N})}{p(\bar x_{1:N})},
\end{align}
where $x_{1:N} = \{x_1,x_2,...,x_N\}$ denotes an entire trajectory of length $N$ and $p(x_{1:N})$ the probability of its occurrence. The overbar denotes time-reversal:
\begin{align}
\bar{x}_{1:N}~=~\{x_N,x_{N-1},...,x_1\}.
\end{align}
Notice that we do not require a time-reversed trajectory probability $\bar{p}[x_{1:N}]$ since there is no external control protocol to be time-reversed.

We will turn to the question of how to compute the average marginal entropy production in Sec.~\ref{sec_hid_estTimeIrreversibility}. Crucially, it underestimates the real entropy production per time step [see Eq.~\eqref{eqn_hid_marginalEPUnderestimates}], which can also be written in terms of time-series irreversibility [Eq.~\eqref{eqn_basics_DissipationKullbackLeibler}]:
\begin{align}
\Delta \Sigma &= \lim\limits_{N \rightarrow \infty} \frac{1}{N} \sum\limits_{z_{1:N}} p(z_{1:N}) \ln\frac{p(z_{1:N})}{p(\bar z_{1:N})}\\
&\geq \Delta\Sigma_x.
\end{align}

\subsection{Dissipation from thermodynamic uncertainty relation}
Another, loosely-related approach is to use the \emph{thermodynamic uncertainty relation}~\cite{Barato2015,Gingrich2016}: From the fluctuations of an observed current one can construct a lower bound on the steady-state entropy production. It has proven useful in the analysis of molecular motors~\cite{Pietzonka2016} and heat engines~\cite{Pietzonka2018a}. Li \emph{et al.}~\cite{Li2019} have recently analyzed its effectiveness in estimating the complete dissipation in systems which are only partially observable.

Importantly, in some way or another all the effective descriptions apart from the marginal entropy production assume some kind of Markovian dynamics which makes them useless in situations in which there is no observable net probability current flowing in the visible part of the system~\cite{Martinez2019}.

\section{Estimation of time-series irreversibility}\label{sec_hid_estTimeIrreversibility}
Mart\'{i}nez \emph{et al.} have found a simple way to calculate time-series irreversibility for masked continuous-time Markov jump processes~\cite{Martinez2019} in the steady state. Here, we adapt this formalism to discrete-time processes. 

\begin{figure}[ht]
	\centering
	\includegraphics[scale=0.5]{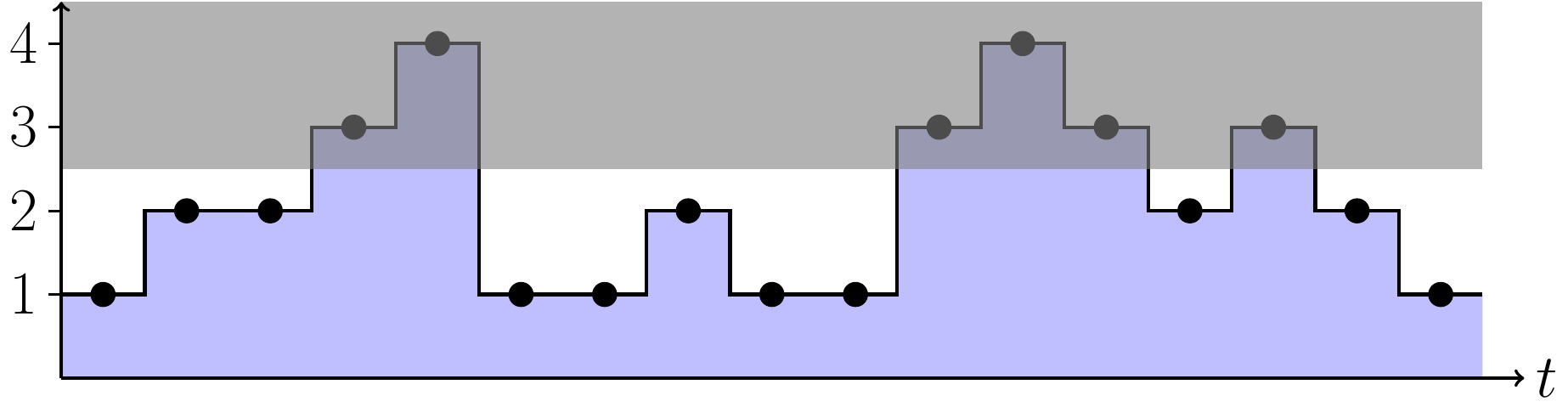}
	\hspace*{-1mm}\includegraphics[scale=0.5]{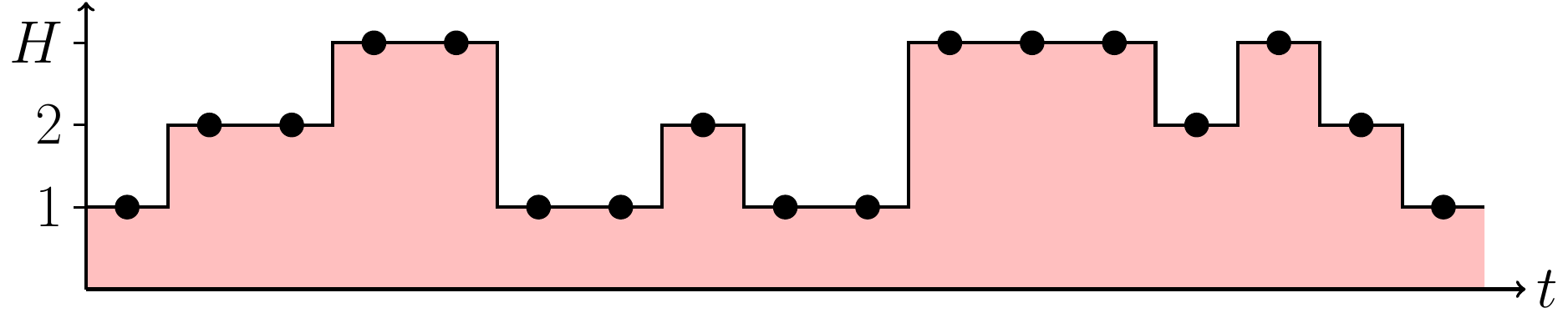}
	\caption{Sample trajectory of the complete (masked) process and the corresponding visible trajectory generated by lumping together the states 3 and 4 into one hidden (H) state.}
	\label{fig_hid_trajectories}
\end{figure}

Figure~\ref{fig_hid_trajectories} shows an example trajectory of the process. Consider the visible trajectory as consisting of jumps between visible states with waiting times $n_i$. A trajectory of length $N$ with $M$ jumps is thus specified by
\begin{align}
x_{1:N} = \left\{(x_1 \rightarrow x_2, n_1) , (x_2 \rightarrow x_3, n_2), ..., (x_M \rightarrow x_{M+1}, n_M)\right\},
\end{align}
where the $x_i \in \{1,2\}$ are elements of the visible set of states only. Note that this restricts us to considering only trajectories starting and ending in the visible part of the network, which can always be achieved by \emph{cutting off} parts of the sampled trajectory.

\begin{figure}[ht]
	\centering
	\includegraphics[scale=0.5]{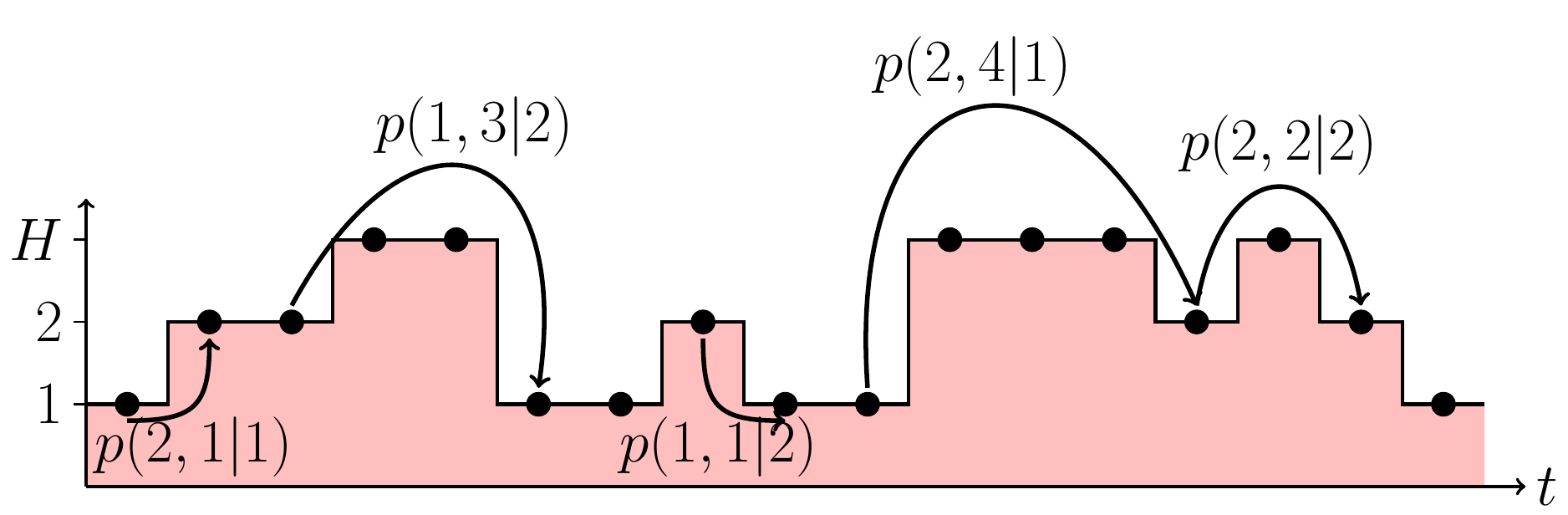}
	\caption{Visible trajectory annotated with jump probabilities between visible states. The jump probability $p(i,n|j)$ denotes the probability that the system jumps from $j$ to $i$ with $n-1$ intermediate hidden states (i.e., in $n$ time steps).}
	\label{fig_hid_trajectoryArrows}
\end{figure}

The probability of such a trajectory reads:
\begin{align}
p(x_{1:N}) = \tilde{\pi}_{x_1}\, p(x_2,n_1|x_1) \, p(x_3,n_2|x_2)... p(x_{M+1},n_M|x_M),
\end{align}
where $\tilde{\pi}_{x_1} := \pi_{x_1}/(\pi_{1} + \pi_{2})$ is the probability that the trajectory started in state $x_1$ and $p(i,n|j)$ denotes the probability that the system jumps from $j$ to $i$ in $n$ time steps (i.e., with $n-1$ intermediate hidden states). Examples for these transition probabilities can be found in Fig.~\ref{fig_hid_trajectoryArrows}. The visible process therefore has the structure of a \emph{semi-Markov chain}~\cite{Martinez2019}, i.e., a discrete-time Markov chain with non-Poissonian waiting times.

The probability of the time-reversed trajectory reads
\begin{align}
p(\bar{x}_{1:N}) = \tilde{\pi}_{x_{M+1}}\, p(x_{M},n_M|x_{M+1}),..., p(x_2,n_2|x_3)\, p(x_{1},n_1|x_2).
\end{align}

We can thus calculate the \emph{average time-irreversibility per jump} between the visible states:
\begin{align}
\delta \Sigma_{x} &= \lim\limits_{M\rightarrow \infty} \frac{1}{M}\sum\limits_{x_1,x_2,...,x_{M+1}} \sum\limits_{n_1,n_2,...,n_M} p(x_{1:N})\ln\frac{p(x_{1:N})}{p(\bar{x}_{1:N})}\\
&= \lim\limits_{M\rightarrow \infty} \frac{1}{M} \sum\limits_{k=1}^M \sum\limits_{x_1,...,x_{M+1}} \sum\limits_{n_1,...,n_M} p(x_{1:N}) \ln\frac{p(x_{k+1},n_k|x_k)}{p(x_k,n_k~|x_{k+1})}\\
&= \lim\limits_{M\rightarrow \infty} \frac{1}{M} \sum\limits_{k=1}^M \sum\limits_{x_k,x_{k+1}} \sum\limits_{n_k} p(x_{k+1},n_k|x_k)\, \tilde{\pi}_{x_k} \ln\frac{p(x_{k+1},n_k|x_k)}{p(x_k,n_k~|x_{k+1})}\\
&= \sum\limits_{i,j=1}^2\sum\limits_{n=1}^\infty p(i,n|j)\, \tilde{\pi}_j \ln\frac{p(i,n|j)}{p(j,n|i)}.
\end{align}

To get the \emph{irreversibility per time step}, we need to divide by the average number $\bar n$ of time steps per jump between the visible states. In a long trajectory this is given by the total jump length $l$ divided by the number of jumps $M$: $ \bar{n} = l/M$. The total jump length is in turn given by the sum of the number of jumps and the total number $n_H$ of hidden symbols in the trajectory: $l = M+n_H$. Finally, if the trajectory length is $N$, we note the following relationship: $N = M + n_H +1$. We thus get:
\begin{align}
\bar n &= \lim\limits_{N \rightarrow \infty} \frac{l}{M} =\lim\limits_{N \rightarrow \infty} \frac{M+n_H}{M}\\
&= \lim\limits_{N \rightarrow \infty} 1+\frac{n_H}{N-n_H-1}= \lim\limits_{N \rightarrow \infty} \frac{1}{1-n_H/N}\\
&= \frac{1}{1-\pi_H} = \frac{1}{\pi_1+\pi_2},
\end{align}
where $\pi_H$ denotes the steady-state probability to observe a hidden state.

Therefore, the time-series irreversibility per time step reads
\begin{align}
\Delta \Sigma_{x} &= \frac{\delta \Sigma_{x}}{\bar n}\\
&=  \sum\limits_{i,j=1}^2\sum\limits_{n=1}^\infty p(i,n|j) \, \pi_j \ln\frac{p(i,n|j)}{p(j,n|i)}. \label{eqn_hid_irreversibilityPerStep}
\end{align}

Importantly, the jump probabilities $p(i,n|j)$ can easily be sampled to the desired accuracy in an experiment. This implies that, at least for masked jump processes, the \emph{average marginal entropy production} is experimentally accessible. This is in contrast to the challenge of obtaining the infinite-point statistics necessary to sample continuous trajectories of diffusive-type processes with continuous state spaces.

\subsection{Calculating the semi-Markov transition probabilities}
In order to apply Eq.~\eqref{eqn_hid_irreversibilityPerStep}, one needs to measure the semi-Markov transition probabilities $p(i,n|j)$. For the purpose of exploring how much irreversibility one can theoretically measure, we proceed to calculate these probabilities from an underlying model.

We first note that the transition matrix $\mathcal{A}$ can be written in a block form:
\begin{align} \label{eqn_hid_blockmatrixForm}
\mathcal{A} = \begin{pmatrix}
\begin{pmatrix}
a_{11}&a_{12}\\
a_{21}&a_{22}\\
\end{pmatrix} & \mathcal{C}\\
\mathcal{B} & \mathcal{H}
\end{pmatrix},
\end{align}
where $\mathcal{B}$ denotes the matrix of probabilities of transitions from the visible part into the hidden part of the network, $\mathcal{C}$ labels those transition probabilities from the hidden to the visible parts, and $\mathcal{H}$ denotes transition probabilities between the hidden states.

The probabilities for the jumps between the visible states can directly be observed from the time-series. In order to jump from one visible state to another in $n\geq 2$ time steps, the process needs to jump from the visible into the hidden part of the network, then jump $n-2$ times among the hidden states, and finally jump into the visible part. Therefore, we obtain:
\begin{align}
p(i,n|j) = \begin{cases}
a_{ij} & \, n = 1\\
\left[ \mathcal{C} \, {\mathcal{H}}^{n-2} \, \mathcal{B} \right]_{ij} & n > 1.
\end{cases} \label{eqn_hid_jumpProbsFromBlockForm}
\end{align}

\subsection{Example network}
\begin{figure}[ht]
	\centering
	\includegraphics[width = 0.6\linewidth]{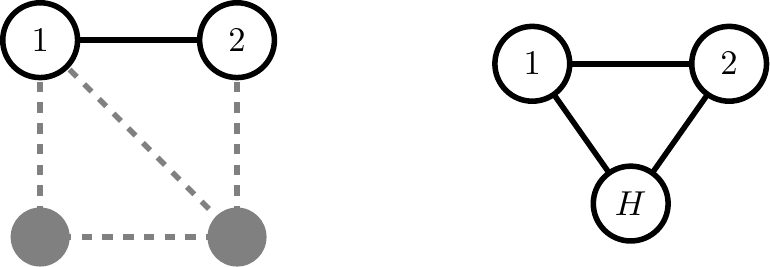}
	\caption{Topology of the example network (left) and reduced visible network (right).}
	\label{fig_hid_exampleNetwork}
\end{figure}

For an explicit example, we consider the network  specified by the transition matrix
\begin{align} \label{eqn_hid_exampleNetwork}
\mathcal{A} = \begin{pmatrix}
0.4 - 0.1 \, e^{\Delta \mu/2} & 0.2 \, e^{-\Delta \mu/2} & 0.3 & 0.3\\
0.1 \, e^{\Delta \mu/2} & 0.9 - 0.2 \, e^{-\Delta \mu/2} & 0.1 & 0\\
0.1 & 0.1 & 0.4 & 0.6\\
0.5 & 0   & 0.2 & 0.1
\end{pmatrix},
\end{align}
where $\Delta \mu$ controls the transition probabilities at the edge $1-2$. In particular, for $\Delta \mu \approx 0.85$ there is no net current flowing over the visible edge. The topology of this example network is depicted in Fig.~\ref{fig_hid_exampleNetwork}.

Assuming perfect statistics, we can analytically calculate the coarse-grained transition probabilities of the corresponding Markov process on the reduced network of states (see Eq.~\ref{eqn_hid_defEffectiveRates} and Ref.~\cite{Esposito2012} for discrete state spaces):
\begin{align}
\tilde{a}_{ij} = \begin{cases}
a_{ij} & i,j =1,2\\
a_{3 j} + a_{4j} & i= H \, ,\, j = 1,2\\
\frac{a_{i3}\,p_3 + a_{i4}\,p_4}{p_3+p_4} & i=1,2\, ,\, j = H\\
\frac{(a_{33}+a_{43})\,p_3 + (a_{34}+a_{44})\,p_4}{p_3+p_4} & i,j = H \;.
\end{cases} 
\end{align}
These can be used in Eq.~\eqref{eqn_hid_defAppEP} to calculate the apparent entropy production.

\begin{figure}[ht]
	\centering
	\includegraphics[width=0.65  \linewidth]{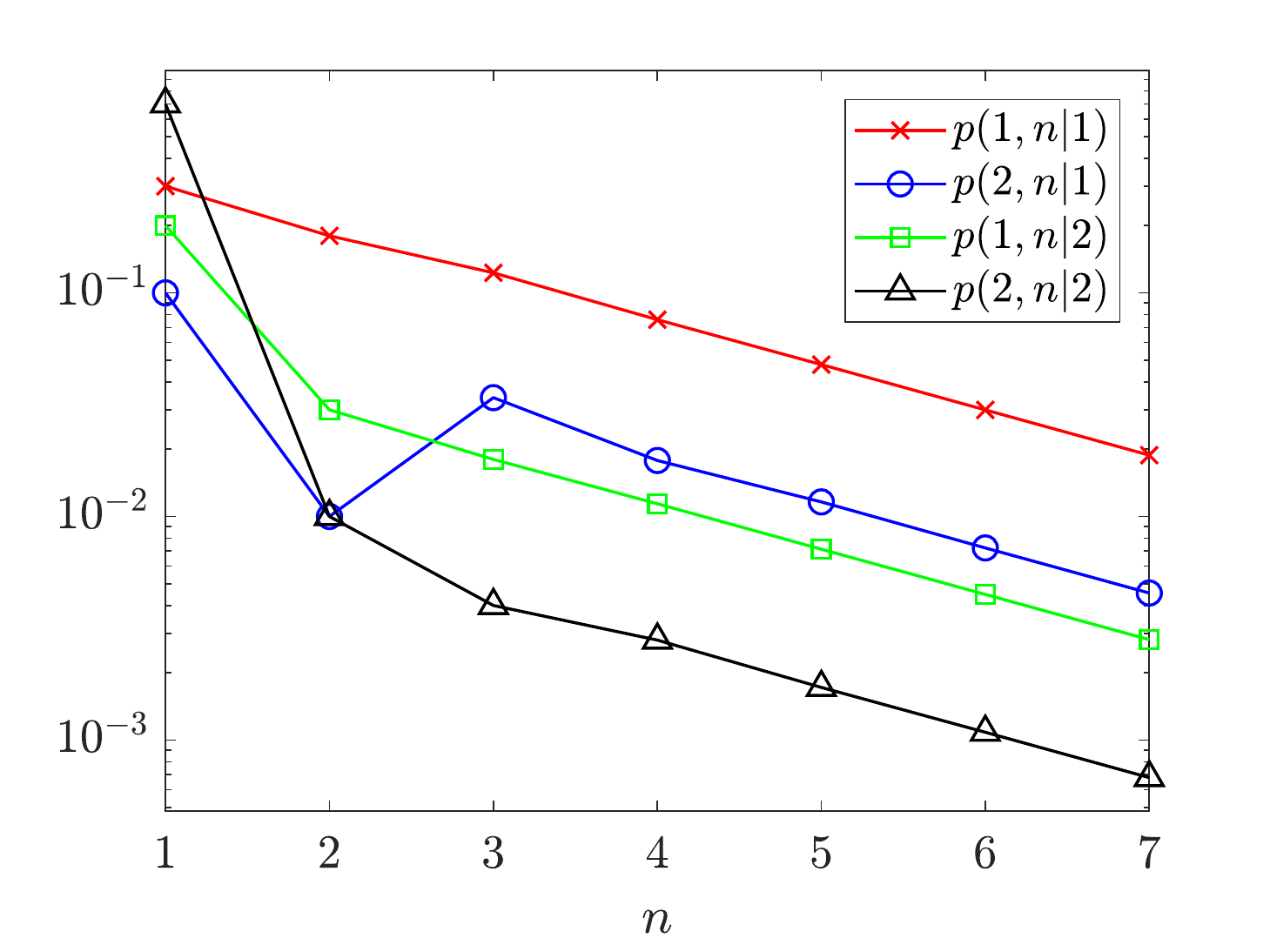}
	\caption{Semi-Markov jump probabilities for the example network.}
	\label{fig_hid_exampleJumpProbs}
\end{figure}

Using Eqs.~\eqref{eqn_hid_blockmatrixForm}~and~\eqref{eqn_hid_jumpProbsFromBlockForm}, we can calculate the matrix $\mathcal{P}(n)$ of semi-Markov jump probabilities for $n>1$:
\begin{align}
\mathcal{P}(n) &:= \begin{pmatrix}
p(1,n|1) & p(1,n|2)\\
p(2,n|1) & p(2,n|2)
\end{pmatrix}\\
&= \begin{pmatrix}
0.123-0.18 \lambda_2 & 0.018-0.03\lambda_2\\
0.034-0.01\lambda_2 & 0.004 - 0.01 \lambda_2
\end{pmatrix} \frac{\lambda_1^{n-2}}{\lambda_1-\lambda_2} \nonumber\\
&\qquad- \begin{pmatrix}
0.033 + 0.18\lambda_2 & 0.003 + 0.03 \lambda_2\\
0.029 + 0.01\lambda_2& -0.001 + 0.01 \lambda_2
\end{pmatrix} \frac{\lambda_2^{n-2}}{\lambda_1-\lambda_2},\label{eqn_hid_exampleTransitionProbs}
\end{align}
where $\lambda_1 \approx 0.63$ and $\lambda_2 \approx -0.13$ are the eigenvalues of the $\mathcal{H}$-submatrix. Fig.~\ref{fig_hid_exampleJumpProbs} shows a plot of these probabilities.

\begin{figure}[ht]
	\centering
	\includegraphics[width = 0.65\linewidth]{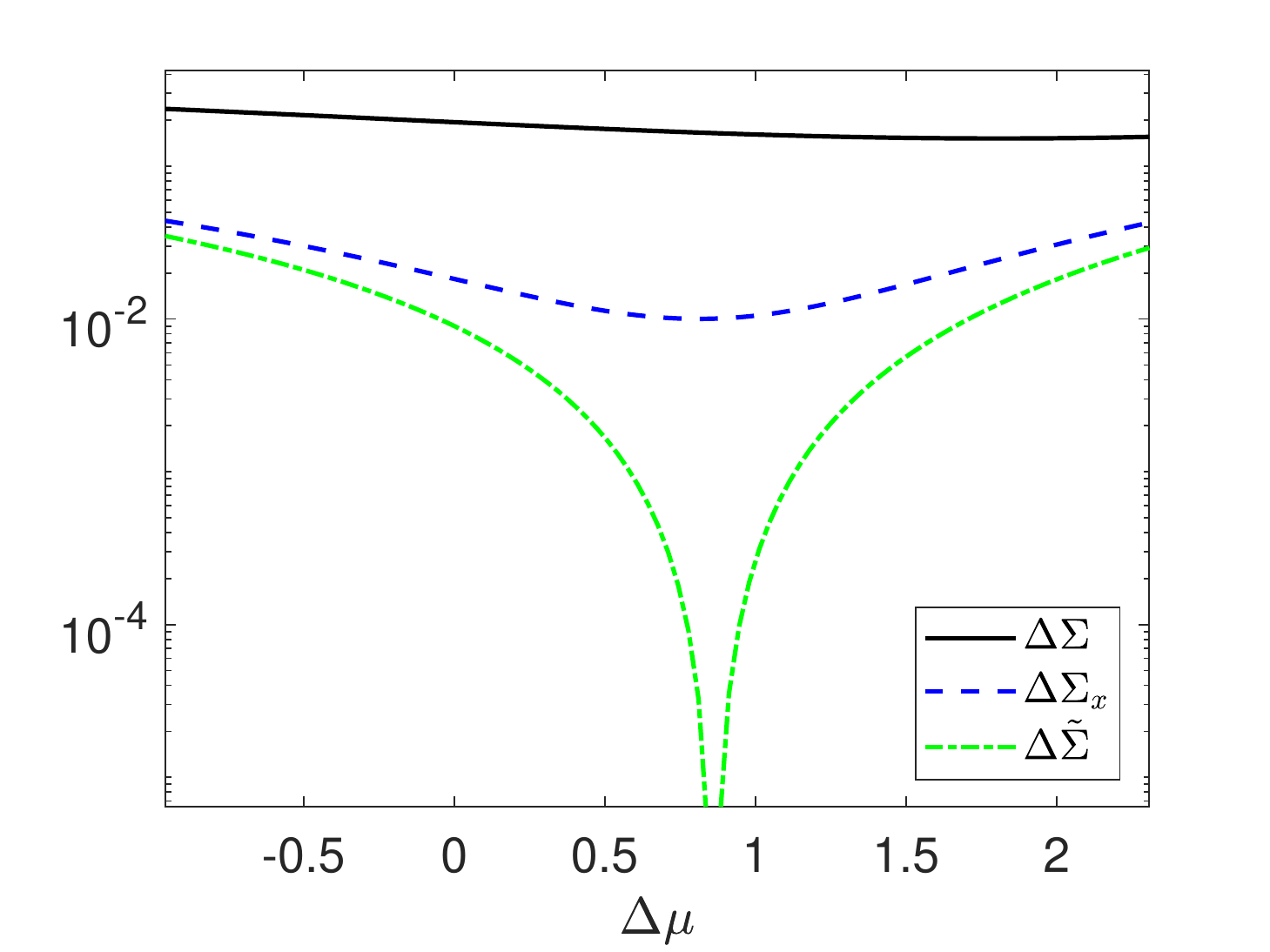}
	\caption{Comparison of irreversibility measures for the example network for different values of $\Delta \mu$. From top to bottom: The real entropy production $\Delta \Sigma$, the time-series irreversibility $\Delta\Sigma_{x}$, and the apparent entropy production  $\Delta\tilde{\Sigma}$.}
	\label{fig_hid_exampleTuneMu_1}
\end{figure}

Using Eq.~\eqref{eqn_hid_irreversibilityPerStep}, we can calculate the time-series irreversibility per time step. Fig.~\ref{fig_hid_exampleTuneMu_1} shows a comparison of the two irreversibility measures together with the real entropy production of the underlying network. Importantly, the time-series irreversibility still captures entropy production, when the apparent entropy production vanishes and thus indicates a reversible process~\cite{Martinez2019}. Nevertheless, both measures significantly underestimate the real entropy production.

\section{Fitting the hidden process}
Let us follow a different approach by finding models of the hidden network that match the available data. By searching the space of all such models, we are able to find refined bounds for the total entropy production of the underlying system.

The situation is well known in the machine learning community and is commonly tackled by training a \emph{hidden Markov model} on the data. This would enable us to find the underlying hidden structure outlined in Sec.~\ref{sec_hid_setupHiddenProcess}.

We could, therefore, use any of the standard algorithms to train a model on the data (see, e.g., Chap.~13.2.1 of Ref.~\cite{Bishop2006}). However, we expect that there are multiple parameter configurations of the underlying network that lead to the observed probabilities. The training algorithm will thus converge to only one solution of the (possibly high dimensional) solution space. Therefore, it seems worthwhile to ask whether one can go further analytically.

First, we need to do a \emph{model selection}, i.e. we need to know how many hidden states the real network has.

\subsection{Bounding the number of hidden states}
We first show that one can in principle use the measured semi-Markov jump probabilities $p(i,n|j)$ to estimate the number of hidden states in the network. From Eq.~\eqref{eqn_hid_jumpProbsFromBlockForm} it is evident that the $n$-dependence of these probabilities is governed by the eigenvalues of the transition matrix $\mathcal{H}$ between the hidden states. This can also be seen in Eq.~\eqref{eqn_hid_exampleTransitionProbs} for the example network.

Thus, although challenging for imperfect statistics, one can in principle fit a sum of exponentials to the measured curves\footnote{This was also pointed out by Amann \emph{et al.}\cite{Amann2010} as a possibility to identify nonequilibrium in a system that has only two observable states.}: 
\begin{align} \label{eqn_hid_fitSumExponentials}
p(i,n|j) = \sum\limits_{k=1}^{H} \alpha_{ij}^{(k)}\, \lambda_k^{n-2}.
\end{align}

The lowest possible number $H$ of terms will then give a lower bound on the number of hidden states. It is only a lower bound since we cannot be certain that the eigenvalues of the $\mathcal{H}$-matrix are not degenerate, although this seems unlikely in realistic scenarios.

\subsection{Unicyclic networks as a special case}
Networks with only one cycle are special: In this case the time-series irreversibility already captures the real entropy production, $\Delta \Sigma_{x} = \Delta \Sigma$. To see why, we start with the marginal entropy production [Eq.~\eqref{eqn_hid_irreversibilityPerStep}] and split out the contribution from the direct jumps:
\begin{align}
\Delta \Sigma_{x} &= \left( a_{12} \,\pi_2 - a_{21}\, \pi_1\right) \ln\frac{a_{12}}{a_{21}}  \nonumber\\
&\quad + \sum\limits_{n=2}^\infty \left[ p(1,n|2) \, \pi_2 - p(2,n|1)\, \pi_1 \right] \ln\frac{p(1,n|2)}{p(2,n|1)}.
\end{align}
Let there be $m$ hidden states. They can be arranged so that the matrix $\mathcal{H}$ is tri-diagonal. Then, the ratio of jump probabilities can be written as:
\begin{align}
\frac{p(1,n|2)}{p(2,n|1)} &= \frac{a_{1,m+2} \left[\mathcal{ H}^{n-2} \right]_{m1} a_{32}}{a_{m+2,1} \left[\mathcal{H}^{n-2} \right]_{1m} a_{23}}.
\end{align}
Due to a special property of tri-diagonal matrices (see the proof in Appendix~\ref{app_triDiagonal}), we can simplify this to:
\begin{align}
\frac{p(1,n|2)}{p(2,n|1)} &= \frac{a_{1,m+2} \, h_{m,m-1}...\, h_{32} \, h_{21}\, a_{32}}{a_{m+2,1} \, h_{m-1,m}...\, h_{23}\, h_{12}\, a_{23}}\\
&= \frac{a_{1,m+2}\, a_{m+2,m+1}...a_{43}\,a_{32}}{a_{m+2,1}\, a_{m+1,m+2}...a_{34}\,a_{23}}\\
&= \frac{p(1,m+1|2)}{p(2,m+1|1)},\quad \text{for } n\geq m+1. \label{eqn_hid_ratioForUnicyclic}
\end{align}

Additionally, we find
\begin{align}
\tilde\pi_1 &= \sum\limits_{n=1}^\infty \left[p(1,n|1)\tilde{\pi}_1 + p(1,n|2)\tilde{\pi}_2\right]\\
\Leftrightarrow 0 &=\sum\limits_{n=1}^\infty \left[ p(2,n|1) \, \tilde{\pi}_1 - p(1,n|2)\, \tilde{\pi}_2\right]\\
\Leftrightarrow 0 &=\sum\limits_{n=1}^\infty \left[ p(2,n|1) \, \pi_1 - p(1,n|2)\, \pi_2\right],
\end{align}
where we used $\sum_{n=1}^\infty \left[p(1,n|1) + p(2,n|1)\right] = 1$ in the second line. Thus,
\begin{align}
-\sum\limits_{n=2}^\infty \left[ p(2,n|1) \, \pi_1 - p(1,n|2)\, \pi_2\right] &=  a_{21}\,\pi_1 - a_{12}\, \pi_2.
\end{align}
Finally, we have
\begin{align}
\Delta\Sigma_{x} &= \left(a_{12}\, \pi_2-  a_{21}\, \pi_1\right) \ln\frac{a_{12}\,a_{23}...\,a_{m+2,1}}{a_{1,m+2}...\,a_{32}\,a_{21}}\\
&= \Delta\Sigma,
\end{align}
where we used the Schnakenberg decomposition~\cite{Schnakenberg1976} (which expresses the entropy production as net flux times affinity summed over all cycles) for the special case of the unicyclic network.

Interestingly, Eq.~\eqref{eqn_hid_ratioForUnicyclic} suggests a way to identify whether a given masked network is unicyclic: The ratio $p(1,n|2)/p(2,n|1)$ has to be constant. Then, the number of hidden states can also be inferred from the smallest $n$ for which this ratio exists.

\subsection{Inferring a hidden two-state network}
We now turn to the key finding: Inferring the hidden transition rates of a two-state hidden network from the measurable jump probabilities $p(i,n|j)$. We thus need to fix the 16 elements of the transition matrix $\mathcal{A}$ most conveniently written in the block-matrix from in Eq.~\eqref{eqn_hid_blockmatrixForm}. We re-introduce the four steady-state probabilities $\pi_i$ (although the are redundant information) and thus need to fix 20 unknowns from the following 21 relations:
\begin{subequations}
	\begin{enumerate}
		\item Four elements for the transitions between the visible states. These are simply the one-step probabilities:
		\begin{align}
		a_{ij} = p(i,1|j), \qquad \text{for } i,j \leq 2.
		\end{align}
		\item Two eigenvalues of the $\mathcal{H}$-submatrix inferred from the fitted sum of exponentials [Eq.~\eqref{eqn_hid_fitSumExponentials}]:
		\begin{align}
		\lambda_1\,\lambda_2 &= a_{33}\,a_{44} - a_{34} \, a_{43}\\
		\lambda_1 + \lambda_2 &= a_{33} + a_{44}.
		\end{align}
		\item Four column sums of one:
		\begin{align}
		1 = \sum\limits_{i=1}^4 a_{ij}.
		\end{align}
		\item Four two-step probabilities $\mathcal{P}(2) = \mathcal{C}\,\mathcal{B}$:
		\begin{align}
		p(1,2|1) &= a_{13}\,a_{31} + a_{14}\,a_{41}\\
		p(2,2|1) &= a_{23}\,a_{31} + a_{24}\,a_{41}\\
		p(1,2|2) &= a_{13}\,a_{32} + a_{14}\,a_{42}\\
		p(2,2|2) &= a_{23}\,a_{32} + a_{24}\,a_{42}.
		\end{align}
		\item Four steady-state equations:
		\begin{align}
		\pi_i = \sum\limits_{j=1}^4 a_{ij}\, \pi_j.
		\end{align}
		\item Two measured steady-state probabilities $\pi_1$ and $\pi_2$.
		\item One sum of probabilities: $\pi_1 + \pi_2 + \pi_3 + \pi_4$ = 1.
	\end{enumerate}
\end{subequations}

Solving these equations simultaneously\footnote{There are 21 equations for 20 unknowns because the eigenvalue equations in point 5 are not linearly independent. The reason for why two free parameters are left in total is that the equations are not independent, since $\sum\limits_{n,i} p(i,n|j) = 1$ for $j=1,2$. This is a complicated relationship between the eigenvalues and the two-step probabilities in point 4.} yields a solution depending on two free parameters chosen as $c_3 := a_{33} + a_{43}$ and $c_4 := a_{34} + a_{44}$:
\begin{subequations}
	\begin{align}
	a_{13} &= \frac{[(c_3-1)\,\tilde{a}_{H2} + p(2,2|2)]\,p(1,2|1)- (c_3-1)\,\tilde{a}_{H1} + p(2,2|1)]\,p(2,2|1)}{\tilde{a}_{H1}\, p(1,2|2) + \tilde{a}_{H1}\, p(2,2|2) -\tilde{a}_{H2}\, p(1,2|1) -\tilde{a}_{H2}\, p(2,2|1)}\\
	a_{14} &= \frac{[(c_4-1)\,\tilde{a}_{H2} + p(2,2|2)]\,p(1,2|1)- (c_4-1)\,\tilde{a}_{H1} + p(2,2|1)]\,p(2,2|1)}{\tilde{a}_{H1}\, p(1,2|2) + \tilde{a}_{H1}\, p(2,2|2) -\tilde{a}_{H2}\, p(1,2|1) -\tilde{a}_{H2}\, p(2,2|1)}\\
	a_{23} &=\frac{[(c_3-1)\,\tilde{a}_{H2} + p(1,2|2)]\,p(2,2|1)- (c_3-1)\,\tilde{a}_{H1} + p(1,2|1)]\,p(2,2|2)}{\tilde{a}_{H1}\, p(1,2|2) + \tilde{a}_{H1}\, p(2,2|2) -\tilde{a}_{H2}\, p(1,2|1) -\tilde{a}_{H2}\, p(2,2|1)}\\
	a_{24} &=\frac{[(c_4-1)\,\tilde{a}_{H2} + p(1,2|2)]\,p(2,2|1)- (c_4-1)\,\tilde{a}_{H1} + p(1,2|1)]\,p(2,2|2)}{\tilde{a}_{H1}\, p(1,2|2) + \tilde{a}_{H1}\, p(2,2|2) -\tilde{a}_{H2}\, p(1,2|1) -\tilde{a}_{H2}\, p(2,2|1)}
	\end{align}
	\begin{align}
	a_{31} &= \frac{(1-c_4)\, \tilde{a}_{H1}-p(1,2|1)-p(2,2|1)}{c_3-c_4}\\
	a_{32} &= \frac{(1-c_4)\, \tilde{a}_{H2}-p(1,2|2)-p(2,2|2)}{c_3-c_4}\\
	a_{31} &= \frac{(c_3-1)\, \tilde{a}_{H1}+p(1,2|1)+p(2,2|1)}{c_3-c_4}\\
	a_{42} &= \frac{(c_3-1)\, \tilde{a}_{H2}+p(1,2|2)+p(2,2|1)}{c_3-c_4},
	\end{align}
	where $\tilde{a}_{H i} = a_{3i} + a_{4i} = 1-a_{1i}-a_{2i}$. The other probabilities read:
	\begin{align}
	a_{33} &= \frac{(-c_4+\lambda_1 + \lambda_2)\, c_3 - \lambda_1\,\lambda_2}{c_3-c_4}\\
	a_{34} &= -\frac{(c_4-\lambda_2)\,(c_4-\lambda_1)}{c_3-c_4}\\
	a_{43} &= \frac{(c_3-\lambda_2)\,(c_3-\lambda_1)}{c_3-c_4}\\
	a_{44} &= \frac{(c_3-\lambda_1 - \lambda_2)\, c_4 + \lambda_1\,\lambda_2}{c_3-c_4}.
	\end{align}
\end{subequations}

Additionally, the solution needs to obey $0\leq a_{ij} \leq 1$ for all $i,j$. This gives 12 additional inequalities which can be checked numerically in a parameter sweep for $0 \leq \{c_3,c_4\} \leq 1$. Because of the arbitrariness in numbering the hidden states, the solution is symmetric with respect to interchanging $c_3$ and $c_4$.

\begin{figure}[ht]
	\centering
	\includegraphics[width=0.65 \linewidth]{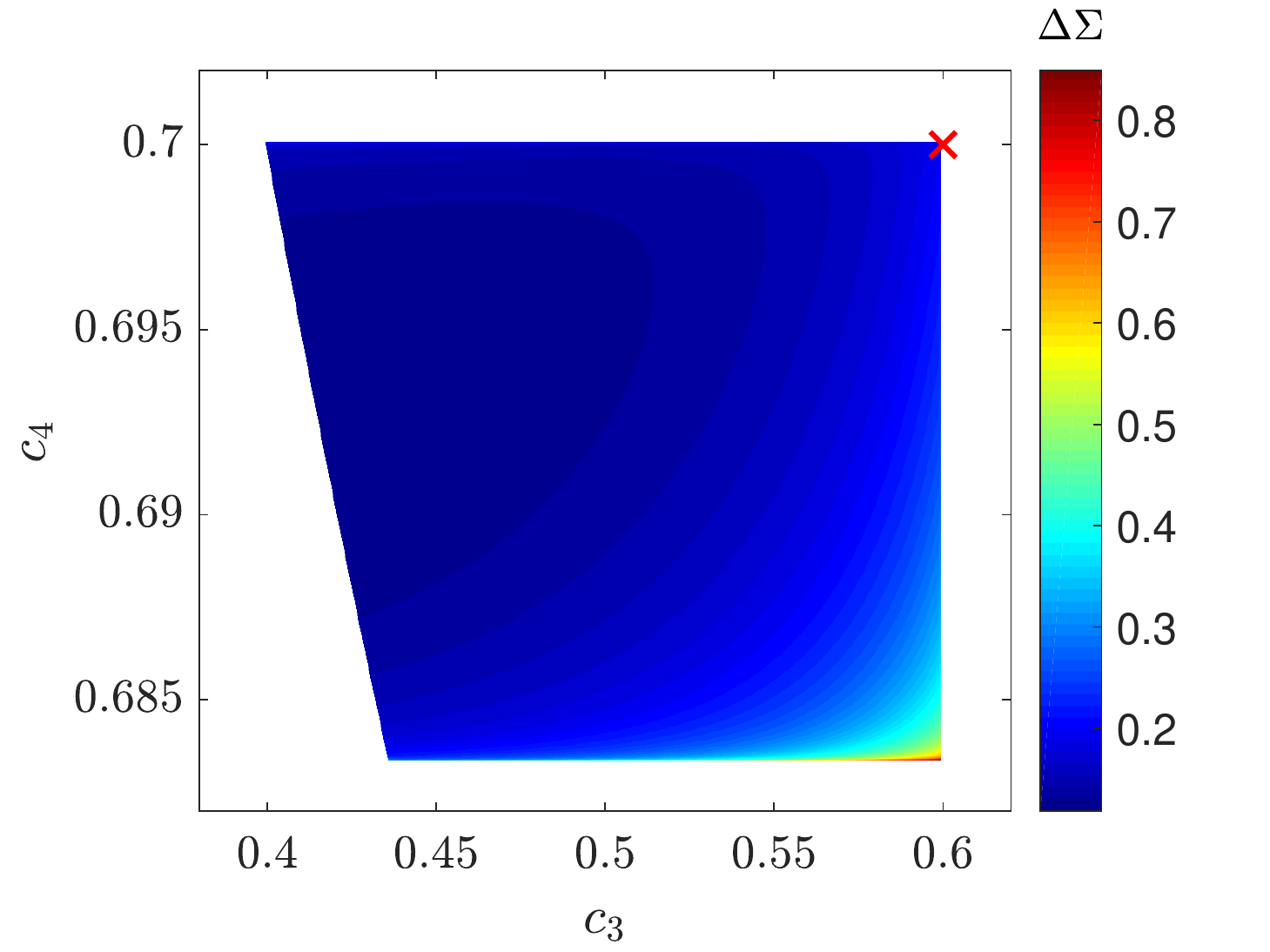}
	\caption{Entropy production $\Delta \Sigma$ for all possible hidden two-state networks capable of generating the visible statistics in Eq.~\eqref{eqn_hid_exampleTransitionProbs} of the example network for $\Delta\mu =0$. The solutions are parametrized by $c_3:=a_{33}+a_{34}$ and $c_4:=a_{34}+a_{44}$. The white regions indicate that no allowed solution is possible. The red cross shows the generating network.}
	\label{fig_hid_exampleSweep}
\end{figure}

Using Eq.~\eqref{eqn_hid_EPProdPerTimeStep}, we calculate the entropy production for all allowed solutions. Figure~\ref{fig_hid_exampleSweep} shows the results for the example network specified by the transition matrix in Eq.~\eqref{eqn_hid_exampleNetwork} for $\Delta\mu=0$. Interestingly, the observed statistics are compatible with a wide range of different entropy productions of the underlying network. The generating network lies at the edge of the solution space. This is probably due to it saturating two of the inequalities since $a_{24} = a_{42} = 0$.

\begin{figure}[ht]
	\centering
	\includegraphics[width=0.65 \linewidth]{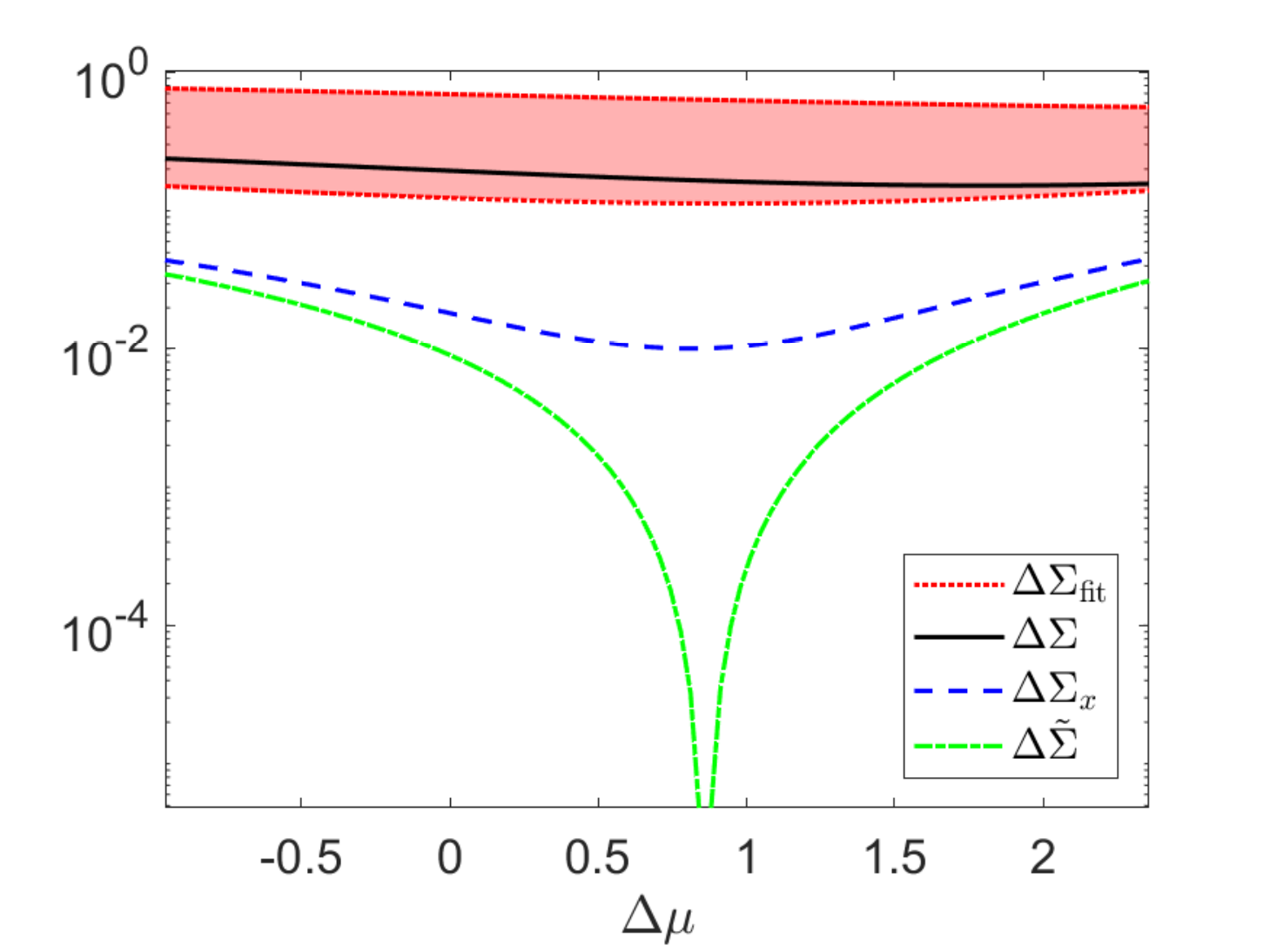}
	\caption{Comparison of irreversibility measures for the example network for different values of $\Delta \mu$. The red band shows the bound $\Delta\Sigma_{\rm fit}$ obtained from finding all possible hidden networks compatible with the observed data. Then, from top to bottom: The real entropy production $\Delta \Sigma$, the time-series irreversibility $\Delta\Sigma_{x}$, and the apparent entropy production  $\Delta\tilde{\Sigma}$.}
	\label{fig_hid_exampleTuneMu_2}
\end{figure}

Thus, remarkably, we obtain upper and lower bounds for the real entropy production. We call this estimate $\Delta\Sigma_{\rm fit}$.

Repeating the above calculations for different values of $\Delta\mu$ allows us to make a plot similar to the one in Fig.~\ref{fig_hid_exampleTuneMu_1}. This is shown in Fig.~\ref{fig_hid_exampleTuneMu_2}. We see that one can significantly improve the previous estimates of entropy production but that there is still a wide range of possible dissipations.

\begin{figure}[ht]
	\centering
	\includegraphics[width=0.65 \linewidth]{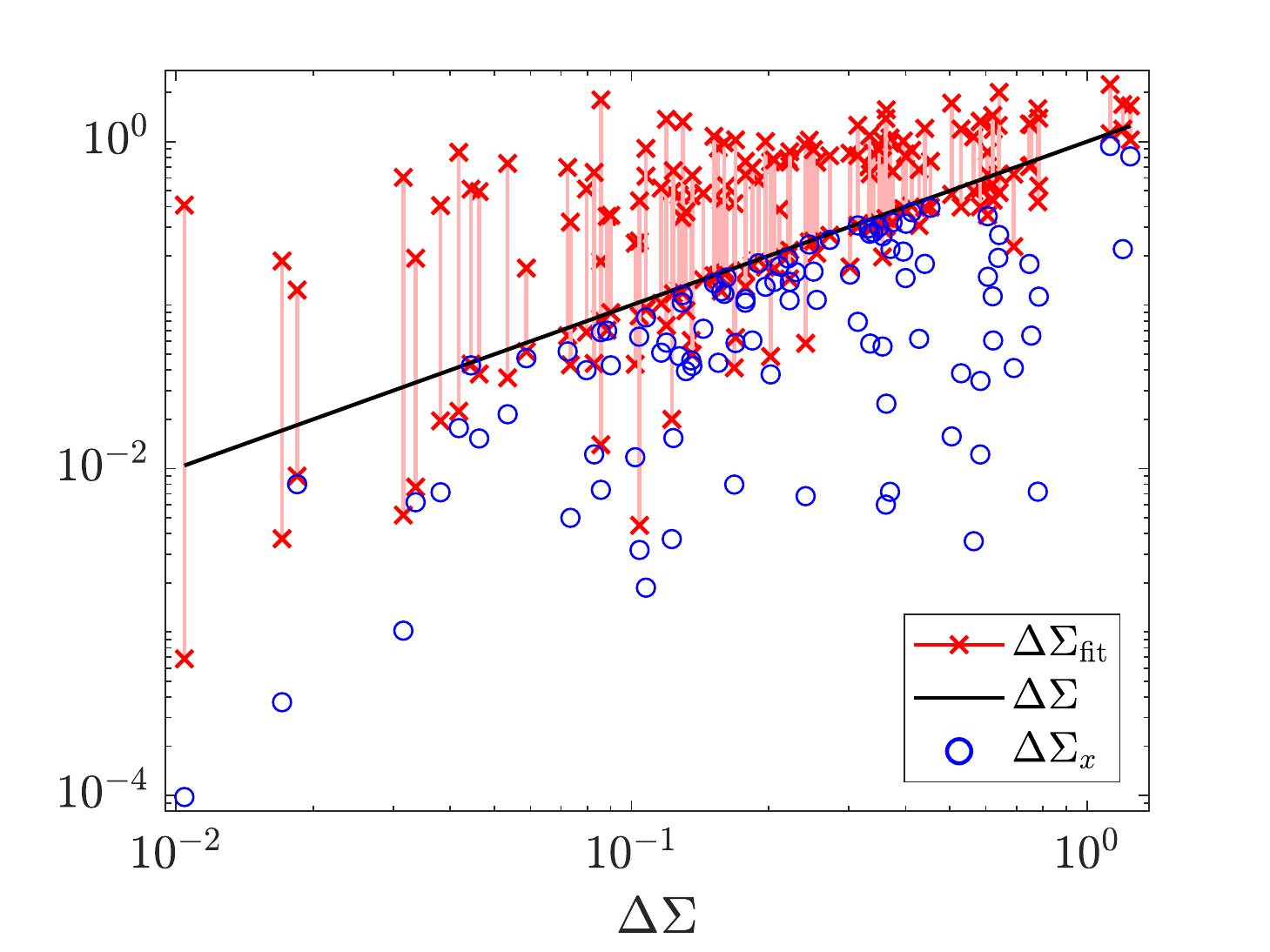}
	\caption{Comparison of entropy productions for $100$ randomly generated fully connected four state networks. We again see that the entropy production $\Delta \Sigma_{\rm fit}$ obtained from fitting the hidden network yields upper and lower bounds for the real entropy production $\Delta \Sigma$. The marginal entropy production $\Delta\Sigma_x$ is also shown.}
	\label{fig_hid_randomNetworks}
\end{figure}

Repeating the above procedure for $100$ randomly generated transition matrices for fully connected four-state networks (i.e., including the missing link in the left part of Fig.~\ref{fig_hid_exampleNetwork}), yields similar results which are shown in Fig.~\ref{fig_hid_randomNetworks}.

Interestingly, the real entropy production $\Delta\Sigma$ tends to cluster around the lower margin of the possible entropy productions $\Delta\Sigma_{\rm fit}$. This indicates that the upper bound is not very tight, which could be a consequence of the fact that it is possible to find solutions with very low transition probabilities $a_{ij}$ on the hidden edge $j \to i$ in one direction only. This would result in very large affinities $\ln\frac{a_{ji}}{a_{ij}}$ and thus large contributions to the total entropy productions from this edge.

\subsection{More hidden states and outlook}
The analysis is straightforward for the case of two hidden states. Next, one could consider an example with three observed and two hidden states. This should further constrain the possible hidden networks. Additionally, the case with an arbitrary number of hidden states can be considered. It is obvious that the solution space will then have a very high dimension, which makes a parameter sweep numerically expensive. 

Instead, one should opt for a numeric optimization method (e.g., simulated annealing, see, e.g., Ref~\cite{Press2007}) that searches for the network with the lowest entropy production under the constrains of producing the observed jump probabilities $p(i,n|j)$.

In general, we cannot expect to find upper bounds on the entropy production: It is always possible to replace one hidden state by a whole cycle (e.g., of three states) with a current between them. This can render the entropy production infinite without affecting the observed jump probabilities. However, it would possibly require degenerate eigenvalues of the hidden matrix, which might be unlikely in real networks (because of the large amount of fine tuning necessary). A study on three hidden states will shed light on this question by showing whether there is an upper bound in a scenario which can produce a cyclical probability current that is entirely hidden.

It is interesting to see whether, in scenarios with more hidden states, the marginal entropy production $\Delta\Sigma_x$ is closer to the least possible entropy production given by $\Delta \Sigma_{\rm fit}$. This could be the case when the solution space is so vast as to allow all kinds of entropy productions. The mere fact that the real entropy production can never be less than marginal entropy production would then guarantee that the marginal entropy production saturates the bound.

Finally, one should consider realistic data sets. We have thus far always assumed perfect statistics to see what bound can be achieved \emph{in principle}. It is a completely different question whether this is also feasible in practice with limited data that result in noisy statistics.

\bookmarksetup{startatroot}
\chapter{Conclusion} 
In this thesis we have investigated several important aspects of the thermodynamic description of microscopic systems. In \textbf{part I} we have presented the main concepts pertaining to the theory of \emph{stochastic processes}, \emph{information}, and \emph{stochastic thermodynamics} (\textbf{chapters~\ref{chap_basics_stochProc}-\ref{chap_stochasticThermodynamics}}). We have shown that from the mathematical formalism of stochastic processes one can construct a robust classical thermodynamic theory for small-scale (biological) systems that correctly accounts for thermal fluctuations relevant on these small scales.

In \textbf{part II} we considered the setup of systems in which multiple degrees of freedom interact with each other. To this end, in \textbf{chapter~\ref{chap_int_reservoirSubs}}, we investigated the role of \emph{micro-reversibility} in ensuring \emph{thermalization} of microscopic systems. We have applied this framework to \emph{collisional baths}, in which a system exchanges energy with reservoir particles in equilibrium, and found that conservation of energy alone is not sufficient to ensure thermalization of the system. We have considered an explicit example of seemingly plausible collision rules and used the breaking of micro-reversibility to prove that they are unphysical. It was shown that this fact can be exploited to break the second law.

In \textbf{chapter~\ref{chap_informationThermodynamics}} we reviewed the thermodynamics of information and given a short historical account of the apparent paradox around \emph{Maxwell's demon}. It was shown how information thermodynamics result from the interplay of two parts: the system and the memory. We shortly reviewed how biological sensors are constrained by thermodynamics. Finally, we presented the framework of \emph{causal conditioning} and \emph{detached path probabilities} with which one can study these setups from one common perspective and recover many results which have previously been obtained separately. Crucially, the entropy productions defined from this formalism split the second law in such a way that the causal influences of individual subsystems are preserved.

\textbf{Part III} presents a continuation of the study of interacting degrees of freedom with the twist that some of them are \emph{hidden} from the observer. We studied the dissipation of \emph{microswimmers} in \textbf{chapter~\ref{chap_hid_microswimmers}} and introduced a concrete model system. It was shown that the dynamics of a microswimmer can be approximated by \emph{active Brownian motion} when one cannot resolve the swimming mechanism itself, thus rendering it a hidden variable. We then showed that one can massively underestimate the energy dissipation of microswimmers when inferring it from their trajectories alone, as is commonly done.

In \textbf{chapter~\ref{sec_hid_hiddenDegreesInSTD}} we formalized the setting of interacting degrees of freedom and presented two \emph{effective descriptions} of the visible parts of the system using different stochastic processes which produce two kinds of effective dissipation measures: The \emph{apparent} and the \emph{marginal entropy production}. We have introduced a simple model system that illustrates how the fluctuation theorems for these effective entropy productions are modified. It was found that the apparent entropy production violates standard fluctuation theorems which can be traced back to the observed process not being Markovian. In contrast, the marginal entropy production, which takes all non-Markovian effects into account, fulfills the usual fluctuation theorems.

Finally, in \textbf{chapter~\ref{chap_hid_inferringDiss}} we showed how the marginal entropy production can be computed for the setup of \emph{masked Markov networks}. We then used the non-Markovian description to \emph{infer the hidden network} that generates the observed trajectories. We showed how this procedure can be used to \emph{improve estimations of entropy production} and tested it on networks with two visible and two masked states.

\section{Outlook}
A complete understanding of the role of hidden slow degrees of freedom in stochastic thermodynamics is still lacking. It is clear by now that not observing all relevant parts of the system invalidates the usual approaches, but the question remains: \emph{How much} is still knowable about the system? This thesis shows that the issue touches on other open problems, especially the question of how to define heat and work for systems strongly coupled to their environment. 

In that context it is interesting to apply causal conditioning to a system strongly coupled to a heat bath and check whether the thermodynamic quantities thus derived agree with those gained from the approach using the Hamiltonian of mean force~\cite{Seifert2016}. This could lend additional credibility to this strategy and perhaps extend the findings of Strasberg and Esposito~\cite{Strasberg2017}, since their approach is only in agreement with the strong-coupling scheme in the limit of time-scale separation.

Recently, a novel effective description for the visible parts of a system with hidden degrees of freedom has been used~\cite{Herpich2019,Strasberg2019}. Based on the Hamiltonian of mean force, it only works for driven dynamics and only captures the \emph{nonadiabatic entropy production}. It can be inferred from the visible dynamics only when one is able to perform the process quasistatically, so that one can sample the instantaneous equilibrium distribution of the visible degrees of freedom for all protocol values. The upside of these limitations is that the thus defined entropy production can be shown to always \emph{overestimate} the real entropy production.

It is worthwhile to investigate whether this description can be used to infer the real entropy production. Certainly, using this new effective description and coarse-graining, one can give upper and lower bounds for the nonadiabatic entropy production. Additionally, one could study the fluctuations of the entropy production using the new effective process. Perhaps this definition lends itself to \emph{thermodynamic inference} whereby one learns about concealed system properties from fluctuations of the observable variables.

An obvious future study is to use the approach laid out in chapter~\ref{chap_hid_inferringDiss} and investigate how well one can bound the entropy production in more realistic settings, i.e., with more hidden states and realistic amounts of data. Additionally, one can apply the scheme to measured data from experiments to improve entropy production estimates.

\titleformat{\chapter}[hang]{\huge\bfseries}{\thechapter\hsp\textcolor{black!50}{\Huge\textbar\huge}\hsp}{0pt}{\huge\bfseries}
\cleardoublepage
\begin{appendices}
\chapter[Conditional velocity averages]{Conditional velocity averages for Langevin systems}\label{app_conditionalVelocityAverages}

We assume that $x(t)$ is generated from an overdamped Langevin equation of the form of Eq.~\eqref{eqn_basics_overdampedLangevinEquation}:
\begin{align}\label{eqn_appendix_overdampedLangevinEquation}
\dot{x} = \nu F(x,t) + \sqrt{2D}\,\xi(t).
\end{align}

We are interested in ensemble averages of the type $\left\langle \dot{x} \circ g(x,t) \right\rangle$, where the multiplication is to be interpreted in the Stratonovich sense. The product within the brackets has to be discretized according to:
\begin{align}
\dot{x} \circ g(x,t) = \lim_{dt\rightarrow 0} \frac{x_{i+1}-x_i}{dt}\cdot g\left(\frac{x_i+ x_{i+1}}{2},t_i+\frac{dt}{2}\right),
\end{align}
where $x_i \coloneqq x(t_i)$. If we assume~\cite{Seifert2012,Seifert2005} that $x_i \approx \frac{x_{i+1}+x_{i-1}}{2}$, we can similarly write:
\begin{align}
\dot{x}\circ g(x,t) = \lim_{dt\rightarrow 0} \frac{x_{i+1}-x_{i-1}}{2 dt}\cdot g\left( x_i,t_i\right).
\end{align}
Then, the ensemble average is rendered as:
\begin{align}
\left\langle \dot{x}\circ g(x,t) \right\rangle = \left\langle \lim_{dt\rightarrow 0} \left\langle \frac{x_{i+1}-x_{i-1}}{2 dt} \right\rangle_{p(x_{i-1},x_{i+1}|x_i)}\cdot g(x_i,t_i)   \right\rangle_{p(x_i,t_i)}.\label{eqn_appendix_ensembleAverage}
\end{align}

The middle average can be split as follows:
\begin{align}
\left\langle \frac{x_{i+1}-x_{i-1}}{2 dt} \right\rangle_{p(x_{i-1},x_{i+1}|x_i)} = \left\langle \frac{x_{i+1}-x_{i}}{2 dt} \right\rangle_{p(x_{i+1}|x_i)} + \left\langle \frac{x_{i}-x_{i-1}}{2 dt} \right\rangle_{p(x_{i-1}|x_i)}.
\end{align}

The first average follows straighforwardly from the transition probability $p(x_{i+1}|x_i)$ which we calculated in Eq.~\eqref{eqn_basics_transitionProbSmallTimes}:
\begin{align}
p(x_{i+1}|x_{i}) = \frac{1}{\sqrt{4\pi D dt}}\exp{\left[- \frac{dt}{4D}\left(\frac{x_{i+1}-x_{i}}{dt} - \nu F_{i+1}\right)^2 -\frac{1}{2} \nu dt F'_{i+1} \right]}\nonumber\\
+ \mathcal{O}\left( dt^{3/2}\right)
\end{align}
where $F^{(n)}_i := \frac{\partial^n}{\partial x_i^n} F\left(\frac{x_i + x_{i-1}}{2},t+\frac{dt}{2}\right)$.

We thus obtain:
\begin{align}
\left\langle \frac{x_{i+1}-x_{i}}{2 dt} \right\rangle_{p(x_{i+1}|x_i)} =\frac{ \nu F_{i+1}}{2} + \mathcal{O}(dt^{1/2}).
\end{align}

The transition probability in the reverse direction $p(x_{i-1}|x_i)$ can be expressed using Bayes's theorem~\cite{Seifert2012}:
\begin{align}
p(x_{i-1}|x_i) &\coloneqq p(x_{i-1};t_i-dt|x_i;t_i) = p(x_i;t_i| x_{i-1};t_i-dt)\, \frac{p( x_{i-1};t_i-dt)}{p(x_{i};t_i)}\\
&= p(x_i;t_i|x_{i-1};t_i-dt)\times\nonumber\\
&\quad \left[ 1 - \left( x_i-x_{i-1} \right) \cdot \pd{x} \ln{p( x_{i};t_i)} - dt\,\partial_t \ln{p(x_{i};t_i)} + O(dt^2) \right]\\
&= \frac{1}{\sqrt{4\pi D dt}}\exp{\left[- \frac{dt}{4D}\left(\frac{x_i-x_{i-1}}{dt} - \nu F_i\right)^2 -\frac{1}{2} \nu dt F'_i \right]}\nonumber\\
&\; \times \left[ 1 - \left( x_i-x_{i-1} \right) \cdot \pd{x} \ln{p( x;t_i)}\Big|_{x_i} - dt\,\partial_t \ln{p(x_{i};t)}\Big|_{t_i} \right]+ \mathcal O(dt^{3/2}),
\end{align}
which then leads to:
\begin{align}
\left\langle \frac{x_{i}-x_{i-1}}{2 dt} \right\rangle_{p(x_{i-1}|x_i)} =\frac{ \nu F_{i}}{2} - 2 D\, \pd{x} \ln{p( x;t_i)}\Big|_{x_i} + \mathcal{O}(dt^{1/2}).
\end{align}

Finally, we obtain:
\begin{align}
\lim_{dt\rightarrow 0} \left\langle \frac{x_{i+1}-x_{i-1}}{2 dt} \right\rangle_{p(x_{i-1},x_{i+1}|x_i)} &= \nu F(x,t) - D\pd{x} p(x,t)\\
&= \frac{j(x,t)}{p(x,t)},
\end{align}
implying
\begin{align}
\left\langle \dot{x} \circ g(x,t) \right\rangle = \left\langle \frac{j(x,t)}{p(x,t)}\, g(x,t) \right\rangle_{p(x,t)},
\end{align}
where we used the definition of the probability flux in Eq.~\eqref{eqn_basics_defProbCurrentFPE}.

This agrees with the result obtained by Seifert~\cite{Seifert2012} and with the one obtained by Sekimoto~\cite{Sekimoto2010} for $g(x,t) = \pd{x} V(x,t)$.

\chapter[Evolution equation of cumulants]{Transformation of linearly coupled Langevin equations to evolution equation of cumulants}\label{app_linearLangevin}
Consider the system of $n$ linearly coupled Langevin equations with different noise intensities:
\begin{align}
\dot{x}_i = \sum\limits_{j=1}^n a_{ij}\, x_j + b_i + \sqrt{2D_i}\, \xi_i(t),
\end{align}
where
\begin{align}
\left\langle \xi_i(t)\, \xi_j(t') \right\rangle = \delta_{ij}\, \delta(t-t').
\end{align}
According to Eq.~\eqref{eqn_basics_multiDimensionalFokkerPlanck}, the corresponding Fokker-Planck equation reads:
\begin{align}
\pd{t}p(\bx,t) = -\sum\limits_{i=1}^n\pd{x_i}\left( \sum\limits_{j=1}^n a_{ij}x_j + b_i \right)p(\bx,t) + \sum\limits_{i=1}^n D_i\frac{\partial^2}{\partial x_i^2}\, p(\bx,t).
\end{align}

This is a multivariate \emph{Ornstein-Uhlenbeck process}~(cf. Sec. 3.2 of Ref.~\cite{Risken1996}) and is thus solved by a multivariate Gaussian
\begin{align}
\mathcal{N}(\bx;M,\mathbf{\Sigma}) = \frac{1}{\sqrt{(2\pi)^d\,|\mathbf{\Sigma}|}}\exp{\left\{ -\frac{1}{2}(\bx-M)^T\mathbf{\Sigma}^{-1}(\bx-M) \right\}},
\end{align}
with time-dependent means $\mu_i := \langle x_i \rangle$ and covariances $c_{ij} := \langle x_i x_j\rangle - \mu_i\mu_j$:
\begin{align}
M= \begin{pmatrix}
\mu_1\\\mu_2\\\vdots\\\mu_n
\end{pmatrix}\qquad 
\mathbf{\Sigma} = \begin{pmatrix}
c_{11}&c_{12}&\hdots& c_{1n}\\
c_{12}&c_{22}&\hdots& c_{2n}\\
\vdots&\vdots& \ddots & \vdots \\
c_{1n}&c_{2n}&\hdots& c_{nn}\\
\end{pmatrix}.
\end{align}

The evolution equations for the cumulants read:
\begin{align}
\dot{\mu}_i(t) &= \sum_{j=1}^n a_{ij}\,\mu_j(t) + b_i\\
\dot{c}_{ij}(t)&= \sum_{k=1}^n \left[a_{ik}\,c_{jk}(t) + a_{jk}\,c_{ik}(t) \right] + 2D_i\delta_{ij}.
\end{align}

\section*{Proof}
The evolution equation for the first cumulants can be obtained by partial integration and using the fact that boundary terms vanish due to normalization:
\begin{align}
\frac{d}{dt} \mu_k &= \int d^n x\, x_k\, \pd{t} p(\bx,t)\\
&= \bigintsss d^n x\, x_k\, \left[ -\sum\limits_{i=1}^n\pd{x_i}\left( \sum\limits_{j=1}^n a_{ij}x_j + b_i \right)p(\bx,t) + \sum\limits_{i=1}^n D_i\frac{\partial^2}{\partial x_i^2}\, p(\bx,t)\right]\\
&=\bigintsss d^nx\left( \sum\limits_{j=1}^n a_{kj} x_j +b_k\right)p(\bx,t)\\
&= a_{kj}\,x_j + b_k.
\end{align}

Similarly, the evolution equations for the second cumulants require two steps of partial integration:
\begin{align}
\frac{d}{d t} c_{kl} &= \int d^n x\, x_k x_l\, \pd{t} p(\bx,t) - \dot{\mu}_k \mu_l - \mu_k \dot{\mu}_l\\
&= \bigintsss d^n x\, x_k x_l\, \left[ -\sum\limits_{i=1}^n\pd{x_i}\left( \sum\limits_{j=1}^n a_{ij}x_j + b_i \right)p(\bx,t) + \sum\limits_{i=1}^n D_i\frac{\partial^2}{\partial x_i^2}\, p(\bx,t)\right] + \nonumber\\
&\qquad\qquad\qquad - \dot{\mu}_k \mu_l - \mu_k \dot{\mu}_l\\
&= \bigintsss d^nx\, x_l\left[\left( \sum\limits_{j=1}^n a_{kj} x_j +b_k\right)p(\bx,t)- D_k \pd{x_k} p(\bx,t)\right] + \nonumber\\
&\quad+\bigintsss d^nx\, x_k\left[\left( \sum\limits_{j=1}^n a_{lj} x_j +b_l\right)p(\bx,t)- D_l \pd{x_l} p(\bx,t)\right]  - \dot{\mu}_k \mu_l - \mu_k \dot{\mu}_l\\
&=\sum\limits_{j=1}^n a_{kj} \left( c_{lj} + \mu_l\mu_j\right) + b_k \mu_l + D_k \delta_{lk} + \sum\limits_{j=1}^n a_{lj}\left( c_{kj} + \mu_k\mu_j\right) + b_l \mu_k +\nonumber\\
&\qquad+ D_l \delta_{kl} - \left(\sum\limits_{j=1}^n a_{kj} \mu_j + b_k\right)\mu_l - \mu_k \left(\sum\limits_{j=1}^n a_{lj} \mu_j + b_l\right)\\
&=\sum\limits_{j=1}^n a_{kj} c_{lj} + \sum\limits_{j=1}^n a_{lj} c_{kj} +  2 D_k\delta_{kl}.
\end{align}

\chapter{Microswimmer efficiency}\label{app_efficiency}
Let us elaborate on the definition of the swimming efficiency $\eta$ in Eq. (43) of our article in Ref.~\cite{Ehrich2019}. Other definitions of efficiency are possible but eventually lead to the same expression. For example, consider the \emph{energy dissipation per distance traveled} (which is the average friction force when friction is linear):
\begin{align}
\frac{\Delta Q}{\Delta \mu_X} = (L-1)\,\frac{1+\nu}{1-\nu},
\end{align}
where $\Delta Q$ is the heat dissipated per cycle [Eq. (39) in the article] and $\Delta \mu_X$ is the average distance traveled in one cycle [Eq. (11a) in the article]. Note that this quantity is independent of the protocol time, which is another consequence of the linear friction.

Contrast this with the corresponding expression for active Brownian motion:
\begin{align}
\frac{\dot{Q}_{\rm eff}}{\langle \dot{X} \rangle} = f_{\rm eff} = (L-1) \, \frac{1-\nu}{1+\nu},
\end{align}
where $f_{\rm eff}$ is given in Eq. (20) of the article.

The ratio of these two quantities again yields the same swimming efficiency:
\begin{align}
\frac{\dot{Q}_{\rm eff}}{\langle \dot{X} \rangle}\Big/\frac{\Delta Q}{\Delta \mu_X} = \frac{(1-\nu)^2}{(1+\nu)^2} = \eta,
\end{align}
the reason is that, with Stokes friction, speed does not matter for energy dissipation.

This also implies that even for finite $\Delta t$, one cannot infer more dissipation from the center of mass movement than from active Brownian motion. Consider the trajectory of $\mu_X(t)$ for $\Delta t =10$ in Fig. 3 of the article. If one wanted to model this behavior using active Brownian motion, one needs a pushing force $f_{\rm eff}(t)$ that is periodically varying. This, however, results in the same dissipation as that of a process which has a constant pushing force with a magnitude determined by the average of $f_{\rm eff}(t)$.

\chapter[Average dissipation for 2d model]{Average entropy production for a model system with hidden degrees of freedom}\label{app_CompareDiss}
A modification of the model system we proposed in Ref.~\cite{Kahlen2018} reprinted in Sec.~\ref{sec_hid_articleHiddenDegrees} was analyzed by Juliana Caspers in her Bachelor's thesis~\cite{Caspers2019}. She added a second velocity component $v$ in pulling the trapping potential in the $y$-direction. The potential then reads (cf. Eq. (13) in our article~\cite{Kahlen2018}):
\begin{align}
V(x,y;t) = \frac{1}{2} (x-u t)^2 + \frac{1}{2} (y-v t)^2 - b(x- u t)\,(y-v t).
\end{align}
Repeating the calculations in Eqs. (14)-(23), one finds the following effective potential for the visible degree of freedom:
\begin{align}
\tilde{V}(x;t) \overset{\lambda_1 t \gg 1}{\longrightarrow} \frac{1-b^2}{2} (x- u t)^2 + \frac{b v}{\nu}(x-u t).
\end{align} 
The apparent work thus reads:
\begin{align}
\tilde{w} &=\int\limits_0^T dt\,  \pd{t} \tilde{V}(x;t) \\
&= -(1-b^2) u \int\limits_0^T dt\, [x(t)- u t] - \frac{u b v}{\nu}. \label{eqn_app_appWork}
\end{align}
Remarkably, this effective description results in the same detailed fluctuation theorem for the apparent entropy production  [Eq. (33) of our article].

The marginal entropy production can be calculated analogously to the scheme outlined in Sec. 3.3. of the article. The difference is that the total work now reads:
\begin{align}\label{eqn_app_realWork}
w = \int\limits_0^T dt \left[ -u( x - u t) -v ( y - v t )+ u b (y - v t) + v b (x - u t) \right]
\end{align}
instead of
\begin{align}
w = \int\limits_0^T dt \left[-u( x - u t ) + u b y \right].
\end{align}
Again, in the limit of time-scale separation the apparent, the marginal, and the total work agree. Performing the time-asymptotic analysis in Eq. (62) of the article for this modified system yields\footnote{Note that Eq. (62) in our article provides the expression in the co-moving reference frame $x\to x - ut$.}
\begin{align}
w_x[x(\cdot)] \approx - \left(u + \frac{b v}{\nu} \right) (1-b^2) \frac{\nu}{\nu + b^2} \int\limits_0^T dt \, (x(t)- u t). \label{eqn_app_marginalWork}
\end{align}

This allows us to compute the different \emph{average dissipation rates} for long times from Eqs.~\eqref{eqn_app_appWork},~\eqref{eqn_app_realWork}, and~\eqref{eqn_app_marginalWork}:
\begin{align}
\dot{\Sigma} &= (v b - u) (\mu_x(t) - u t) + (u b - v ) (\mu_y(t) - v t)\\
\dot{\tilde{\Sigma}} &= -(1-b^2)u\,(\mu_x- u t) -\frac{u b v}{\nu}\\
&= u^2\\
\dot{\Sigma}_x &= - \left(u + \frac{b v}{\nu} \right) (1-b^2) \frac{\nu}{\nu + b^2} \, (\mu_x(t)- u t)\\
&= \frac{(u \nu + bv)^2}{\nu(\nu + b^2)},
\end{align}
where we used the time-asymptotic mean values of $x$ and $y$,
\begin{align}
\mu_x(t) &= u t - \frac{u \nu + b v}{\nu ( 1-b^2)}\label{eqn_app_mux}\\
\mu_y(t) &= v t - \frac{u \nu b + v}{\nu ( 1-b^2)},
\end{align}
which follow from solving the Fokker-Planck equation corresponding to the process~\cite{Caspers2019}.

\begin{figure}[ht]
	\centering
	\includegraphics[width=0.65  \linewidth]{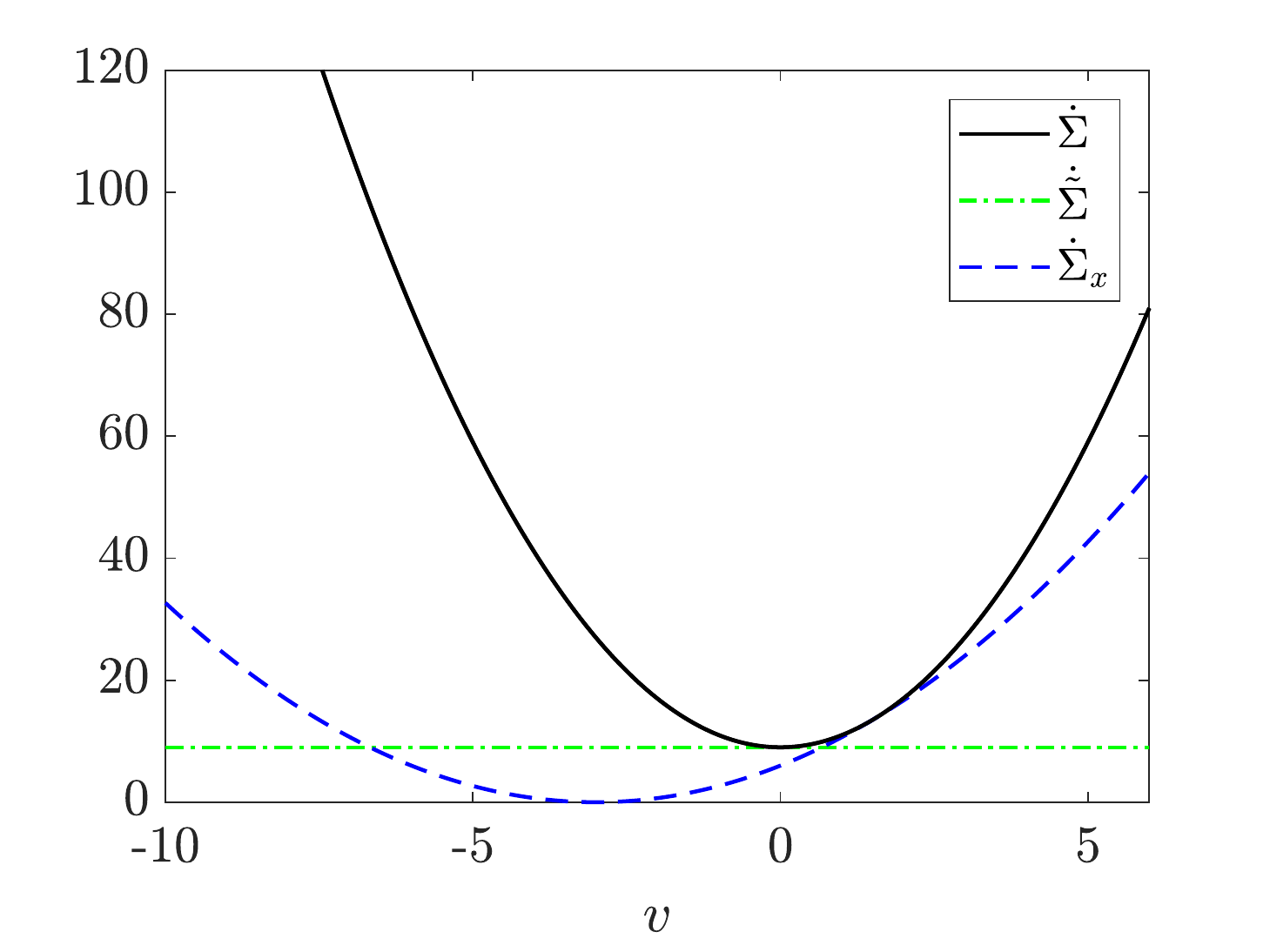}
	\caption{Comparison of average dissipation rates for different velocities $v$ in the $y$-direction for a representative set of parameters ($\nu = b = 1/2$ and $u=3$). The total entropy production $\dot{\Sigma}$ is always greater or equal than the rate of apparent entropy production $\dot{\tilde{\Sigma}}$ and the rate of marginal entropy production $\dot{\Sigma}_x$.}
	\label{fig_app_compAvgDiss}
\end{figure}

The dependence of the entropy production rates on $v$ is plotted in Fig.~\ref{fig_app_compAvgDiss} for a representative set of parameters.

While both effective entropy production rates underestimate the real entropy production rate, as was expected, the quality of the bounds they provide depend on the value of the hidden velocity. Importantly, the marginal entropy production  vanishes completely at $v=-\nu u /b$. Because of the coupling between $x$ and $y$ and the perfectly balanced velocities $u$ and $v$, one can no longer differentiate the forward process from the time-reversed process \emph{by the $x$-trajectories alone}. 

This can be seen from Eq.~\eqref{eqn_app_mux}, which then reads $\mu_x(t) = u t$ with no offset from the center of the potential. Thus, the driving through $v$ forces the $x$-position to be ahead of the center of the potential, which is exactly balanced by the driving through $u$ that causes the particle to lag behind.

It seems odd that the marginal entropy production should be a worse bound than the apparent one, since taking into account more information than provided by a Markovian description should result in a better estimation of the time-irreversibility of the trajectories. Importantly, we are dealing with a \emph{driven process}, i.e., the time-reversed version also includes a time-reversed driving that generates different trajectories for the reverse process.

The apparent entropy production rate $\dot{\tilde{\Sigma}}$ does not depend on the velocity $v$ because it essentially only captures the net probability flow in the $x$-direction, which is independent of how the potential is pulled in the $y$-direction.

The intersection between the apparent and the total entropy production rates lies at $v = u b$ and results from the fact that the apparent work $\tilde w$ and the total work $w$ become the same function.

\chapter{Ratio of elements of powers of tri-diagonal matrices}\label{app_triDiagonal}
\section*{Step 1}
Consider an $m \times m$ tri-diagonal matrix $\mathcal{A}$ with elements $a_{ij} \neq 0, |i-j| \leq 1$. For $m \geq j > i$ and $n \in \mathbb{N}$ the following holds:
\begin{align} \label{eqn_step1_1}
\left[\mathcal{A}^n\right]_{ij} = \begin{cases}
\prod\limits_{k=i}^{j-1} a_{k,k+1} & n = j-i\\
0 & n < j-1\,.
\end{cases}
\end{align}

\subsection*{Proof}
The statement is true for $n=1$. By induction:
\begin{align}
\left[\mathcal{A}^{n+1}\right]_{ij} &= \theta[i-1]\, a_{i,i-1}\, \left[\mathcal{A}^{n}\right]_{i-1,j} + a_{ii}\,[\mathcal{A}^n]_{ij} + a_{i,i+1}\, [\mathcal{A}^n]_{i+1,j}\\
&= \begin{cases}
a_{i,i+1}\, [\mathcal{A}^n]_{i+1,j} = \prod\limits_{k=i}^{j-1} a_{k,k+1} &\qquad n+1 = j-i\\
0 &\qquad n+1 < j-i\,,
\end{cases}
\end{align}
where $\theta[i] = \begin{cases}
0 & n < 1\\
1 & n\geq 1\, .
\end{cases}$

Similarly, for $m\geq j > i $:
\begin{align}\label{eqn_step1_2}
\left[\mathcal{A}^n\right]_{ji} = \begin{cases}
\prod\limits_{k=i}^{j-1} a_{k+1,k} & n = j-i\\
0 & n < j-1\,.
\end{cases}
\end{align}

\section*{Step 2}
For $m \geq j >i$ and $n \geq j-i$ the following holds true:
\begin{align}
\frac{[\mathcal{A}^n]_{ij}}{[\mathcal{A}^n]_{ji}} = \prod\limits_{k=i}^{j-1}\frac{a_{k,k+1}}{a_{k+1,k}}.
\end{align}

\subsection*{Proof}
Following step 1, the statement holds true for $n=j-i$. Using Eqs.~\eqref{eqn_step1_1}~and~\eqref{eqn_step1_2}, we find by induction:
\begin{align}
[\mathcal{A}^{n+1}]_{ij} &= [\mathcal{A} \, \mathcal{A}^n]_{ij} =  \theta[i-1]\, a_{i,i-1}\, [\mathcal{A}^n]_{i-1,j} + a_{ii} \, [\mathcal{A}^n]_{ij} + a_{i,i+1} \, [\mathcal{A}^n]_{i+1,j}\\
[\mathcal{A}^{n+1}]_{ji} &= [\mathcal{A}^n \, \mathcal{A}]_{ji} = \theta[i-1]\,[\mathcal{A}^n]_{j,i-1}\, a_{i-1,i} +[\mathcal{A}^n]_{ji} \, a_{ii} + [\mathcal{A}^n]_{j,i+1}\, a_{i+1,i}\\
&= \theta[i-1]\,[\mathcal{A}^n]_{i-1,j} \left( \prod\limits_{k=i-1}^{j-1} \frac{a_{k+1,k}}{a_{k,k+1}} \right)a_{i-1,i} +[\mathcal{A}^n]_{ji} \left( \prod\limits_{k=i}^{j-1} \frac{a_{k+1,k}}{a_{k,k+1}} \right) a_{ii} \nonumber\\
&\qquad\qquad\qquad+ [\mathcal{A}^n]_{i+1,j} \left( \prod\limits_{k=i+1}^{j-1} \frac{a_{k+1,k}}{a_{k,k+1}} \right) a_{i+1,i}\\
&= \left( \prod\limits_{k=i}^{j-1} \frac{a_{k+1,k}}{a_{k,k+1}} \right) \left( \theta[i-1]\, a_{i,i-1}\, [\mathcal{A}^n]_{i-1,j} + a_{ii} \, [\mathcal{A}^n]_{ij} + a_{i,i+1} \, [\mathcal{A}^n]_{i+1,j} \right)\\
&= \left( \prod\limits_{k=i}^{j-1} \frac{a_{k+1,k}}{a_{k,k+1}} \right) \, [\mathcal{A}^{n+1}]_{ij}.
\end{align}
From this, the statement immediately follows. As a special case, the following holds true:
\begin{align}\label{eqn_proofTriDiagonal}
\frac{[\mathcal{A}^n]_{1m}}{[\mathcal{A}^n]_{m1}} = \frac{a_{12} \, a_{23}...\,a_{m-1,m}}{a_{m,m-1}... \, a_{32}\, a_{21}}.
\end{align}
\end{appendices}

\bookmarksetup{startatroot}
\cleardoublepage
%%%%%%%%%%%%%%%%%%%%%%%%%%%%%
{\small
\nocite{apsrev41Control}
%\addcontentsline{toc}{chapter}{Bibliography}
\bibliography{referencesMaster,controlBib}
\bibliographystyle{apsrev4-1}
}

\end{document}